\documentclass[aps,prd,nofootinbib,10pt,tightenlines,notitlepage, twocolumn,floatfix,eqsecnum,superscriptaddress,showkeys]{revtex4-1}
    
\usepackage[utf8]{inputenc}          
\usepackage{graphicx}         
\usepackage[usenames,dvipsnames]{xcolor}
\usepackage{array,dcolumn,longtable} 

\usepackage{amsmath,amssymb,amsfonts,slashed} 
\usepackage{mathtools}          

\usepackage[linktocpage,breaklinks]{hyperref}

\usepackage{txfonts}
\usepackage{bm}
\usepackage{stmaryrd}
\usepackage{tensor}
\usepackage[utf8]{inputenc}

\usepackage{epsfig}
\usepackage{epstopdf}

\usepackage{natbib}
\usepackage{cleveref}

\definecolor{mred}{RGB}{127,0,25}
\definecolor{mdgr}{RGB}{51,51,51}
\definecolor{mag}{RGB}{211, 54, 130}
\definecolor{verm}{RGB}{164, 25, 0}
\definecolor{purp}{RGB}{153,50,204}

\hypersetup{colorlinks=true,
            citecolor=NavyBlue,
            linkcolor=NavyBlue,
            urlcolor=NavyBlue}

\usepackage{siunitx}                                  
\DeclareSIUnit{\fm}{\femto\metre}                     

\begin{document}
\title{Extreme Matter meets Extreme Gravity: Ultra-heavy neutron stars with crossovers and first-order phase transitions}
\date{\today}

\author{Hung Tan}
\affiliation{Illinois Center for Advanced Studies of the Universe, Department of Physics, University of Illinois at Urbana-Champaign, Urbana, IL 61801, USA}
\author{Travis Dore}
\affiliation{Illinois Center for Advanced Studies of the Universe, Department of Physics, University of Illinois at Urbana-Champaign, Urbana, IL 61801, USA}
\author{Veronica Dexheimer}
\affiliation{Department of Physics, Kent State University, Kent, OH 44243 USA}
\author{Jacquelyn Noronha-Hostler}
\affiliation{Illinois Center for Advanced Studies of the Universe, Department of Physics, University of Illinois at Urbana-Champaign, Urbana, IL 61801, USA}
\author{Nicol\'as Yunes}
\affiliation{Illinois Center for Advanced Studies of the Universe, Department of Physics, University of Illinois at Urbana-Champaign, Urbana, IL 61801, USA}

\begin{abstract}

The speed of sound of the matter within neutron stars may contain non-smooth structure related to first-order phase transitions or crossovers. 
Here we investigate what are the observable consequences of structure in the speed of sound, such as bumps, spikes, step functions, plateaus, and kinks. 
One of the main consequences is the possibility of ultra-heavy neutron stars (with masses larger than 2.5 solar masses) and mass twins in heavy (with masses larger than $2$ solar masses) and ultra-heavy neutron stars. 
These stars pass all observational and theoretical constraints, including those imposed by recent LIGO/Virgo gravitational-wave observations and NICER X-ray observations. 
We thoroughly investigate other consequences of this structure in the speed of sound to develop an understanding of how non-smooth features affect astrophysical observables, such as stellar radii, tidal deformability, moment of inertia, and Love number. 
Our results have important implications for future gravitational wave and X-ray observations of neutron stars and their impact in nuclear astrophysics. 
\end{abstract}

\maketitle

\section{Introduction}
\label{sec:intro}
Exotic degrees of freedom, such as baryons made of strange quarks (hyperons), have been the topic of intense discussion since they were first studied in neutron stars in the beginning of the eighties \cite{Glendenning:1982nc}. The new degrees of freedom, in spite of being energetically favoured at intermediate densities, create new channels to redistribute the Fermi energy and, as a consequence, inevitably remove a source of pressure, making the matter equation of state (EoS) softer. When the EoS model-builders try to compensate for this by changing some of their nuclear physics interaction parameters, and, thus, producing an EoS that is stiffer in the presence of hyperons, they very often generate stars with larger radii. This apparent conflict, when comparing predictions with observations, has been referred to as the ``hyperon puzzle" \cite{Bednarek:2011gd,Fortin:2014mya,Buballa:2014jta}.

The possibility of hybrid neutron stars, containing an inner core with deconfined quark matter, an outer core with bulk hadronic matter, and a crust with nuclei, has drawn a considerable amount of attention (see e.g. Ref.~\cite{Annala:2019puf}). Hybrid stars were first proposed in 1965 \cite{Ivanenko:1965dg} and their stability carefully investigated in the seventies \cite{Keister:1976gc} and eighties \cite{Grassi:1987bu}. These stars have recently resurfaced in light of new astrophysical data from gravitational wave observations of possibly the most massive neutron stars ever detected \cite{Abbott:2020khf,Tan:2020ics,Dexheimer:2020rlp,Demircik:2020jkc,Christian:2020xwz,Blaschke:2020vuy,Ayriyan:2021prr,Li:2020dst,Otto:2020hoz,Nandi:2020luz,Ferreira:2020kvu}. 
The GW190814 event \cite{Abbott:2020khf} has been interpreted as being produced by the coalescence of a compact binary, composed of a heavy black hole and a mysterious compact object of mass $M = 2.5-2.67$~M$_\odot$, for which an electromagnetic counterpart was not detected. An argument has been made that the mysterious compact object was a very low-mass black hole~\cite{Most:2020bba,Broadhurst:2020cvm,Fishbach:2020ryj,Nathanail:2021tay}. However, this argument is not based on a smoking gun observable derived from the GW190814 data itself, like measuring that the tidal deformability was zero to sufficient accuracy to exclude neutron stars, which have non-zero deformabilities. Rather, this black hole argument is based on other more ``indirect'' reasons. One such reason is that the two past LIGO events consistent with neutron stars (GW170814 and GW190425), and the distribution of pulsars in our galaxy suggest a neutron star mass distribution that disallows ultra-heavy neutron stars (with masses above $2.5 M_\odot$). This conclusion, however, hinges on small number statistics from the two LIGO observations, and the assumption that all neutron stars in the universe look like those observed through radio observations in our galaxy. Another reason for  the argument presented above is based on the electromagnetic emission coincident with the GW170817 event, which suggests a maximum mass of $\sim 2.3 M_\odot$~\cite{Rezzolla:2017aly,Khadkikar:2021yrj,Demircik:2020jkc}. This conclusion, however, hinges on details of numerical relativity simulations that currently do not contain all of the physics that may be relevant in binary neutron star mergers (like viscosity, neutrino transport, and precise knowledge of the EoS)~\cite{Alford:2017rxf,Alford:2019kdw,Routray:2020zkf}. 

Given these caveats, we are inspired here to investigate what nuclear physics conclusions we would be led to {\emph{if}} we assumed the light companion of GW190814 was actually a slowly-rotating neutron star.  What EoSs can produce (slowly-rotating) neutron stars in the mass range of $M = 2.5-2.67$~M$_\odot$, or heavier? Does the GW190814 event exclude the possibility of a first-order phase transition inside neutron stars, and thus the existence of mass twins~\cite{Christian:2020xwz}, i.e.~neutron stars with very different radii but the same mass ? If one were to require that EoSs allow for (slowly-rotating) ultra-heavy neutron stars, with masses $M > 2.5 $~M$_\odot$, what would the implications be for the stellar radii and tidal deformability? How does the information we have learned about neutron stars from this event translate to information about the EoS and the speed of sound of dense matter? Can we understand in detail what the generic features of neutron stars that contain first-order phase transitions in quark cores are?

The above questions are of great interest to the broader nuclear physics community as they would influence our understanding of the transition from hadronic (neutrons, protons, hyperons, etc) to deconfined quark matter. The nature of \emph{how} heavy hadrons transition into light quarks varies with temperature and baryon density, as well as with other properties of the system, such as excess of strangeness and chemical equilibration (or lack thereof). Therefore, let us carefully define the different types of phase transitions that are possible. An nth-order phase transition is where the nth-susceptibility (i.e.~the nth derivative of the pressure with respect to the chemical potential, $\chi_n \equiv \partial^n P/\partial\mu_B^n$) becomes non-differentiable. For a first-order phase transition, $\chi_1$ has a jump at a given chemical potential, while for a phase transition of order $n>1$, then $\chi_{n>1}$ has a divergence at some chemical potential. This behavior has important implications for the speed of sound, which through thermodynamic identities can be written as $c_s^2 \equiv dP/d\epsilon = \chi_1/(\chi_2 \mu_B)$ at zero temperature, where $\epsilon$ is the energy density. In particular, during a first-order phase transition, the speed of sound drops to zero in some finite range of baryon densities, while during a second-order phase transition, the speed of sound spikes to zero at a single value of the baryon density (e.g. see \cite{Parotto:2018pwx,Grefa:2021qvt,Stafford:2021wik}). In contrast, a crossover transition is one in which all susceptibilities remain differentiable, and the speed of sound therefore does not go to zero at all. Because phase transitions of order higher than three are very hard to measure, the nuclear community has sometimes referred to high-order phase transitions as crossovers. 

A related concept that is important for our analysis is that of a critical point. This quantity is defined as the boundary between a first-order and a crossover phase transition, and therefore a critical point occurs at a second-order phase transition.  A crossover phase transition is well-established to exist at finite temperatures (above $10^{12}$ K) and in a finite range of baryon chemical potentials that includes zero  \cite{Aoki:2006we} but does not include the chemical potentials relevant to neutron stars \cite{Borsanyi:2021sxv}.  Therefore, if a first-order phase transition exists at low temperatures (specifically, inside neutron stars and/or in neutron star mergers), then a critical point must also exist at finite temperatures and somewhat lower baryon chemical potential to separate the cross-over from the first-order phase transition \cite{Asakawa:1989bq,Berges:1998rc,Halasz:1998qr}. Such a critical point has been dubbed the {\emph{QCD critical point}}, and current efforts are underway using heavy-ion collisions to search for it, where the STAR experiment has measured the most promising signals so far (kurtosis of net-proton fluctuations) at $3.1\sigma$ significance \cite{Adam:2020unf}. For further discussions see recent reviews \cite{Dexheimer:2020zzs,Ratti:2018ksb,Bzdak:2019pkr,Monnai:2021kgu}.

An important consequence of a first-order phase transition is the existence of mass twin stars. Twin stars were first proposed in~\cite{Gerlach:1968zz} for the general case of first-order phase transitions, and in~\cite{Kampfer:1981yr} for the specific case of deconfinement to quark matter. Recently, twins have been thoroughly studied in the case of a constant speed of sound for quark matter \cite{Christian:2020xwz,Blaschke:2020vuy,Han:2020adu}, polytropes \cite{Pang:2020ilf} or using two different EoSs, one of which is easily adjustable~\cite{Jakobus:2020nxw}. The second stable-branch family appears when (i) the central stellar density reaches the threshold for a strong first-order phase transition, creating an unstable region in the mass-radius diagram, and when (ii) there is a large enough portion of the star made up of matter that reaches densities beyond the transition threshold and described by a stiff enough EoS. Such first-order phase transitions, in fact, are motivated from the studies of deconfinement of quark matter. Reference~\cite{Keister:1976gc} argued that a phase transition to non-interacting 4-flavor quark matter was unlikely inside stable stellar systems. Reference~\cite{Grassi:1987bu} showed that a phase transition to a 3-flavor interacting quark phase could generate stable stars. Finally, Ref.~\cite{Glendenning:1992vb} showed in detail how to achieve stability for the case of two conserved charges, baryon number and electric charge. In this paper, although we do not focus on the nature of first-order phase transitions themselves, we do investigate density regimes of several times nuclear density, in which deconfined quark matter is expected to appear \cite{Baym:1979etb}.

Given this, in this paper we attempt to address the questions listed above by building a large number of EoSs based on the functional form of the speed of sound, following the prescription of~\cite{Tan:2020ics}. These configurations allow for one or more ``bumps" in the speed of sound, with different low density (below nuclear saturation) crusts, shapes, widths, heights, wells, and intervals in which the speed of sound falls to zero (associated with the strength of a first-order phase transition), which are consistent with features found in realistic nuclear physics EoSs \cite{Dexheimer:2014pea,Dutra:2015hxa,McLerran:2018hbz,Jakobus:2020nxw,Alford:2017qgh,Zacchi:2015oma,Alvarez-Castillo:2018pve,Li:2019fqe,Wang:2019npj,Fadafa:2019euu,Xia:2019xax,Yazdizadeh:2019ivy,Shahrbaf:2019vtf,Zacchi:2019ayh,Zhao:2020dvu,Lopes:2020rqn,Blaschke:2020qrs,Duarte:2020xsp,Rho:2020eqo,Marczenko:2020wlc,Minamikawa:2020jfj,Hippert:2021gfs,Pisarski:2021aoz,Sen:2020qcd,Stone:2021ngh,Ferreira:2020kvu,Kapusta:2021ney,Somasundaram:2021ljr}. We investigate in detail which conditions generate ultra-heavy stars (hereon defined as $M \gtrapprox 2.5$~M$_{\odot}$) and which conditions lead to a second family of stable stars with or without twin configurations.

\subsection*{Executive summary}

The purpose of this paper is to test the possible functional forms of the neutron star EoS using known gravitational wave and X-ray constraints in order to understand qualitative features of the speed of sound.  There is a specific focus on bumps in the speed of sound that can arise due to new hadronic degrees of freedom \cite{Stone:2021ngh},  deconfinement crossover phase transitions \cite{Motornenko:2019arp,Baym:2019iky}, quarkyonic matter \cite{McLerran:2018hbz,Zhao:2020dvu,Sen:2020qcd,Duarte:2020xsp,Sen:2020peq}, breaking of chiral symmetry into a gapped Fermi surface \cite{Hippert:2021gfs},  certain vector interactions \cite{Pisarski:2021aoz}, and first-order phase transitions that produce either a continuous mass-radius sequence or mass twins. These features are of interest because they can signal deconfined quark degrees of freedom within the core of neutron stars, which are of broad interest to a number of fields, including high-energy physics. Below, we provide an executive summary of our main conclusions:

{\bf Result 1: The maximum mass of neutron stars and observational constraints from LIGO/Virgo and NICER observations}. When considering the maximum possible mass of a neutron star, the radius measurement from the GW170817 event~\cite{TheLIGOScientific:2017qsa} (deduced from its tidal deformability) place the tightest constraints. The radius measurement of PSR J0740+6620 from NICER observations~\cite{Riley:2019yda,Miller:2019cac} currently does not have sufficiently small enough uncertainties to add further constraints. This is because, in general, EoS models that lead to large maximum masses possess very stiff EoSs, resulting in larger stellar radii. The impact of LIGO/Virgo mass-radius constraints on the maximum mass, however, depend on the analysis method employed (spectral EoS versus universal relations). The two methods lead to different radius posteriors (differing by about $0.5$~km) because they employ different priors. Even such a small discrepancy in radius posteriors can affect the limits by about $10\%$ on the maximum possible mass of causal neutron stars.  Combining these discrepancies with uncertainties in the crust of neutron stars, we estimate that the absolutely maximum mass of a neutron star could be in the range of $M=2.5-3.6$~M$_\odot$ and $R=11-16$~km without violating any observational constraints. 

{\bf Result 2: Functional form of $c_s^2$ to reach $M>2.5 ~$M$_\odot$ from the GW190814 event}. Even though there are a number of constraints on the neutron star EoS, there is still a wide parameter space available when constructing EoSs that can reach masses as high as the light component of GW190814 i.e. reproduce $M\geq 2.5~ $M$_\odot$. The only generic feature that all of these EoSs share is a steep rise in $c_s^2$ at low densities $n_B\lesssim 3~n_{sat}$.  Nevertheless, with the existing data, one cannot conclude that a bump in the $c_s^2$ is present, because similar results are also consistent with a plateau.  We find that certain features in $c_s^2$ can either shift the maximum mass up or down without affecting the radius (e.g.~the width of a peak in $c_s^2$ or the end point in $c_s^2$) or can shift both the mass and radius, i.e.~ the location of the peak (peaks at large $n_B$ are inversely correlated with the mass and radius).  Finally, much more complicated $c_s^2$ structures are entirely possible within a neutron star, such as double bumps, so \textit{a priori} assumptions of relatively smooth EoSs would eliminate a number of reasonable and realistic EoSs.  This then allows us determine the maximum possible central baryon density reached within our generated EoSs. If one assumes that the light component of GW190814 was a neutron star, then we are not able to generate an EoS with a central baryon density greater than $\sim 6~n_{sat}$. Without this assumption, we find that the maximum central baryon density that we were able to generate in our EoS model was less than $\sim 8~n_{sat}$. 

{\bf Result 3: Neutron star EoSs that are consistent with the binary companion to V723 Mon}. Recently, a dark object of a mass $M\geq 2.91\pm0.08 $~M$_\odot$  was measured \cite{Jayasinghe:2021uqb}. Although this companion is likely a black hole, we consider what the implications on the EoS would be if it were a neutron star. We construct a number of EoSs that are able to reach up to such large masses with radii in the range of $R\sim 12.5-15.5$~km.  If such a massive neutron star exists, then its tidal deformability would be as low as $\Lambda \sim 2.5$. This value is very small, and in fact quite close to that of a black hole. This implies that (a) one does not require an exotic compact object to reach these small tidal deformabilities, and (b) it may be quite difficult to measure such small tidal deformabilities accurately enough to exclude the black hole limit with gravitational wave observations. 

{\bf Result 4: Influence of the crust on the stellar maximum mass and radius}. While the EoS at low densities is better constrained due to a better understanding of nuclear physics in that regime combined with symmetry energy constraints, uncertainties on the order of $10\%$ still remain \cite{Baym:2017whm,Rather:2020gja,Ferreira:2020zzy}. To take that into account, we vary the description of our EoSs in this ``low'' density regime (below $1$--$3$ nuclear saturation density) using 3 different crust models (QHC19, SKa, and SLy). We find that the low density part of the EoS plays a significant role for neutron stars of mass $M\sim 1.4$~M$_{\odot}$, and also, the crust affects the maximum possible causal EoS that one can produce.  Interestingly enough, the crust affects the radius of neutron stars as massive as $M\sim 2$~M$_\odot$. Therefore, future constraints on the radius coming from the NICER observation of PSR J0740+6620 \cite{Riley:2021pdl,Miller:2021qha} with smaller uncertainties would also be important for learning about the neutron star EoS at low densities.

{\bf Result 5: First-order phase transitions in heavy neutron stars $M\geq 2$~M$_\odot$}. We are able to generate a number of EoSs that lead to disconnected twins that reach a maximum mass larger than $M\geq 2$~M$_\odot$, in contrast to \cite{Pang:2020ilf}, due to non-trivial structure in $c_s^2$. The large maximum mass can occur in either the first or second stable branch. Even more massive stars, potentially as massive as the low-mass component of GW190814, do not preclude the possibility of a first-order phase transition within a neutron star. Applying first all known LIGO/Virgo/NICER and saturation density constraints, we are able to construct mass-radius diagrams that have kinks or connected twins but not disconnected branches. The only way we can produce mass-radius diagrams with disconnected twin branches is if we ignore the radius constraint from GW170817. Of course, as shown in \cite{Blaschke:2020vuy}, one can produce disconnected twin branches at very low mass, but this requires a first-order phase transition at potentially too low baryon density (below or close to $n_{sat}$).

{\bf Result 6: First-order phase transitions and bumps in the $c_s^2$}. We construct an assortment of mass twins that also contain a range of structure in the $c_s^2$, such as bumps, spikes,  kinks, and slants, and we determine qualitative observational features that are of interest. For instance, if a narrow bump is placed before a first-order phase transition well, then the mass-radius curve presents disconnected mass twins; but if the bump is widened, then the mass-radius curve only has a single stable branch. 
A more extended phase transition (in density) switches from having a connected branch, to a kink, to disconnected or connected twins, and, finally, to only one stable branch. In the case of twins, the more extended the phase transition, the larger the radius in which they are distributed.
Additionally, after the phase transition, one does not need a step function to a constant $c_s^2$ to produce disconnected mass twins; a slanted $c_s^2$ that goes to a plateau suffices.  A softer slope leads to a flatter second stable branch, whereas a stiffer slope turns the second stable branch more vertical. A similar effect is seen if one increases the height of the plateau in $c_s^2$ after the phase transition (a larger value turns the second branch more vertical), although values that are too small prevent a second stable branch entirely.  Varying the width of the phase transition, one can switch between a disconnected mass twin, or a single stable branch, but this also depends on other features of the EoS, like the position and width of the bump.

{\bf Result 7: Approximately universal relations for EoSs with structure}. We have calculated the I-Love-Q relations (inter-relations between the moment of inertia, the Love number and the quadrupole moment of rotating stars) for a large set of EoSs with various kinds of structure in the speed of sound (such as kinks, spikes, bumps, plateaus, first-order phase transitions, etc). We find that no matter what we do to the speed of sound, the I-Love-Q relations continue to be roughly the same. The insensitivity to the wide variations in the speed of sound that we considered is such that the relative fractional variability is below $1.5\%$. This implies that inferences drawn on any two quantities in the I-Love-Q trio from the measurement of the third are robust to structure in the speed of sound. 

The remainder of this paper presents the details of the conclusions described above. Section~\ref{sec:constraints} discusses constraints from gravitational wave observations, astrophysics, causality, and nuclear physics.  Section~\ref{sec:modelingEoS} outlines our formalism to create structure in the speed of sound and how that connects to EoSs. Section~\ref{sec:25msol} provides explicit examples of EoSs that lead to mass-radius sequences that have maximum masses of $M\geq 2.5$~M$_\odot$, tests causality constraints, and verifies the influence of the crust. Section~\ref{sec:NS-with-1st-O} incorporates first-order phase transitions into non-trivial speed of sound models and studies the possibility of creating heavy mass twins. Section~\ref{sec:comparison} compares the properties of stars with first-order phase transitions and crossover structure in the speed of sound, including the I-Love-Q relations. Section~\ref{sec:future-dir} concludes and points to future work. 

\section{Constraints on the EoS}
\label{sec:constraints}

In this section, we discuss the observational and theoretical constraints one can place on the EoS. We begin by describing how to compute observables given an EoS. We then proceed with a description of observational constraints, and conclude with theoretical constraints.  
\subsection{Observable quantities}

Let us begin by discussing how we connect EoSs to astrophysical observables in a general sense. We provide here only a short summary and refer the interested reader to Ref.~\cite{Yagi:2016bkt} for further details. 

Consider an isolated neutron star that rotates uniformly with low enough angular velocity $\Omega$ that the Einstein equations can be expanded in powers of $\Omega$. Angular velocity is a dimensionful quantity, so by ``slow-rotation'' here we mean slow relative to the mass-shedding limit. This approximation is valid for most stars, even millisecond pulsars, with the exception perhaps of proto-neutron stars shortly after their birth from a supernova explosion~\cite{Martinon:2014uua}.

At $\mathcal{O}(\Omega^0)$, the $(t,t)$ and $(r,r)$ components of the Einstein equation, together with the conservation of the stress-energy tensor, yield the well-known Tolman-Oppenheimer-Volkoff (TOV) equation and the continuity equation. Given an EoS and a central energy density $\epsilon_c$, the solution to these equations yields the spacetime metric for a star of a certain mass and radius. By varying the central density, one can then produce a mass-radius curve, until the sequence becomes unstable, i.e.~$dM/d\varepsilon_c<0$. When considering twin starts, there may be more than one stable branch, so it is important to continue to solve the TOV equations passed the first stable branch. Note that the second branch is only stable if the mass-radius curve
rotates clockwise with increasing central
pressure at the extremum that gives raise to it \cite{Alford:2017vca}. Alternatively, a slow conversion from hadronic to quark phase inside stars could turn $dM/d\varepsilon_c<0$ stars dynamic stable due to how the phase interface moves under stellar pulsation (see Ref.~\cite{Pereira:2017rmp} and references therein) but, so far, there is no strong evidence that this is the case \cite{Alford:2014jha}.

With the ${\cal{O}}(\Omega^0)$ solution under control, one can then proceed to obtain the ${\cal{O}}(\Omega)$ and ${\cal{O}}(\Omega^2)$ corrections.  At first-order in $\Omega$, one finds a correction to the gravito-magnetic sector of the spacetime metric, whose exterior asymptotic behavior is controlled by the moment of inertia $I$. This quantity can be obtained once the interior and exterior solutions at linear order in $\Omega$ are matched at the stellar surface. At second order in $\Omega$, the diagonal sector of the metric is modified, leading to a coupled set of differential equations. The solutions to these equations in the interior and exterior of the star are then matched at the stellar surface through the appropriate choice of integration constants, and this determines the quadrupole moment of the star. The latter controls how the mass-energy re-distributes into an oblate spheroid due to the rotation of the star. As in the case of the mass-radius relation, the moment of inertia and the quadrupole moment can be computed once an EoS and central density are selected. Repeating the calculation over a set of central densities then leads to the $I$--$C$ and $Q$--$C$ curves, where $C = M/R$ is the stellar compactness, or equivalently the $I$--$M$ and $Q$--$M$ curves. 

With the mass-radius, moment of inertia and quadrupole moment of an isolated star computed, one can then shift gears to a star that is no longer rotating but is also no longer in isolation. When in the presence of a companion, a neutron star will tidally deform. The re-distribution of the mass energy inside the star into an oblate spheroid due to the external perturbation can be captured through a multipolar decomposition. At leading order in perturbation theory, the quadrupole moment dominates the deformation. How much of a quadrupolar deformation is excited, given an external quadrupolar tidal field, is controlled by the (electric-type, $\ell=2$) Love number, or its dimensionless counterpart, the (electric-type, $\ell=2$) tidal deformability $\Lambda$\footnote{Technically, there are two types of tidal deformabilities --an electric-type and a magnetic-type-- that are characterized by how they transform under a parity transformation. The electric-type tidal deformability dominates, affecting the diagonal sector of the spacetime metric. The magnetic-type enters at higher order in post-Newtonian theory, affecting first the gravito-magnetic sector.}. The calculation of the tidal deformability requires the solution to the linearized Einstein equations, which in turn requires the solution of a differential equation in the interior and exterior of the star that must be made continuous and differentiable at the stellar surface through the appropriate choice of integration constants. With this done, the tidal deformability can be computed from this solution and its derivative evaluated at the stellar surface. As in the case of the moment of inertia or the quadrupole moment, one choice of central density yields one value of the tidal deformability for a given star. Repeating the calculation over various central densities, one can then obtain the $\Lambda$-C (sometimes called Love-C), or $\Lambda$-M (sometimes called Love-M) relations.  

\subsection{Observational constraints}

\subsubsection{Gravitational wave observations}\label{sec:GWobs}

The LIGO/Virgo Collaboration (LVC) used LIGO and Virgo data of the GW170817~\cite{TheLIGOScientific:2017qsa} event to place the first gravitational wave constraints on the EoS~\cite{Abbott:2018exr}, which we now review. The GW170817 event consisted of gravitational waves emitted in the inspiral, late inspiral and merger of a compact binary composed of compact objects with masses $(m_1,m_2) \sim (1.5,1.3) $~M$_\odot$~\cite{TheLIGOScientific:2017qsa}. The gravitational waves encode the tidal deformabilty of the compact object in the waveform phase, and a non-zero posterior on these quantities (together with an assumption of small spins) ensured the compact objects were neutron stars and not black holes. The question then is how to go from information about the tidal deformabilities to the EoS of matter. 

The LVC followed two approaches\footnote{Recently, there have been new studies of LVC data that use an EoS model-agnostic approach, in terms of Gaussian processes~\cite{Landry:2018prl,Essick:2019ldf,Landry:2020vaw}, which leads to somewhat different confidence regions, though given the degree of information contained in the data, all of these are technically statistically consistent with each other.}. In the first approach, which we shall call the ``universal relations approach,'' they used approximately  EoS-insensitive relations to perform EoS inferences. These are universal relations between the two tidal deformabilities of compact objects in a binary (the so-called binary Love relations)~\cite{Yagi:2015pkc,Yagi:2016qmr}, as well as between the tidal deformability of any one of the compact objects and its compactness (the so-called Love-C relations)~\cite{Yagi:2016qmr}. The LVC first used the binary Love relations to express the waveform phase entirely in terms of one of the two tidal deformabilities (thus analytically breaking parameter degeneracies in the waveform phase). They then used the binary Love relations again to infer the second tidal deformability from their measurement of the first one. Finally, they used the Love-C relations to infer the compactnesses of the two stars. This last inference, together with their measurements of the component masses, yielded an inference on the stellar radii of the two stars. 

The second LVC approach used a specific model of the EoS to carry out parameter estimation, which we shall call the ``spectral EoS approach.'' The phenomenological spectral EoS is constructed from $c_s^2 := dp/d\epsilon = \Gamma(p) p/(\epsilon + p)$, where $p$ and $\epsilon$ are pressure and energy density respectively, while $\Gamma$ is the adiabatic index for a polytrope, which is expanded via $\Gamma = \exp[\sum_k \gamma_k \log(p/p_0)^k]$, where $\gamma_k$ are parameters to be fitted and $p_0$ is the smallest pressure at which the spectral EoS is used (below that pressure, the LVC used a SLy EoS). The LVC then sampled in the $\gamma_k$ (including up to $k=4$) and in the central densities directly, when comparing the waveform model to the data. Importantly, the priors the LVC used for $\gamma_k$ do not allow for the possibility of first-order phase transitions or any other sharp feature in the speed of sound. In this way, they were able to construct posterior probability distributions for the $\gamma_k$ and directly for the EoS. With this information, one can also infer a posterior distribution on the M-R plane, as well as on other inferred quantities, like in the I-M place and the Love-M plane. 

Each of the two approaches have advantages and disadvantages. The obvious advantage of the universal relations method is that one never has to specify a functional form for the EoS, but since the universal relations are not exact, one has to marginalize over the residual EoS sensitivity. Conversely, the advantage of the spectral EoS approach is that one can infer a posterior on the EoS directly, so all EoS-dependent quantities can be easily computed. Specifying a functional form for the EoS, however, has its own problems, since it can bias inferences related to nuclear physics for the cases in which Nature's EoS cannot be well-represented by the spectral EoS model. The posterior distributions in the mass-radius plane one obtains in these two approaches, however, are consistent with each other, with the spectral EoS posteriors being tighter than the universal relations. 

Let us also mention, in passing, that most of the information about the tidal deformabilities come from the very, very late inspiral part of the coalescence. Indeed, the analysis of the GW170817 that led to posteriors on the tidal deformabilities is somewhat sensitive to (i) the accuracy of gravitational wave models near but right before merger, and (ii) any calibration of the detector at those high frequencies.  The LVC did carry out a careful analysis, extracting the signal and analyzing it with various waveform models and noise representations. These various analysis suggest that the posteriors on the tidal deformability (and thus on the mass-radius plane) are robust to these systematics, yet the fact that these systematics exist should be kept in mind.

The LVC has recently observed another interesting event, GW190814, which consisted of the gravitational waves emitted in the inspiral and merger of a compact  binary with component masses $m_1 \approx 25.6$~M$_\odot$ and $m_2 \approx 2.59$~M$_\odot$. For this event, however, the signal-to-noise ratio was not large enough to measure the tidal deformability to be zero with enough accuracy to exclude neutron stars, which have non-zero deformabilities.  Lacking this, and any electromagnetic counterpart, it is not possible to determine whether the lighter object was an ultra-light black hole or an ultra-heavy neutron star from that data alone. In this paper, we will investigate whether it is possible or not to form such ultra-heavy neutron stars.

\subsubsection{Astrophysical observations}

The NICER collaboration used x-ray data on PSR J0030+0451 to place the first NICER constraints on the EoS~\cite{Riley:2019yda,Miller:2019cac}, which we now review. PSR J0030+0451 is an old (roughly $7.8$ Gyr old), isolated millisecond pulsar with a period of about $4.87$ ms. X-rays emitted by the hot spots on the surface of the star (hotter relative to the average surface temperature) are then recorded by the NICER instrument as a function of time, allowing for the resolving of the X-ray pulse profile. This profile encodes information about a variety of properties of the system, such as the hot spot shape, but in particular also information about the compactness of the star and its mass. The question here is how to model these pulses and how to analyze this data set in practice. 

The collaboration followed two approaches, which we shall refer to as the ``Amsterdam analysis'' (AM) and the ``Illinois-Maryland analysis'' (IL/MD). The two approaches differ in the way the hot spots are modeled, and in the way the parameter space is explored (e.g.~Amsterdam used nested sampling, while Illinois-Maryland used Markov-Chain Monte Carlo sampling) and their choice of priors. These differences led to different confidence regions in the mass-radius plane, although the differences are statistically consistent with each other~\cite{Riley:2019yda,Miller:2019cac}. A recent comparison paper~\cite{Riley:2021pdl} suggests that the use of MultiNest may have led to non-convergent results with live-point number, suggesting an explanation for the discrepancy in the IL/MD and the AM posteriors in the mass-radius plane. 

The NICER collaboration has also recently presented their analysis of the millisecond pulsar, PSR J0740+6620, with a period of about 2.9 ms, in a binary system with a white dwarf \cite{Riley:2021pdl,Miller:2021qha}. Because of the companion, the mass of the pulsar was known from pulsar radio observations to be about $2.14 $~M$_\odot$. For this pulsar, the NICER team was able to use X-ray measurements from XMM Newton to estimate the background. As before, two analysis were carried out on this data set, leading to two posteriors in the mass-radius plane that are statistically consistent with each other~\cite{Miller:2021qha,Riley:2021pdl}. 

Another set of observations that we will refer to in this paper are those of the dark companion to the bright nearby red giant V723 Mon. Due to the geometrical orientation of the binary system's orbit, the components eclipse each other. From the light curve of this eclipsing binary, researchers were able to infer that the companion to V723 Mon has a mass of $M\geq 2.91\pm0.08$~M$_\odot $ \cite{Jayasinghe:2021uqb}. This object is, therefore, either a black hole in the so-called mass gap, or an ultra-heavy neutron star (ie.~a neutron star with mass larger than $2.5 $~M$_\odot$). 

\subsubsection{Comparison between LIGO/Virgo and NICER observations}

\begin{figure}
\centering
\includegraphics[width=0.9\linewidth]{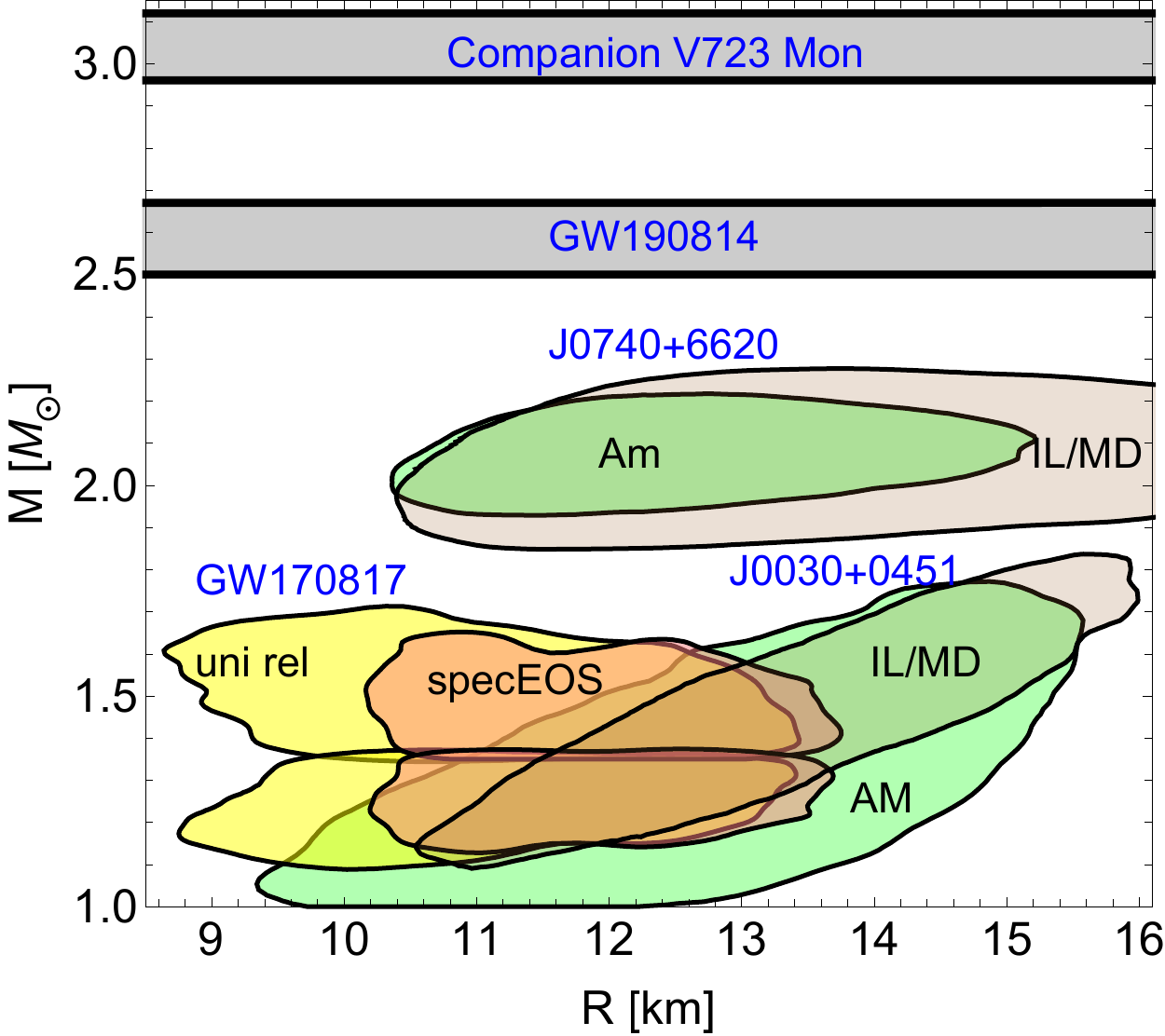} 
\caption{(Color online) Observational constraints on the neutron star mass-radius plane from LIGO/Virgo and NICER data. The yellow and pink (looks orange when overlapped) regions correspond to 90\% confidence regions, obtained from the GW170817 event using the universal relations and spectral EoS approaches, respectively. The brown and green regions correspond to 90\% confidence regions obtained from NICER data on PSR J0030+0451 and J0740+6620, using the Illinois-Maryland and the Amsterdam analysis, respectively. Finally, the horizontal regions are 90\% confidence regions for the mass of the lighter object in GW109814 and the Companion V723 Mon object. 
}
\label{fig:allCON}
\end{figure}

Figure~\ref{fig:allCON} shows the various observational constraints we have discussed so far in the mass-radius plane (see caption for details). The radius constraints for GW170817 differ between the two LVC analysis by roughly $0.5$~km, and it stems from the fact that the spectral EoS approach included the prior that only EoSs that support a $2$~M$_\odot$ star or heavier are allowed (a prior that was not imposed in the universal relations analysis). For J0030+0451, The IL/MD analysis extends more, in the direction of wider radii, which implies that it is the least constraining between IL/MD and AM (the maximum mass and the lowest radius of the AM posterior do not strongly affect our results because more massive neutron stars tend to be produced by EoSs that reproduce intermediate mass stars that are wider). 

The real impact of different constraints on nuclear physics is strongly dependent on which range and which quantity the observations constrain. For example, many standard EoSs (assuming they do not describe strange quark stars, dark matter, or mirror matter) predict that a radius in $R\sim 11$-$17$~km for a star with mass $M\sim 1.4$~M$_\odot$
\cite{Bednarek:2011gd,Fortin:2014mya}. Thus, the low end of the radius constraints from LIGO/Virgo and NICER data do not affect the vast majority of EoSs. On the other hand, the high-end of the radius constraints from GW170817 does provide a significant constraint on many EoSs, especially when one considers the possibility of very massive neutron stars. Since we assume that the overlap region between the J0030+0451 NICER and GW170817 LIGO/Virgo constraints is the most likely for stars of about $1.4$~M$_\odot$, then this implies the maximum radius is constrained from the GW170817 data.  Henceforth, for the most part, we will choose to show our results in comparison to constraints obtained from the universal relations analysis, together with the Illinois-Maryland analysis. However, as discussed above, some gravitational wave constraints are susceptible to the functional form of the EoS used, so in Sec.\ \ref{sec:LIGOcheck} we explore the approximately $0.5$~km difference in the constraints obtained from the use of an spectral EoS and the use of universal relations.

\subsection{Theoretical constraints}

\subsubsection{Causal limit and pQCD}

A theoretical bound that all neutron star EoSs must not violate is causality.  This means that the speed of sound of a neutron star cannot be greater than the speed of light, i.e.~$0\leq c_s^2\leq 1$ (in natural units). A way to probe the absolute limits of the largest possible neutron stars is to generate an EoS that transitions from a crust (at low densities) to the causal limit, i.e.~$c_s^2\rightarrow 1$. Typically, this is done with almost a step function. In this paper, however, when we explore the causal limit, we always incorporate a slant to avoid a divergence in the derivative of $c_s^2$. We discuss the effects of varying the slant inclination in section \ref{sec:NS-with-1st-O}. This kind of study can place bounds on the absolute maximum neutron-star mass for a specific crust EoS and specific stellar radius constraints.

A somehow opposite constraint comes from perturbative QCD techniques, which have found that $c_s^2\rightarrow 1/3$ from below at large baryon densities \cite{Kurkela:2009gj,Kurkela:2014vha,Hebeler:2013nza}.  Currently, the lowest estimate for the applicability of this argument is at $n_B\gtrsim 40~n_{sat}$, which is well beyond the maximum central density of all stars shown in this paper.  However, it is still interesting to study the consequences of the final value that  $c_s^2 $ reaches at its maximum central density in neutron stars, which we also explore in this paper. Furthermore, if quark matter is indeed reached with the neutron star densities, the only explanation for a further significant decrease in speed of sound is the appearance of new degrees of freedom, further phase transitions to other quark phases with different symmetries \cite{Alford:2017qgh,Li:2019fqe}, or drastic changes in the interactions \cite{Pisarski:2021aoz}.

\subsubsection{Nuclear Physics Constraints}

Assuming that the constituents of the neutron star in some region are baryons, the hadronic matter in this region must present properties that are compatible with the corresponding reliable theoretical predictions and laboratory experimental data. The former comprehends \textit{ab initio} approaches to model the attractive and repulsive components of the strong interaction between hadrons. In particular, Chiral Effective Field theory produces reliable calculations for pure neutron matter for densities that extend a little beyond \textit{saturation density} (see, for example, Fig.~2 in Ref.~\cite{Hebeler:2013nza}), which we define next.

Hadrons are bound together inside nuclei by the residual strong force, which is short-ranged. This gives rise to the phenomenon of saturation, imposing a limit for the size of nuclei and producing an average nuclear baryon density of $\rho_{\rm sat} \sim 2.3\times10^{14}$~g/cm$^3$, or equivalently an average nuclear number density of $n_{\rm{sat}}\sim0.16$~fm$^{-3}$; we will refer to the latter as the nuclear saturation density \cite{Dutra:2012mb}. This value can be calculated as the density that corresponds to the minimum in the binding energy per hadron. Several properties have been measured for nuclei at this density, including binding energy, incompressibility, and effective mass of nucleons. Hadronic models are constructed in such a way that they reproduce saturation properties at saturation density for isospin symmetric matter (i.e.~containing the same amount of neutrons and protons).

Since neutron-star matter is \textit{not} isospin symmetric due to the neutronization process that takes place in supernova explosions (combined with forbidden decays due to Pauli blocking), hadronic models that are used in astrophysics must in addition reproduce asymmetric properties. This is done by modelling the energetic cost of producing asymmetric matter (much more incompressible than symmetric matter) to be in agreement with experimental data of symmetry energy at saturation density and its slope \cite{Tsang:2012se}. The latter is a very important quantity, as it guides models beyond saturation. A lower slope of the symmetry energy implies a less incompressible (more soft) EoS at intermediate densities. Such a slope leads to neutron stars in better agreement with NICER data and LIGO/Virgo data\cite{Fattoyev:2012uu}.

At larger densities (beyond $~2$~n$_{\rm sat}$) heavier hadrons are expected to appear. The uncertainties in their properties only increases the uncertainties in the modelling of the neutron star core, independently of those being hyperons \cite{Weissenborn:2011kb}, Delta resonances \cite{Li:2018qaw}, or negative parity states \cite{Dexheimer:2007tn}.

\section{Modeling neutron star EoSs with crossover and first-order phase transitions}
\label{sec:modelingEoS}

In the previous section we discussed constraints that have been imposed on the EoS of neutron stars from observations and theoretical considerations. Now we discuss how to build EoSs that satisfy these constraints. One key feature of our model-building, however, is that we will only consider EoSs that contain one of two types of structure: first-order phase transitions or crossover transitions. An EoS with a {\bf first-order phase transition} corresponds to one in which there exists a region in energy density $\epsilon$ inside which the speed of sound $c_s^2 := dp/d\epsilon$ vanishes (forming a well), or equivalently, in which pressure $p$ is a constant across a range of $\epsilon$. By {\bf crossover} structure, we mean regions in $p$--$\epsilon$ in which the speed of sound behaves non-monotonically, for example presenting kinks, bumps, jumps, spikes, or plateaus. A crossover transition could be triggered by a new state of matter or by new degrees of freedom (or simply by different interactions, as discussed above). First-order phase transitions, on the other hand, are usually related o the appearance of a new state of matter, such as deconfined quarks. In either case, simple relations among central speed of sound and stellar masses \cite{Moustakidis:2016sab} do not apply.

\begin{figure}[t]
\centering
\includegraphics[width=0.9\linewidth]{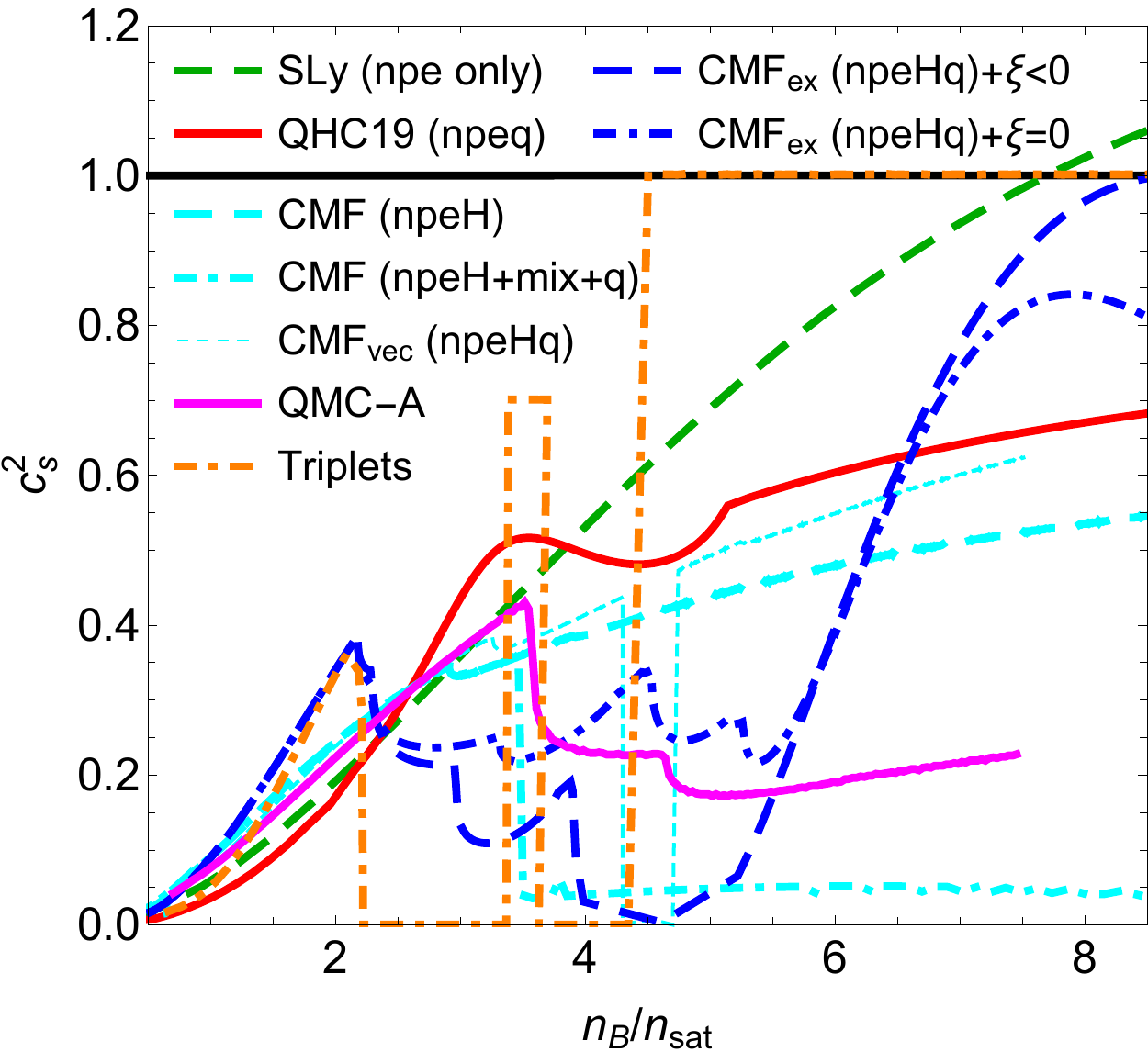} 
\caption{(Color online) Comparison of the speed of sound squared, calculated from the SLy EoS (n+p+e), the QHC19 EoS (n+p+e+quarks), the CMF EoSs (n+p+e+$\mu$+Hyperons+quarks), the QMC-A EoS (n+p+e+$\mu$+Hyperons), and the Tiplets (n+p+e+$\mu$+Hyperons+quarks) EoS as a function of baryon number density in nuclear saturation units. EoSs constructed from state-of-the-art models generically lead to non-trivial, non-monotonic structure in the speed of sound.
}
\label{fig:known}
\end{figure}

Why do we focus on such EoSs? EoSs containing only nucleonic degrees of freedom within the core of a neutron star generally have speeds of sound $c_s^2$ that monotonically increase with increasing baryon density $n_B$. However, at large enough $n_B$, nucleonic EoSs often become acausal ($c_s^2>1$), as  demonstrated in Fig.\ \ref{fig:known} with the SLy EoS, where $n_{\rm sat} := 0.16\; {\rm{fm}}^{-3}$ stands for nuclear saturation density. The inclusion of more exotic degrees of freedom at large densities, such as hyperons or quarks naturally creates a non-monotonic structure in the speed of sound, with kinks, bumps, jumps, spikes, wells, or plateaus \cite{Bedaque:2014sqa,Dutra:2015hxa,McLerran:2018hbz,Jakobus:2020nxw,Annala:2019puf,Zacchi:2019ayh,Zhao:2020dvu,Pisarski:2021aoz,Ferreira:2020kvu,Sen:2020qcd,Malfatti:2020onm,Minamikawa:2020jfj,Hippert:2021gfs,Stone:2021ngh,Kapusta:2021ney,Somasundaram:2021ljr}.  This is also demonstrated in Fig.\ \ref{fig:known} with the QHC19, QMC-A, Triplets, and various versions of the CMF model. 
QHC19 is an EoS constructed with a percolation to quark matter described by a 2+1 NJL model \cite{Baym:2019iky}, which leads to a bump followed by a dip in $c_s^2$. 
The QMC-A is an EoS that includes valence quarks inside nucleons and hyperons interacting self-consistently \cite{Guichon:1995ue,Stone:2021ngh}. The kinks associated with the appearance of hyperons combine into a larger bump that brings the speed of sound below $c_s^2=1/3$.
The Triplets EoS is a combination of a density-dependent RMF model with nucleons and hyperons, a two-flavor color-superconducting (2SC) phase, and a colorflavor-locked (CFL) phase \cite{Alford:2017qgh}. The two distinct phase transitions reproduce two EoS wells.
CMF is an EoS constructed from a chirally symmetric Lagrangian that describes hadrons and quarks within the same formalism. Figure \ref{fig:known} shows five realizations of the CMF EoS -- the original description with hyperons \cite{Dexheimer:2008ax} and a first-order phase transition to quark matter connected through a mixture of phases \cite{Dexheimer:2009hi}, a new parametrization with extra vector interactions and a first-order phase transition to quark matter (but now without a mixture of phases) \cite{Dexheimer:2020rlp}, and, finally, two EoS that include an excluded volume for the hadrons, resulting in a crossover to the quark phase (shown for two different parametrization of the strange vector quark couplings) \cite{Dexheimer:2014pea}. 
The appearance of hyperons and crossover transition to quarks appear as kinks or bumps of different sizes. The first-order phase transition reproduces a small well with $c_s^2=0$ and the mixture of phases presents very small $c_s^2$ for an extended range of densities. The excluded volume of hadrons generates a large raise in $c_s^2$, which can generate stable twin stars (for the CMF$_{\rm exc}$ with $\xi<0$).
This figure clearly shows that realistic neutron star EoSs contain non-monotonic structures in the speed of sound that would be impossible to model accurately with a spectral EoS (unless a very large number of $\gamma_k$ constants are included). 
 
\begin{figure}[t]
\centering
\includegraphics[width=\linewidth]{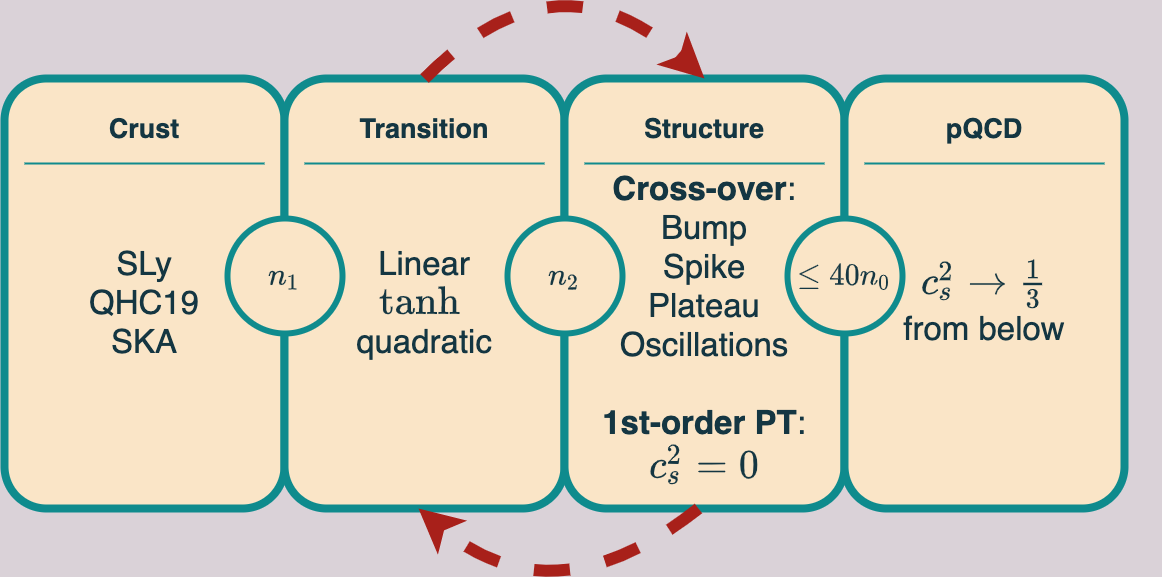} 
\caption{(Color online) Flow chart of the piecewise functional forms of the speed of sound used to create EoSs. More complicated structures in the speed of sound can be created with multiple transition and structure functions. 
}\label{fig:flow}
\end{figure}

The approach we will follow in this paper is to create functional forms of the EoS using the speed of sound, $c_s^2$, directly.  At low densities we use well-established crust EoSs (SLy is our default, but we study the consequence of the QHC19 \cite{Baym:2019iky} and SKa \cite{Fortin:2016hny} crusts as well), and at a transition baryon density $n_1 \in (1.5$--$3)~n_{\rm sat}$, we switch to a new set of functional forms for the EoS, as sketched in Fig.\ \ref{fig:flow}.  The functional forms have a \textit{transition} piece that is either a linear function, a quadratic function, or a hyperbolic tangent (smooth transition) function of the energy density. This transition piece connects the crust to a \textit{structure} piece once a baryon density $n_2$ is reached.  The structure piece can model a smooth bump, a spike, a flat plateau, oscillations, or a plateau at $c_s^2=0$, which defines a first-order phase transition. We restrict the, in principle infinite, space of structure functions as follows:
\begin{itemize}
\setlength\parskip{-0.1cm}
\setlength\parsep{-0.1cm}
    \item Causality: $0\leq c_s^2\leq 1$,
    \item Switching density: $n_1\geq 1.5~n_{sat}$,
    \item Mass-radius relationship must fit within known constraints by LIGO/VIRGO/NICER \cite{Abbott:2018exr,Abbott:2018wiz,Riley:2019yda,Miller:2019cac}. 
\end{itemize}
Regarding the last point, we here use the Illinois-Maryland analysis of the NICER data for J0030+0451 and the universal-relations analysis of the LIGO/Virgo data for the GW170817 event.  Additional constraints arising from the analysis of PSR J0740+6620, from the analysis of GW170817 with a spectral EoS, from GW190814, or from the companion of V723 Mon are also shown in a few figures for comparison.  At very large densities, i.e.~$n_B\sim 40~n_{\rm sat}$, one can also require that $c_s^2\rightarrow 1/3$ from below \cite{Kurkela:2009gj,Hebeler:2013nza,Kurkela:2014vha}, which is sometimes referred to as the \textit{conformal limit}.  However, since all of our EoSs have maximum central densities below $n_B^{max}\leq 10~n_{\rm sat}$, this requirement has no effect on our functional forms. We have created a \texttt{Mathematica} notebook that can reproduce these EoSs, and is available at GitHub~\cite{jakigit}

\subsection{Low density (crust) EoS}

As a default, we use the SLy EoS \cite{Chabanat:1997un,Douchin:2000kad,Douchin:2000kx,Douchin:2001sv} for our low density regime, $n_B \lesssim 2~n_{sat}$, although two other prescriptions are also used for comparison.  The SLy EoS employs a nuclear interaction of the Skyrme type and generates a unified EoS from crust to a liquid core of neutrons, protons, electrons, and muons. The crust has an effective Hamiltonian with the Argonne two-nucleon interaction AV18 and it includes the
Urbana model of three-nucleon interactions.

The second prescription we use is that of the QHC19 EoS. It contains purely hadronic degrees of freedom in the crust, whose free energy is calculated using a variational method \cite{Togashi:2017mjp}. We employ this framework up to a baryon number density of $2~n_{sat}$, where a transition to a quark matter phase takes place (described by a 2+1 NJL model) via percolation, which is comparable to a crossover.

A third prescription we explore is that used in the SKa EoS \cite{KOHLER1976301,DANIELEWICZ200936,Gulminelli:2015csa}. The physics at the microscopic level is very similar to that of the SLy EoS, in which a Skyrme type interaction is adopted to generate a unified EoS. However, this single nucleon approach is extended to incorporate the effects of non-trivial clustering via a nuclear statistical equilibrium method. 
Additionally, the outer crust employs a variational method, making use of available experimental mass tables of nuclei. 

\subsection{Transition function}

We often need to connect different pieces of the EoS to describe different regions of the star.  This could be the crust and the core or just two difference structures like a bump and a phase transition.  Let us say that the low density function that we need to connect is $f_1(n_B)$ and the high density function is $f_2(n_B)$. Typically,  we start our connection from $f_1(n_B)$ to the transition piece at $n_1$ and we end our connection at $n_2$.  In this work, we have defined three different transition functions.  

The first function we use is a linear transition function:
\begin{equation}
    f_{\rm lin}(n)\equiv mx+b \quad n_1<n_B<n_2\,,
\end{equation}
where
\begin{eqnarray}
m &=&\frac{f_1(n_1)-f_2(n_2)}{n_2-n_1} \qquad
b =  f_1(n_1)-m\cdot n_1\,,
\end{eqnarray}
This gives a kink in $c_s^2$ when connected to $f_1(n_B)$ and $f_2(n_B)$. 

The next function provides a smooth transition between two functions by using a hyperbolic tangent:
\begin{equation}\label{eqn:endpoint}
 f_{\rm tanh}(n_B)\equiv S(n_B) f_2
 (n_B)+\left[1-S(n_B)\right]f_1(n_B)\,,
  \end{equation}
where $S(n_B)$ is a smoothing function defined as
\begin{equation}
S(n_B) := 0.5+0.5\tanh \left[(n_B/n_{sat}-a)/b\right]\,,
\end{equation}
where $b$ determines the width of the smoothing region and $a$ is an offset parameter. Here, $n_1$ and $n_2$ have some arbitrariness as the two functions smoothly connect to each other.

Lastly, we use a quadratic function to transition.  The primary reason for this is to change a function from concave to convex (or visa versa). In particular, we employ 
\begin{equation}
    f_{\rm quad}(n)\equiv (n_B-x_0)^2+y_0 \quad n_1<n_B<n_2\,,
\end{equation}
where we have defined
\begin{eqnarray}
x_0 &=&\frac{f_2(n_2)-f_1(n_1)+\left(n_1^2-n_2^2\right)}{2\left(n_1-n_2\right)} ,\\
y_0 &=&  f_2(n_2)-(n_2-x_0)^2\,.
\end{eqnarray}
This is used when we want to create a kink in the $c_s^2$, for instance.  With these there different transition functions, we are able to connect the crust to the core (or multiple layers of structure) quite easily. 

\subsection{Structure function}

In this paper, we build in multiple different types of structure functions that we detail below.  We turn on our structure function at the baryon density $n_2$, and in some cases, connect multiple structure functions together, e.g.~ to another function $f_2(n_B)$ that could be connected to another $f_3(n_B)$ at $n_3$, where $f_3$ may be either a transition function or another structure function.

Most commonly, we build bumps or spikes into our EoS through a structure function of the form
\begin{equation}\label{eqn:bump}
     f_{\rm bump}(n_B)\equiv f_3(n_3)+d \exp\left[-\frac{(n_B-n_{peak})^2}{w^2}\right] \quad n_2\leq n_B\leq n_3\,,
\end{equation}
where $d$ defines the magnitude of the bump, $w$ is the width of the bump, and $n_{peak}$ is any offset that places the peak at a specific location in $n_B$.  Often times we define $n_2=n_{peak}$ if we want to have a sharper rise to the peak (i.e. something like a spike), rather than a simple Gaussian, which can be accomplished through a transition function. We have studied a number of different $f_3(n)$ possibilities, such as just a constant value (a plateau after the bump), oscillations, a first-order phase transition, or secondary bumps. Additionally, we sometimes build in valleys or spikes to lower $c_s^2$, and this is accomplished by choosing $d$ to be negative.

We have also tested oscillations after a bump to test the sensitive of EoS to complicated structures after a large bump. The format is somewhat similar to Eq.\ \ref{eqn:bump}.   This includes damped oscillations as well, such as
\begin{equation}
f_{\rm osc}(n_B)\equiv f_3(n_3)+ d\exp\left[-w\frac{n_B}{ n_{sat}}\right]\cos\left[f\cdot n_B/n_{sat}\right] \quad n_B\leq n_3\,,
\end{equation}
where $d$ describes the magnitude and $w$ describes the width (as with Eq.\ \ref{eqn:bump}) and $f$ describes the frequency of the oscillations.  To study just oscillations (no dampening) then on can set $w=0$.

A plateau can occur simply by using a constant value of $c_s^2$, i.e.
\begin{equation}
     f_{\rm plat}(n_B)\equiv {\rm const} \quad  n_2 \leq n_B\,.
\end{equation}
For the case of first-order phase transitions, we use a plateau of a constant value that is nearly zero e.g. $c_s^2=0.001$ across a width of $n_B$.  Typically we have transition functions that are either linear or quadratic that connect the first-order phase transition to the rest of the functional form.  Finally, we often end at another plateau of a different value by the time we reach the stellar central baryon density.

\subsection{Reconstructing the EoS from $c_s^2$}

The low density EoS is taken directly from our crust table.  At the baryon number density $n_1$ at which one switches between the crust and a functional form of $c_s^2$, we use the initial values from the crust for the pressure $p_1$ and the energy density $\varepsilon_1$ to reconstruct the EoS iteratively. Taking small step sizes in the baryon density $\Delta n_B$, it is possible to solve the following equations
\begin{eqnarray}
  n_{B,i+1}&=&n_{B,i}+\Delta n_{B}\,, \\
  \varepsilon_{i+1}&=&\varepsilon_i+\Delta\varepsilon=\varepsilon_i+\Delta n_B \left(\frac{\varepsilon_i+p_i}{n_{B,i}}\right)\,, \\
  p_{i+1}&=&p_i+c_s^2(n_{B,i}) \, \Delta \varepsilon\,, 
\end{eqnarray}
as was shown in Ref.~\cite{Tews:2018kmu}
, assuming that
\begin{align}
  c_s^2 &= \frac{\partial p}{\partial \varepsilon}\,, \qquad
  p+\varepsilon =n_B \mu_B\,,
\end{align}
for zero temperatures, and 
assuming electric charge neutrality and chemical equilibrium with leptons (otherwise, there would be a $n_Q\mu_Q$ term on the right-hand side that would require another constraint). 
When considering twin EoSs, it is extremely important to take very small steps sizes in $\Delta n_B$ in order to avoid numerical error. A good consistency check when determining $\Delta n_B$ that we have employed is to obtain the EoS from this method, and then recalculate $c_s^2$ to ensure that it is compatible with the original functional form.

\begin{figure*}[t!]
\centering
\begin{tabular}{c c c c}
\includegraphics[width=0.245\linewidth]{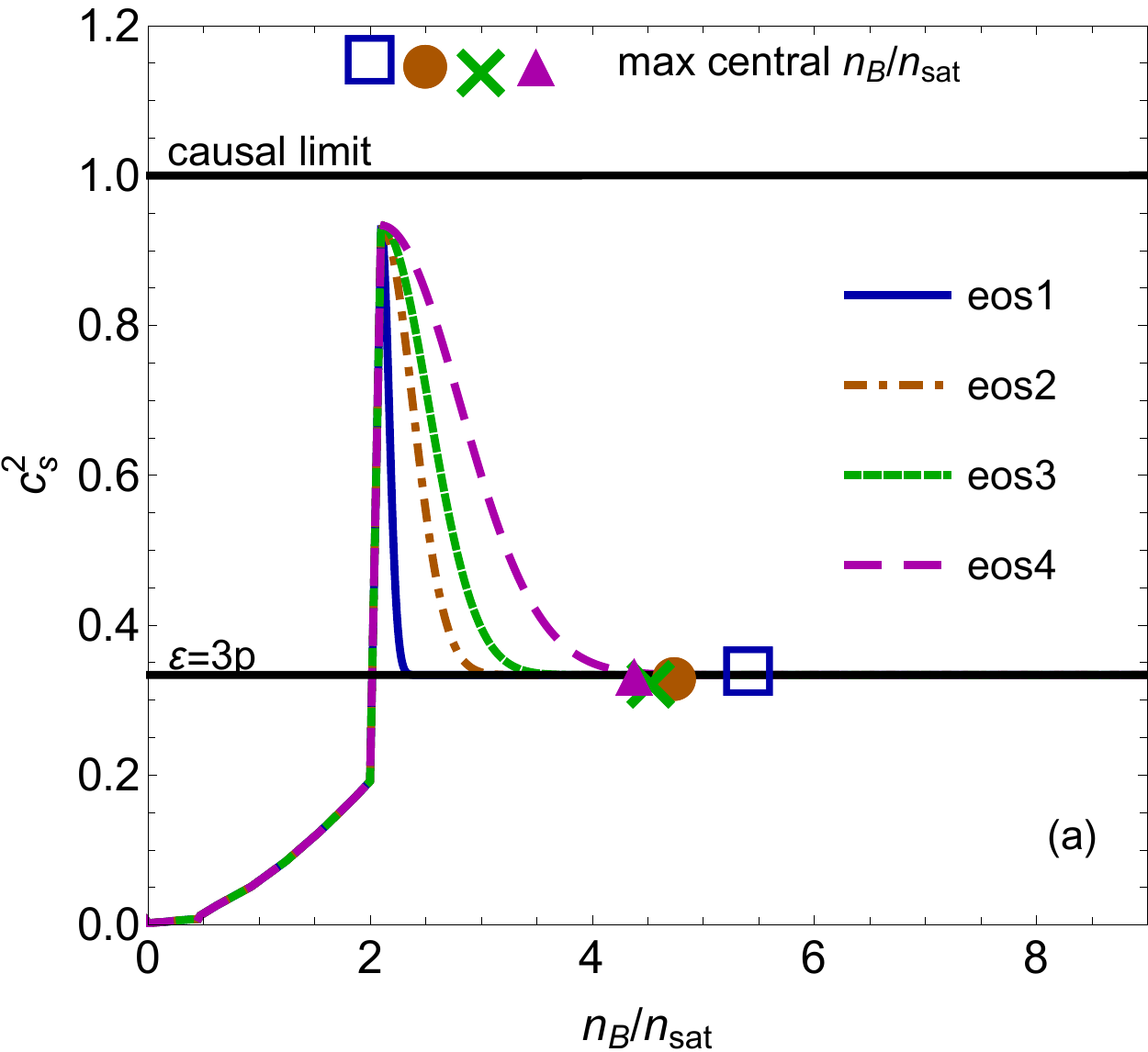} &
\includegraphics[width=0.245\linewidth]{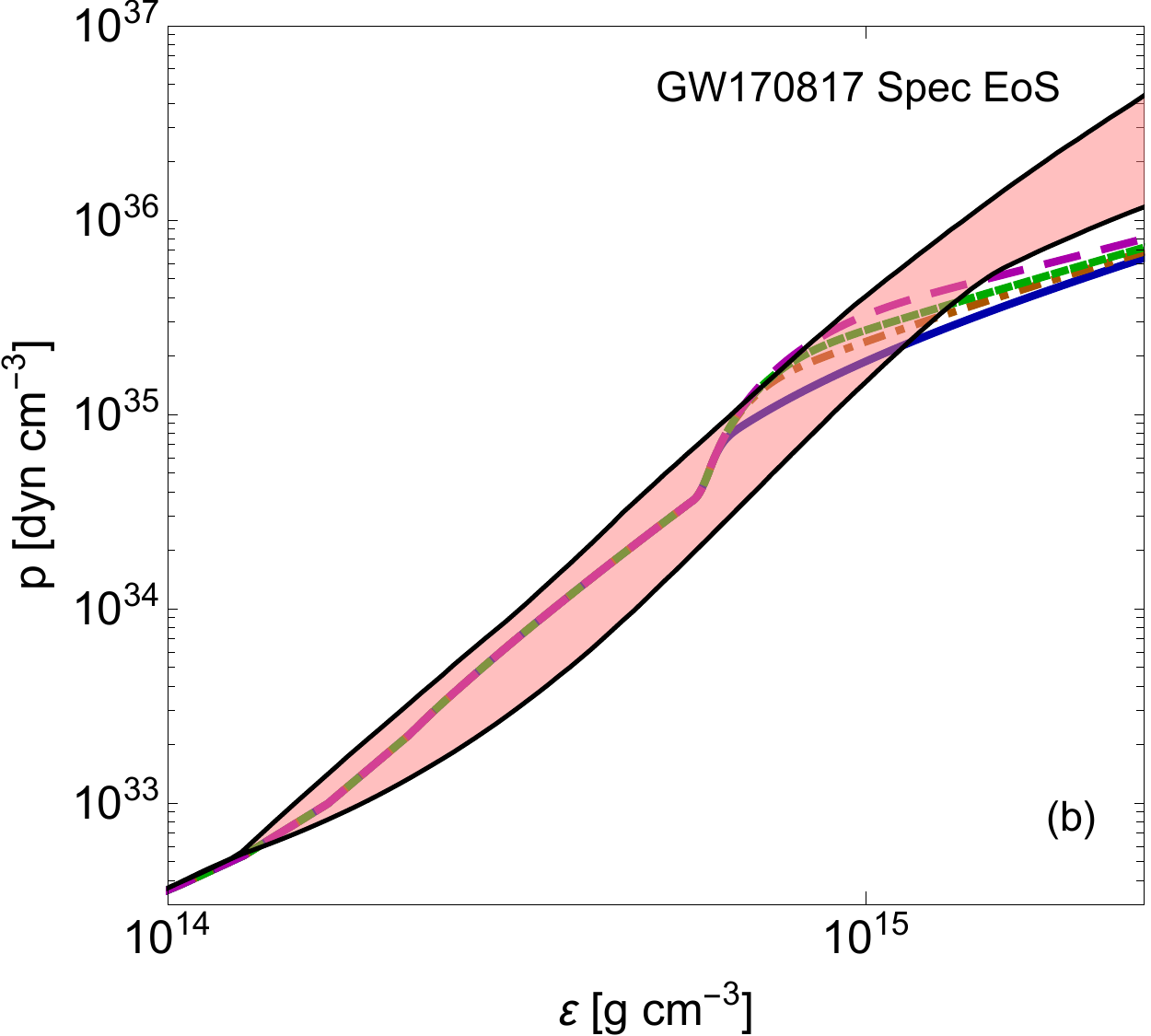} &
\includegraphics[width=0.245\linewidth]{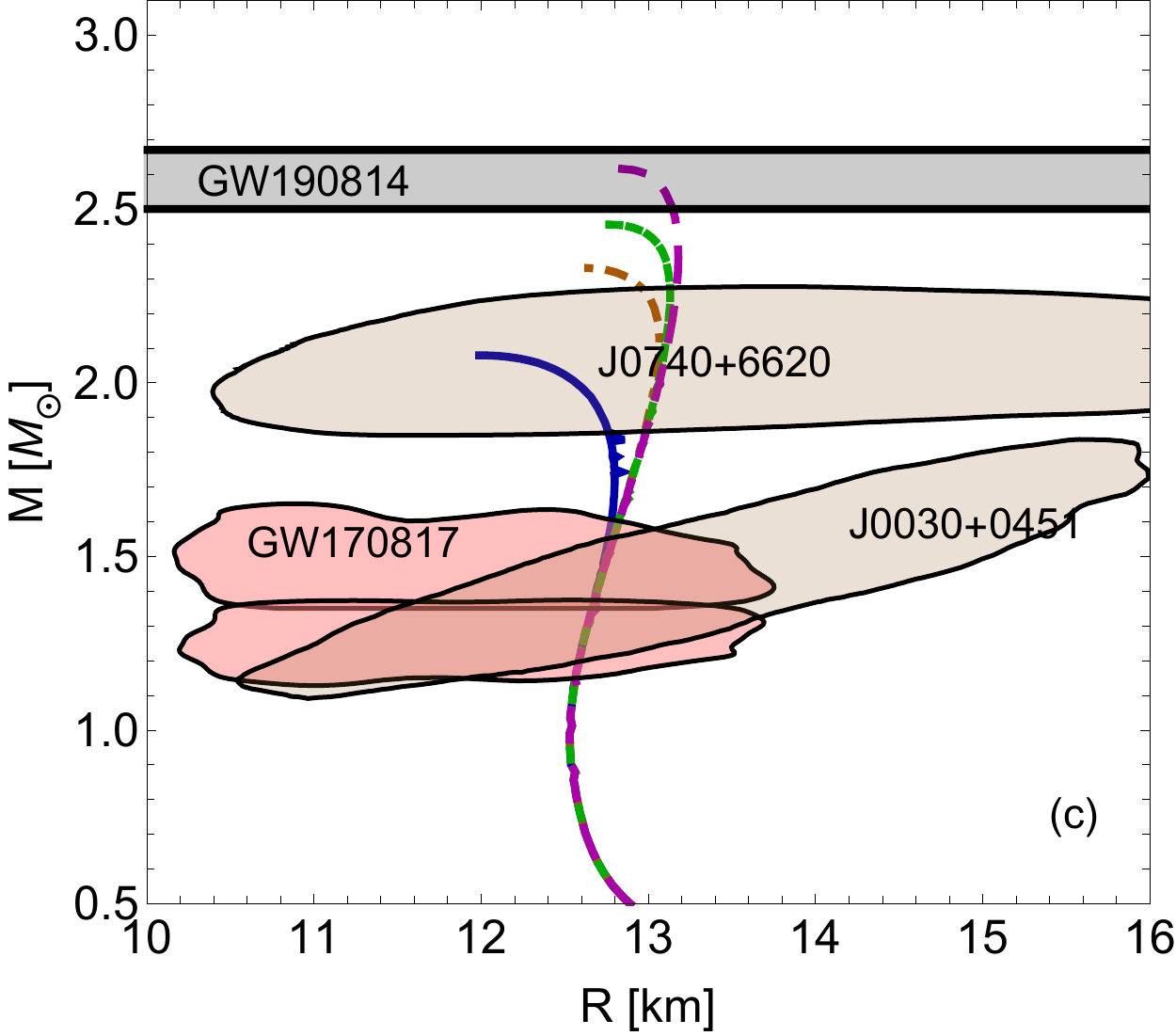} & 
\includegraphics[width=0.245\linewidth]{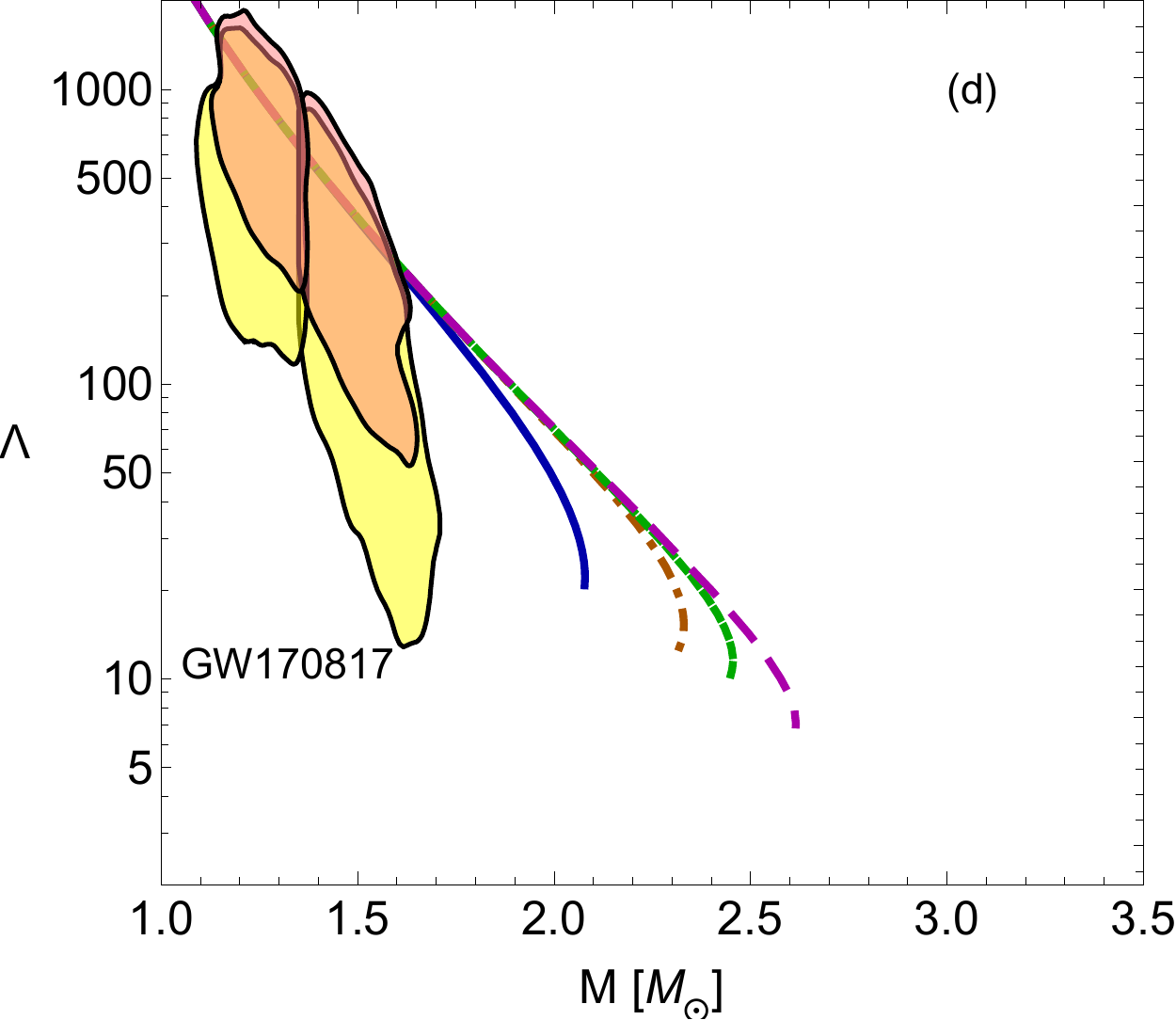}  \\
\includegraphics[width=0.245\linewidth]{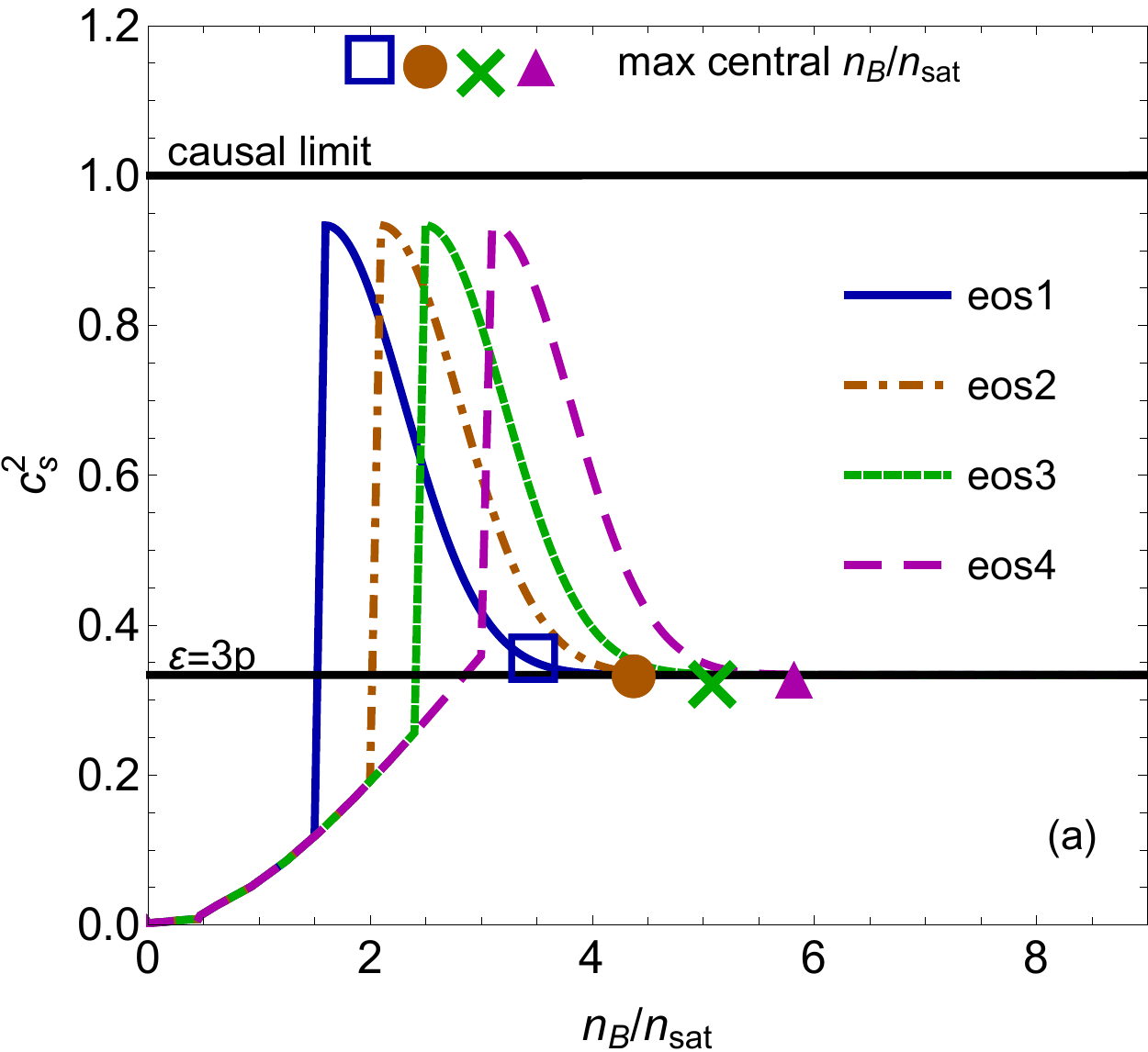} &
\includegraphics[width=0.245\linewidth]{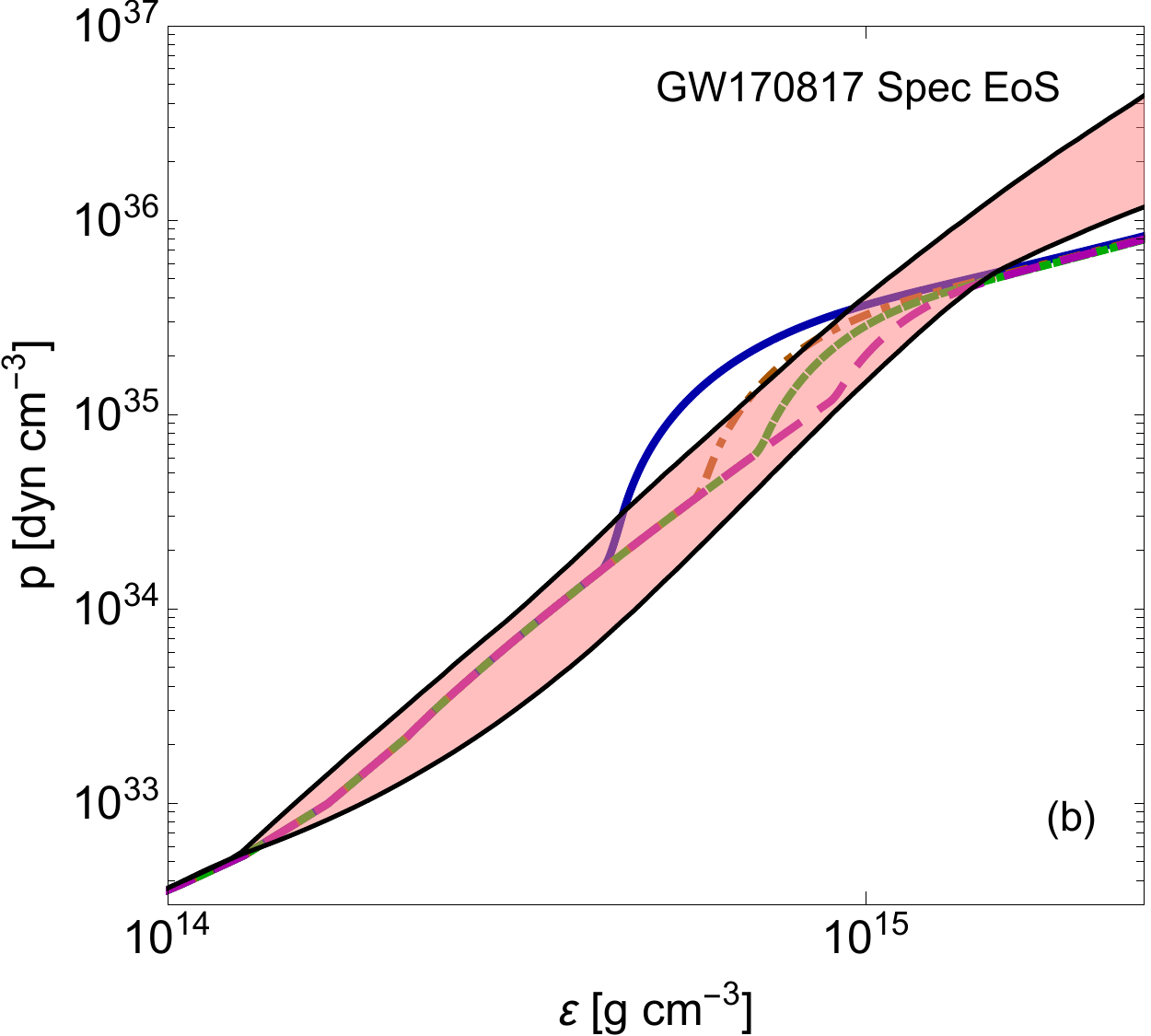} &
\includegraphics[width=0.245\linewidth]{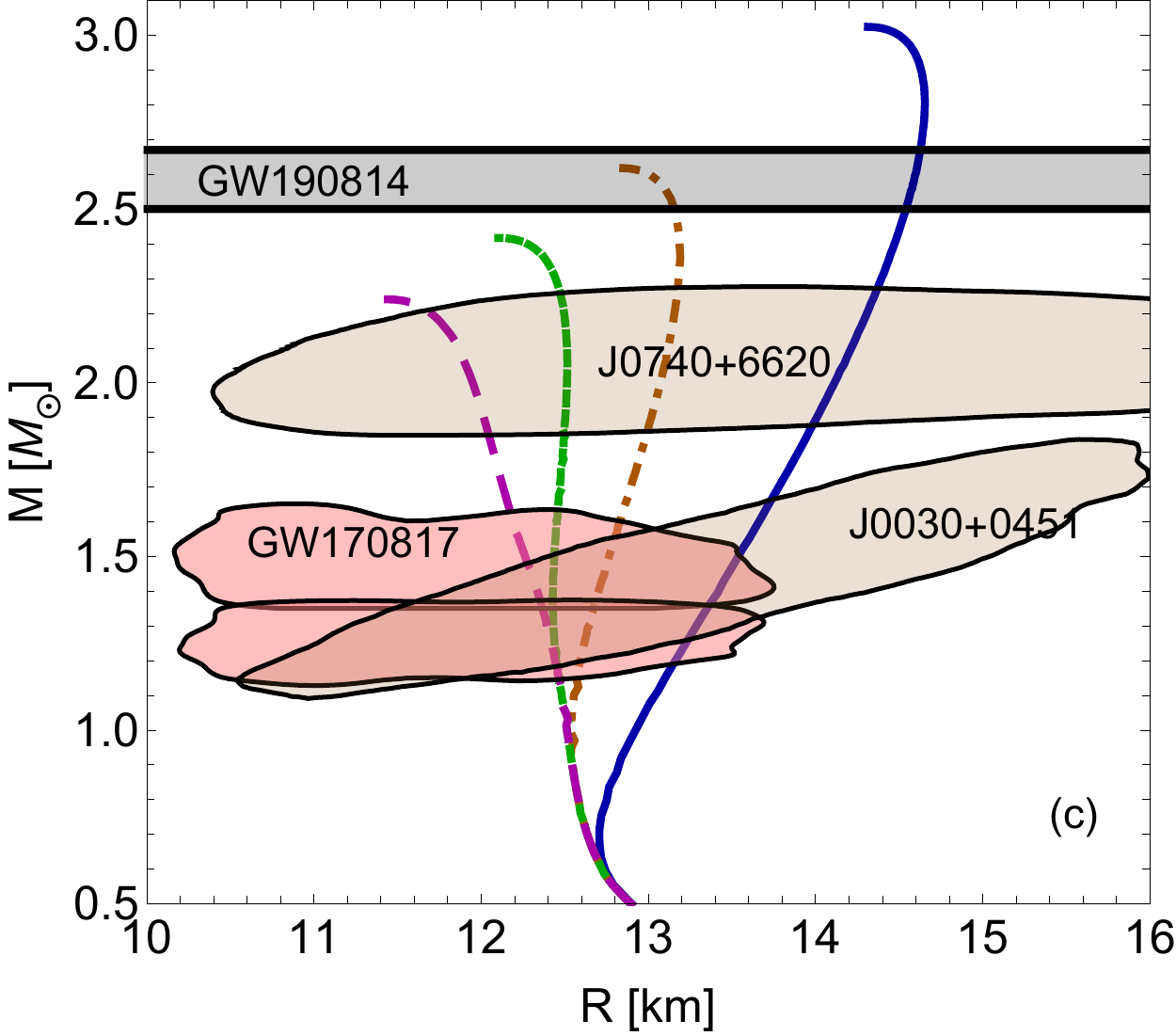} & 
\includegraphics[width=0.245\linewidth]{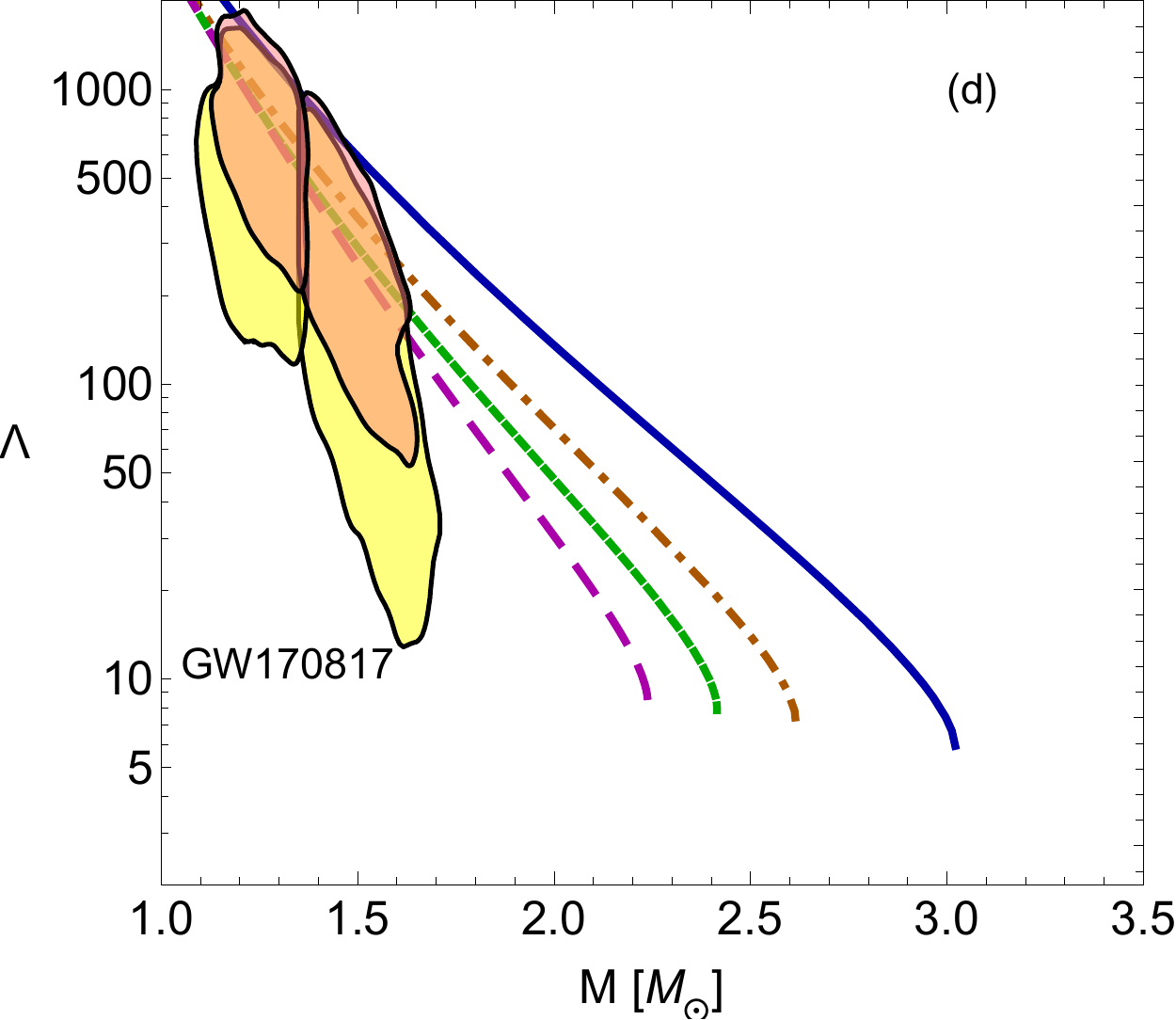} 
\end{tabular}
\caption{(Color online) Speed of sound (first or far left panel), EoS (second panel), mass-radius diagram (third panel), and tidal deformability (fourth or far right panel) for a sub-family of equations of state with peaks in the speed of sound of different widths at the same location (top) and peaks of the same width at different locations (bottom). Symbols show the central density for the most massive star of the sequence. The wider the bump or the lower the density at which it occurs, the larger the maximum mass of neutron stars, but only the low-density bump generates larger intermediate-mass stars with larger radius and tidal deformability.
}
\label{fig:width_peak}
\end{figure*}

\section{Neutron stars with crossover transitions}\label{sec:25msol}

In this section we build EoSs with crossover transitions, as described in the previous section, focusing on EoSs that have kinks, bumps, jumps, spikes and plateaus in the speed of sound. Of these structures, a rapid rise in the speed of sound at some baryon number density could be indicative of a rapid change in the degrees of freedom of the system (such as quark matter), and has been dubbed by some an ``un-phase transition''~\cite{McLerran:2018hbz}. We will investigate the impact that such un-phase transitions, as well as other crossover structures, have on observable quantities, such as the mass-radius and tidal deformability curves. In particular, we will see how these structures can lead to extremely heavy neutron stars with small deformabilities.  

\subsection{How to construct ultra-heavy neutron stars}

One of the main consequences of crossover structures in the speed of sound is the possibility of creating extremely heavy neutron stars. In this paper, we refer to non-spinning stars with a maximum TOV mass larger than $2.5$~M$_\odot$ as ``ultra-heavy.'' As mentioned in the introduction, one can obtain even heavier systems by allowing for rigid or differential rotation, which increases the maximum TOV mass by about ${\cal{O}}(10\%)$ for millisecond periods rotation rates~\cite{Yagi:2014bxa}, but we do not discuss such systems here. As we shall see, the universal feature in the speed of sound that allows for the construction of ultra-heavy neutron stars is a steep rise in $c_s^2$ at a relatively low transition density.  

\subsubsection{Bump in $c_s^2$: Width and location}

Let us begin by studying the consequences of a single bump in the speed of sound of a fixed height but various widths and transition densities in the speed of sound. The motivation for this study is the fact that a steep rise in the speed of sound at intermediate densities has been associated with higher-order repulsive terms in the description of the strong force among nucleons and hyperons~\cite{Dexheimer:2020rlp,Pisarski:2021aoz}. Moreover, such a bump in the speed of sound has also been shown to improve the agreement with the observation of neutron stars above $M\geq 2$~M$_\odot$ \cite{Bedaque:2014sqa,Tews:2018kmu,Tan:2020ics,Jimenez:2021wil}. The structure functions we use to model these cases are either $f_{bump}$ or $f_{osc}$ with a variety of transition functions. In all cases considered below, we ensure that the resulting mass-radius and mass-tidal deformability curves satisfy the constraints discussed in Sec.~\ref{sec:constraints}.

The results of this analysis are presented in Fig.\ \ref{fig:width_peak}, which shows the effect that a single bump in $c_s^2$ has on the EoS  and the resulting mass-radius and mass-tidal deformability relations. The top panels of this figure show the effect of a bump of varying width (but fixed transition baryon number density $n_1$), so let us discuss these first. The wider the bump in the $c_s^2$, the taller the raise in p vs $\varepsilon$ in the EoS and the larger the maximum stellar mass (while maintaining a nearly constant low and intermediate mass-radius relation), because a larger portion of the star is described by a more stiff EoS. Furthermore, the wider the speed of sound peak, the less compressed the maximum mass stars are, and the lower the central densities they present (symbols in far left panel). Finally, the tidal deformability of low and intermediate mass stars are not affected by differences in the peak, but for a given massive star, the wider peak EoSs produce a larger tidal deformation.

We now focus on the influence of the location of the peak in the speed of sound bump, while keeping the width and height fixed, which is shown in the bottom panels of Fig. \ref{fig:width_peak}. The lower the baryon number density $n_1$ of the transition, the larger the corresponding stellar masses and radii, as intermediate mass stars are strongly affected by this shift. Additionally, as $n_1$ shifts to larger baryon densities, the central stellar density also shifts to larger central densities. More compressed matter translates into a much lower tidal deformability and Love number for a given mass, as shown also in the figure. Therefore, our results indicate that one can easily create a family of EoSs that reach $M\geq 2.5$~M$_\odot$, either by implementing a narrow peak at low transition baryon densities $n_1$ or a wide peak at higher $n_1$.   

\begin{figure*}
\centering
\begin{tabular}{c c c c}
\includegraphics[width=0.245\linewidth]{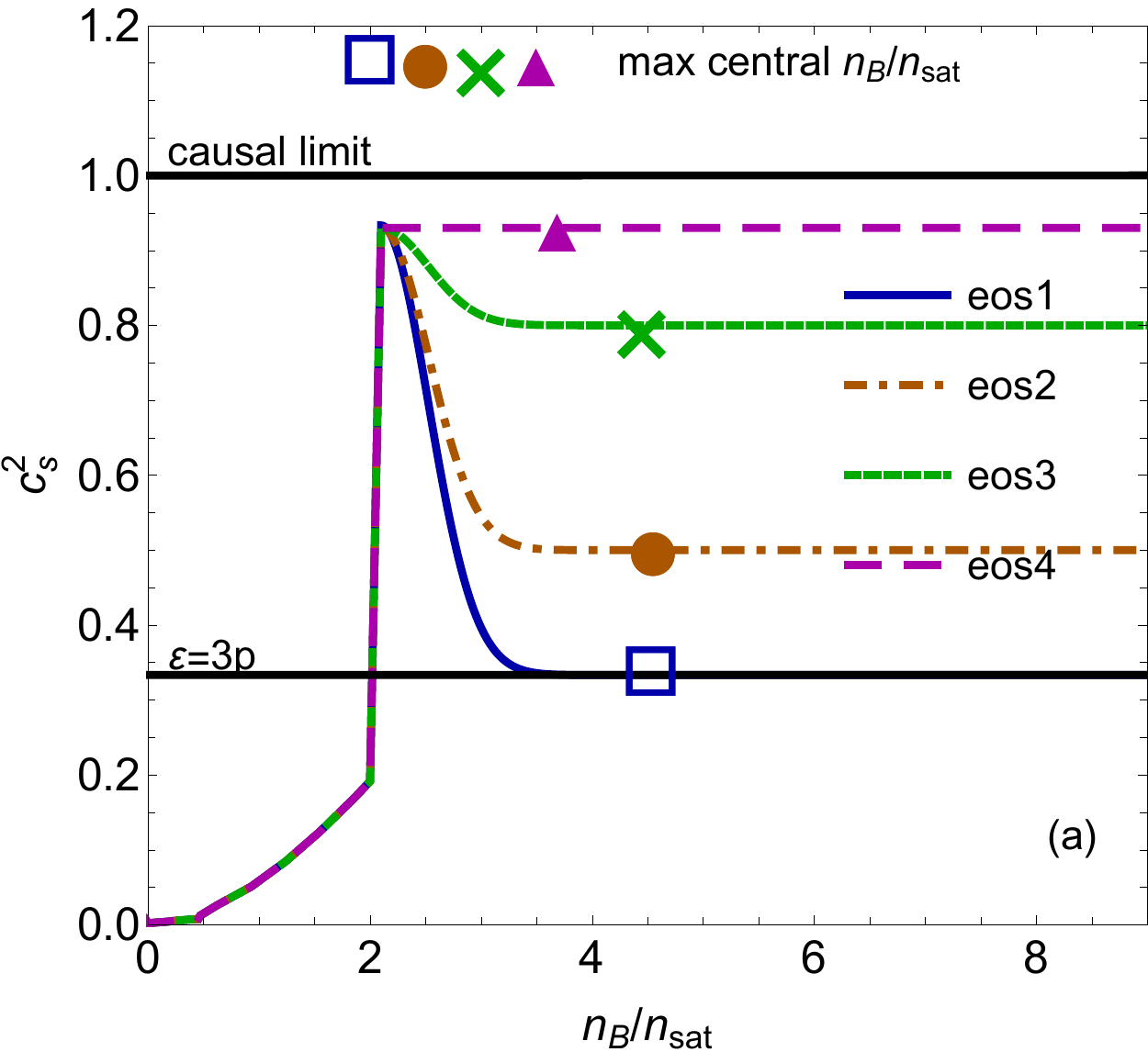} &
\includegraphics[width=0.245\linewidth]{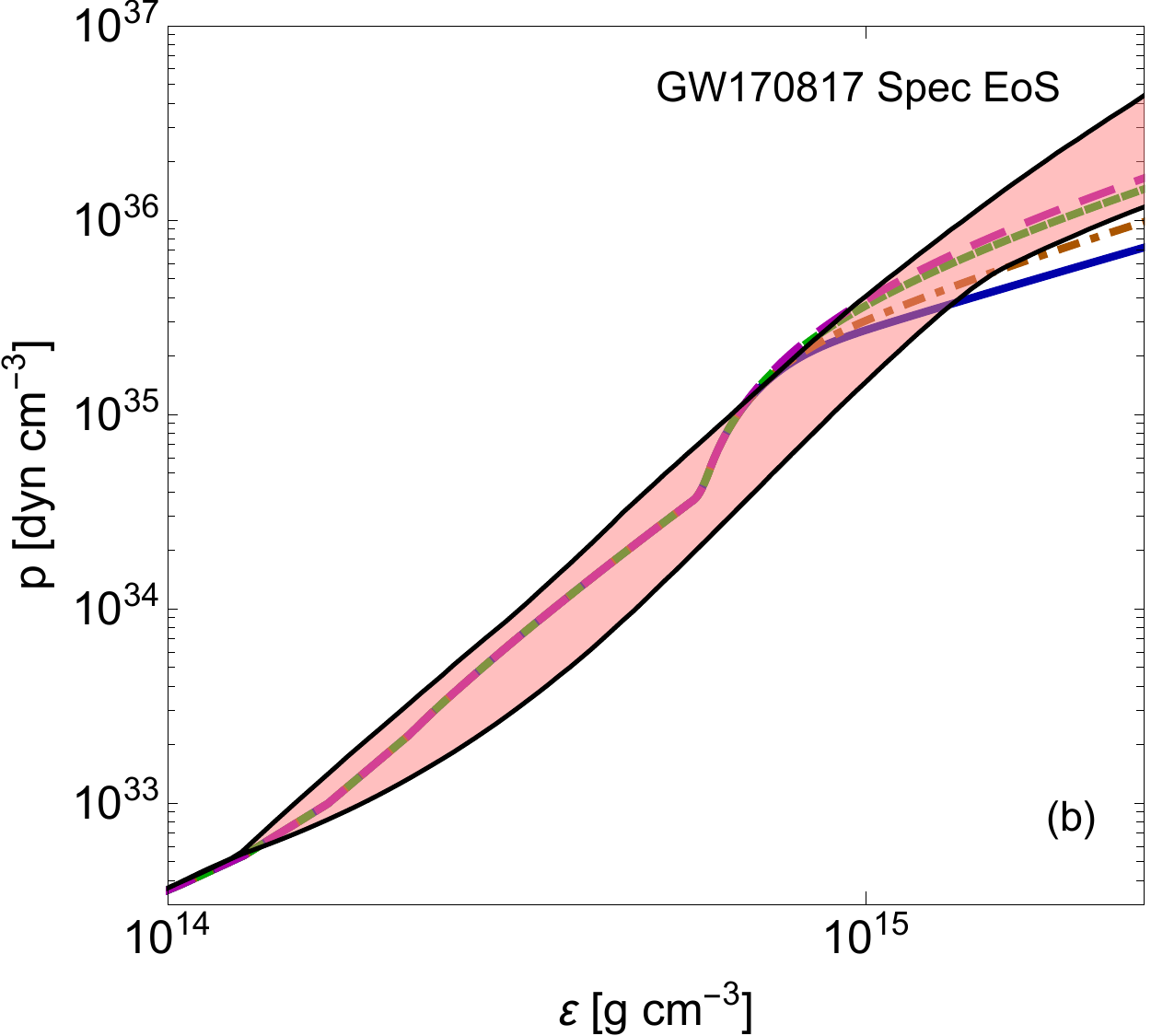} &\includegraphics[width=0.245\linewidth]{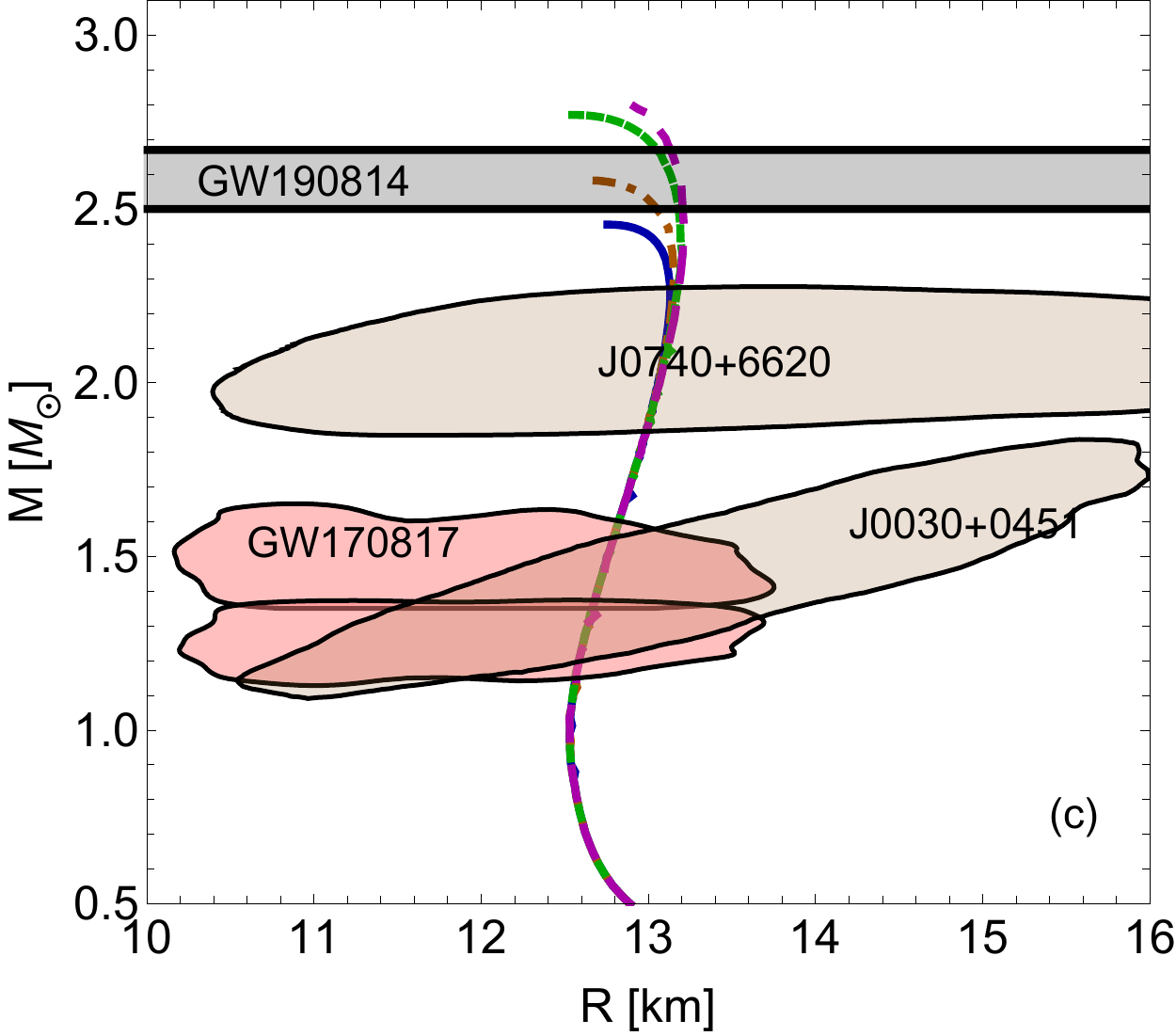} & \includegraphics[width=0.245\linewidth]{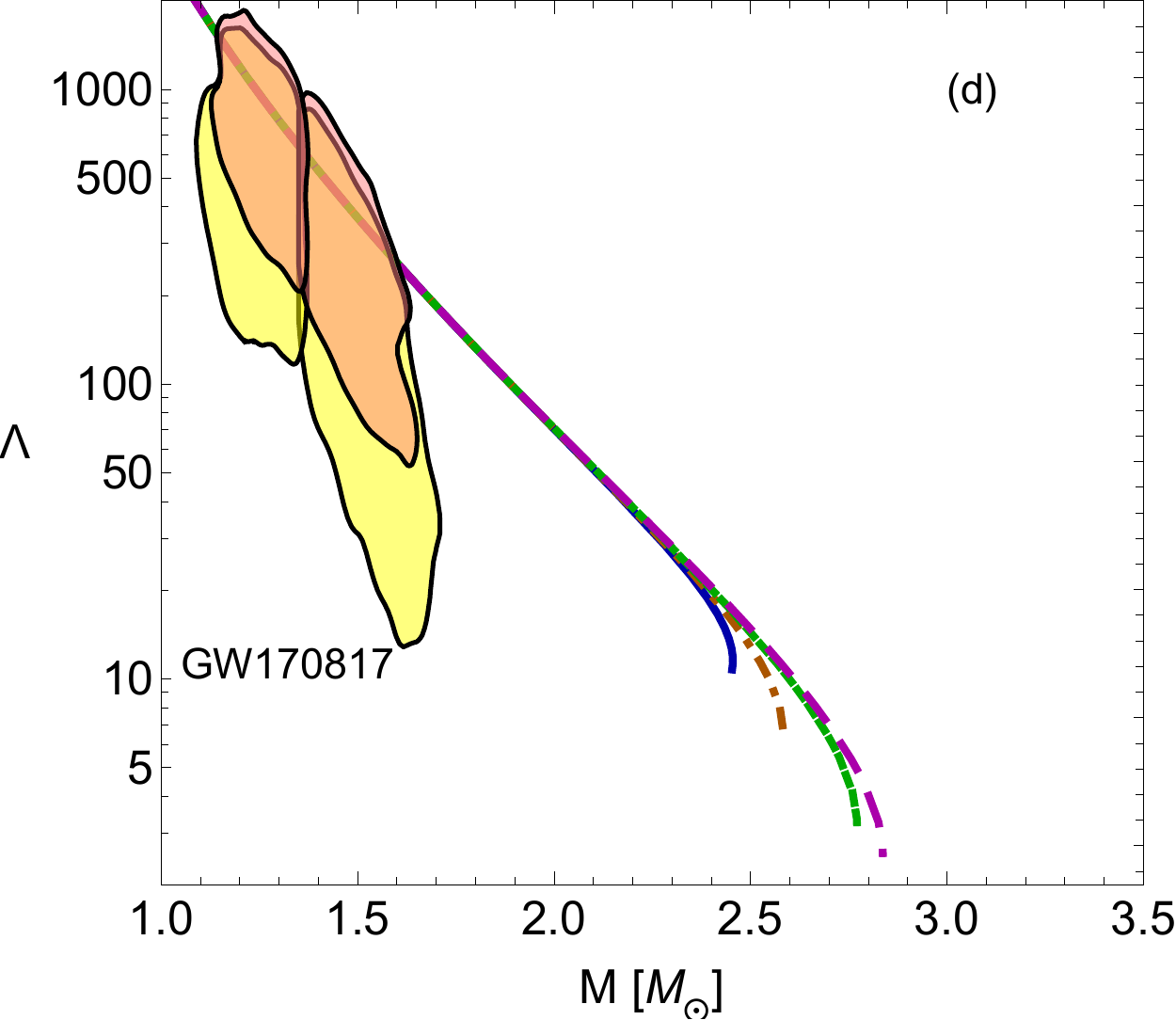}
\end{tabular}
\caption{(Color online) Same as Fig.~\ref{fig:width_peak}, but for a sub-family of EoSs with different asymptotic values for the speed of sound at large densities (i.e.~different plateau heights). The plateau height has a moderate effect in the maximum mass and the large-mass portion of the mass-radius diagram and mass-tidal deformability curves.} 
\label{fig:slant_end}
\end{figure*}

Interestingly, the effects of increasing the width of the bump is opposite to the effect of increasing the bump location, i.e.~widening the bump increases the maximum mass, while increasing the bump location decreases it. Could one then tune the widening of the mass and the increase of the bump location so that these two effects approximately cancel each other, leading to a roughly constant maximum mass? The answer to this question is yes. Indeed, one can widen the bump so much so that it becomes a plateau, and as long as this occurs at the right transition density $n_1$ (below $n_1\sim 3~n_{sat}$), the resulting mass-radius curve will have the same maximum mass as that resulting from an EoS with a single narrow bump at low $n_1$. While the maximum mass will be roughly the same, however, the radii will be different, with larger $n_1$ leading to stars with smaller radii. We therefore conclude that a star with a large radius at $M\sim 2.5 $~M$_\odot$ implies a larger raise in the $c_s^2$ at low transition densities. Similarly, a star with a smaller radius at $M\sim 2.5 $~M$_\odot$ is much more likely arising from a transition that occurs at $n_1\sim 3~n_{sat}$ or below, and that has a wide peak or plateau. 

\subsubsection{Secondary structure in $c_s^2$ at high densities}

\begin{figure*}
\centering
\begin{tabular}{c c c  c}
\includegraphics[width=0.245\linewidth]{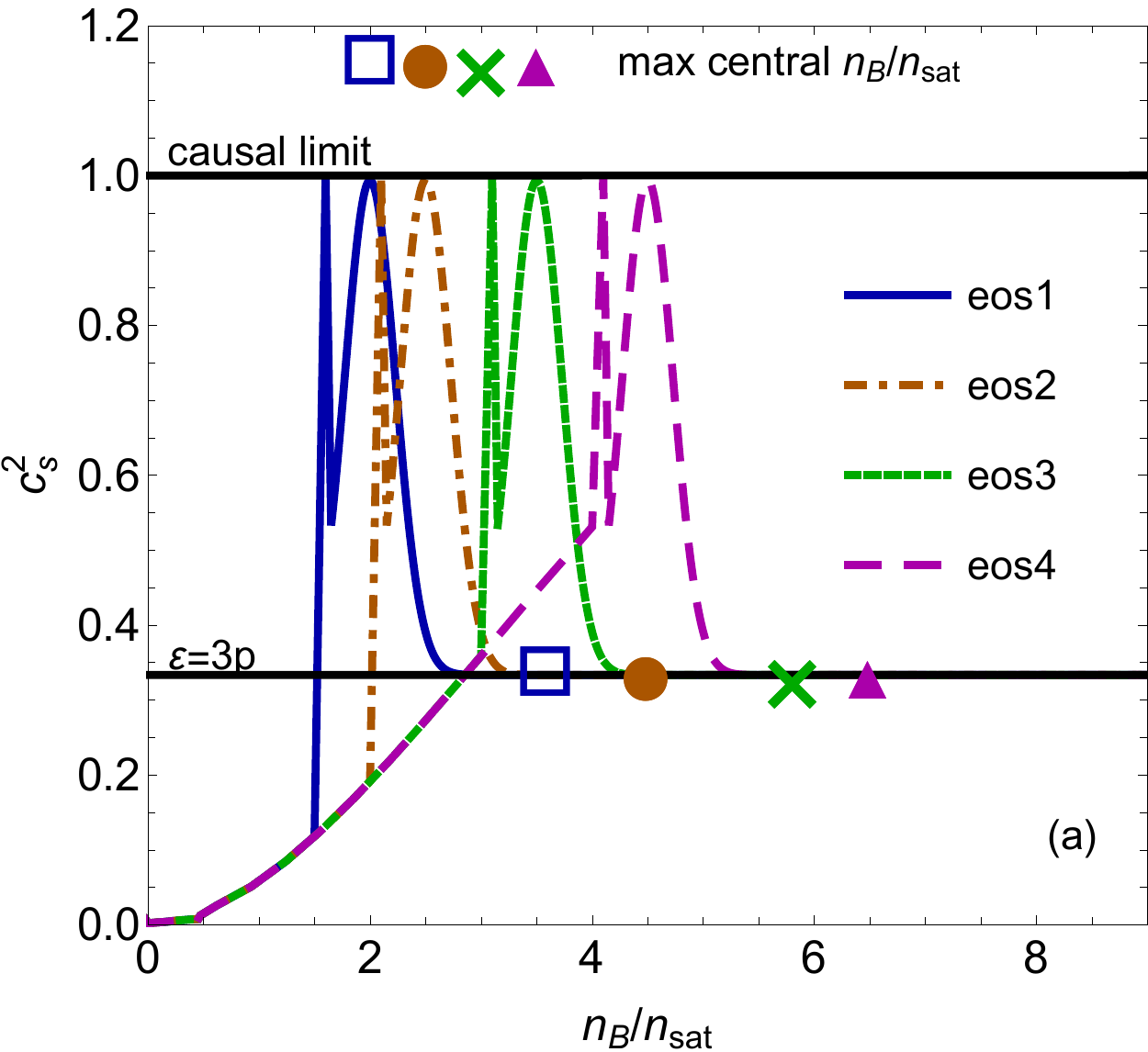} &
\includegraphics[width=0.245\linewidth]{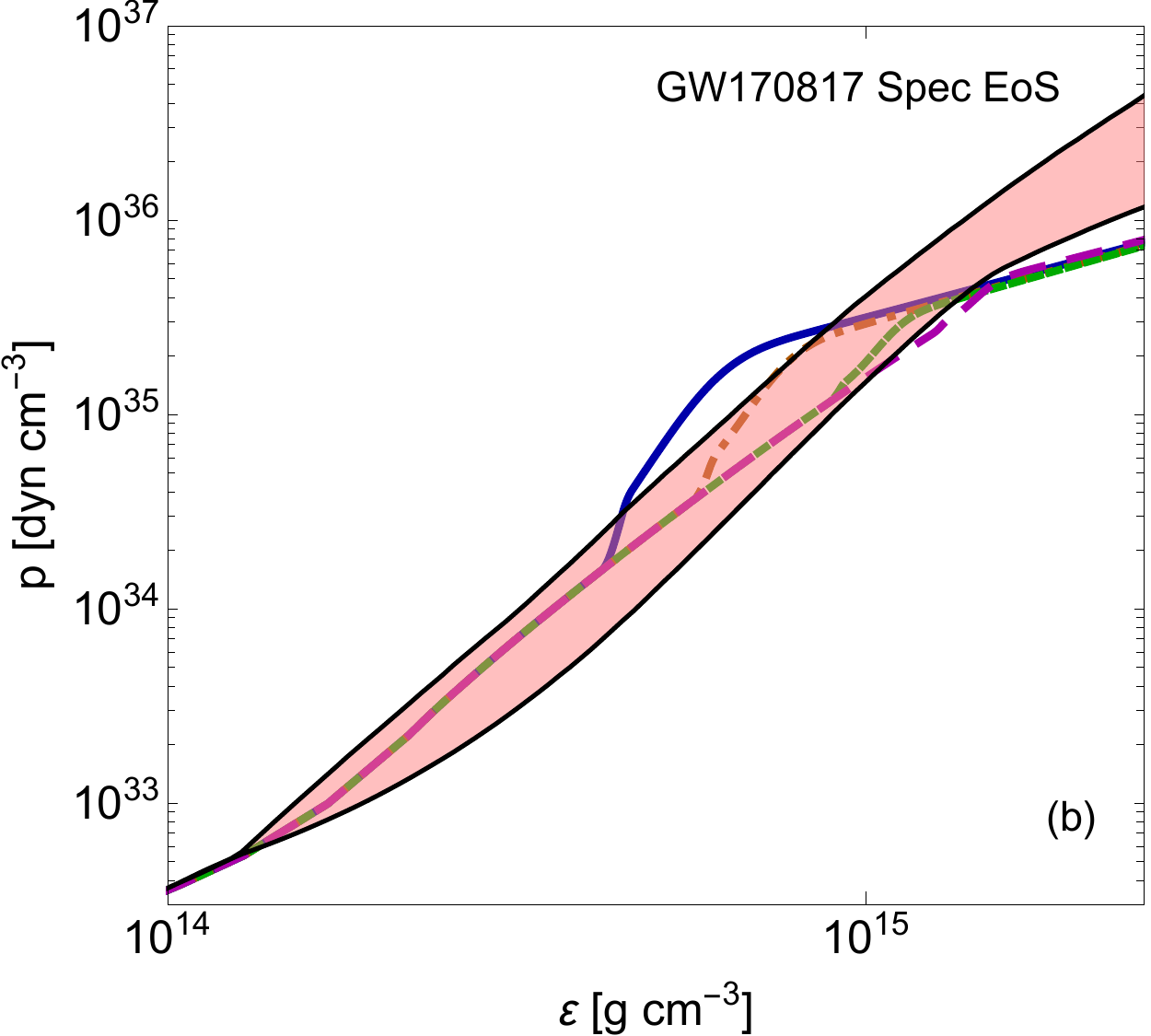} &\includegraphics[width=0.245\linewidth]{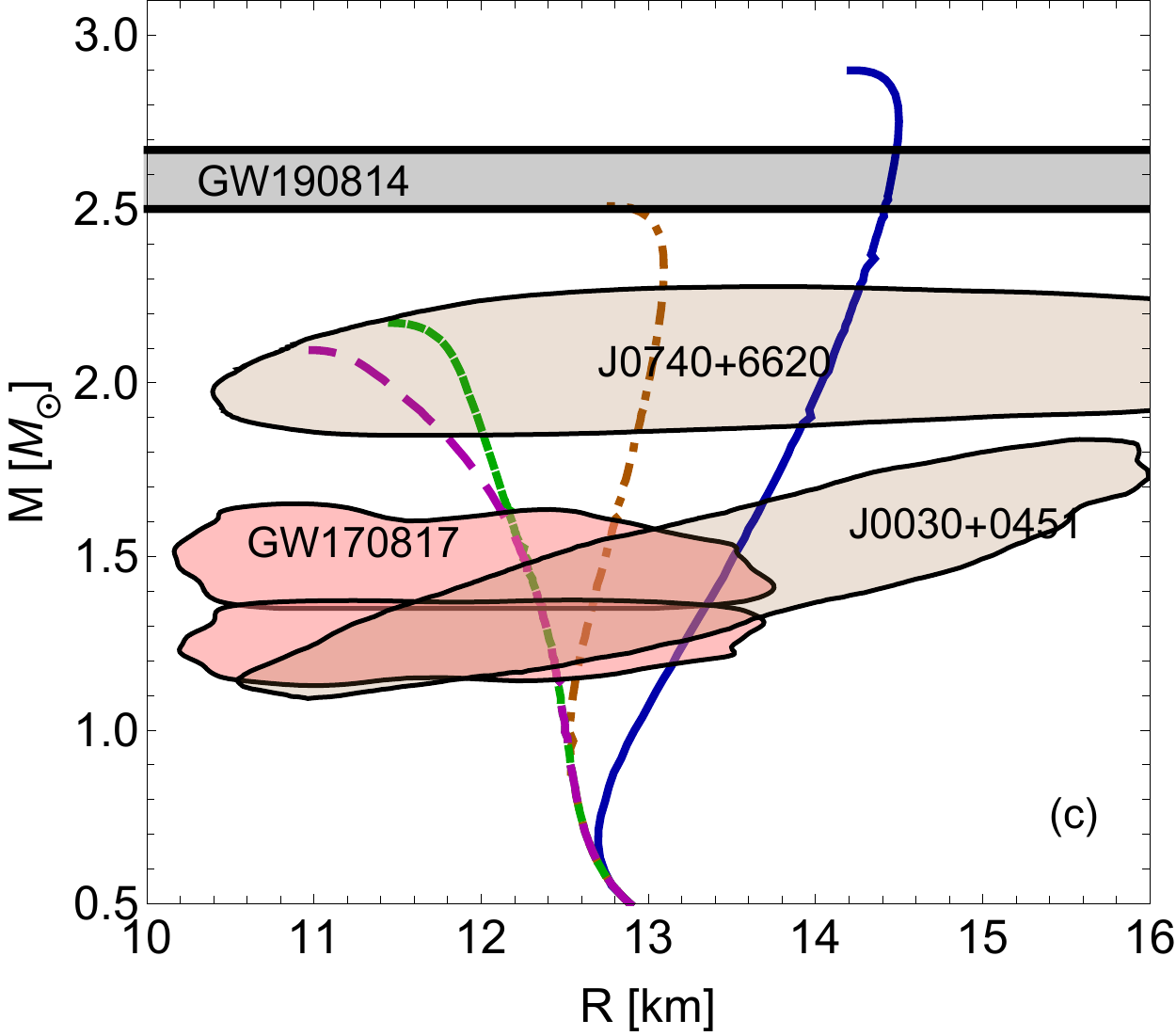} & \includegraphics[width=0.245\linewidth]{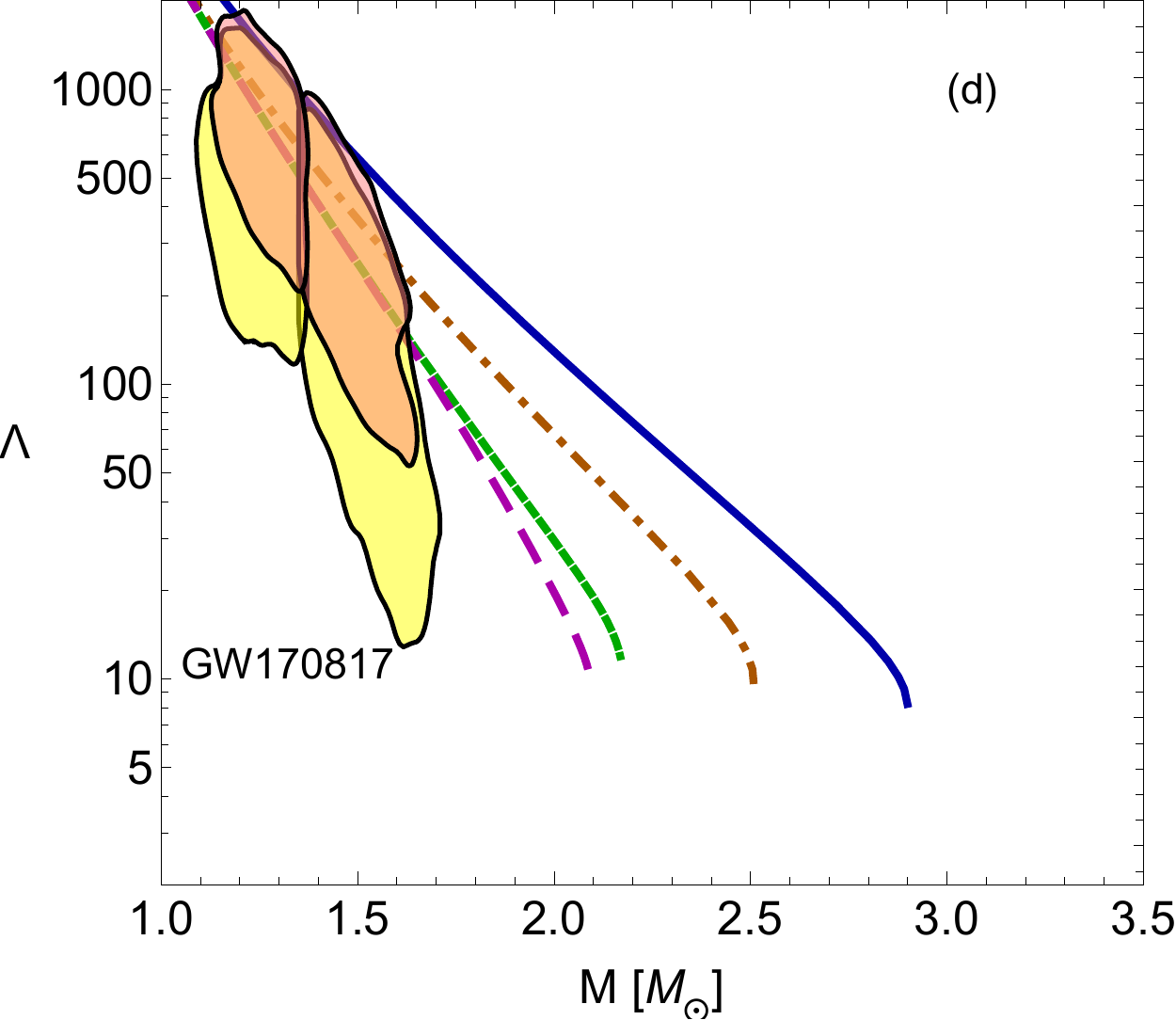}
\end{tabular}
\caption{(Color online) Same as Fig.~\ref{fig:width_peak}, but for a sub-family of EoSs with double peaks in the speed of sound at different locations. The double peak structure allows for ultra-heavy neutron stars, with larger radii and large $\Lambda$ (for the same mass) for bumps at larger densities.  }
\label{fig:2bumps}
\end{figure*}

Let us now study the consequences of the behavior of $c_s^2$ at large baryon densities. We first consider a scenario in which $c_s^2$ increases and then plateaus at various $c_{end}$, so we can additionally test the consequences of  the pQCD limit $c_s^2\rightarrow 1/3$ at very large densities.   The structure function we use in this case is just a constant controlled by $c_{end}$, and thus, it is this parameter that determines the asymptotic value of the speed of sound at large densities.  The effect of changing $c_{end}$ then leads to either a stiffer high density behavior (for a large value of $c_{end}$) or a softer high density behavior (for low values of $c_{end}$).

Figure\ \ref{fig:slant_end} shows the impact of this crossover structure in the EoS, as well as in astrophysical observables. The variation of $c_{end}$ between $1/3$ and $9/10$ has an effect only on massive stars, which are the ones able to have a sizable dense matter core, both in the mass-radius diagram and the $\Lambda$--$M$ relation. The central stellar density is barely affected. Therefore, given current uncertainties, we are not strongly sensitive to a return of $c_s^2\rightarrow 1/3$, even if it were to occur at densities reached by massive stars (especially when one considers other uncertainties in the functional form of $c_s^2$ as detailed earlier). While not shown here, we have also studied the inclusion of oscillations (using multiple trigonometric functions) and a number of other bumps and wiggles occurring \textit{after} the large initial peak.  We find no significant effect on the maximum mass, because the large initial peak appears to soak up any detailed structure in the $c_s^2$ at larger baryon densities.

\subsubsection{Primary bump structure in $c_s^2$}

Let us now consider the possibility that the first initial bump has additional substructure in it, such as a double peak. This is different from the previous case studied, because in those cases the speed of sound had a single and simple bump, which was then followed at larger densities by wiggles or additional bumps. Here, we are concerned with the structure of the first dominant bump itself. A physical motivation for this is that  expected new degrees of freedom could open up at intermediate baryon densities (e.g.~hyperons) and this could produce multiple bumps or kinks in the speed of sound at the dominant bump (see Fig.~\ref{fig:known}). We test this possibility through an initial spike followed by a large bump, which we model with two structure functions of bumps followed by a plateau.

Figure~\ref{fig:2bumps} shows the resulting speeds of sound, their impact on the EoS, and their corresponding mass-radius and $\Lambda$--$M$ curves. A spike followed by a bump can still sustain ultra-heavy neutron stars, provided the spikes occur at $n_B=1.5~ n_{sat}$ and $n_B=2~n_{sat}$.  One can, in principle, obtain a similar maximum mass for the other curves if one were to further adjust the width of their second bump.  In fact, results from the double bump peaks look fairly similar to the ones from single bump peaks shown in Fig.\ \ref{fig:width_peak}. For the heavier neutron stars, the double bumps decrease the maximum mass only about a few percent, while holding the radius nearly fixed, whereas for the lighter neutron stars a double bump could decrease the maximum mass and the radius at the maximum mass slightly more.

\subsection{How to construct the heaviest neutron stars}

\begin{figure*}[htb!]
\centering
\begin{tabular}{c c c c}
\includegraphics[width=0.24\linewidth]{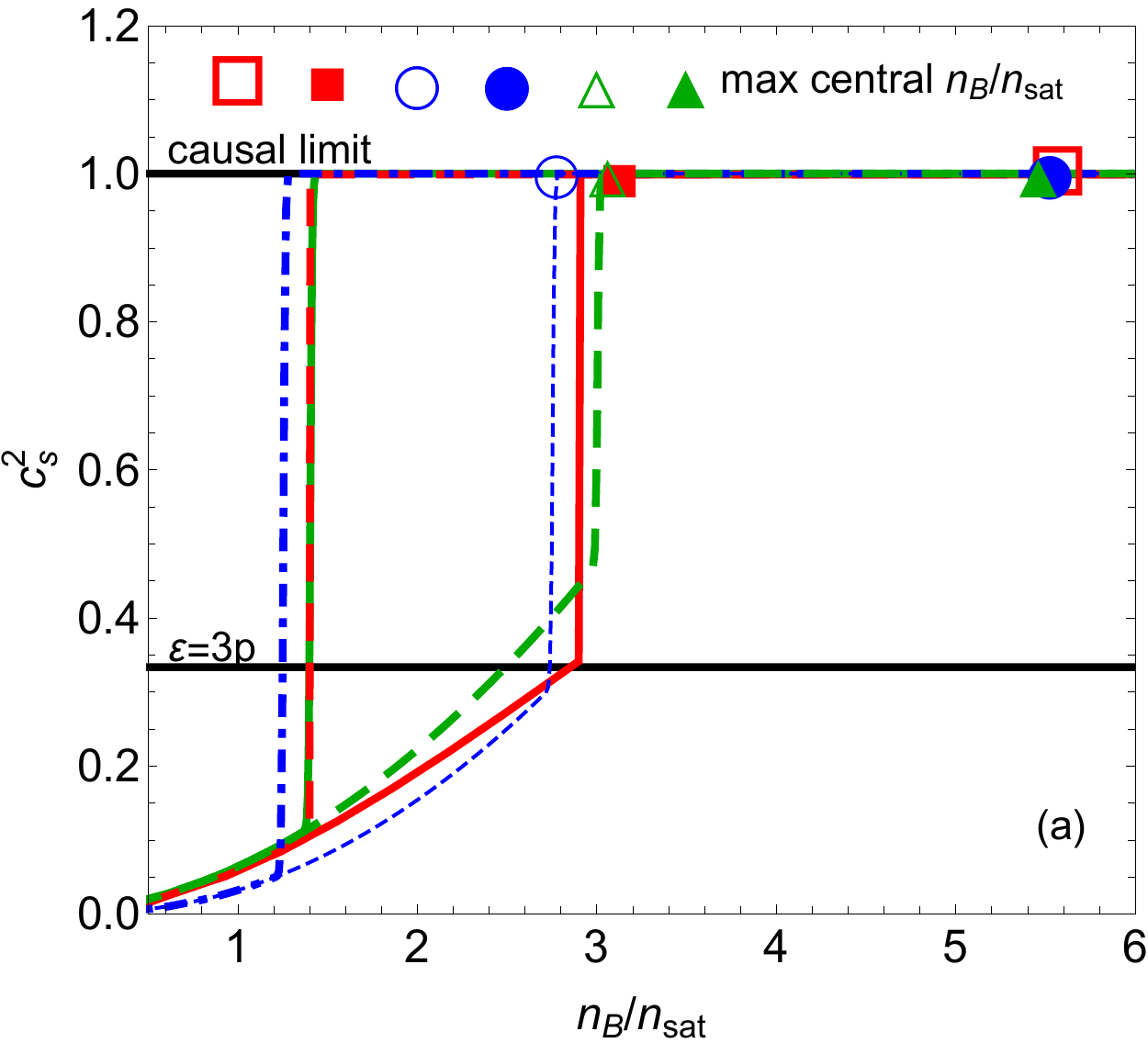} & \includegraphics[width=0.24\linewidth]{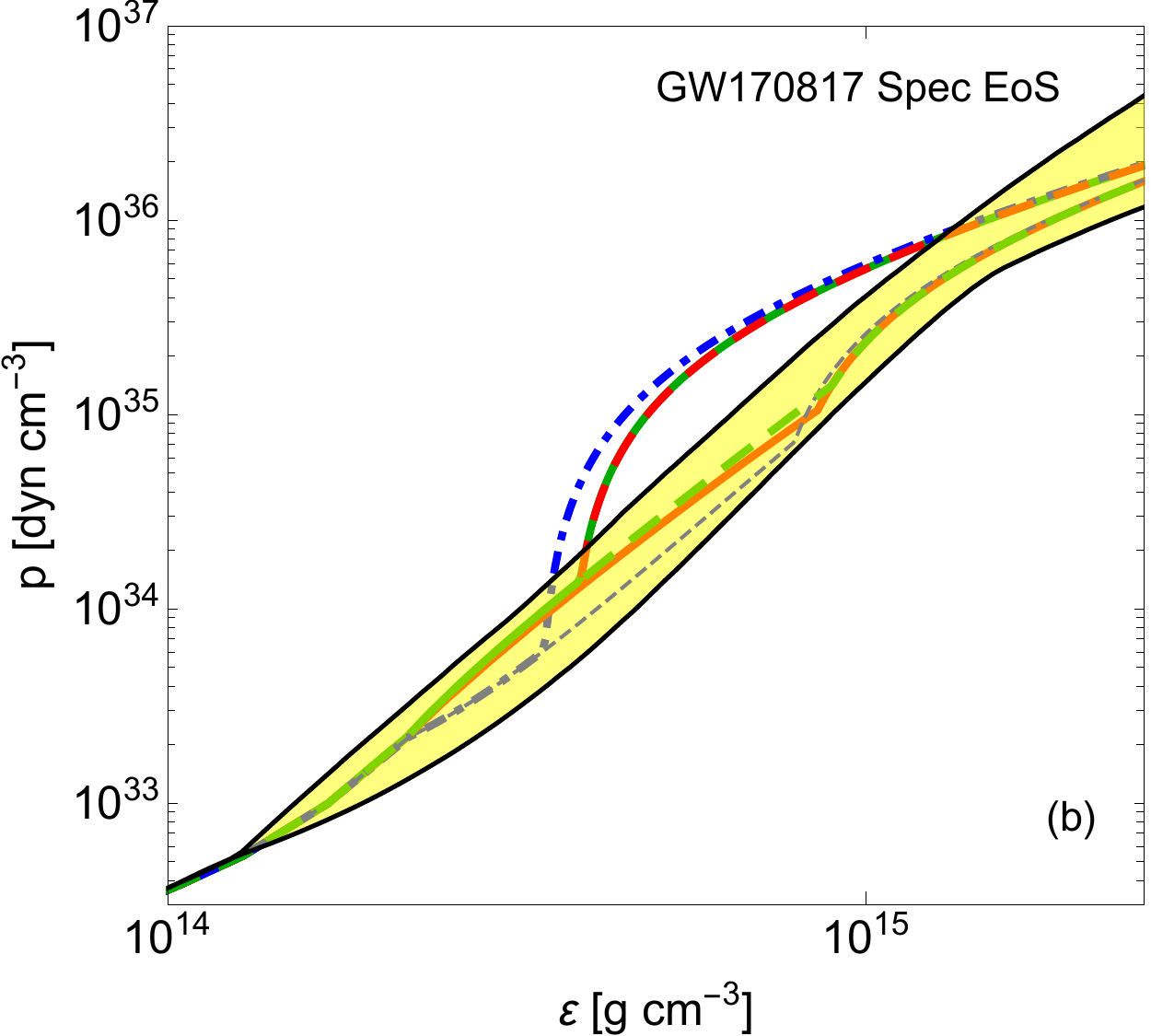} & \includegraphics[width=0.24\linewidth]{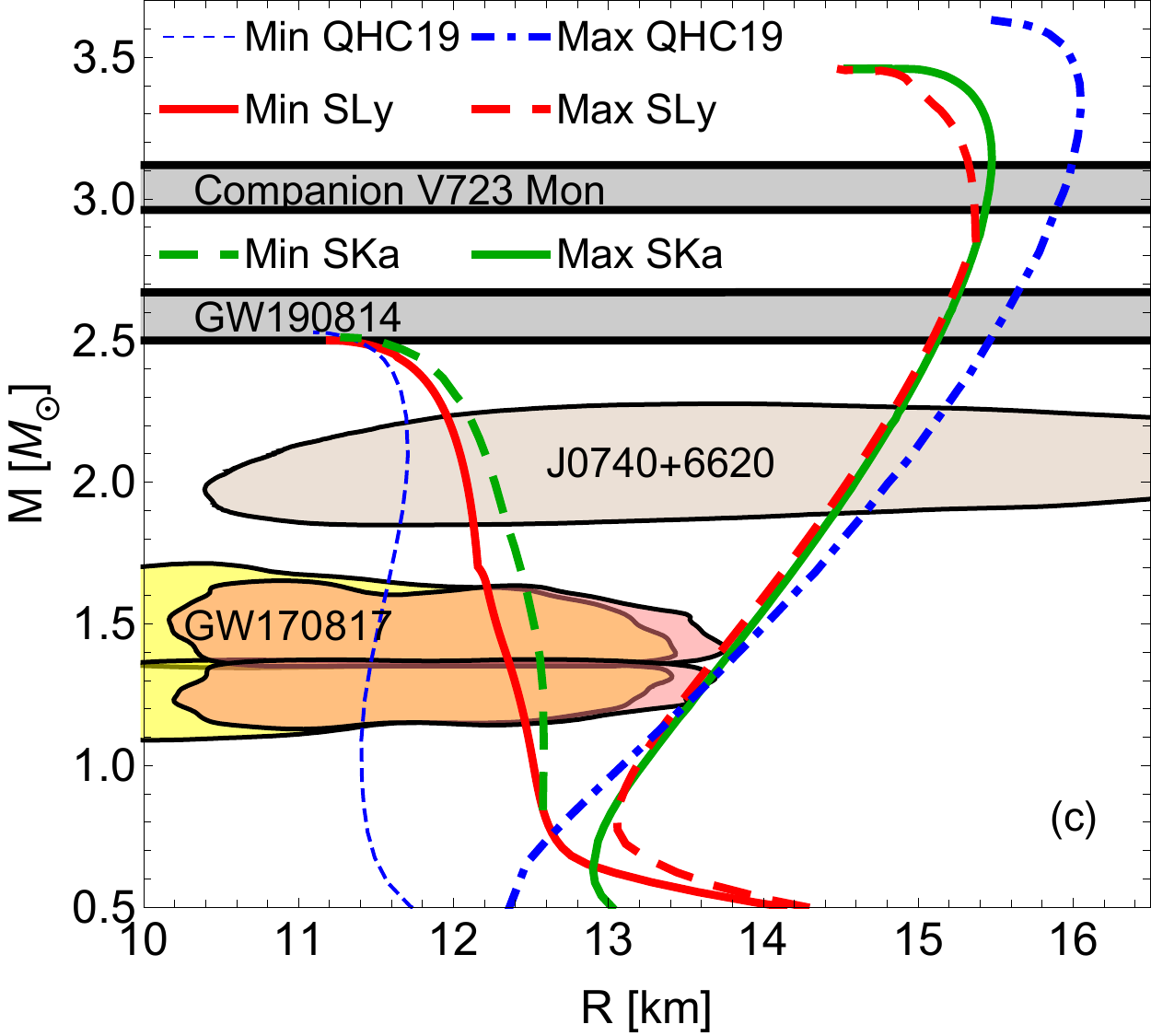} & \includegraphics[width=0.24\linewidth]{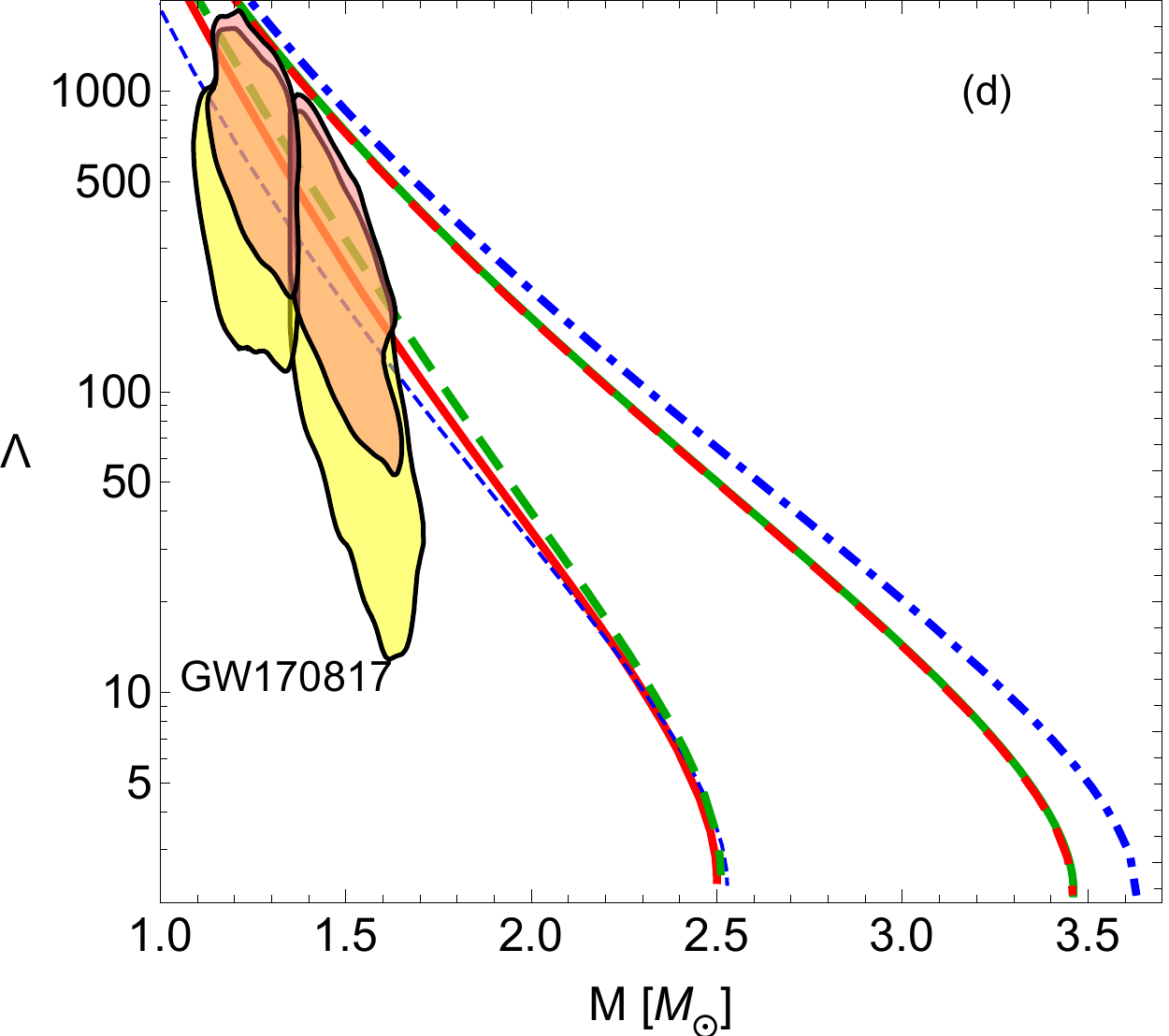}
\end{tabular}
\caption{(Color online) Same as Fig.~\ref{fig:width_peak}, but for a sub-family of EoSs that incorporates three different crusts: QHC19, SLy, SKa coupled to a core that transitions to the causal limit at a transition density $n_1$. We vary the latter in order to find either the minimum possible maximum mass that fits within the GW190814 error bands or the maximum possible mass that fits within all radii constraints. Crust model impacts significantly the mass-radius relation, and also impacts the value of $n_1$ needed to achieve an extremely heavy neutron star and agree with observational constraints. 
}
\label{fig:crusts}
\end{figure*}

A major challenge for realistic but still standard neutron star EoS models is producing a large maximum mass. Reaching masses as high as that of the light component of the GW190814 event is very difficult, specially when including hyperonic degrees of freedom (without resorting to fast rotation \cite{Nathanail:2021tay,Dexheimer:2020rlp,Zhang:2020zsc,Li:2020dst,Rather:2021yxo,Hernandez-Vivanco:2021urj,Sen:2021bms,Demircik:2020jkc}, or strong magnetic fields \cite{Rather:2021azv}). Reaching masses as high as the companion of V723 Mon would not be possible at all with most standard EoS models.  However, it is also entirely possible that new and yet unexplored states of matter exist within the core of a neutron stars that lead to structures in the speed of sound, which can allow for such large masses. Indeed, the preliminary study of~\cite{Tan:2020ics} showed that bumps in the EoS can lead to neutron stars as massive as that of the light component of the GW190814 event. In this section, we will investigate this structure systematically to find the heaviest neutron stars allowed by crossover structure in the speed of sound. We will focus first on the impact of the crust and the causal limit on the maximum mass, and then discuss the impact of particular crossover structures.

\subsubsection{Influence of crust on maximum mass}

\begin{figure*}[htb]
\centering
\begin{tabular}{c c c c}
\includegraphics[width=0.24\linewidth]{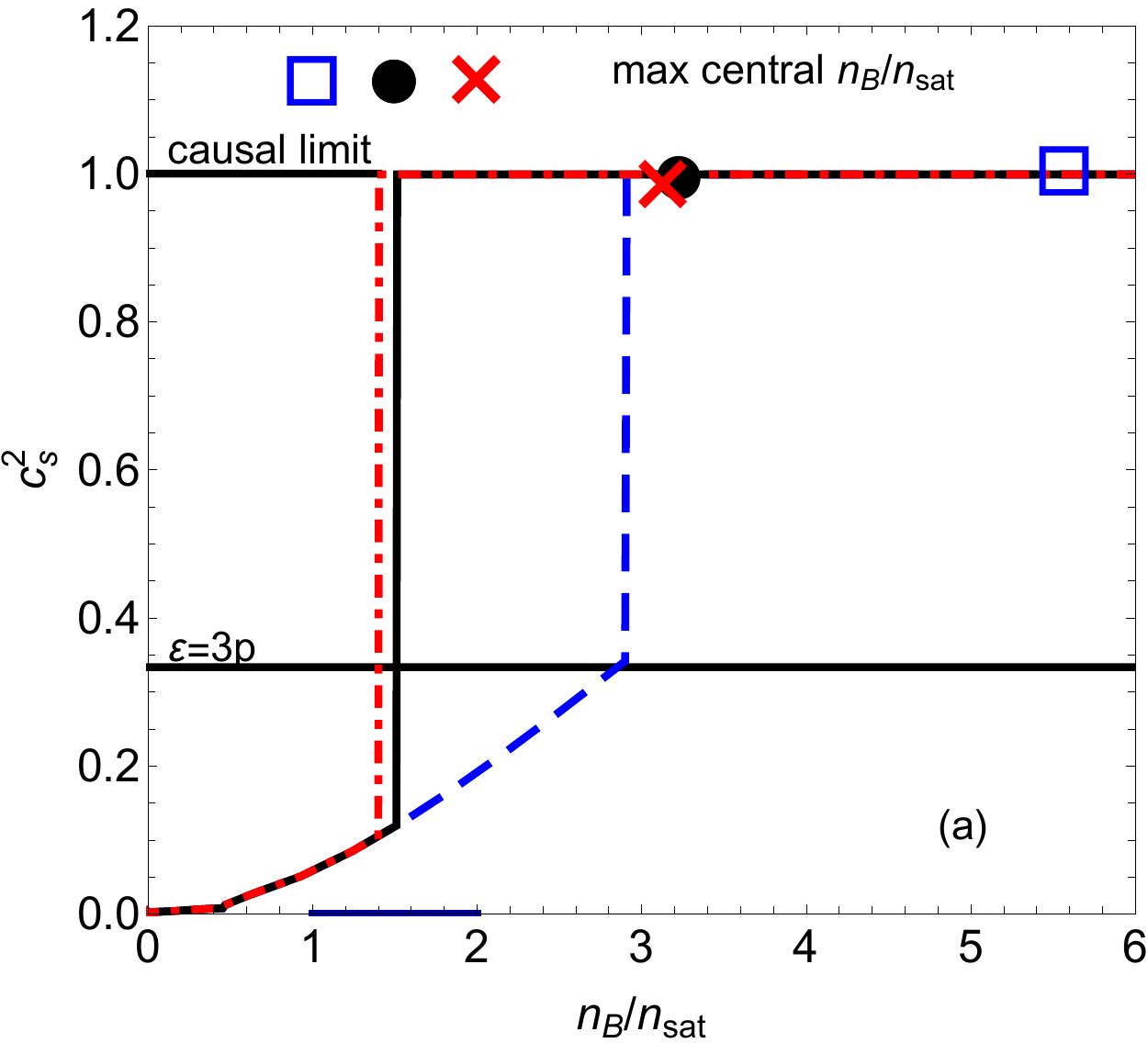} & \includegraphics[width=0.24\linewidth]{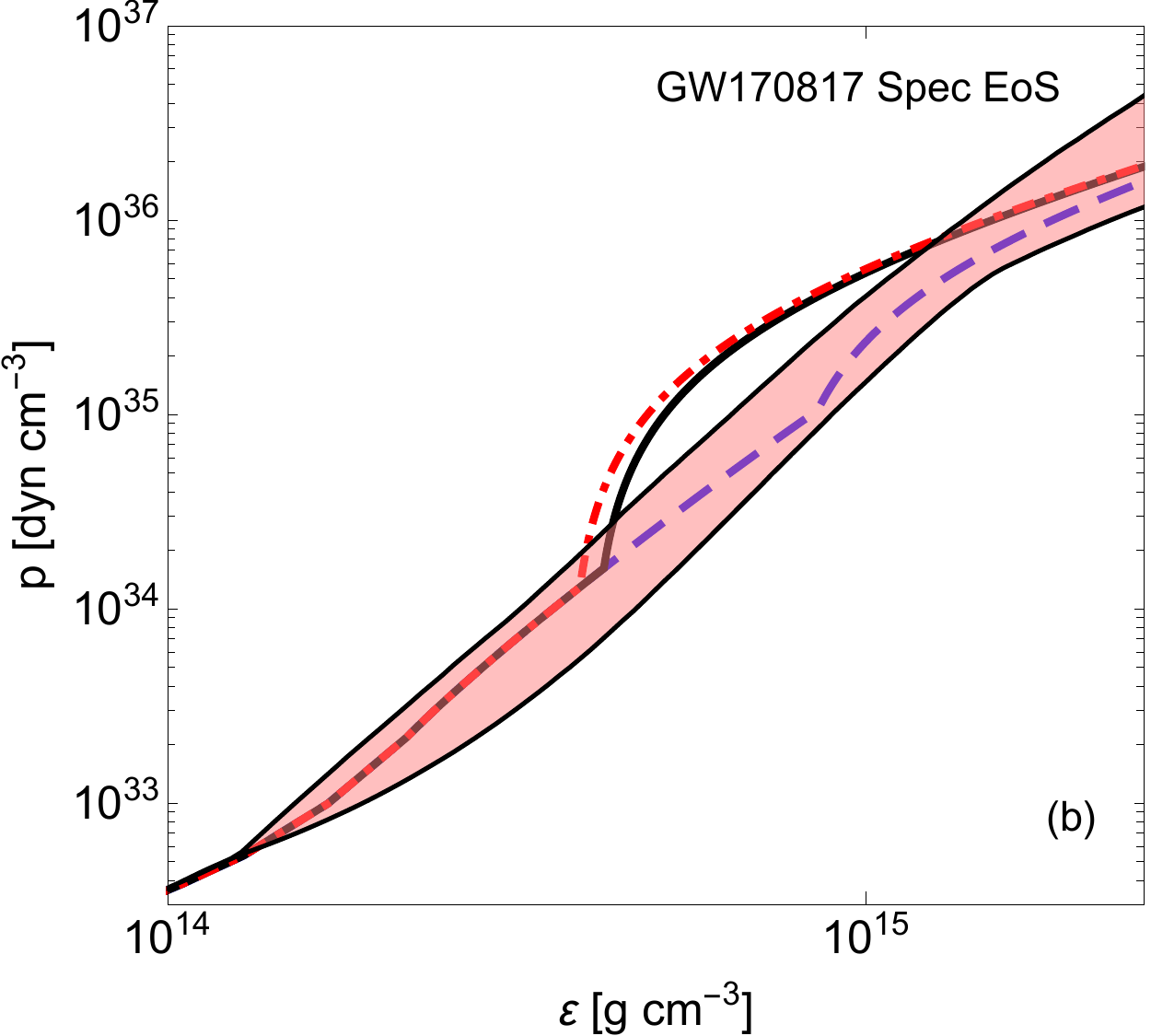} & \includegraphics[width=0.24\linewidth]{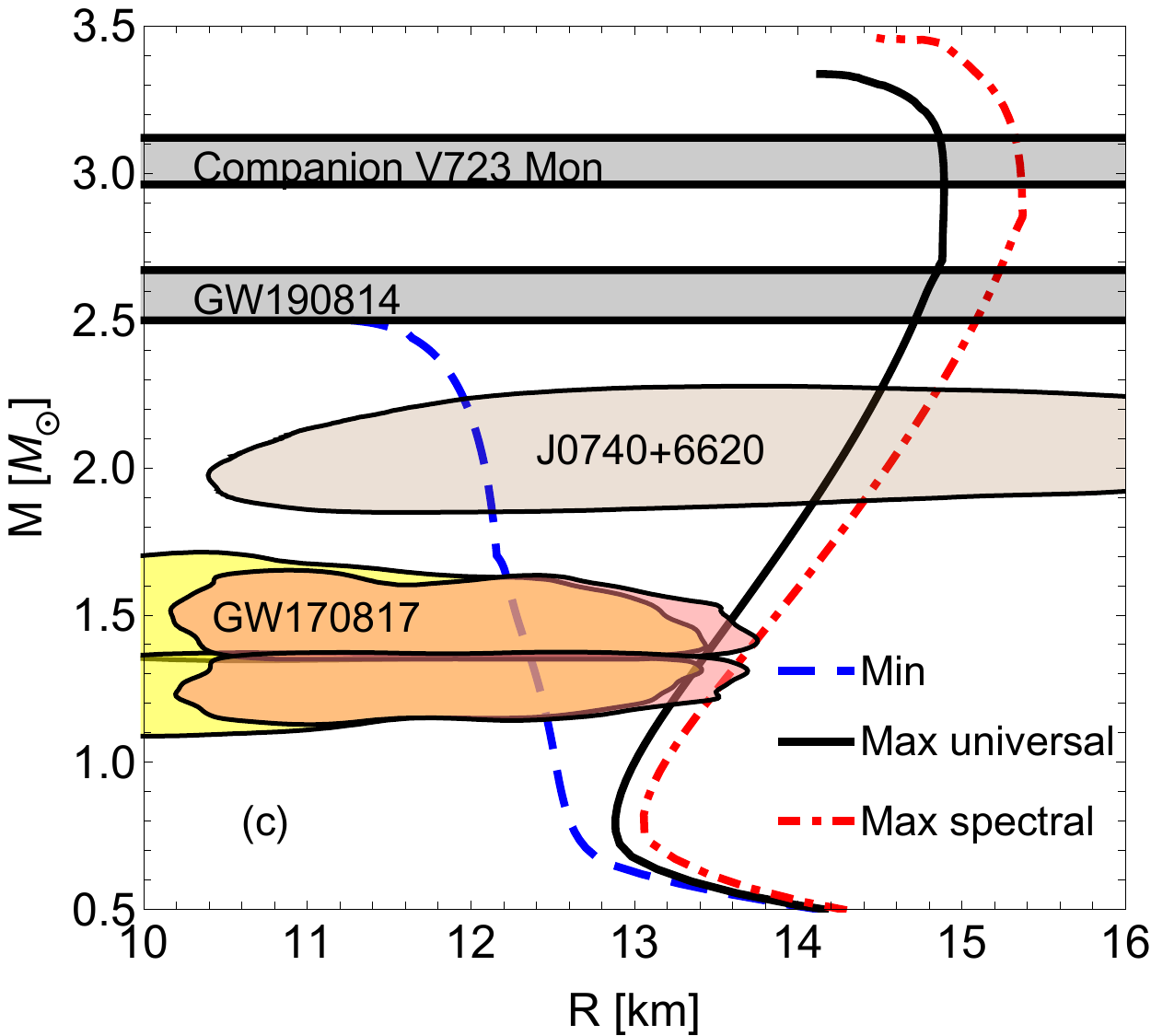} & \includegraphics[width=0.24\linewidth]{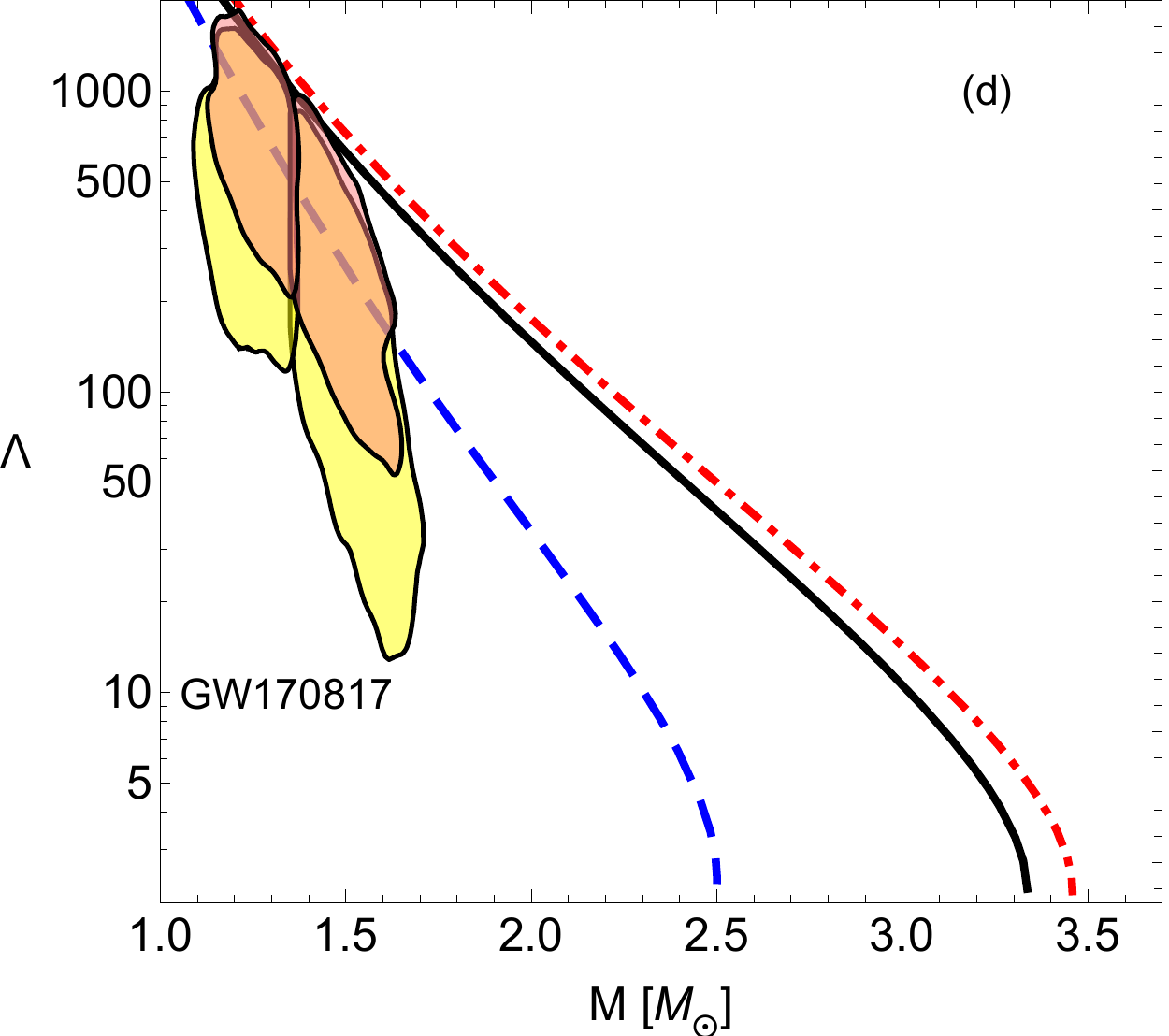}
\end{tabular}
\caption{(Color online) Same as Fig.~\ref{fig:width_peak}, but for a sub-family of EoSs that lead to the most extreme mass-radius curves  that remain causal and fit within the two distinct GW170817 bounds. The right-most edge of the posterior in the mass-radius diagram, derived from the two analysis (universal relations and spectral EoS) of the GW170817, can impact the largest maximum mass achievable.}
\label{fig:extreme}
\end{figure*}

As we mentioned before, we here use the word ``crust'' in a non-rigorous way to refer to the low density portion of the EoS all the way until the fast rise in the speed of sound takes place. Since this rise can happen anywhere from just after saturation density until $\sim 4~n_{sat}$, this comprises regions described by both nuclei and bulk nucleonic matter.
While the low density regime of a neutron star is relatively well constrained from a variety of techniques \cite{Tews:2018iwm,Han:2021kjx}, still some uncertainty remains \cite{Ducoin:2011fy,Baym:2017whm,Gamba:2019kwu,Ferreira:2020zzy}.  Up until this point, we have considered only the SLy EoS to model the low density portion of our EoSs.  In this subsection, we relax that assumption to test how a different choice in crust would affect our maximum mass and range of radii for an ultra-heavy neutron star of $M\geq 2.5$~M$_{\odot}$.  

Unfortunately, it is not possible to systematically check all known crusts because not all of them are available up to the density we require for matching to our functional forms of $c_s^2$.  Thus, we choose two other EoSs\footnote{In a recent work \cite{Li:2021crp}, the QMF EoS \cite{Zhu:2018ona} and the DD2 EoS \cite{Fortin:2016hny} were also studied to find their respective maximum masses.} to study beyond the SLy EoS: the QHC19 EoS \cite{Baym:2019iky} and the SKa EoS \cite{Fortin:2016hny}.   In order to test the effect of these on the mass-radius relation, we fit second-order polynomials to their speeds of sound and connect them to the causal limit at a transition density $n_{1}$. We then vary $n_1$ so that the resulting mass-radius relation is in agreement with the LIGO/Virgo and NICER constraints. 

Figure~\ref{fig:crusts} shows our results for this study where we find that behavior of the mass-radius relation depends strongly on the crust model, even if the EoSs behave the same after switching at more dense regions. Because the QHC19 EoS has the softest EoS in the low density region that we consider here, it requires the lowest value of  $n_{1}$ in order to support heavy neutron stars. Consequently, also because of the low value of $n_{1}$ and the soft low density EoS, this leads to a larger maximum mass and radius that fits within known constraints (and a larger tidal deformability for a fixed mass). Note that, this is the only part of our analysis where we allow $n_1$ to be smaller than $1.5\ n_{sat}$. We see the opposite effect for the SLy and SKa that has the stiffest crusts, which then can have a larger values of $n_1$, which leads to the smallest maximum mass that still fits the constraints. Overall, varying the crust model can change the maximum mass and radius up to $6\%$. See Ref.~\cite{Hu:2020ujf} for a discussion of the relation between different symmetry energy slopes in crusts models and the tidal deformability.
 
By increasing $n_{1}$, we are able to produce mass-radius relations that allow for the minimum radius of stars with masses as high as $2.5$~M$_{\odot}$ and pass all observational constraints. The minimum and maximum values of $n_1$ that can support a neutron star of $2.5$~M$_{\odot}$ and still fit LIGO/Virgo constraints are in the range $n_{1,\rm{min}} \approx (1.25,1.5) ~n_{\rm sat}$ and $n_{1,\rm{max}} = (2.75,3) ~n_{\rm sat}$, which of course depends on the crust EoS. The stiffness or softness of the crust determines the value of $n_1$ that is required; a softer crust typically requires lower values of $n_1$. 

Let us finally make a comment about the findings in Ref.~\cite{Blaschke:2020vuy}. This reference argues that if the NICER marginalized posteriors on the radius using observations of PSR J0740+6620 were to exclude $R\gtrsim 11.5$ at some given confidence, then it must be that what NICER has observed was a hybrid star with a quark core, whose EoS contains a first-order phase transition. This hypothesis was arrived at by investigating the mass-radius curve of a phenomenological EoS model composed of a hadronic crust coupled to constant-speed-of-sound, quark matter. Our findings in Fig.\ \ref{fig:crusts} do not support this hypothesis because there is still an important influence of the crust in the mass-radius relation, even up to neutron star masses $M\sim 2.1 $~M$_\odot$.  Essentially, a soft crust can lead to a star with a small radius that falls below the $R\lesssim 11.5$~km cutoff of Ref.~ \cite{Blaschke:2020vuy}. One could even reach smaller radii without requiring a first-order phase transition if one did not require the EoS to also allow for neutron stars of $2.5 $~M$_\odot$. 

The above discussion highlights the importance of including uncertainties in the crust when considering hypotheses about high-density nuclear physics. Astrophysical data can and should of course be used to constrain crust uncertainties, as done recently in \cite{Han:2021kjx}. The knowledge gained from such constraints can then be used as priors in future analysis, but such priors will still contain some degree of uncertainty. Ignoring uncertainties in the crust can lead to incorrect inferences about nuclear physics at higher densities. 

\subsubsection{Impact of uncertainty in gravitational wave constraints}\label{sec:LIGOcheck}

The two-dimensional marginalized 90\% confidence regions obtained by the LVC when analyzing the GW170817 event vary depending on whether one uses the universal relations method or the spectral EoS method \cite{Abbott:2018exr,Abbott:2018wiz}, see e.g.~Fig.~\ref{fig:allCON}. Indeed, the maximum radius allowed at 90\% confidence for a $\sim1.4 $~M$_\odot$ star is larger when one considers spectral EoSs than one considers universal relations by about $0.5$~km. Although this difference is small, it is enough to impact our understanding of the EoS (and consequently what we consider the possible maximum mass of a neutron star). Let us then investigate how this uncertainty in the LVC analyses impacts our maximum mass conclusions. 

Let us consider a set of EoSs designed to create the largest allowed radii neutron stars with a mass of $2.5$~M$_{\odot}$.  To do this, we build a jump in $c_s^2$ at as small of a $n_B$ as possible. In this case, a transition at low $n_1$ leads to a very stiff EoS that has a large radius and a large maximum mass. With this family in hand, we then find the minimum $n_1$ possible to still fit within each of the GW170817 bounds (using the Sly crust).

Figure\ \ref{fig:extreme} shows our results. The two different EoSs that fit at the very edge of each radius bound from the GW170817 event lead to stars with a radii of about $14.5$-$15.5$~km.  The spectral EoS bound allows for a slightly larger maximum radius and mass ($M_{max}\sim 3.5 $~M$_\odot$) and, therefore, requires a smaller $n_1=1.4 ~n_{\rm sat}$.  The most extreme EoS that fits through the universal relations bound on mass and radius leads to a slightly smaller maximum mass and radius ($M_{max}\sim 3.3 $~M$_\odot$) and transitions at $n_1=1.5 ~n_{\rm sat}$. Moreover, these extreme EoSs lead to stars with tidal deformabilities that are much larger at a fixed mass than more standard EoSs. This is because these extremely massive stars are not very compact (see the small stellar central density in the far left panel), since their radius is very large.   

With our EoS family, we can also find the largest possible value of $n_1$ that still produces a star with a maximum mass of $M\geq 2.5 $~M$_{\odot}$. At large  $n_1$, it is quite difficult to produce a large $M_{\rm max}$ and, therefore, only an extreme jump works. We find that $n_1=2.9 ~n_{\rm sat}$  produces a $M_{\rm max} = 2.5 $~M$_{\odot}$ star with radius $R\sim 11$~km. Overall, we find that an approximately $10\%$ extension of the mass-radius posterior allows for a $10\%$ change in the neutron-star maximum mass and maximum radius (comparing universal posterior to spectral posteriors). The minimum radius of the posterior region does not seem to matter much. Thus, future gravitational wave detectors with more precise measurements on the radius will allow for better constraints on the EoS and will help to constrain the maximum possible mass of a neutron star.

\subsubsection{Influence of structure in $c_s^2$}

\begin{figure*}
\centering
\begin{tabular}{c c c c}
\includegraphics[width=0.24\linewidth]{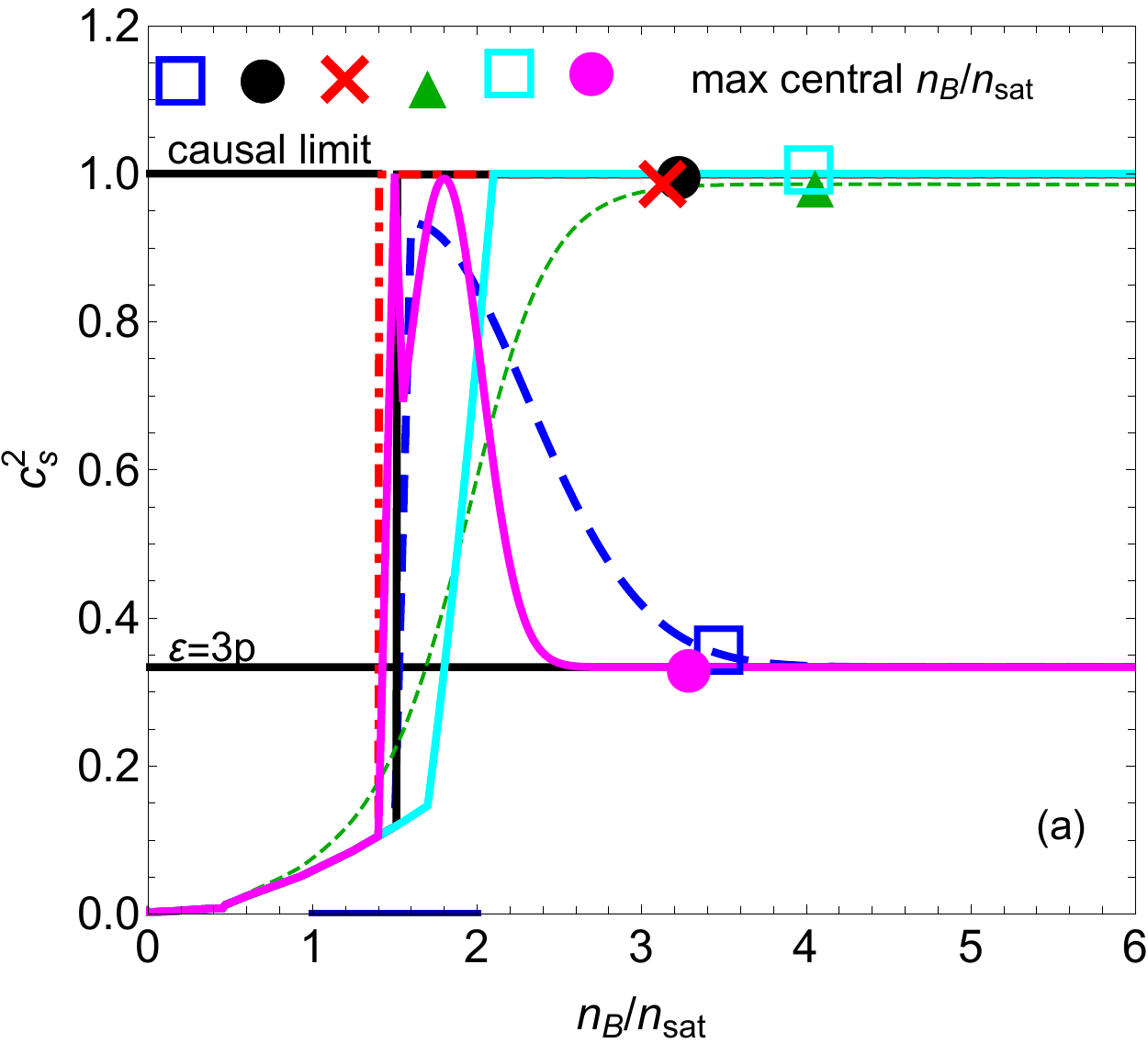} & \includegraphics[width=0.24\linewidth]{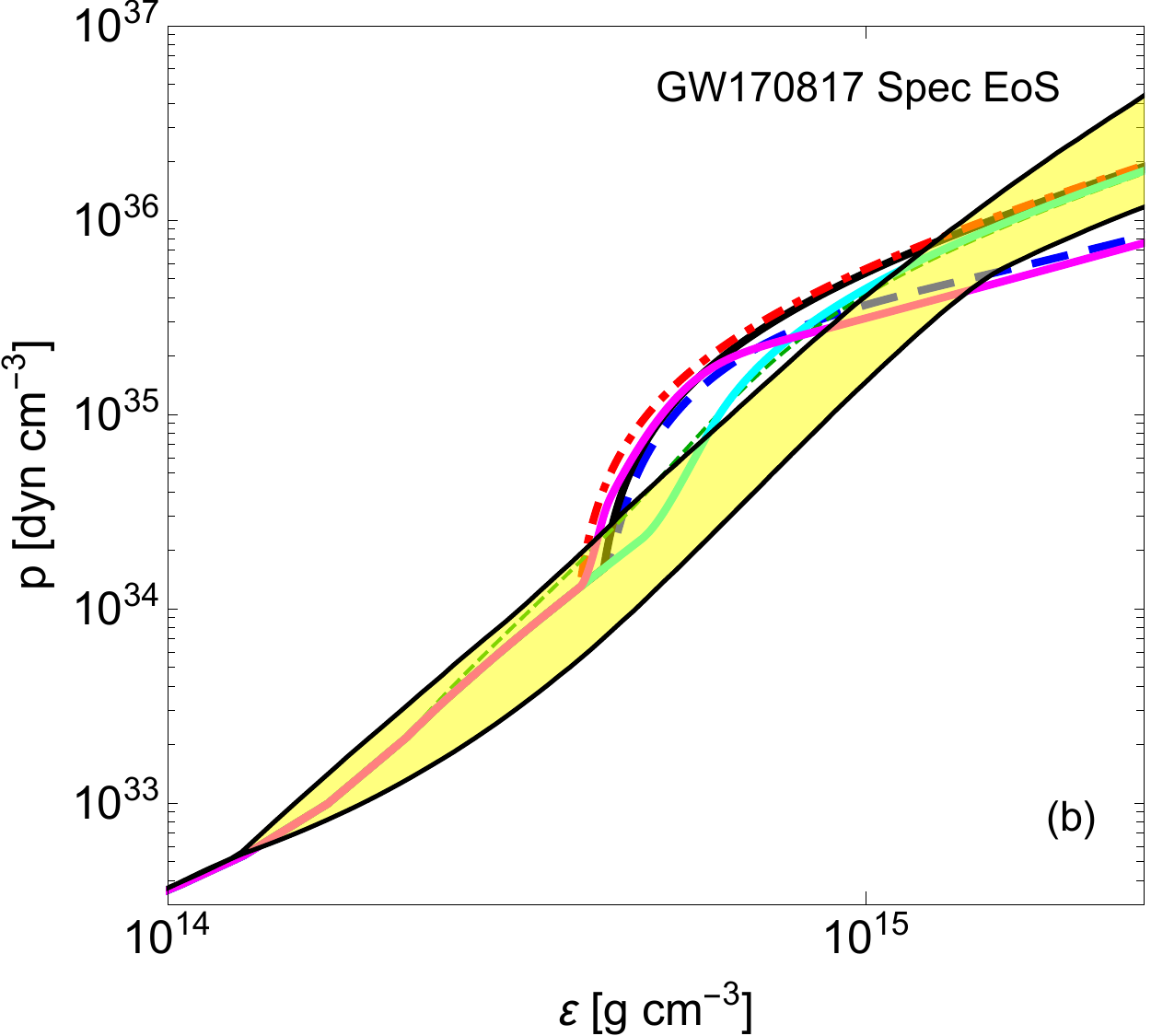} & \includegraphics[width=0.24\linewidth]{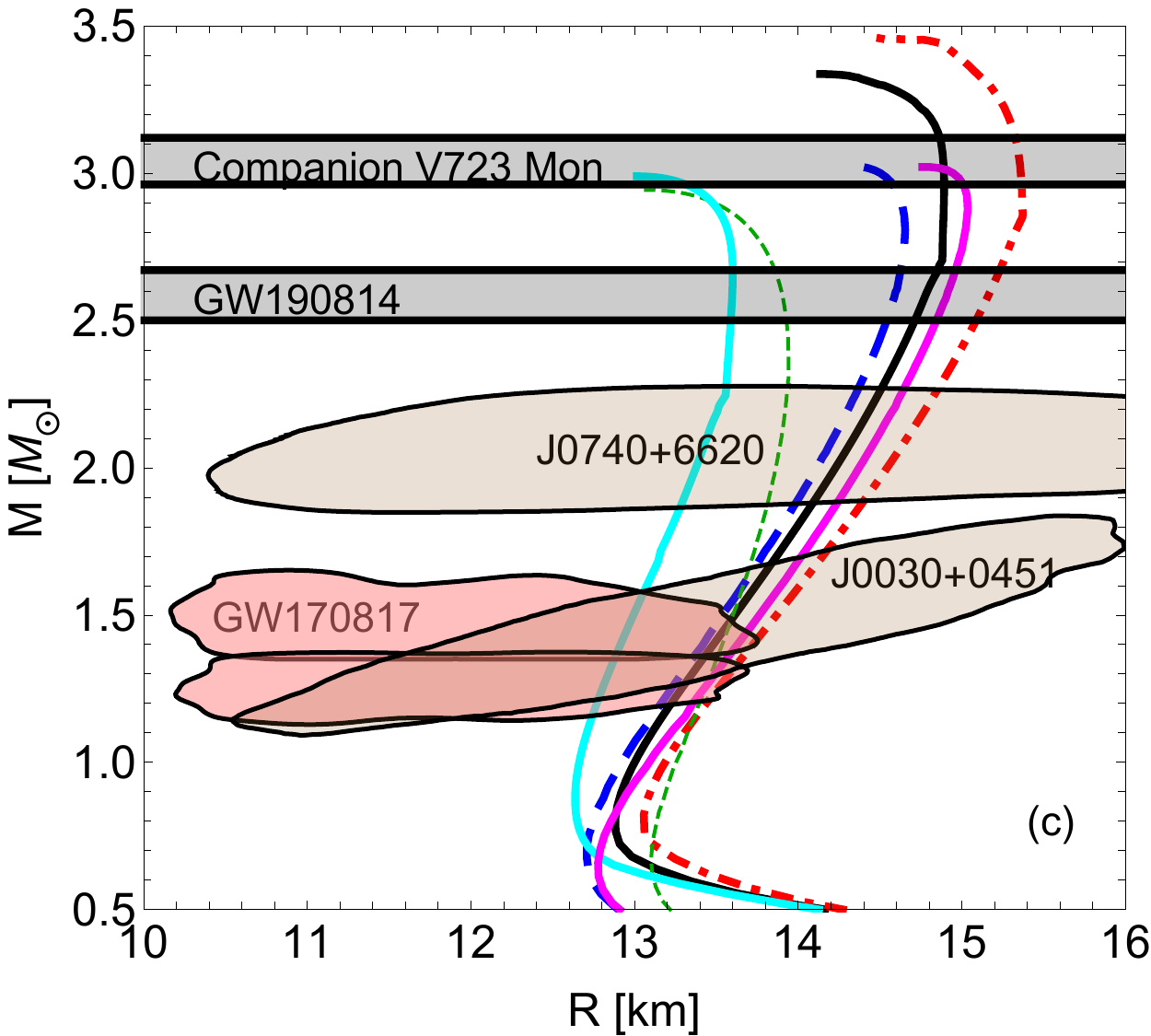} & \includegraphics[width=0.24\linewidth]{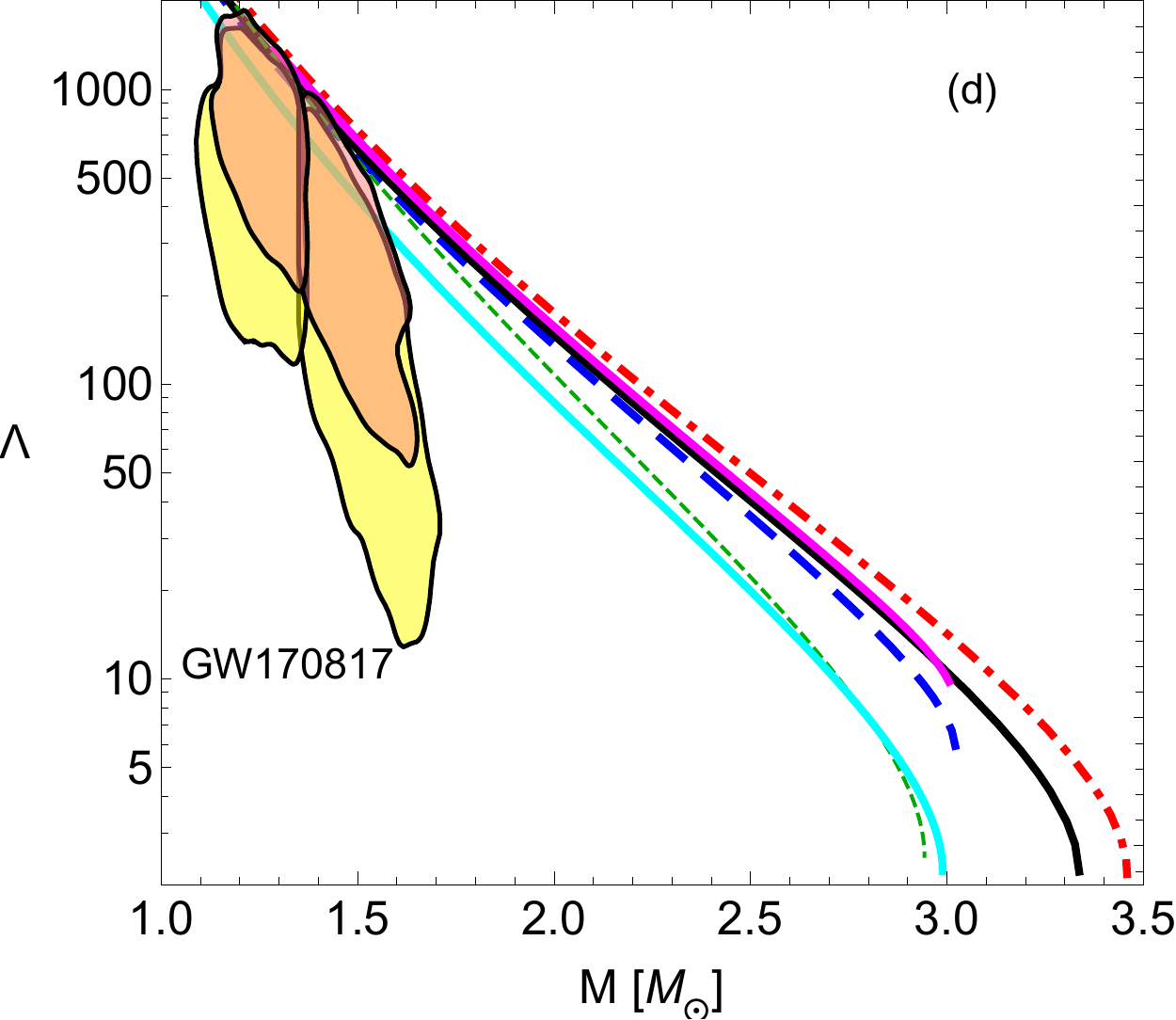}
\end{tabular}
\caption{(Color online) Same as Fig.~\ref{fig:width_peak}, but for a sub-family of EoSs that can produce neutron stars with masses above that of the companion of V723 Mon, yet still fit within all other observational bounds, and remain causal. There is a variety of ways in which such extremely heavy neutron stars could be achieved, leading to drastically different $\Lambda$--$M$ curves.}
\label{fig:3Msun}
\end{figure*}

We have so far in this subsection considered only EoSs which transition to the causal limit with a very sharp slope at a given transition density, but these are not the only EoSs that allow for extremely heavy neutron stars. Indeed, cross over structure in the speed of sound can easily produce extremely heavy stars, as we discuss next. Figure~\ref{fig:3Msun} shows only a characteristic subset of the structures we have considered (all with a SLy crust), including plateaus, single bumps and double bumps. All EoSs shown lead to stars that are massive enough to allow the companion of V723 Mon~\cite{Jayasinghe:2021uqb} to be described as a neutron star, while also satisfying all other observational constraints. The radius of these stars ranges between $12.5$ and $15$~km, leading to various stellar compactnesses and tidal deformabilities. As expected, the most massive neutron stars have exceedingly small tidal deformabilities ($\Lambda \lesssim 10$, and in particular $\Lambda_{\rm min}\sim 2.5$). A measurement of such a small tidal deformability would probably require third-generation gravitation wave detectors. 


\section{Neutron stars with first-order phase transition}
\label{sec:NS-with-1st-O}

Whenever new fermionic degrees of freedom appear inside a neutron star, for example hyperons in a hadronic phase or strange quarks in a quark phase, the energy levels of each of the existing species decrease, softening the EoS. 
The change of slope cannot be directly seen in the EoS, but it can clearly be seen in its derivatives, such as the compressibility, the adiabatic index and the speed of sound \cite{Balberg:1998ug,Haensel:2002qw,Dexheimer:2007mt,CASALI2010,Miyatsu:2017teh,Stone:2019abq,Motta:2020fvm}.
A similar effect appears when light meson condensation takes place, eliminating a source of pressure \cite{Schneider:2019vdm}.
Depending on the microscopic model description of the new degrees of freedom, the EoS may present either a first-order phase transition, where $c_s^2$ abruptly drops to zero in some density region, or crossover structures, where $c_s^2$ presents bumps, spikes, kinks or plateaus over some density region. In the previous section, we discussed the impact of crossover structures on neutron stars, so here we will focus on the impact of first-order phase transitions. As we will see, the size of the first-order phase transition jump, its extension, and the baryon density at which it occurs all have interesting consequences for the structure and stability of neutron stars. 

\begin{figure}
    \centering
    \includegraphics[width=\linewidth]{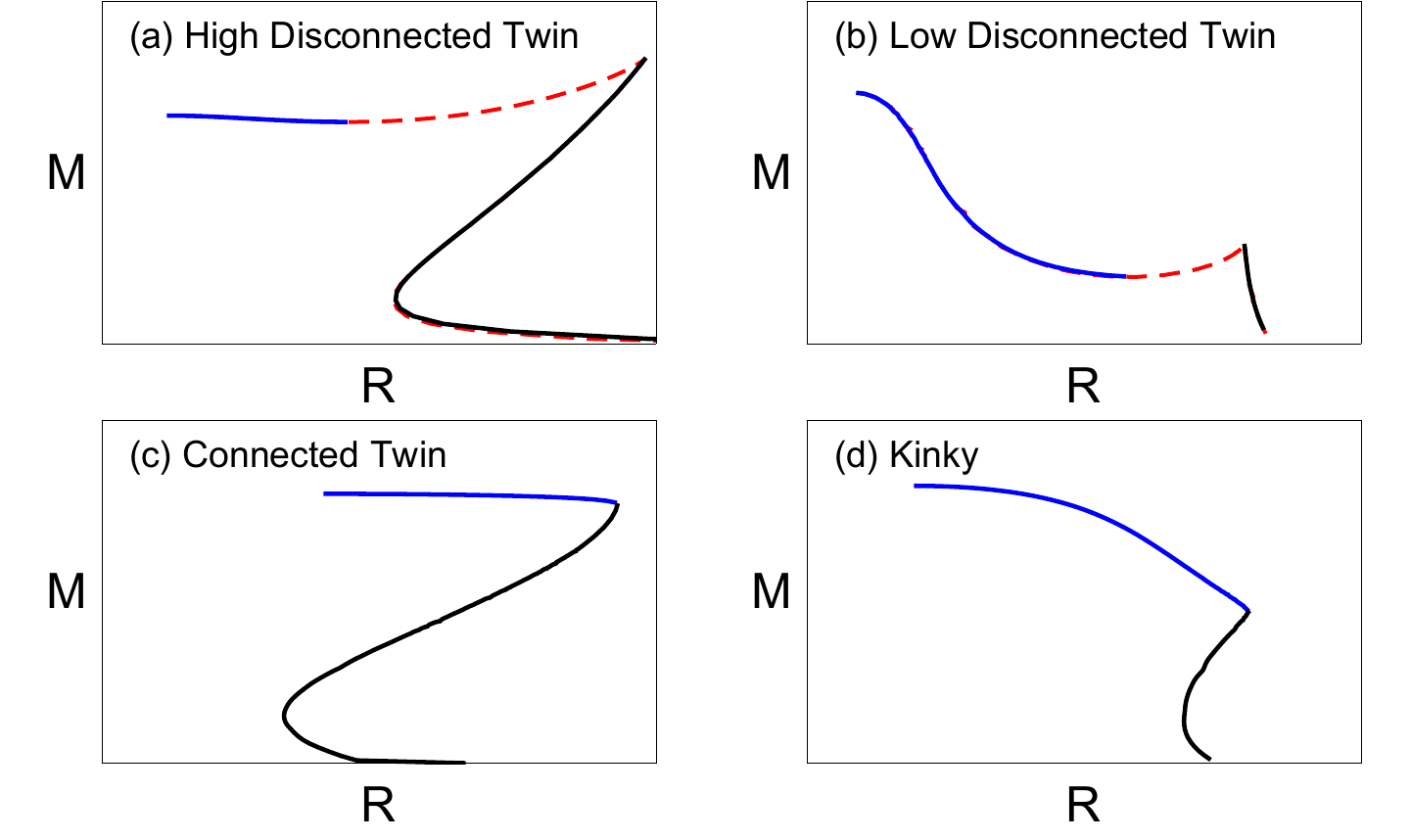}
    \caption{Cartoon of different mass radius diagrams when first-order phase transitions are present. The black lines indicate the first stable branch (at $n_B$ below the phase transition), the blue lines indicate the second stable branch (at $n_B$ above the phase transition), and the red dashed lines indicate the unstable branches.  The right panels show phase transitions at low-densities, while the left panels show phase transitions at high-densities. Only very specific conditions can generate a stable twin branch, as shown in the top panels.}
    \label{fig:cartoon}
\end{figure}

Before we embark on a deep exploration of first-order phase transitions, it is useful to first define a classification and terminology. Depending on the strength, duration and location of the first-order phase transition, different neutron star sequences are possible, which we classify as follows:
\begin{itemize}
    \item \textbf{Disconnected Twins}: Sequences where there can exist stars with different radii but the same mass that lie on two or more stable branches separated by an unstable branch.
    \item \textbf{Connected Twins}. Sequences where there can exist stars with different radii but approximately the same mass that lie on two or more stable branches that are connected to each other. 
    \item \textbf{Kinky}. Sequences where stars of a given mass have a unique radius, yet present two or more stable branches that are connected non-smoothly to each other at a point.
\end{itemize}
Of course, the existence of a first-order phase transition does not guarantee a twin or kinky mass-sequence, as it is also possible that there will only be a single stable branch, and any other branch will be unstable; we will not focus on such cases here, since those mass-radius sequences are degenerate with crossover EoSs. A cartoon of characteristic examples of the mass-radius diagram for each of these classes is shown in Fig.~\ref{fig:cartoon}. 

Examples of all three classes have appeared previously in the literature from both phenomenological EoS models~\cite{Alford:2013aca,Alford:2015dpa,Alford:2015gna,Ranea-Sandoval:2015ldr,Han:2018mtj,Chatziioannou:2019yko,Blaschke:2020vuy,Blaschke:2020vuy,Pang:2020ilf,Ayriyan:2021prr}, as well as more realistic nuclear physics models~\cite{Dexheimer:2014pea,Mishustin:2002xe,Jakobus:2020nxw,Alford:2017qgh,Zacchi:2015oma,Alvarez-Castillo:2018pve,Li:2019fqe,Wang:2019npj,Fadafa:2019euu,Xia:2019xax,Yazdizadeh:2019ivy,Shahrbaf:2019vtf,Lopes:2020rqn,Blaschke:2020qrs,Marczenko:2020wlc,Jokela:2020piw,Li:2020dst,Malfatti:2020onm,Kojo:2020ztt}. For example, a disconnected mass twin sequence arises from a nuclear-physics-based EoS driven by strangeness, as shown in Fig.\ \ref{fig:known} for the CMF$_{ex}$ $\xi<0$ EoS (see the particle population in Fig.~7 of Ref.~\cite{Dexheimer:2014pea}). This EoS contains an enormous amount of structure in terms of bumps and kinks, but what allows for twins (and also kinky) neutron star sequences is the abrupt drop in the speed of sound to zero. Whether one obtains a (connected or disconnected) mass twin or a kinky sequence depends on the interplay of the bump structure that appears before the first-order phase transition, the properties of the phase transition itself, and the dense matter EoS behavior, as we shall see in this section. 

A note of caution about terminology is due at this point. For example, \cite{Alford:2004pf} (and many others) use the term ``hybrid'' neutron stars to refer to those that contain a quark core, whether that is reached from a crossover or a first-order phase transition. However, because a first-order phase transition is often implied in such models, the connection to the mass-radius sequence becomes ambiguous. As we have seen in the previous section and we will continue to see in this section, the structure of neutron stars can be quite different depending on whether there is a first-order phase transition or crossover structure, even if they possess a quark core. Therefore, to avoid this confusion, we avoided using the word ``hybrid'' in this paper.  

With that out of the way, let us now discuss each of the classes in more detail, beginning with disconnected twins. Recall that these are stellar sequences in which there are two disconnected stable branches that are separated by an unstable branch, i.e. a region in the mass-radius relation where ${dM}/{d\varepsilon_c}<0$. In Fig.\ \ref{fig:cartoon} the unstable branch is shown as a red dashed line, the first stable branch is in solid black (this branch is for central densities below the first-order phase transition), and the second stable branch is shown in blue (this is for densities above the phase transition).

We can further separate disconnected twins into two sub-classes: {\bf high disconnected twins} and {\bf low disconnected twins}.  The high/low term refers to the location of the first-order phase transition. A high disconnected twin has a phase transition at high $n_B$, often after some sort of structure has occurred at lower $n_B$. A low disconnected twin, instead, has a phase transition at low $n_B$, followed by structure in the $c_s^2$ at higher $n_B$. Figure\ \ref{fig:cartoon} shows a characteristic mass-radius diagram for low and high disconnected twins.  Of course, the definition of low/high is somewhat arbitrary because in realistic EoSs there may be structure both before and after a phase transition, yet we still find it useful to distinguish these two subclasses in this paper.   

Let us now discuss the other two classes of neutron star sequences: connected twins and kinky stars. An EoS with a first-order phase transition may lead to two stable branches (identified with the EoS before and before/after the phase transition) that are connected to each other. We can separate sequences with connected branches into two distinct classes: those that lead to stars with approximately the same mass but different radius (connected twins) and those that do not (kinky stars). In Fig.\ \ref{fig:cartoon}, we show a characteristic mass-radius diagram for both connected twins and kinky sequences. Unlike in the disconnected twin cases, there are no unstable branches here, and thus no red dashed curve.

\begin{figure*}[t]
\centering
\begin{tabular}{c c c c}
\includegraphics[width=0.24\linewidth]{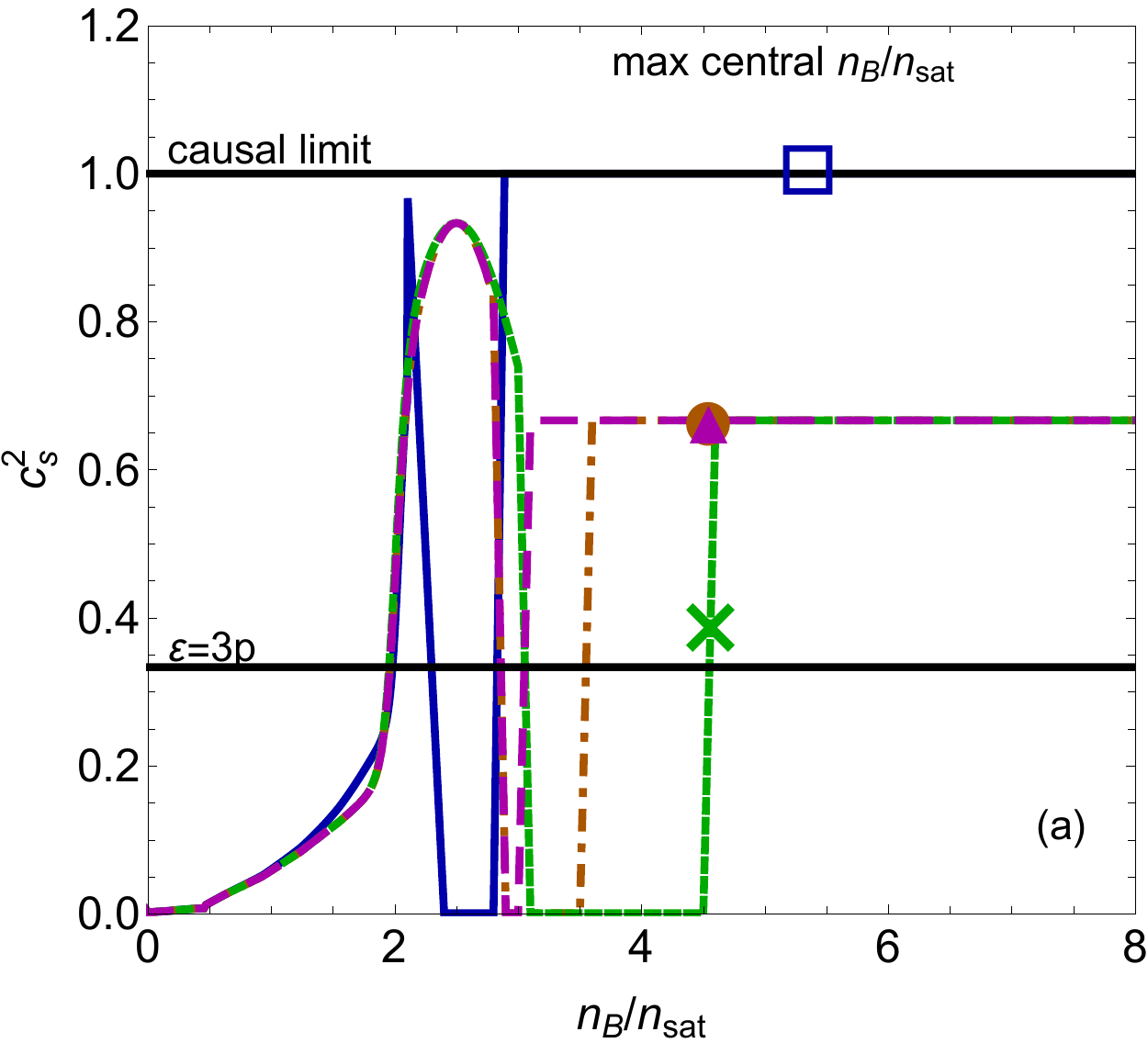} &
\includegraphics[width=0.24\linewidth]{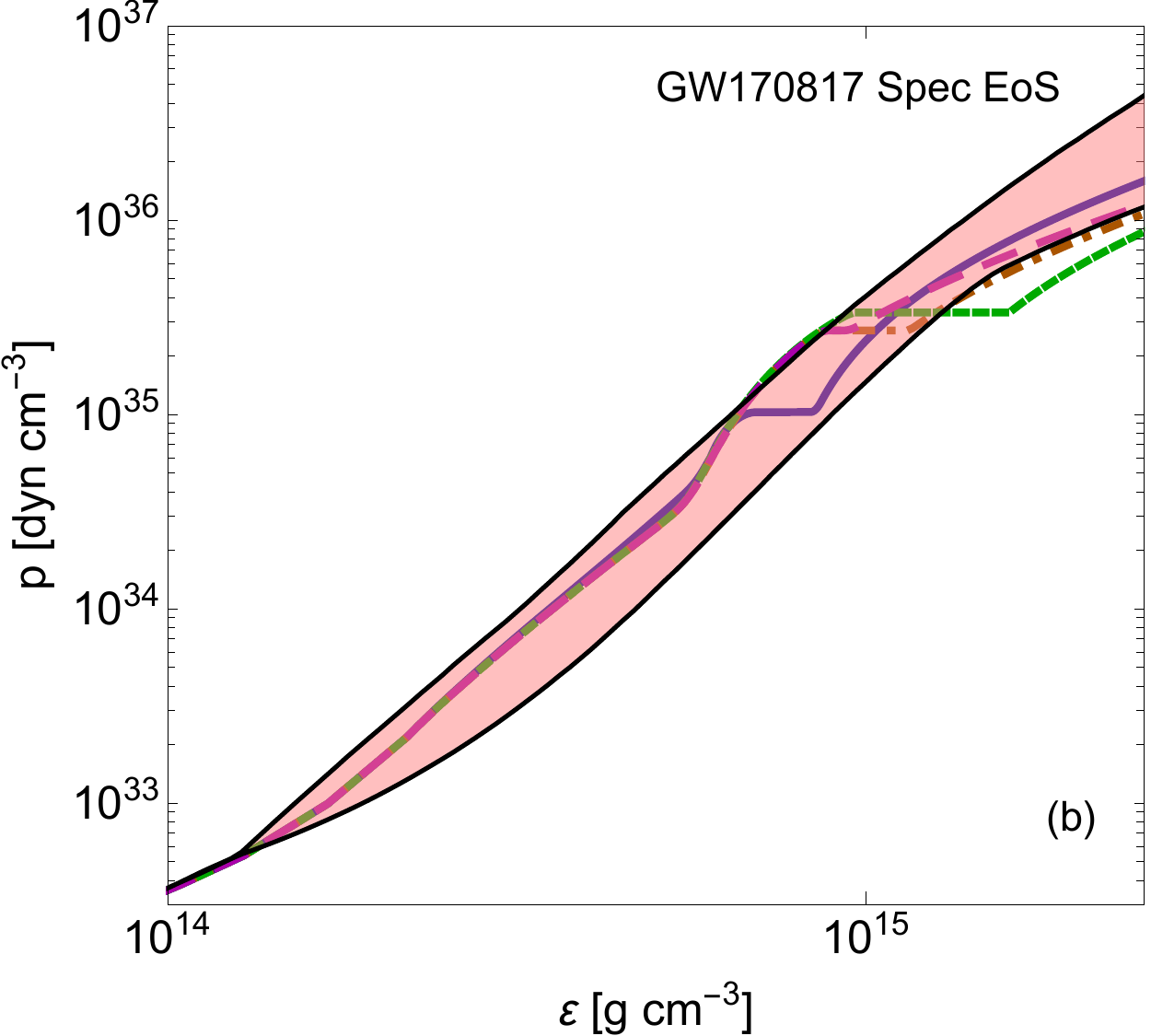} & \includegraphics[width=0.24\linewidth]{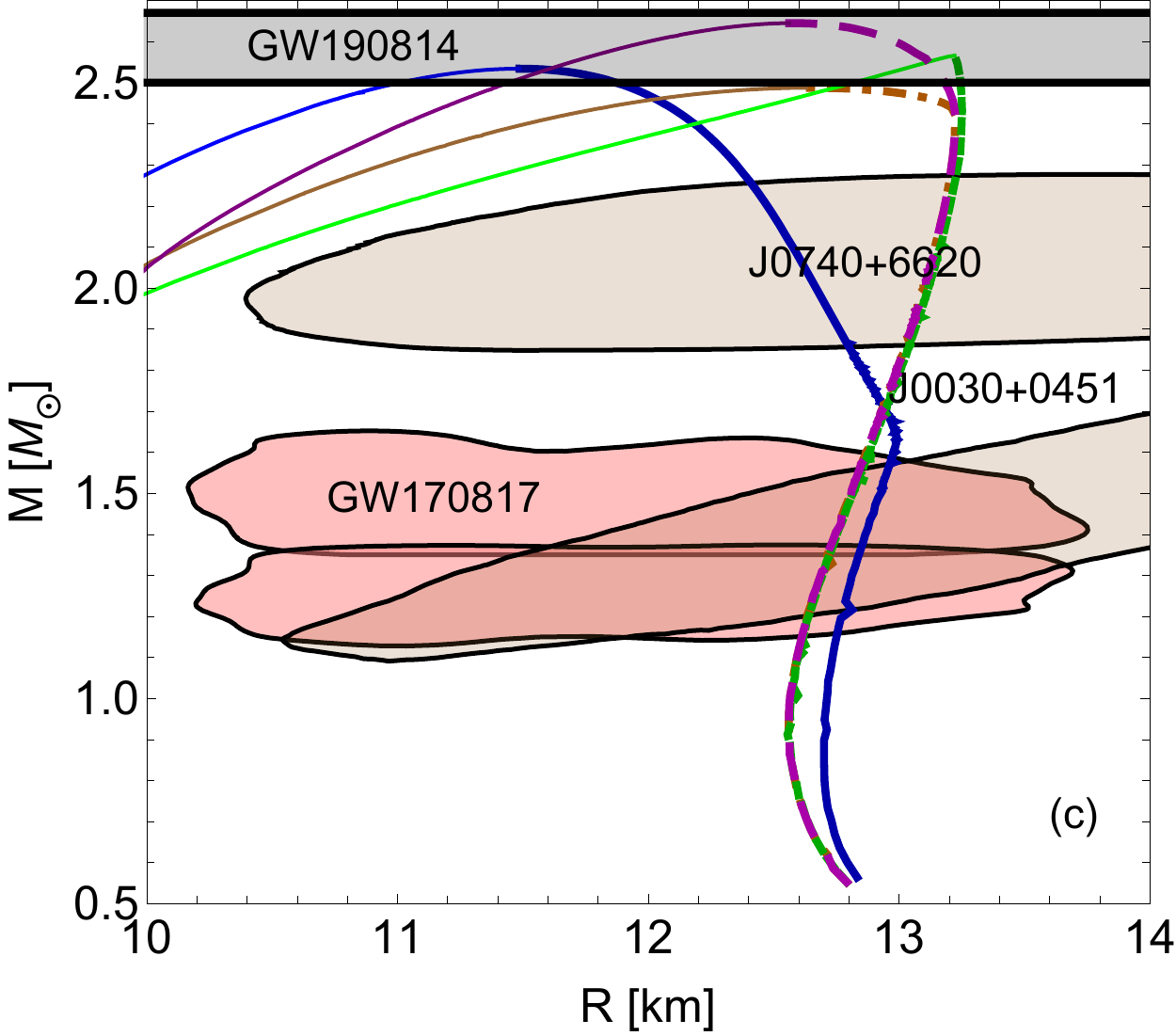} & \includegraphics[width=0.24\linewidth]{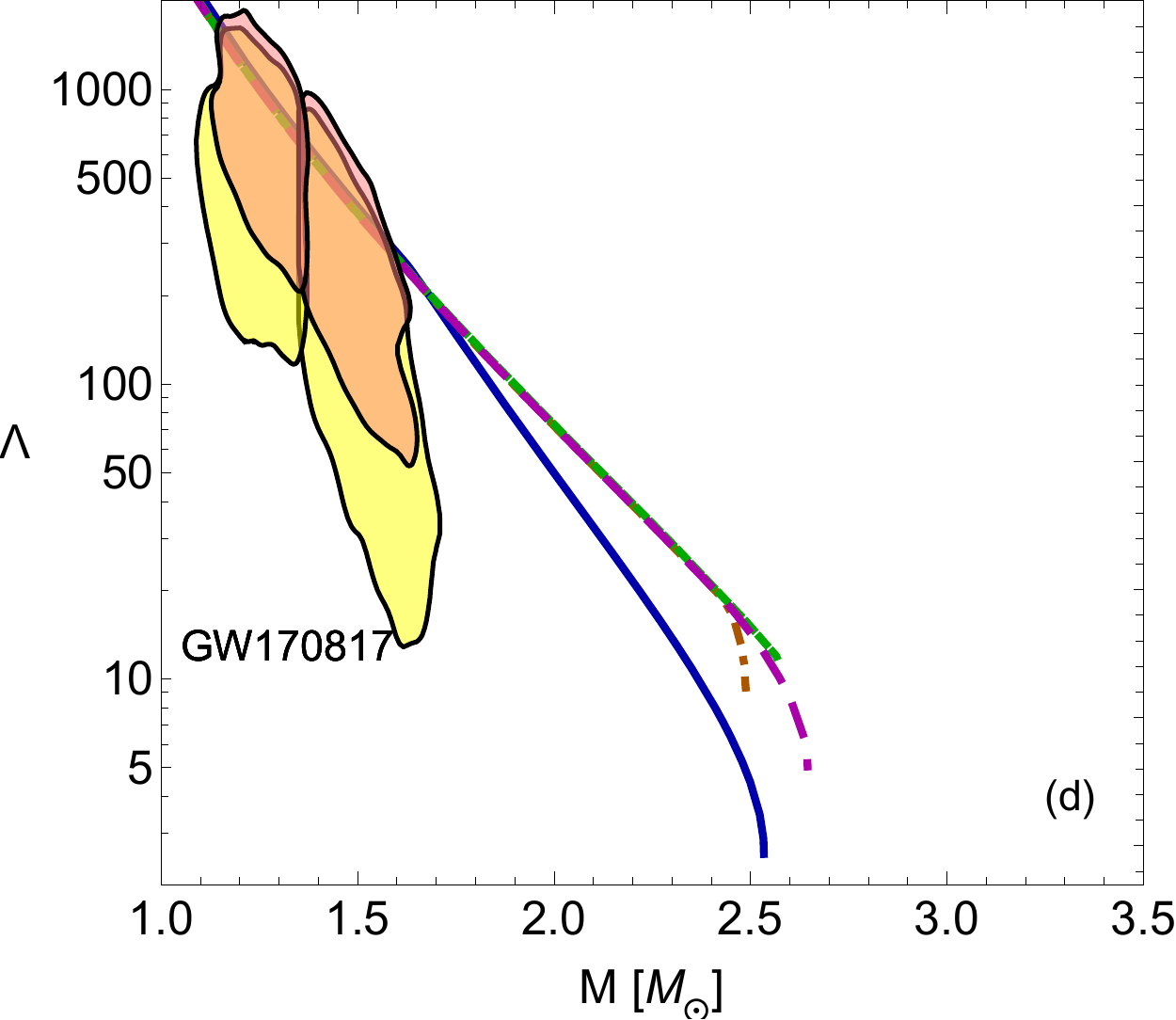}
\end{tabular}
\caption{(Color online) Same as Fig.~\ref{fig:width_peak}, but for a sub-family of EoSs with first-order phase transitions. Thick curves in the mass-radius panel correspond to stable branches, while thin curves are unstable branches; in the $\Lambda$--$M$ panel, all curves correspond to stable branches. Observe that one can easily produce ultra-heavy neutron stars with masses larger than $2.5$~M$_\odot$ with either a connected twin sequence or a kinky sequence, while still satisfying all other observational constraints. }
\label{fig:TWsummary} 
\end{figure*}

We have to be careful with the definition of connected twins though. Mathematically speaking, this class of EoSs does not formally lead to twins because their mass-radius curves are such that stars of different radii do have different masses. What distinguishes connected twins from kinky stars is that there is a stable branch in which stars of different radius have very similar (though technically not identical) masses. Observationally, however, it would be very difficult to distinguish two stars in this stable branch due to statistical or systematic uncertainties in the observations. All data comes with uncertainties, and currently, systems for which we can measure the radius of a neutron star have errors in the mass of about 10\% or more. In the future, these uncertainties will decrease, but it seems unlikely that we will be able to distinguish two stars with radii differing by $1.5$~km but mass differing by less than $\sim 10^{-1} $~M$_\odot$. If the existence of connected twins is ignored and they were to exist in nature, researchers may be led to believe that they have detected a disconnected twin instead. 

Another possibility for EoSs that lead to mass-radius diagrams with connected stable branches is through a kinky sequence. We define those as mass-radius curves with more than two stable branches that are connected at a point in the mass-radius plane in a non-differentiable way (i.e.~leading to a mathematical ``kink'' in the mass-radius relation). At that kink, the first derivative is discontinuous and the second derivative (the ``curvature'') becomes formally infinite. Such a mass-radius sequence would look similar to that of a neutron star without a first-order phase transition, yet they would differ in the behavior of the mass-radius curve in the neighbourhood of the kink. The detection of this kink would require the observation of several stars with varying masses and the precise measurement of their radii, so as to map the kink structure in the mass-radius plane.

In the rest of this section, we study all of these classes in detail through the phenomenological EoS model described in Sec.~\ref{sec:modelingEoS}. Just as a reminder, the first-order phase transitions are implemented by setting $c_s^2=0$ inside a region in baryon density $n_B$ between $n_B^{1\rm PT,\rm low}$ and $n_B^{1\rm PT,\rm hi}$.  The broader the $c_s^2=0$ region across $n_B$, the stronger the phase transition. After the end of the phase transition, $c_s^2$ will be typically a plateau of a given height. Before the start of the phase transition, $c_s^2$ will typically have a bump of a certain width and height. We will study here how all of these choices affect the structure of neutron stars, and in particular the mass-radius and $\Lambda$--$M$ curves. Importantly, when showing mass-radius diagrams in this section, we will always include all stable and unstable branches (unlike in Sec.~\ref{sec:25msol} where only stable branches were shown).   

\begin{figure*}[t]
\centering
\begin{tabular}{c c c c}
\includegraphics[width=0.24\linewidth]{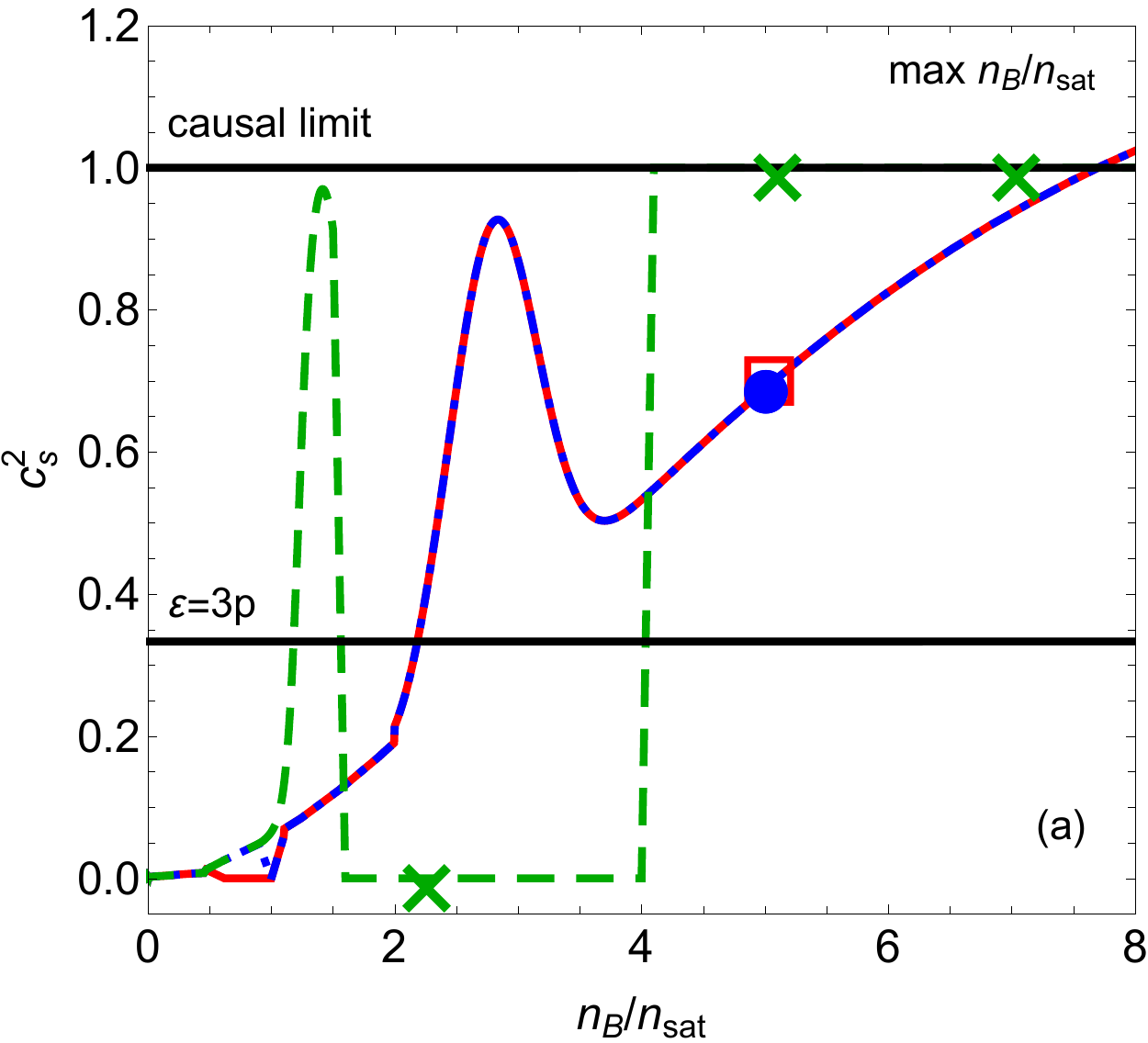} & \includegraphics[width=0.24\linewidth]{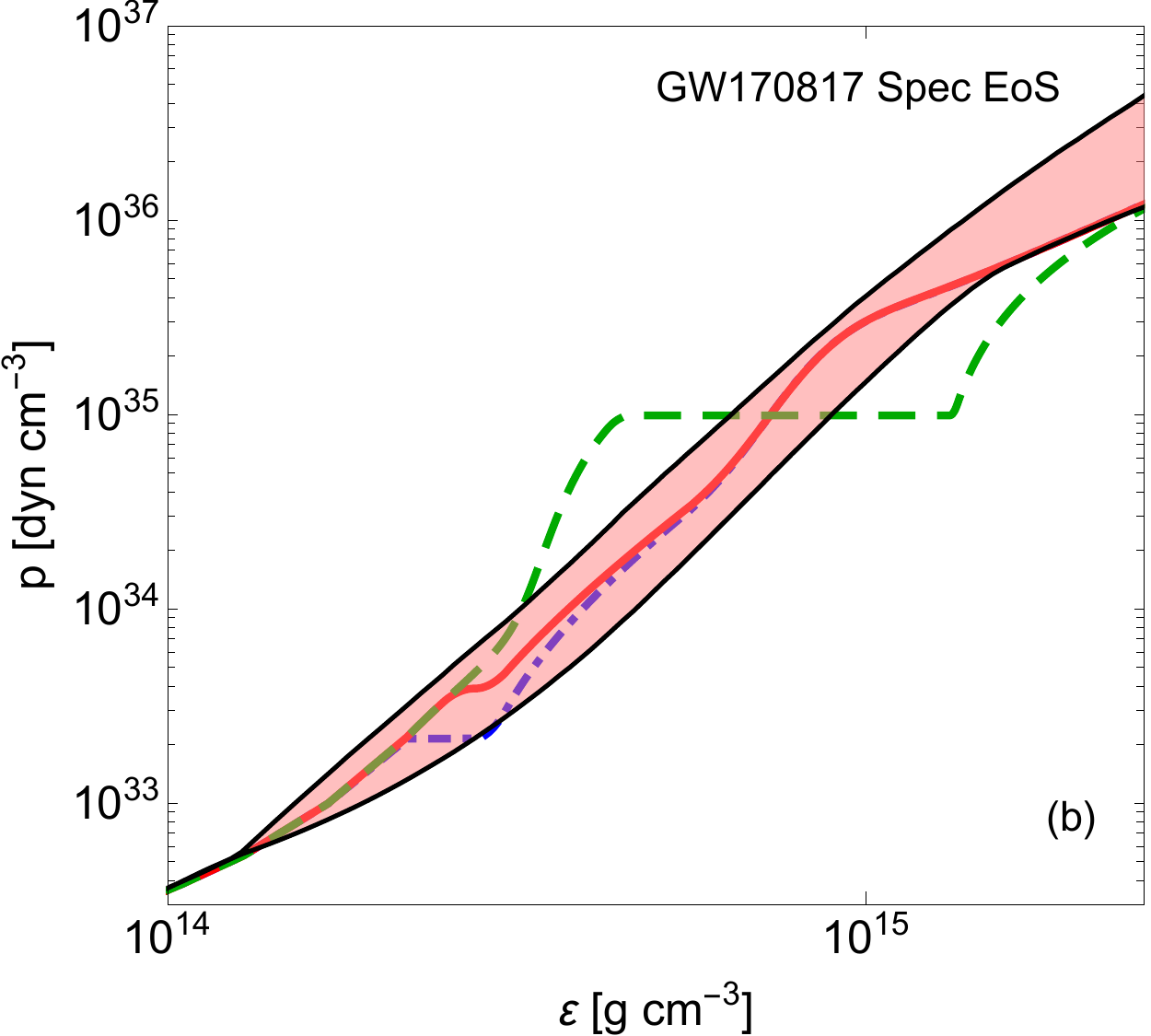} & \includegraphics[width=0.24\linewidth]{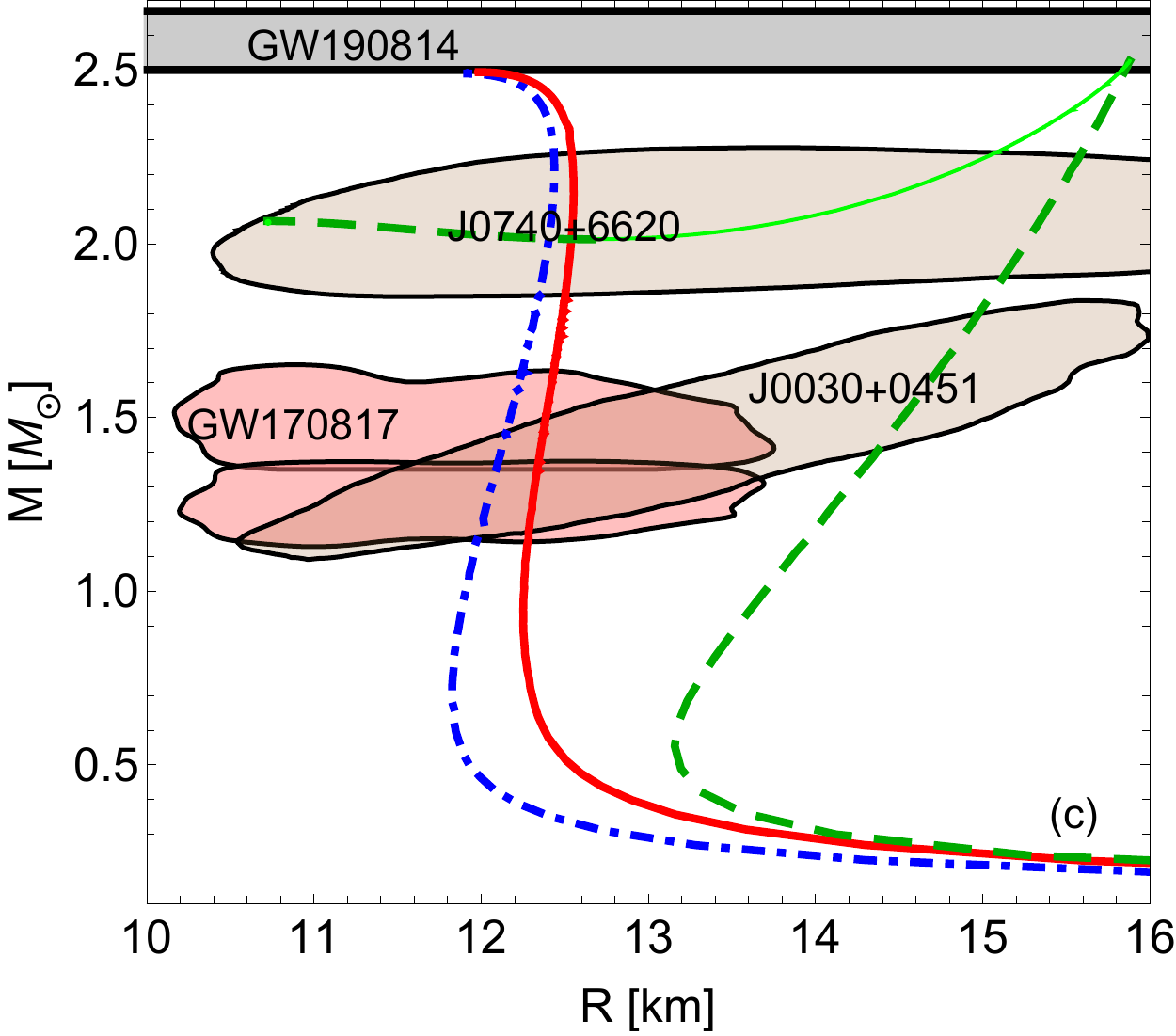} & \includegraphics[width=0.24\linewidth]{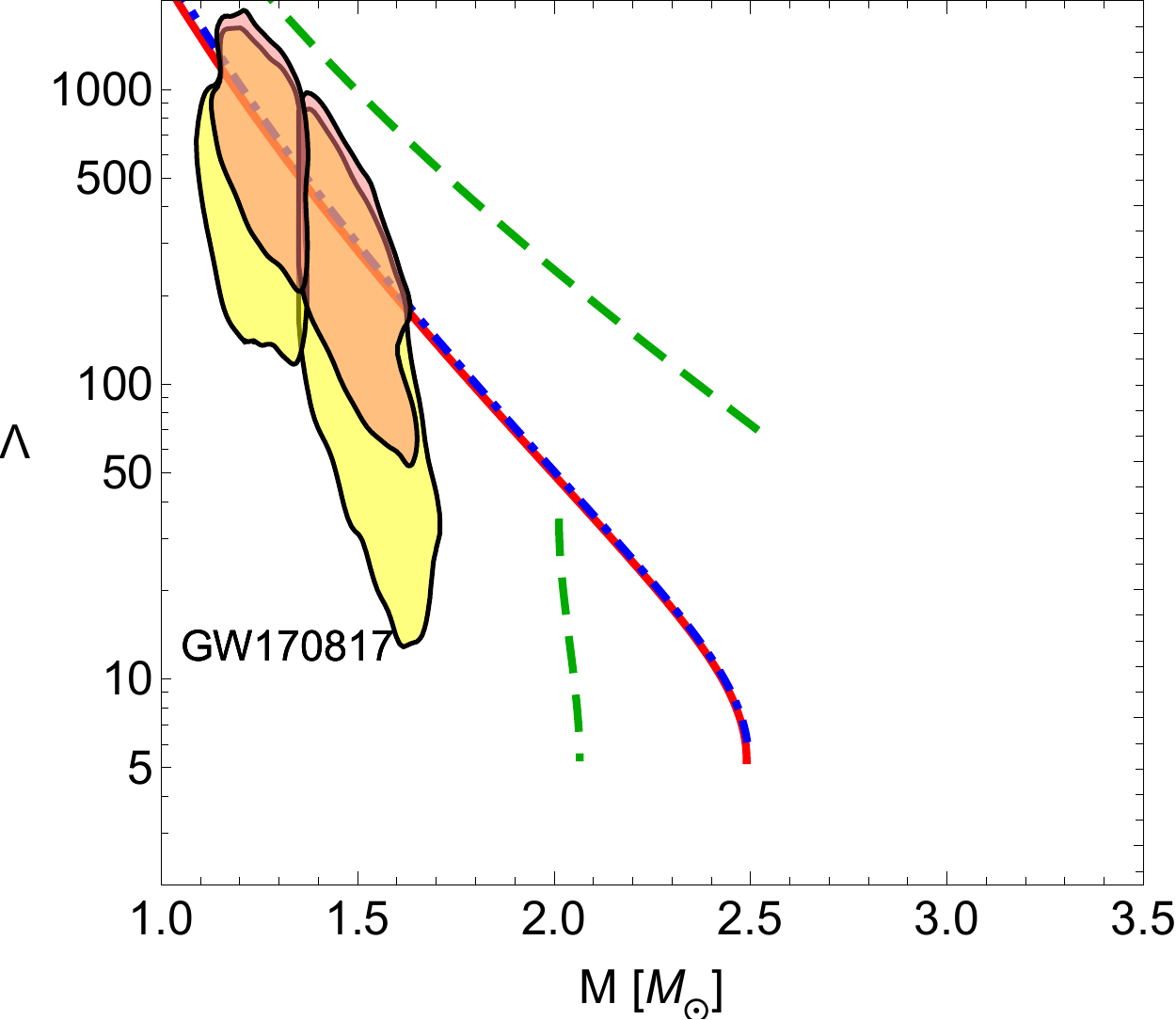}
\end{tabular}
\caption{(Color online) Same as Fig.~\ref{fig:TWsummary}, but for a sub-family of EoSs with first-order phase transitions that lead to disconnected twins. It is now not possible to create ultra-heavy neutron stars with disconnected twins, while simultaneously satisfying LVC constraints. }
\label{fig:Blatsummary} 
\end{figure*}

\subsection{Can there be first-order phase transitions in ultra-heavy neutron stars?}

Now that we have established a dictionary of potential neutron stars with first-order phase transitions, the natural question that arises is if all of these different types of sequences can support ultra-heavy neutron stars (ie.~stars with masses larger than $2.5 $~M$_\odot$). A recent paper \cite{Christian:2020xwz} argued that if  the light component of the GW190814 was a neutron star, then it would exclude the possibility of mass twins. Another work suggested a possible deconfinement transition at very low baryon densities \cite{Blaschke:2020vuy} could produce mass twins at $M<1 $~M$_\odot$ but with the hybrid branch reaching $M\geq 2.5 $~M$_\odot$. As we will see below, it is actually possible to produce ultra-heavy neutron stars with a twin sequence, provided it is connected, in addition to kinky sequences.  

\subsubsection{Connected twins and kinky sequences are possible for $M\geq 2.5 $~M$_\odot$}

Figure\ \ref{fig:TWsummary} shows that indeed it is possible to produce ultra-heavy stars with either connected twins or kinky sequences. Recall that in this section we include both the stable (thick lines) and unstable branches (thin lines) in the mass-radius diagram, but now the right most panel showing $\Lambda$--$M$ curves. The key general features needed for these results are a very high peak in the $c_s^2$ bump at relatively low densities, and a first-order phase transition that occurs at not too high of a density. There is an interplay between the width of the bump and how strong is its effect on the mass-radius diagram. For instance, the blue curve in Fig.\ \ref{fig:TWsummary} has a very thin bump that we place at low density, and which leads to a kinky mass-radius sequence at about $M\sim 1.6 $~M$_\odot$. In contrast, the other EoSs have a wider bump at a slightly larger densities, which pushes the phase transition to larger densities. In this second scenario, either the first-order phase transition makes the neutron star immediately unstable (green dashed line), or only a small region extends past the first-order phase transition (purple and brown EoSs). The stars within this small region still contain a first-order phase transition that generates a either a flattening in the mass-radius curve (brown curve) or a very small kink (purple curve) that cannot be discerned by eye. 

Figure\ \ref{fig:TWsummary} also demonstrates the strong interplay between the strength of the first-order phase transition, which is controlled by its density width, and the observable consequence in the mass-radius sequence.  Comparing the purple dashed and brown dot-dashed curves, one can see that they both have identical EoSs at $n_B$ below the first-order phase transition, and the only discernible difference is the width of the phase transition. A stronger first-order phase transition leads to a more pronounced kink in the mass-radius curve, which then leads to a connected mass twin sequence. In contrast, a milder first-order phase transition leads to a much milder kink (which is present in the purple mass-radius curve but cannot be discerned by eye), leading to a non-twin mass-radius sequence.  

\subsubsection{Disconnected twins are probably not possible for $M\geq 2.5 $~M$_\odot$}

Recent Ref.~\cite{Blaschke:2020vuy} has established that it is possible to produce ultra-heavy disconnected mass twins, as long as the second stable branch occurs at very low masses, i.e.~a low disconnected mass twin.  In order to produce such a branch, it was shown that one must introduce structures in the EoS at very low densities. Until now, we had not considered introducing such structure, but in this subsection we will do so to study this possibility further.

\begin{figure*}[t]
\centering
\begin{tabular}{c c c c}
\includegraphics[width=0.24\linewidth]{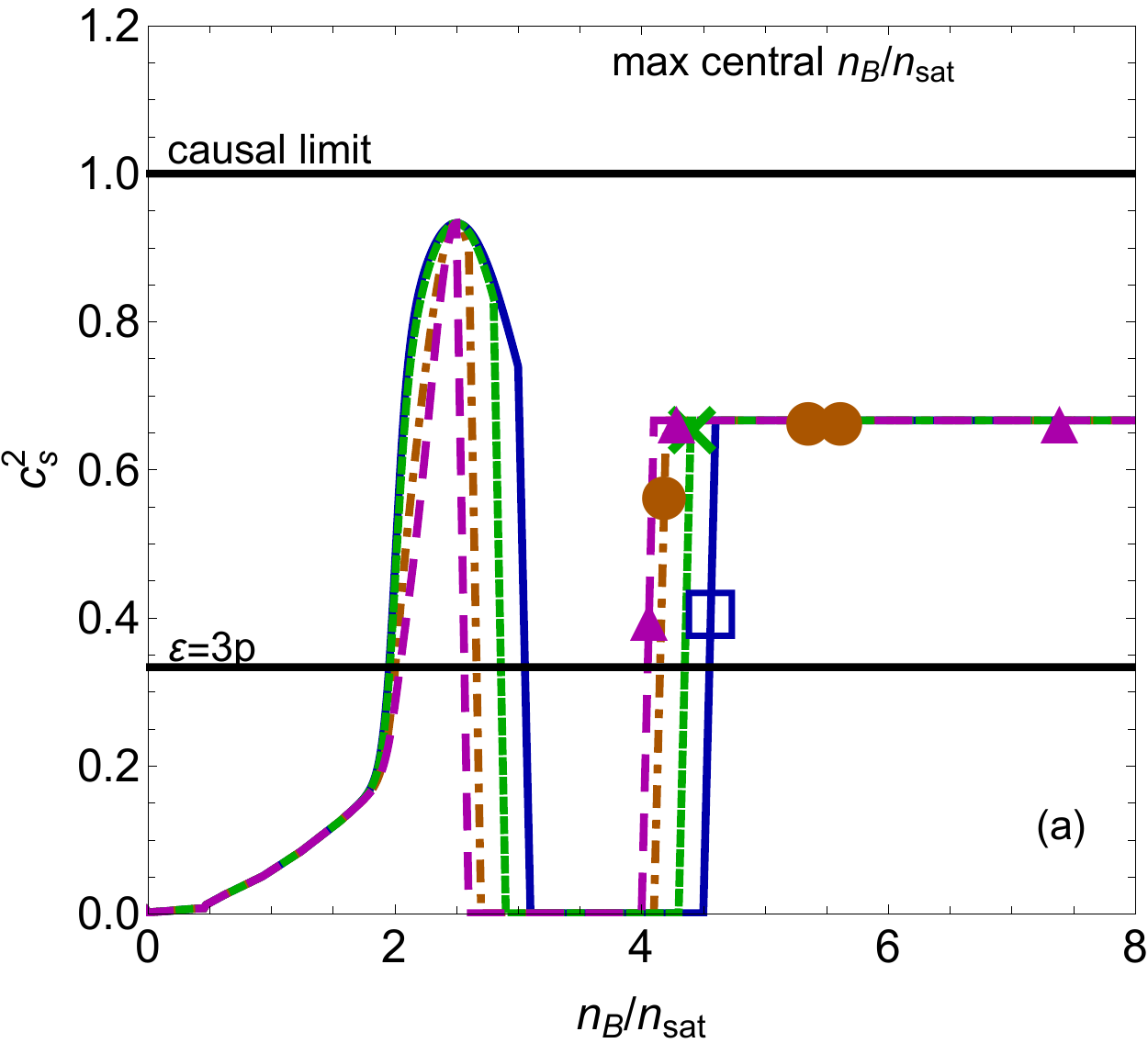} & \includegraphics[width=0.24\linewidth]{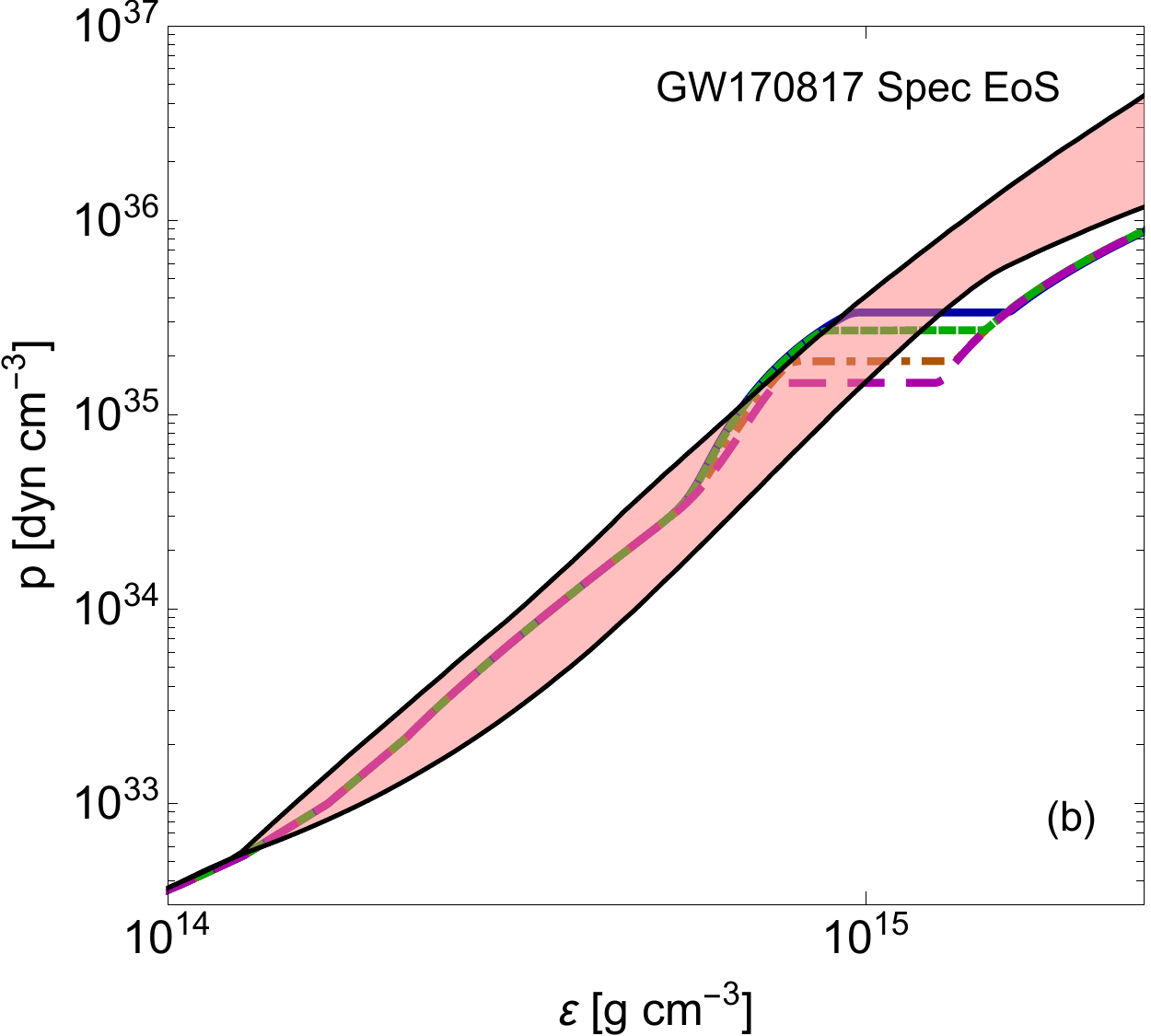} & \includegraphics[width=0.24\linewidth]{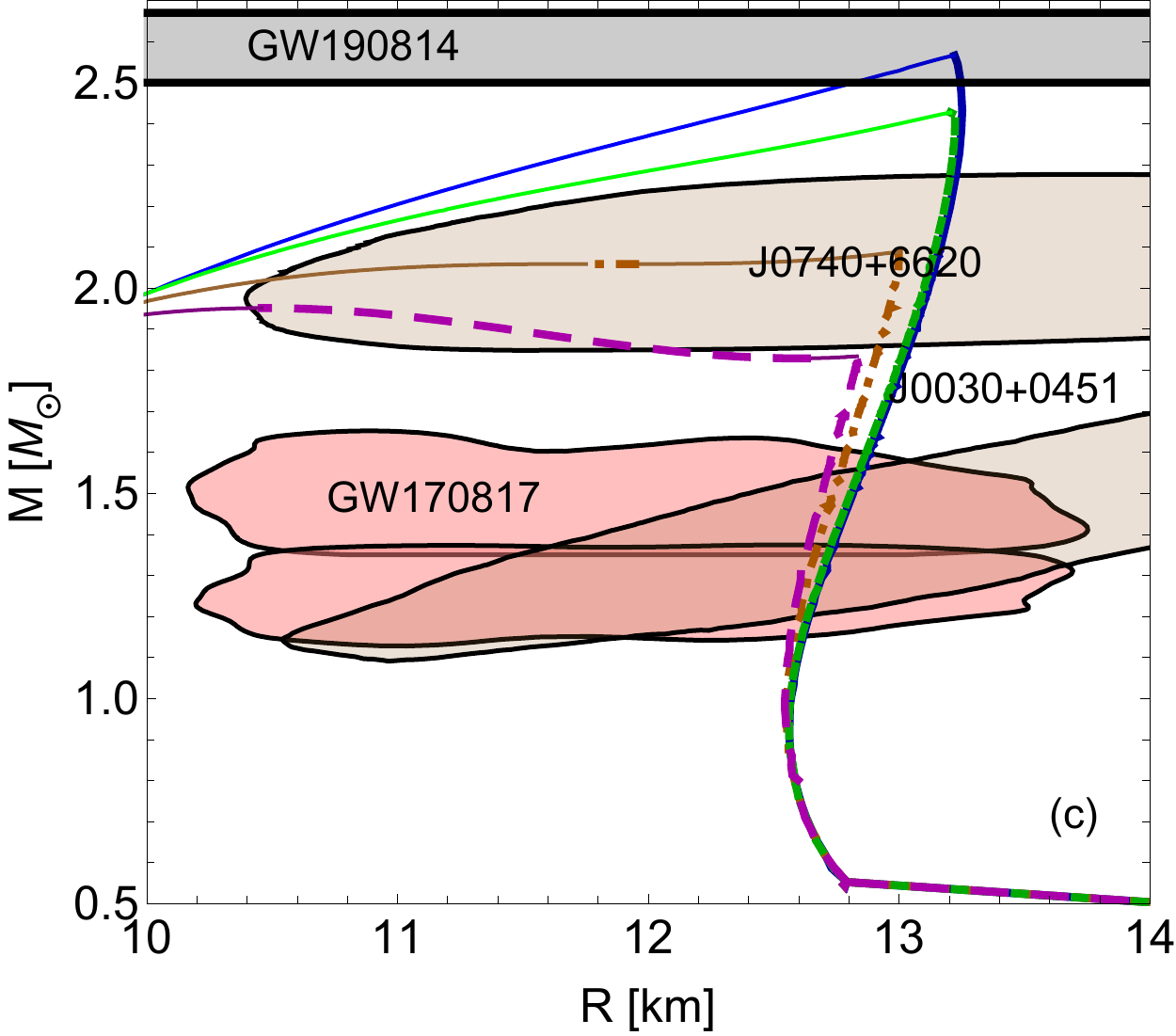} & \includegraphics[width=0.24\linewidth]{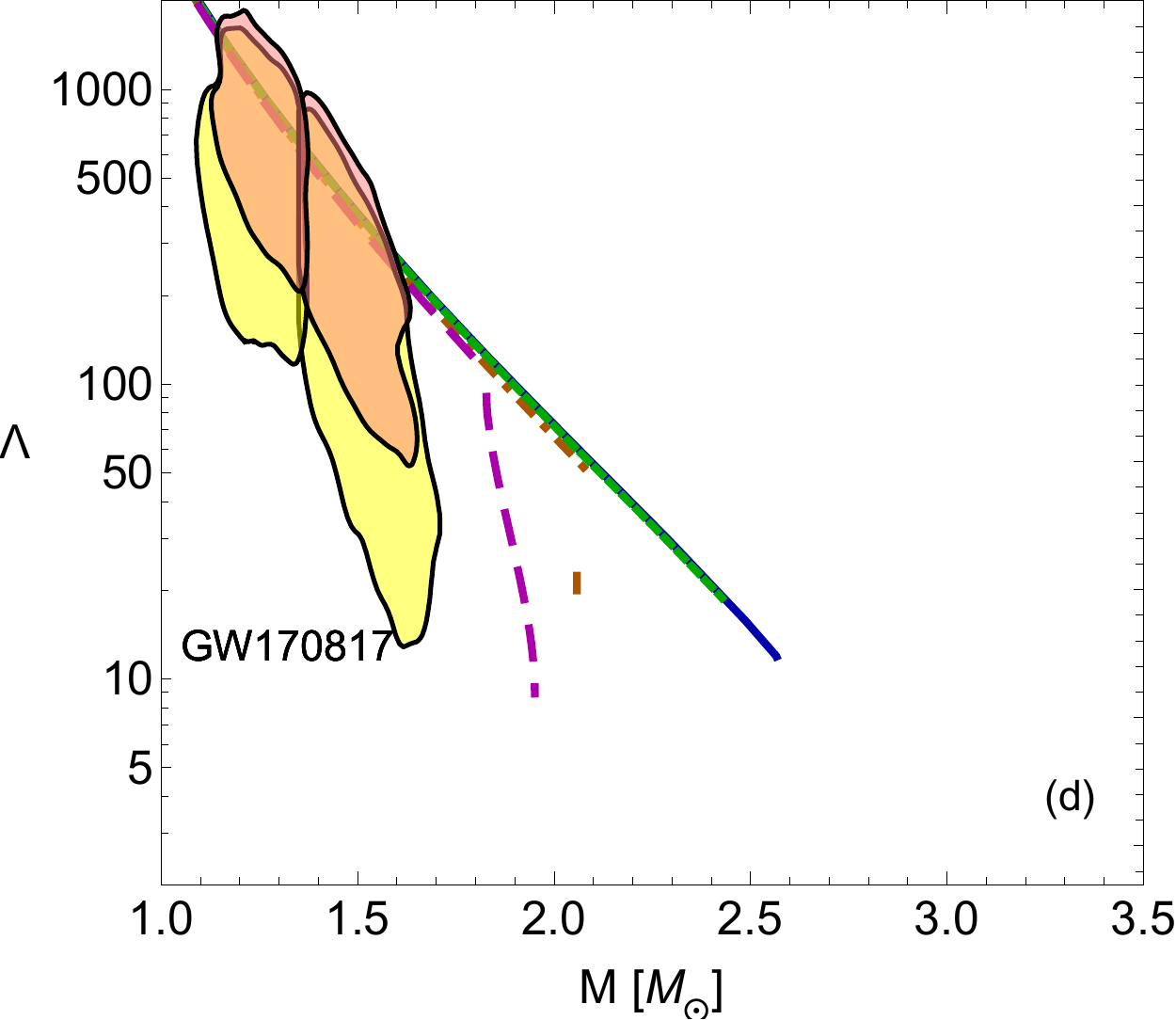}
\end{tabular}
\caption{(Color online) Same as Fig.~\ref{fig:TWsummary}, but for a sub-family of EoSs that contain bumps in the speed of sound of different widths at the same location, followed by a first-order phase transition of the same width in density, and a region of the same constant speed of sound (i.e.~a plateau). As the bump widens, the mass-radius curves transition from disconnected mass twins to a single stable branch, while the tidal deformability goes from nearly vertical to slanted.
}
\label{fig:TWpeakwidth}
\end{figure*}

We first consider EoSs with a weak first-order phase transition right around $n_{sat}$, followed by a bump. A bump is required to reach up to mass of $M\geq 2.5 $~M$_\odot$. Figure~\ref{fig:Blatsummary} shows the result of this exercise, where the red solid and blue dot-dashed curves represent EoSs with very weak and weak first-order phase transitions respectively. The mass-radius sequences satisfy all observational constraints, but there is no discernible disconnected mass twin behavior in the mass-radius curve. Depending on the strength of this low-$n_B$ first-order phase transition, one could produce a disconnected mass twin at very low masses, but the discontinuity would occur below the Chandrasekhar mass and it would be so small to be essentially un-observable. 

Let us now investigate EoSs with a large, thin bump in $c_s^2$ slightly above $n_{sat}$, followed by a wide first-order phase transition. The skinny bump is needed so that the first-order phase transition occurs below the maximum central density probed by these neutron stars. As Fig.~\ref{fig:Blatsummary} shows (green curves), this leads to a very clear high disconnected mass twin that satisfies all NICER constraints, but leads to stars with radii that is outside the LVC constraints. We investigated a few variations of this model, and were not able to produce a high disconnected mass twin that led to ultra-heavy neutron stars. A thorough statistical investigation of this EoS model would be required to confirm these findings, but this is outside the scope of this paper. 

Thus, we are forced to conclude the following.  If the light component of the GW190814 event is a neutron star, then a first-order phase transition is possible, but this most likely not lead to a disconnected mass twin sequence. This is in agreement with the findings from Ref.~\cite{Christian:2020xwz}.

\subsection{Properties of heavy twins with first-order phase transitions}

Let us now no longer constrain the EoS to reproduce a large maximum mass star, but rather seek to understand qualitatively how certain features in the $c_s^2$, combined with a first-order phase transition, affect the mass-radius relation and the existence of connected and disconnected twins, and kinky mass-radius sequences for heavy neutron stars (i.e.~for masses $M\geq 2$~M$_{\odot}$ consistent with NICER observations of PSR J0740+6620~\cite{Riley:2021pdl,Miller:2021qha}). We focus on stars with a speed of sound that combines a large initial peak followed by a strong first-order phase transition.  Throughout this exercise, we vary the width of the peak, the width of the phase transition, the slope of the rise to a constant $c_s^2$ after the first-order phase transition, and the height of the $c_s^2$ plateau at large densities.  This is not meant as a thorough analysis, since for that one would require a statistical study with a significantly larger number of EoSs; rather the aim of this investigation is to present a qualitative understanding of how a first-order phase transition would manifest in both the mass-radius and the $\Lambda$--$M$ diagrams for EoSs more complex than what is typically considered for mass twins. 

\subsubsection{From disconnected twins to unstable sequences through bumps in the speed of sound}

\begin{figure*}[t]
\centering
\begin{tabular}{c c c c}
\includegraphics[width=0.24\linewidth]{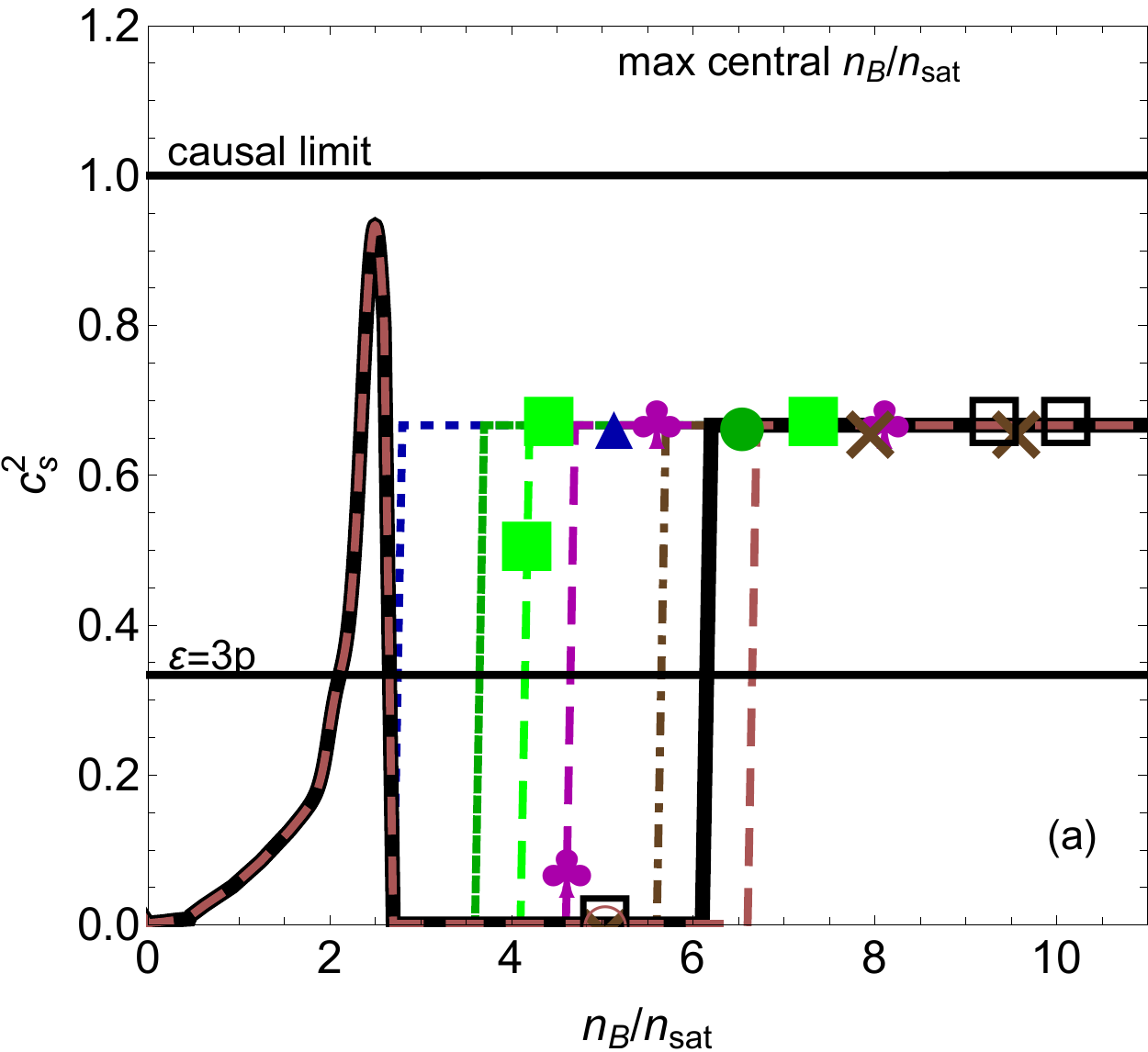} & \includegraphics[width=0.24\linewidth]{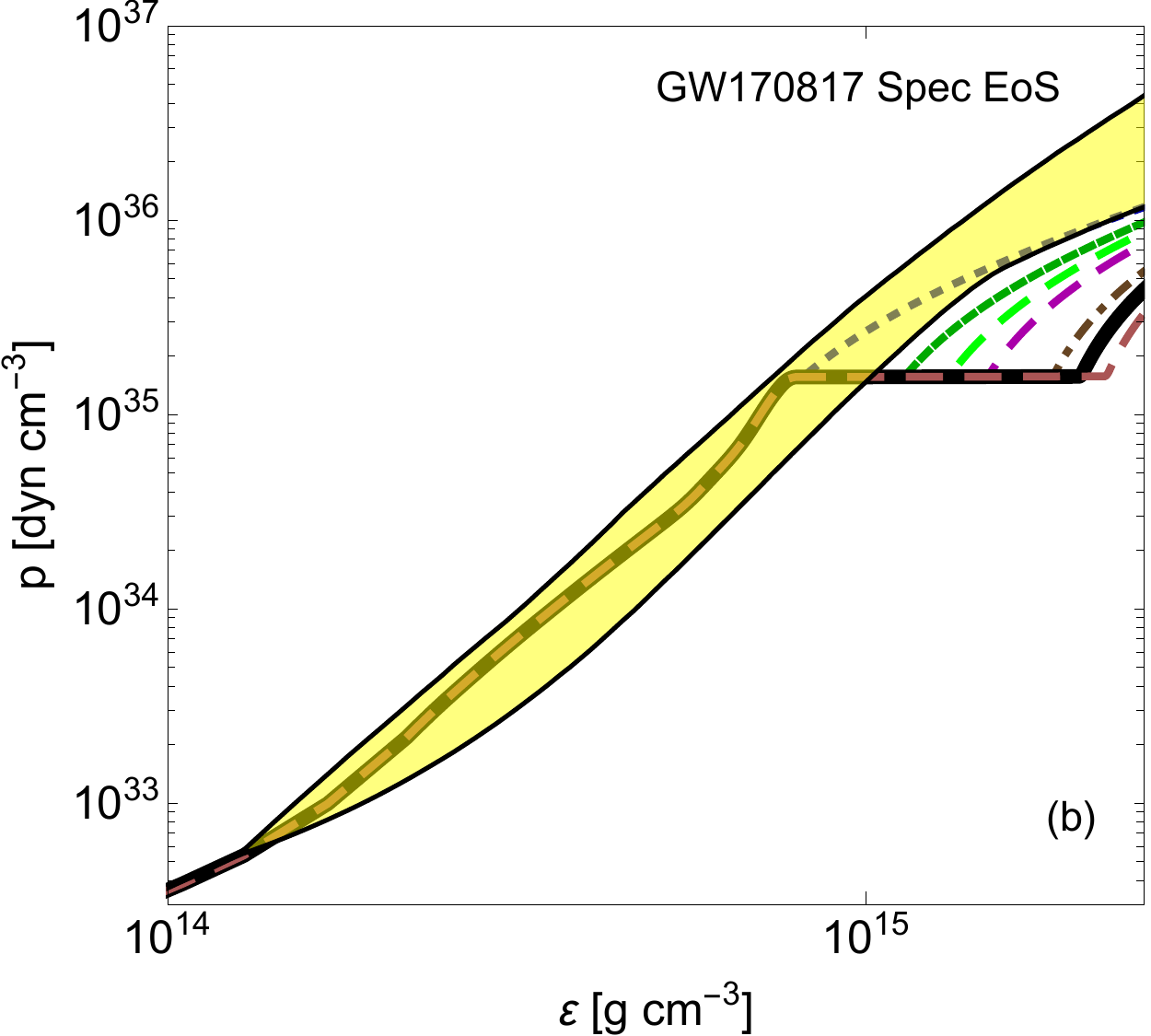} & \includegraphics[width=0.24\linewidth]{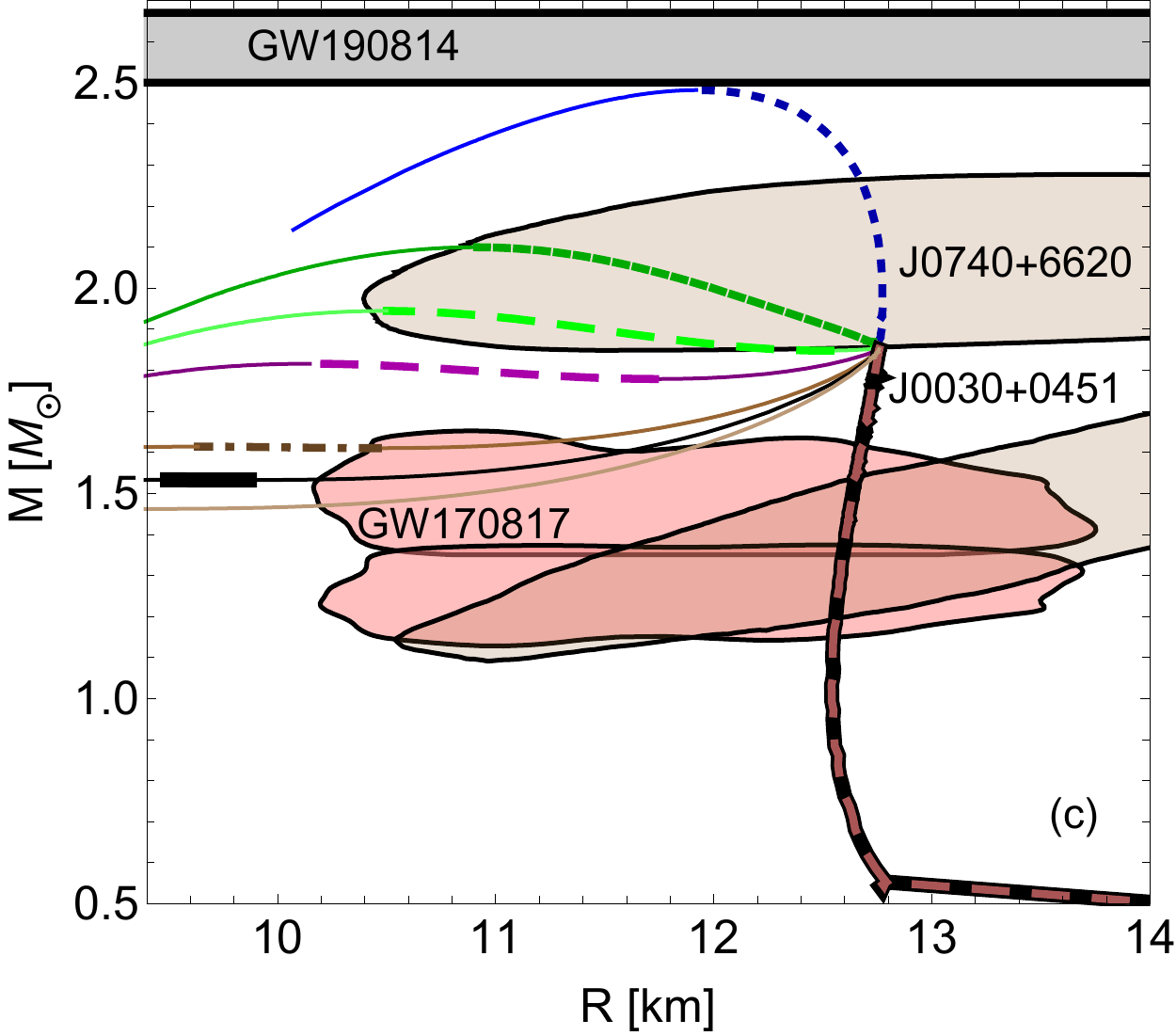} & \includegraphics[width=0.24\linewidth]{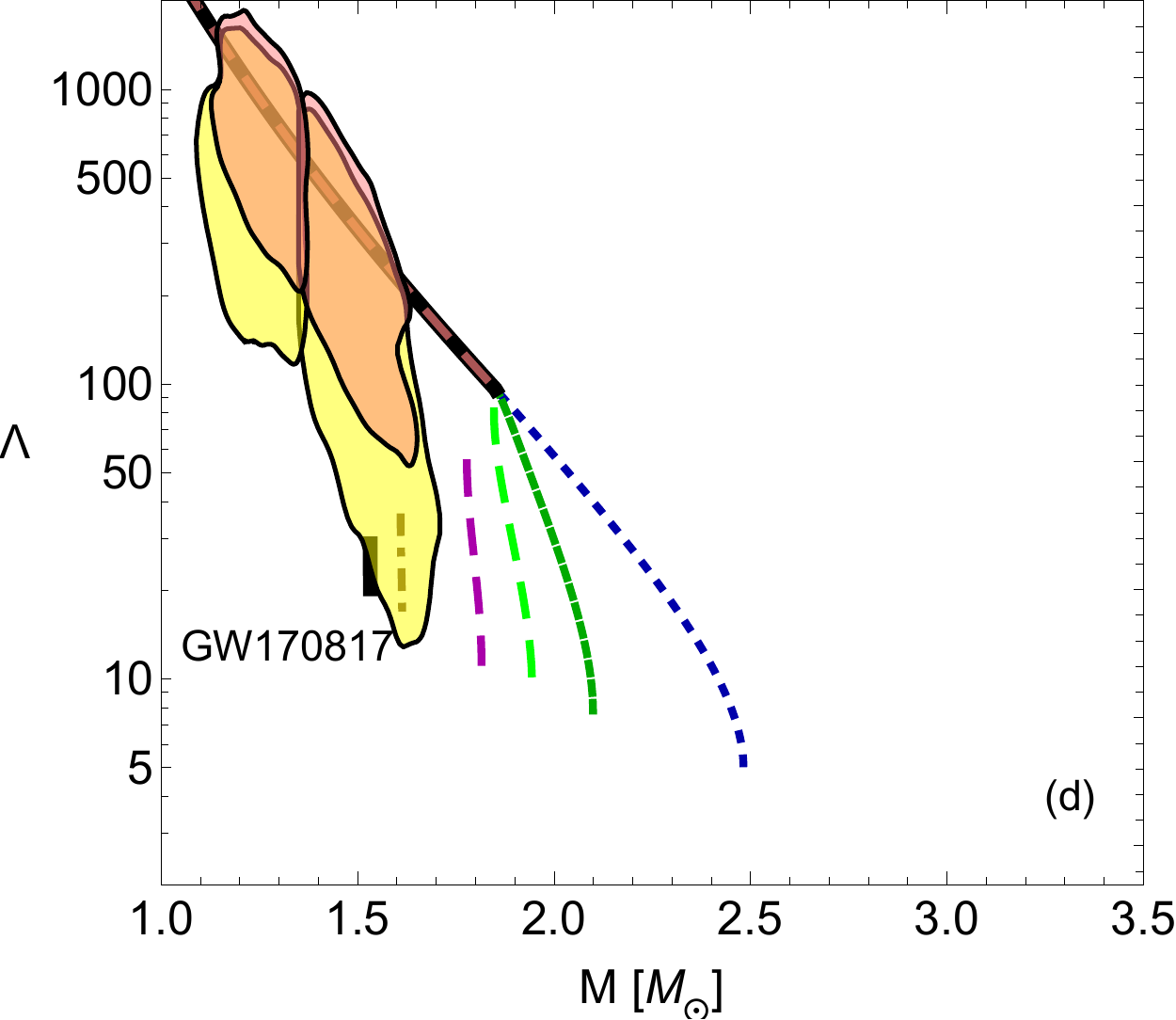}
\end{tabular}
\caption{(Color online) Same as Fig.~\ref{fig:TWsummary}, but for a sub-family of EoSs with different strengths of the first-order phase transition. The strength is related to having a connected branch, kinky, disconnected or connected twins, in addition to the radius range for twin stars.}
\label{fig:TWPTwidth}
\end{figure*}

Our first step is to look into a variety of bump widths in the speed of sound. Figure\ \ref{fig:TWpeakwidth} varies the width of the bumps while keeping the width of the phase transition ($n_B^{1PT,low}-n_B^{1PT,hi}=const$) and the location of the bump fixed; following the phase transition, we use a constant $c_s^2=0.7$. Therefore, bumps that are wider present a first-order phase transition at slightly higher densities, as shown in the far left panel of Fig.~\ref{fig:TWpeakwidth}. As one increases the width of the $c_s^2$ bump, the mass-radius sequence transitions from a disconnected mass twin to a single stable branch. Just as in Sec.\ \ref{sec:25msol}, the wider the peak, the larger the maximum mass of the stars. Finally, the far right panel shows that the second stable branch of the disconnected mass twin sequence presents an almost vertical fall in tidal deformability, thus, leading to much smaller $\Lambda$ than their equivalent main branch counterpart of the same mass. This could also be seen for the green line in Fig.~\ref{fig:Blatsummary}. See Refs.~\cite{Han:2018mtj} for a discussion of the tidal deformability in the presence of first-order phase transitions and twin stars \cite{Christian:2018jyd,Miao:2020yjk,Benitez:2020fup,Bauswein:2020xlt}.

Let us now discuss the maximum central densities that these sequences achieve. The maximum central density of the first stable branch and the central densities of the second stable branch (the minimum and maximum) are shown in the far left panel of Fig.~\ref{fig:TWpeakwidth} with symbols (circles and triangles for the mass-twins and a square and an $\times$ for EoSs with only one stable branch). All EoSs reach their maximum central density of the first stable branch right after the first-order phase transition (around $n_B\sim 4~n_{\rm sat}$). Thus, we see that the effect of the first-order phase transition is to make stars unstable. After the phase transition, the star either remains unstable or a second stable branch appears at even larger densities, depending on the width/strength of the phase transition.  The mass twins' second stable branch begins at a slightly larger baryon density and allows the stars to reach a larger maximum central baryon density (as high as $n_B \sim 7~n_{\rm sat})$.

As we have clearly demonstrated in Fig.~\ref{fig:TWpeakwidth}, we are easily able to produce mass twins that reach $M\sim 2.1 $~M$_\odot$, and thus are consistent with the NICER observations of PSR J0740+6620~\cite{Riley:2021pdl,Miller:2021qha}.  In fact, due to the large posterior distribution in radius emerging from the PSR J0740+6620 observations, it is not possible to determine whether this pulsar belongs to the first or second stable branch, if it were to be a twin. For this particular example (the brown EoS in Fig.\ \ref{fig:TWpeakwidth}), the first stable branch reaches a maximum mass of $M\sim 2.09$~M$_\odot$ and a radius of $R\sim 13$~km. The second stable branch begins at $M\sim 2.06$~M$_\odot$ and $R\sim 11.9$~km and then ends at $M\sim 2.06$~M$_\odot$ and $R\sim 11.7$~km.  Thus, if one knew somehow that this pulsar was a mass twin and one wished to determine whether it belonged to the first or the second stable branch, one would require a much higher precision in the measurement of the radius, i.e.~the posterior distribution of radius at 90\% confidence would need to have a width in radius no larger than $\sim 1.3$~km. 

What would be required for a LIGO/Virgo observation to distinguish between stable branches of twin stars? Of course, this depends on the particular mass twin sequence considered. For the GW170817 event, which is the only one for which the LVC was able to measure the tidal deformability, the $\Lambda$--$M$ curves below $1.7 $~M$_\odot$ are all approximately the same because the phase transition happens at larger densities. Therefore, one cannot use this event to distinguish between stable branches of the mass twin family considered here. One also cannot use the PSR J0740+6620 event that NICER observed to place constraints with LIGO/Virgo data, because this pulsar was not observed by gravitational wave detectors. We can, however, determine the precision that LIGO/Virgo would have to measure $\Lambda$ to in order to distinguish between the two stable branches. Let us then assume that LIGO/Virgo has observed the gravitational waves emitted by a binary system, where one of them has a mass of $\sim 2.06$, and at a sufficiently high signal-to-noise ratio so that $\Lambda$ could also be estimated. If so, the LIGO/Virgo posterior would have to measure a $\Lambda$ as low as $20$ with less than a 100\% error; at 90\% confidence, this could be achieved with a sufficiently loud event when LIGO/Virgo reach design sensitivity. 

\subsubsection{Impact of the strength of first-order phase transition on the mass-radius sequence}

\begin{figure*}[t]
\centering
\begin{tabular}{c c c c}
\includegraphics[width=0.24\linewidth]{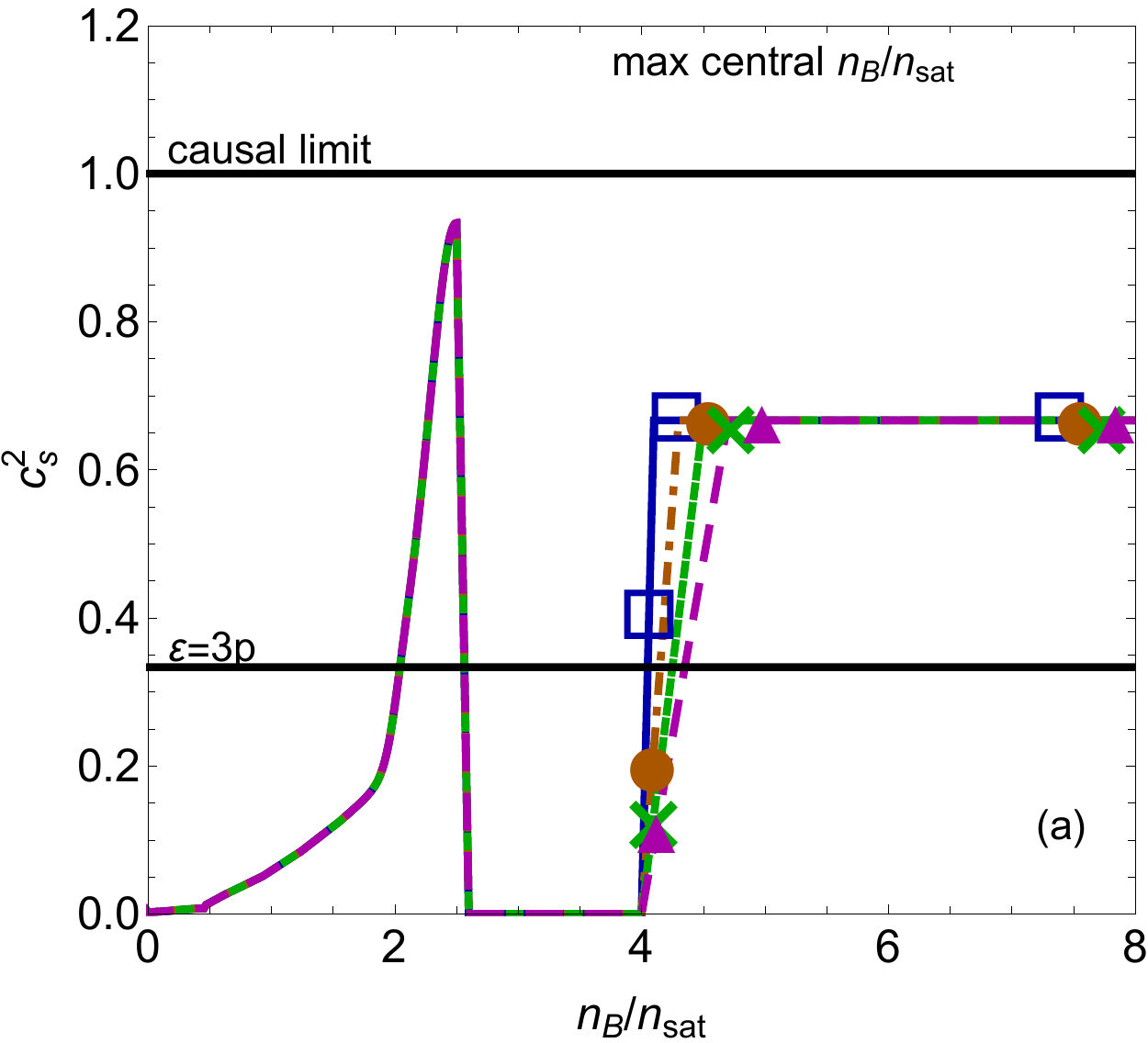} & \includegraphics[width=0.24\linewidth]{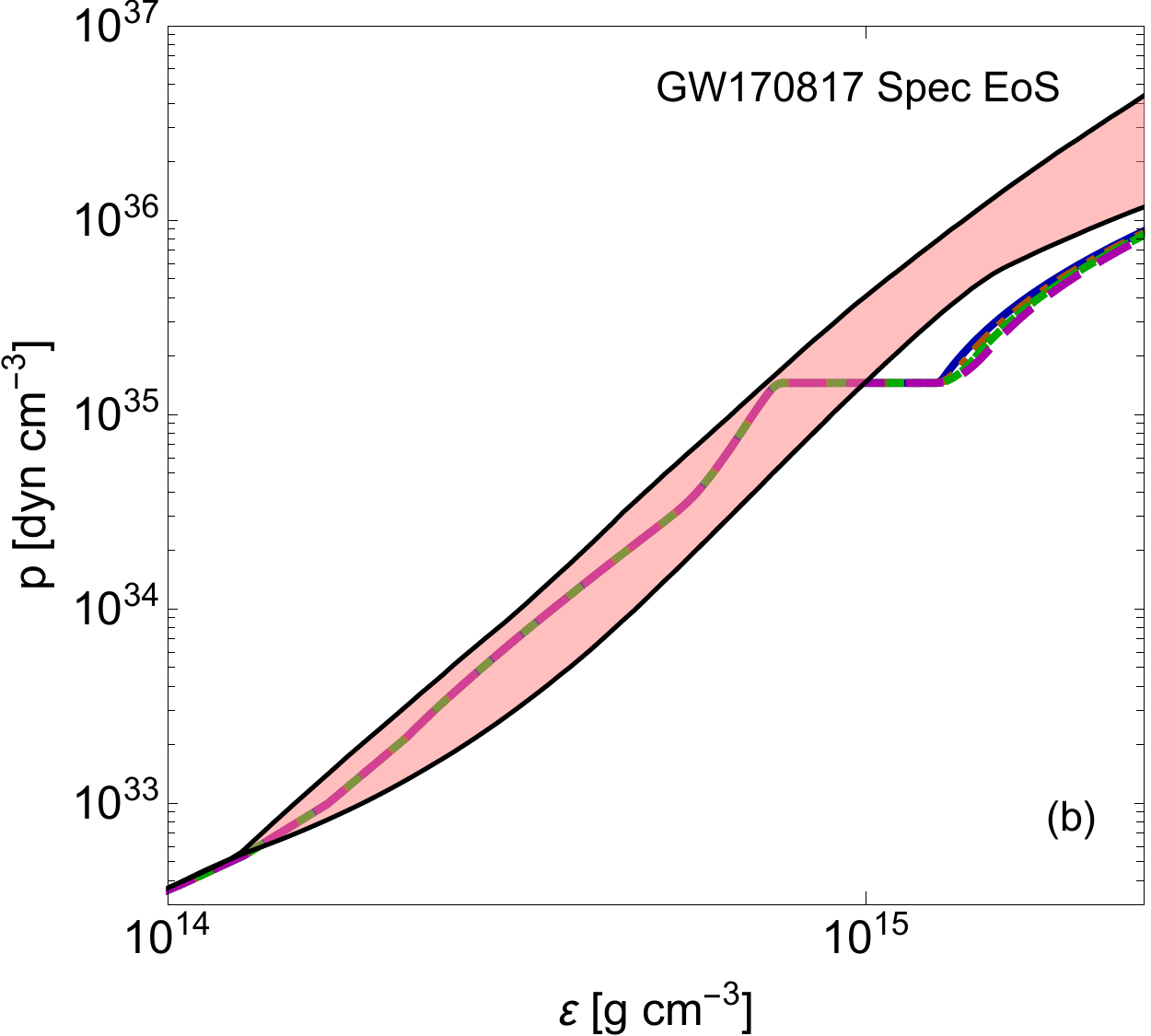} &  \includegraphics[width=0.24\linewidth]{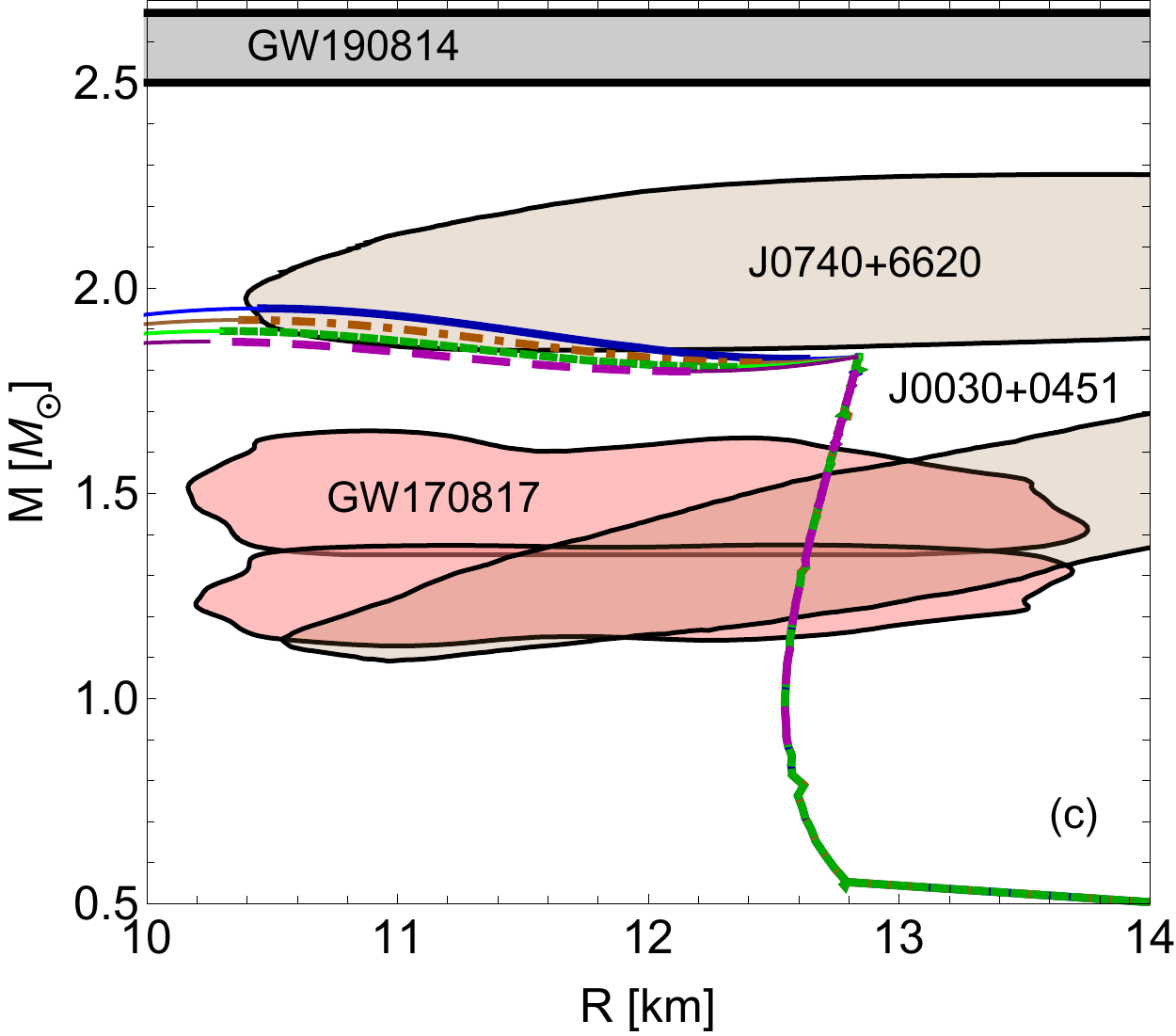} & \includegraphics[width=0.24\linewidth]{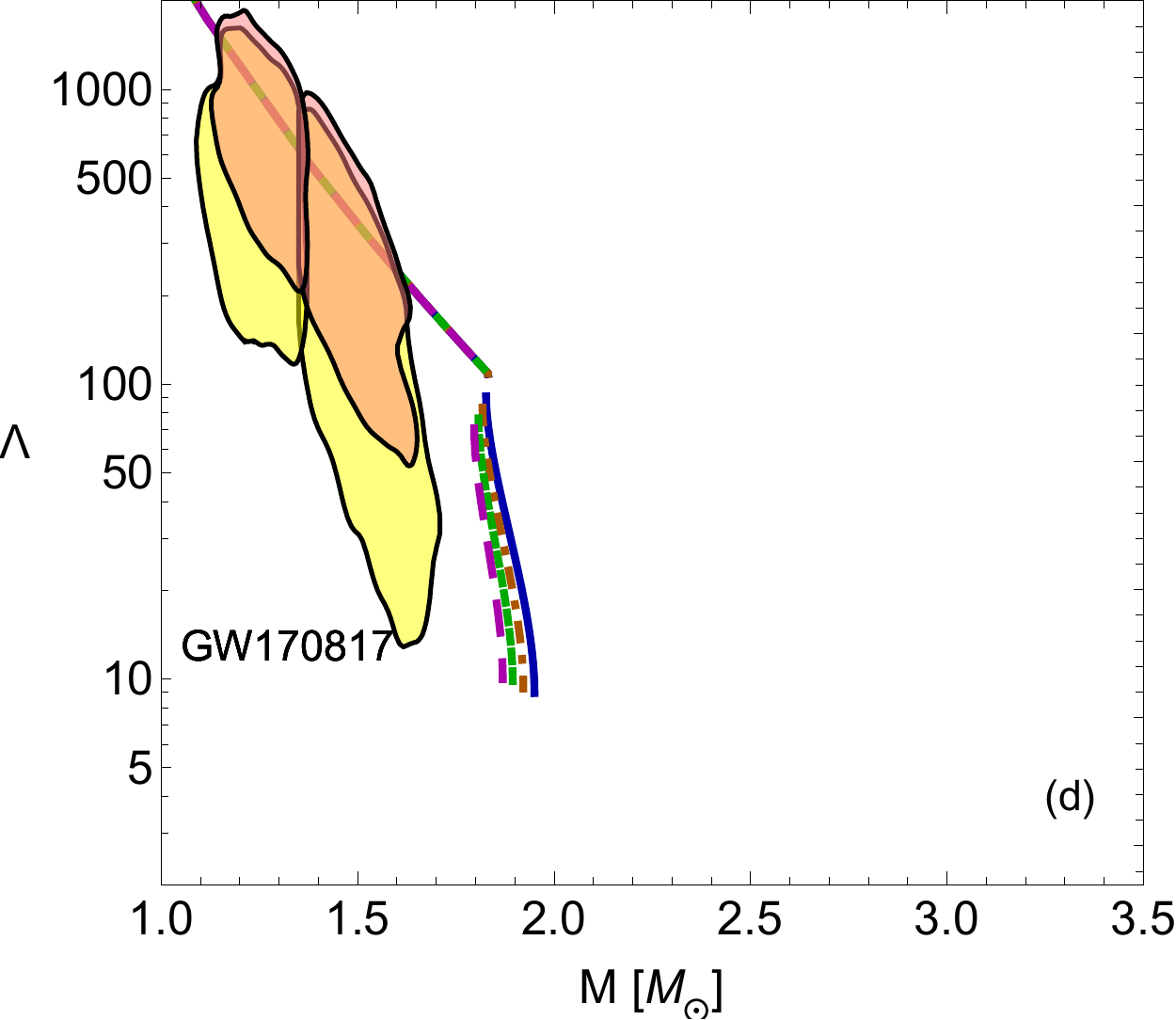}
\end{tabular}
\caption{(Color online) Same as Fig.~\ref{fig:TWsummary}, but for a sub-family of EoSs that have bumps in the speed of sound of the same width,  followed by a first-order phase transition of the same length in density, but different rises to a region with constant speed of sound (i.e.~plateaus). The steeper the rise of the speed of sound after the first-order phase transition, the (slightly) higher the maximum mass of the second stable branch.}
\label{fig:TWslant}
\end{figure*}

If a first-order phase transition exists within neutron stars, it is not yet clear how strong of a phase transition it is. What are the consequences of a strong vs.~weak first-order phase transition on the mass-radius sequence and tidal deformability? Let us begin with a very weak first-order phase transition that is essentially only a spike to $c_s^2\rightarrow 0$, and then systematically increase the width across $n_B$ where $c_s^2= 0$ until we reach such a large first-order phase transition that the second branch is no longer stable. The resulting speeds of sound and EoS are shown in Fig.\ \ref{fig:TWPTwidth}. Note that for very strong first-order phase transitions the pressure is well outside the LIGO 90\% posterior obtained from their spectral EoS analysis using the GW170817 event.

\begin{figure*}[t]
\centering
\begin{tabular}{c c c c}
\includegraphics[width=0.24\linewidth]{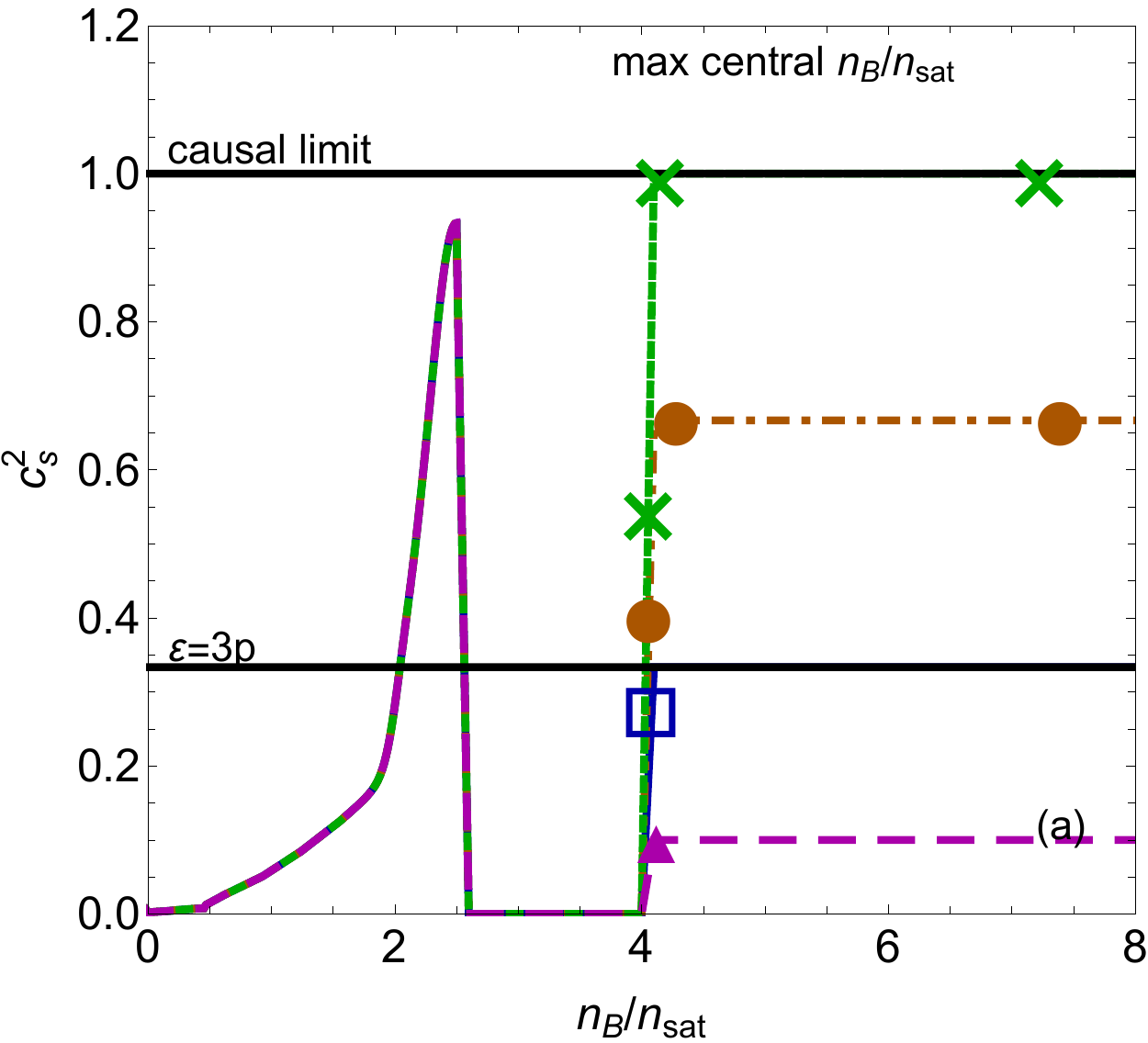} & \includegraphics[width=0.24\linewidth]{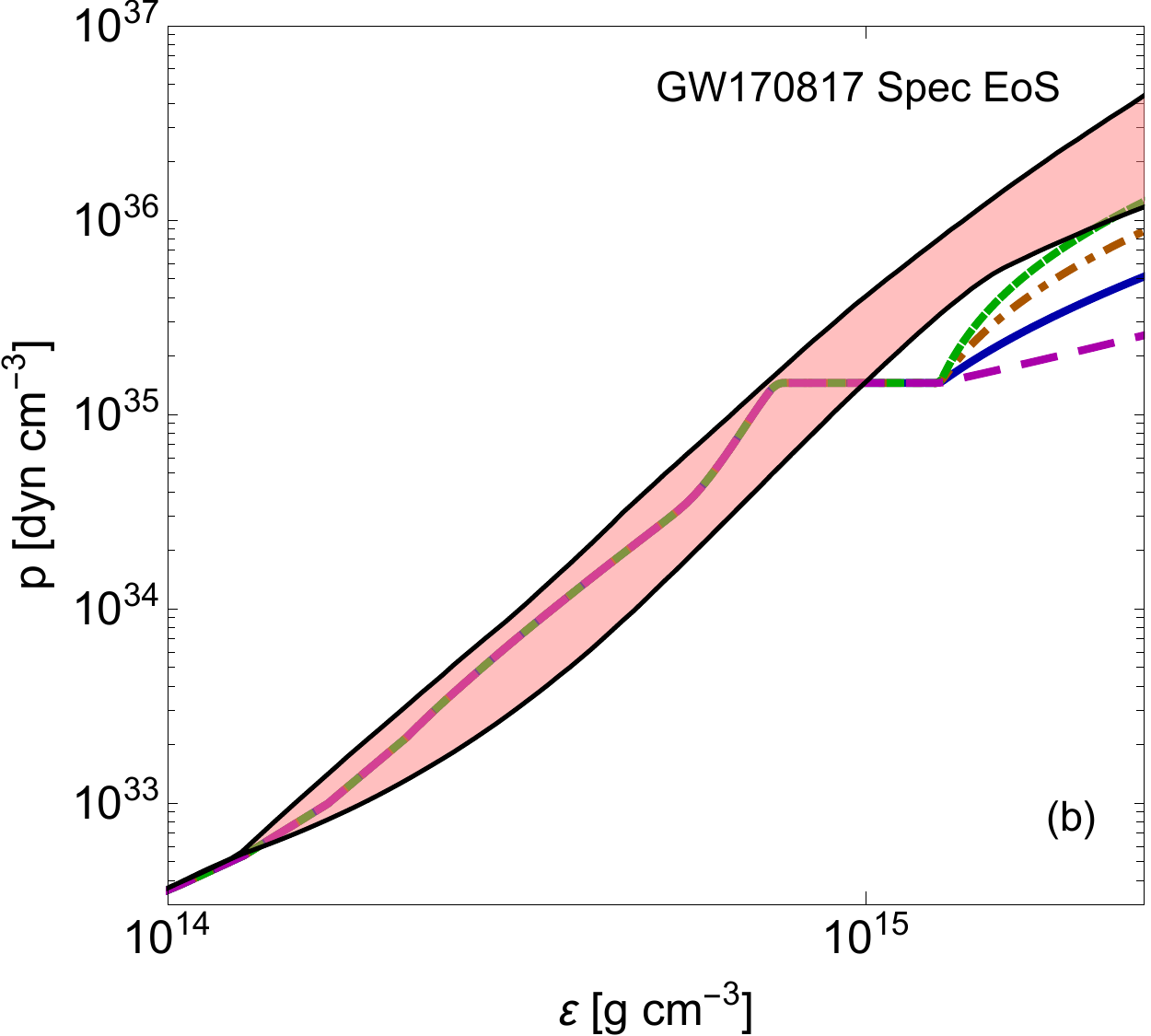} & \includegraphics[width=0.24\linewidth]{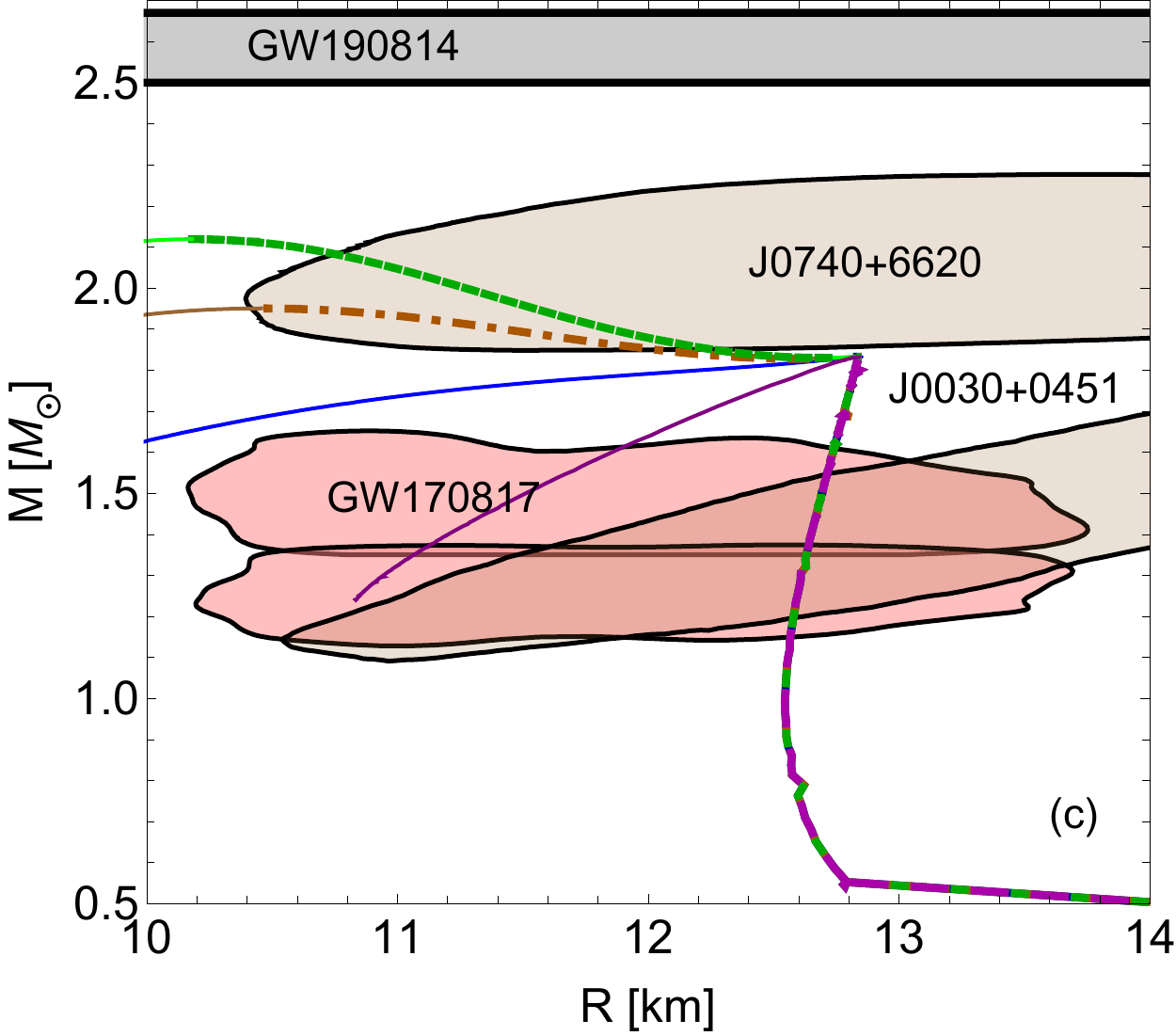} & \includegraphics[width=0.24\linewidth]{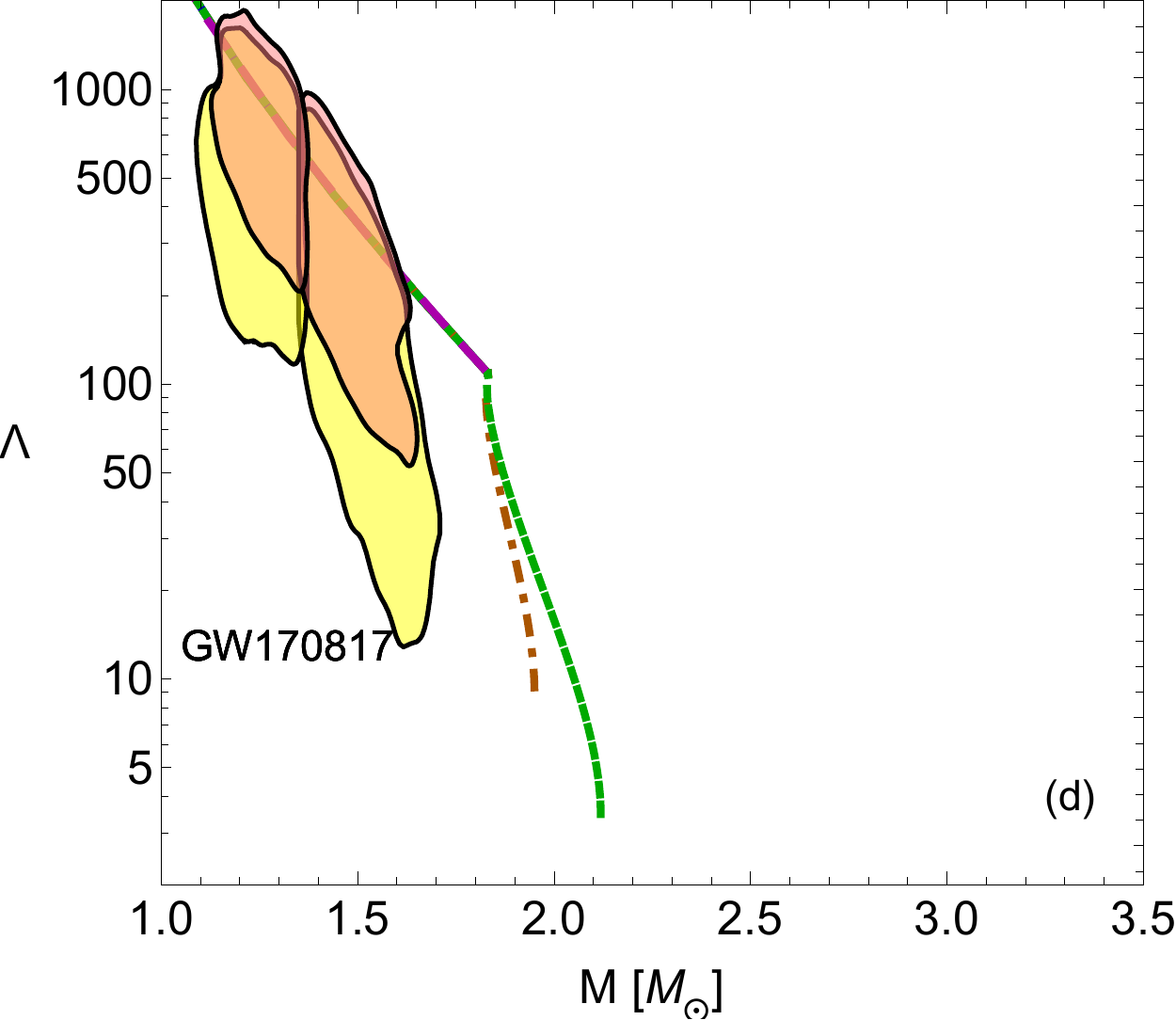}
\end{tabular}
\caption{(Color online) Same as Fig.~\ref{fig:TWsummary}, but for a sub-family of EoSs that have bumps in the speed of sound of the same width,  followed by a first-order phase transition of the same length in density, that rise rapidly to a plateau in the speed of sound of different heights. For sufficiently high plateaus, a disconnected twin branch appears.}
\label{fig:TWend}
\end{figure*}

Weak first-order phase transitions lead to two connected branches with a kink so small that it is not visible, as shown in the dark blue dashed curve in Fig.\ \ref{fig:TWPTwidth}.  As the phase transition strengthens, a kink in the mass-radius sequence is formed while the maximum mass lowers, as shown in the dark green solid curve in Fig.\ \ref{fig:TWPTwidth}.  Eventually, for a strong enough first-order phase transition, the second stable branch disconnects from the first stable branch and disconnected twins appear, as shown in the light green dashed curve in Fig.\ \ref{fig:TWPTwidth}. Continuing to increase the strength of the first-order phase transition past this point still maintains a disconnected mass twin sequence, but the second stable branch shortens and is shifted to lower masses and radii, as shown in the maroon long dashed, brown dot-dashed, and solid black curves in Fig.\ \ref{fig:TWPTwidth}.  Eventually, a too strong of a first-order phase transition shrinks the stable branch to the point where it disappears entirely, as shown in the light brown curve in Fig.\ \ref{fig:TWPTwidth}.  Analogously, strengthening the first-order phase transition forces the minimum $\Lambda$ to be larger and to occur at lower masses, with a stronger disconnect from the first stable branch. 

The effects discussed in Fig.~\ref{fig:TWPTwidth} are in agreement with the investigation of the strength of the phase transition and its influence on twin stars from Ref.~\cite{Ayriyan:2017nby}. In this reference, the authors investigated the effect of introducing a mixed phase in the EoS. They found that weakening the phase transition (in this case by having a softer dip in the speed of sound) changes the dense matter configurations from twins to connected branches.

\subsubsection{Impact of the speed of sound past the first-order phase transition}

Let us now consider the impact of the speed of sound past the first-order phase transition in the properties of these heavy neutron stars. First, we consider adding a constant slope in the speed of sound to transition between $c_s^2 = 0$ to some constant value of $c_s^2$, 
as shown in the far left panel of Fig.\ \ref{fig:TWslant}. Why should one consider a steep rise in the speed of sound after a first-order phase transition?  Rapid raises in speed of sound at large density can be consistently introduced in the EoS of dense matter through, for example, an excluded volume prescription, which is used to account for the finite volume of hadrons~\cite{Baacke:1976jv,Steinheimer:2010ib,Satarov:2014voa}. This has been used, for example, to allow for new degrees of freedom, such as deconfined quarks, to appear in the modelling of dense matter in the core of neutron stars~\cite{Dexheimer:2014pea,Dexheimer:2012eu,Mukherjee:2017jzi,Motornenko:2019arp} (see e.g.~blue dashed line in Fig.~\ref{fig:known}).

The third panel of Fig.\ \ref{fig:TWslant} shows that a steeper slope after the phase transition increases the maximum mass of stars, specifically raising slightly the maximum mass of the second branch. Additionally, the steepness of the slope is directly connected to the slope in mass versus radius of the second stable branch.  What this implies is that a softer slope after the phase transition leads to a flatter second stable branch, whereas a steep slope leads to a second stable branch that increases more in mass as the radius decreases. This same effect is seen in the $\Lambda$--$M$ curves of the far right panel.  Furthermore, this panel demonstrates that one does not require a step transition to the causal limit in order to produce a mass-twin; rather, there is some flexibility in the form of $c_s^2$ after the first-order phase transition that still allows for mass twins.

Let us now consider the influence of the high density part of $c_s^2$, beyond the first-order phase transition. Figure\ \ref{fig:TWend} demonstrates how a higher speed of sound plateau after a first-order phase transition largely affects the mass-radius and $\Lambda$--$M$ relations.  A lower plateau leads to no stable second branch, while higher plateaus generate either a connected or a disconnected twin branch, depending on how large the constant value of $c_s^2$ is after the phase transition. In this scenario, we are able able to produce a twin mass star that is consistent with the NICER observations of PSR J0740+6620~\cite{Riley:2021pdl,Miller:2021qha}. Unlike the brown EoS in Fig.\ \ref{fig:TWpeakwidth}, where both the fist branch and twin branch reached masses consistent with PSR J0740+6620, in this scenario only the high-density regime of the EoS would reach the mass-radius range of PSR J0740+6620.  

\section{Comparison between neutron stars with crossover and first-order phase transitions}
\label{sec:comparison}

Up until now, we have discussed EoS models with crossover structure separately from models with first-order phase transitions. In this section, we bring both of these features together so that we can compare and contrast them. We will begin with a discussion of how central baryon densities are affected by these structures in the speed of sound. We will then describe whether these structures affect the I-Love-Q relations. 

\subsection{Central baryon density and maximum masses}

EoSs that produce mass twins reach much larger central densities in stable stars when compared to nearly equivalent EoSs that only produce one stable branch \cite{Alford:2013aca}.  Figures~\ref{fig:TWpeakwidth} to~\ref{fig:TWend} demonstrate this quite well. The EoSs that only have one stable branch reach baryon central densities of about $n_B\sim 4~n_{\rm sat}$, whereas the mass twins in those figures reach almost up to $n_B\sim 10~n_{\rm sat}$; note that these figures consider EoSs that lead to neutrons stars with maximum masses that are consistent with PSR J0740+6620.  Fig.\ \ref{fig:TWPTwidth} demonstrates nicely that as the phase transition strengthens the maximum central baryon density reaches as well.  

However, if we allow neutron stars to reach higher maximum masses, as is the case in Fig.\ \ref{fig:TWsummary} where all neutron stars reach $M\geq 2.5 $~M$_\odot$, then their maximum central densities are now similar to what we found in Sec.\ \ref{sec:25msol}, i.e.~the maximum central density is $n_B\leq 6~n_{sat}$. Thus, if the light component of the GW190814 event is a neutron star, then the maximum central density within the core for such a sequence could be $n_B\leq 6~n_{\rm sat}$.  If even heavier neutron stars are measured, then this would further restrict the maximum central density to even lower values, as in see in Fig.~\ref{fig:3Msun}, where the constraint of producing stars with $M\geq 2.9 $~M$_\odot$ leads to a maximum central density of $n_B\leq 5~n_{\rm sat}$. 
If we fix the maximum mass, and look at a variety of neutrons stars that reach that maximum mass (or surpass it), the ones that reach the largest maximum central baryon density are also the ones that have the smallest radii at that fixed mass. Thus, information about the radius of heavy neutron stars can help us to also determine the maximum central density reached within these stars (when constrained to reach the same minimum maximum mass).

Before proceeding, a small caveat: the crust model does affect the maximum central baryon density reached in a neutron star sequence, as shown in Fig.~\ref{fig:crusts}. In particular, a softer crust (e.g.~QHC19) results in a larger rise to the causal limit, which results in the lowest central baryon densities. However, the stiffest crust we considered (SKA), which has a shorter rise to the causal limit, leads to a larger maximum central baryon density. 

\subsection{I-Love-Q relations}

\begin{figure}[t]
\centering
\begin{tabular}{c c }
\includegraphics[width=0.48\linewidth]{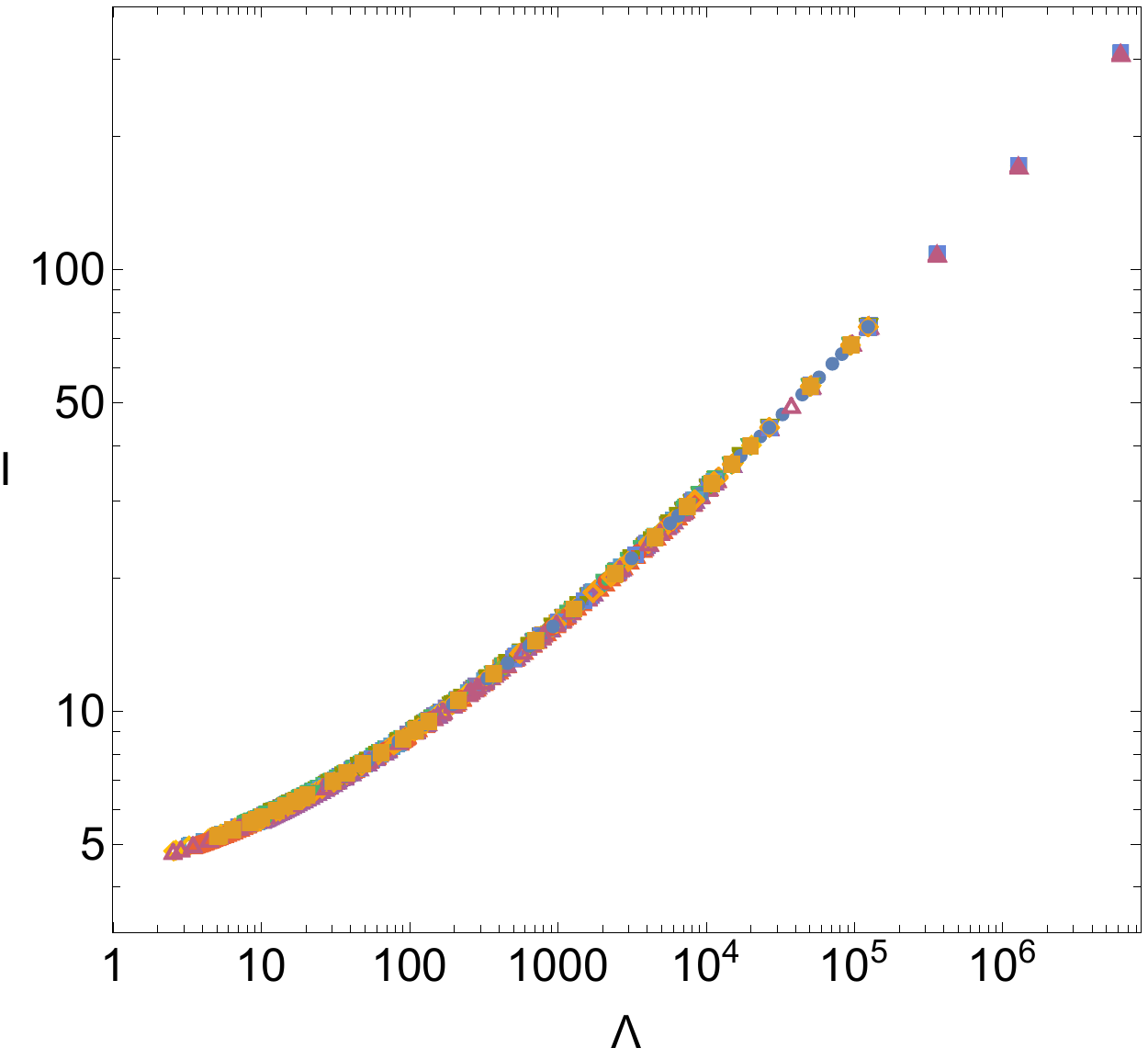} & \includegraphics[width=0.48\linewidth]{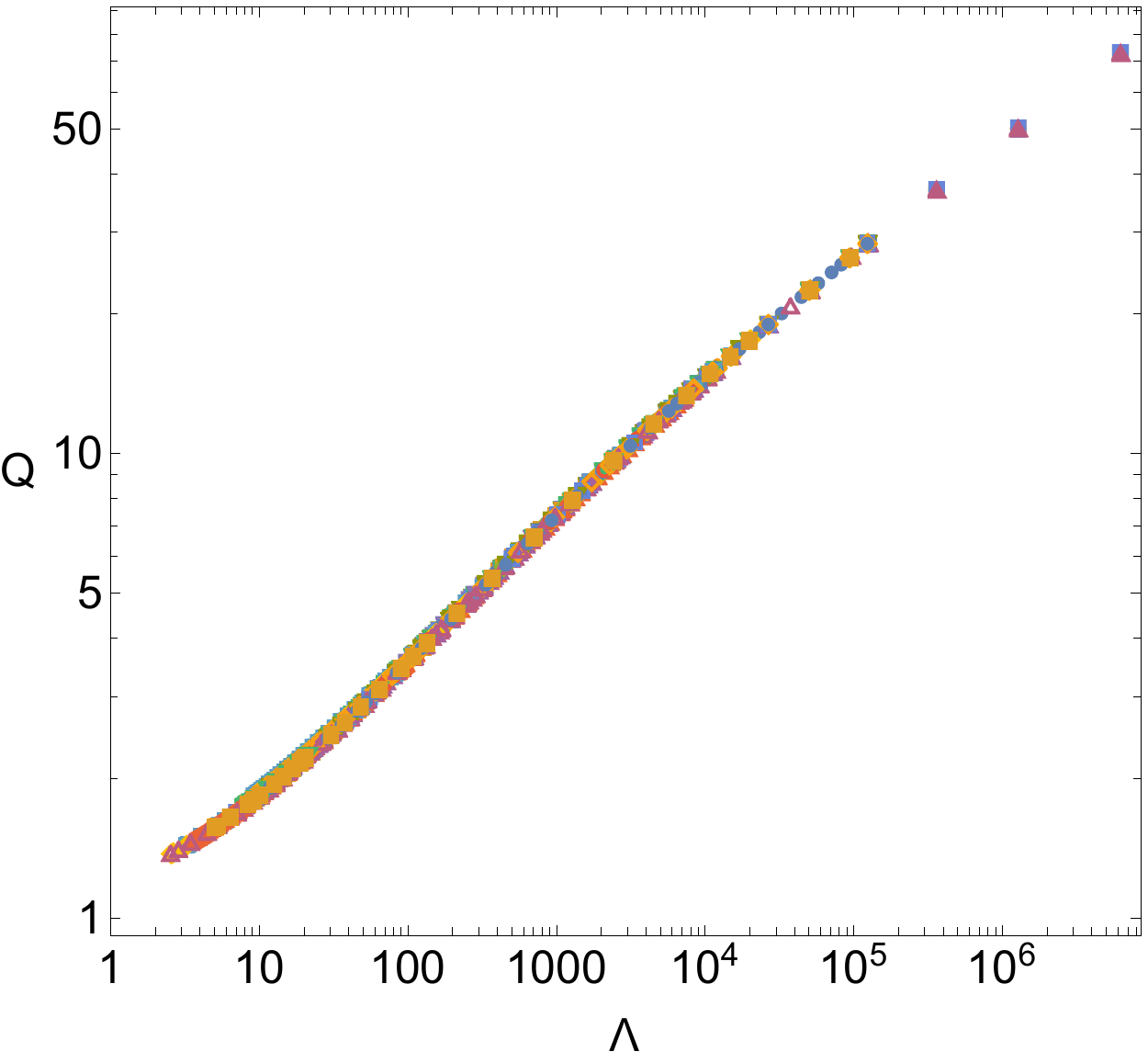} 
\end{tabular}
\caption{(Color online) I-Love and Love-Q relations for all the EoSs shown in this paper. Observe that regardless of the inclusion of crossover structure or first-order phase transitions in the speed of sound, the I-Love-Q relations remain EoS insensitive, with a relative fractional variability smaller than $1.5\%$ 
}
\label{fig:ILoveQ}
\end{figure}

Let us now consider whether the I-Love-Q relations remain approximately EoS insensitive when one includes non-smooth structure in the speed of sound, such as crossover features or first-order phase transitions. As described in Sec.~\ref{sec:GWobs}, the I-Love-Q relations are those that connect a dimensionless version of the moment of inertia, the Love number and the quadrupole moment. For EoSs that do not contain the speed of sound structure considered here, these relations have been shown to be EoS insensitive to better than $1\%$ in relative fractional difference~\cite{Yagi:2013awa,Yagi:2013bca}. Figure~\ref{fig:ILoveQ} shows the I-Love-Q relations for all of the EoSs considered in this paper. In spite of the non-smooth structure introduced in the speed of sound, the I-Love-Q relations remain EoS insensitive. In particular, the relative fractional variability between all of the data is less than $\sim 1.5\%$. This means that inferences made from measurements of any of these quantities can still be made, in spite of the possibility of the presence of non-smooth structure in the speed of sound. 

\section{Future directions}
\label{sec:future-dir}

We have investigated the properties of neutron stars that result from equations of state (EoSs) with structures in the speed of sound, including bumps, spikes, kinks, plateaus, kinks, and first-order phase transitions. We find that this structures can easily lead to ultra-heavy neutron stars, i.e.~those with masses larger than $2.5 $~M$_\odot$. In addition, this structures allows for disconnected and connected mass twins, as well as mass-radius sequences with kinks at least up to  $\sim 2$~M$_\odot$ ($>2.5$~M$_\odot$ in some cases). How heavy a neutron star can be depends sensitively on the structure in the speed of sound introduced, the crust model, and precise marginalized radius posteriors derived from observations. Whether mass twins or kinky mass-sequences appear in the mass-radius sequence also depends sensitively on the particular details of the speed of sound structure (see Sec.~\ref{sec:intro} for an executive summary).  

One of our findings suggests that future observations could lead to stringent constraints on the speed of sound. In particular, an accurate measurement of the tidal deformability could constrain the density at which the speed of sound changes rapidly. Similarly, an accurate measurement of the mass and radius of very massive neutron stars could lead to a constrain on the value of the speed of sound at large densities, and possible secondary structures. Current data, however, is not sufficiently informative to accurately map the entire behavior of the speed of sound at densities above nuclear saturation in neutron stars. 

However, the association of the light companion of the GW190814 event with a neutron star would have strong implications for the speed of sound. In particular, neutron stars as massive as $2.5$~M$_\odot$ would require a steep rise in the speed of sound, after which more complicated structure could be possible. These conclusions, however, must be taken with a pinch of salt because the crust model also has a strong influence, even at these high stellar masses. In particular, a softer crust model can support stars with a larger maximum mass. This suggests that it may be interesting to carry out a Bayesian inference study of the properties of the speed of sound and crust using the GW190814 event. 

But what if even heavier neutron stars existed, such as the companion of V723 Mon? Such extremely heavy neutron stars would possess exceedingly small tidal deformabilities (dimensionless $\Lambda \sim 2$). A very high signal-to-noise ratio event would be required to unambiguously measure such a small tidal deformability with a posterior that does not have support at the black hole limit of zero deformability, i.e.~that the error bars of the measurement in $\Lambda$ at some confidence interval does not overlap zero. Such a measurement may be possible with third-generation detectors. 

Ultra-heavy neutron stars ($M\geq 2.5 $~M$_\odot$) can indeed be in a mass twin configuration, but it must be connected. Disconnected mass twins (i.e.~where the two stable branches are disconnected) are not possible with any of the structures we studied in the speed of sound. The only possibility for ultra-heavy and disconnected mass twins would be to introduce a first-order phase transition at very small low baryon densities. The disconnected branches would then occur at exceedingly small masses, and would thus not be relevant to astrophysical observations. 

Our results also imply that it may be exceedingly difficult to infer with certainty the existence of a \textit{heavy} mass twin from an observation in the mass-radius plane. This is because although heavy mass twins have roughly the same mass, their difference in radius is not that large. The difference in radius, in fact, can be less than half a kilometer, depending on the EoS model. Marginalized posterior distributions on the radius would then have to be smaller than this at some confidence interval to guarantee that a twin star was observed. Once more, this may be possible in the future, with the advent of third-generation detectors, but it seems unlikely with current instruments. 

Our results also imply that future observations of massive neutron stars may place important constraints on the maximum baryon densities achievable at the center of neutron stars. For example, if the light companion in the binary that produced GW190814 was indeed a neutron star, then the maximum baryon density could not exceed six times nuclear saturation density. As more gravitational wave observations are made in the future, it becomes a real possibility to further constraint the maximum central baryon density, and thus the analysis of data should not be restricted to necessarily small central baryon densities \textit{a priori}.  

Note that a combination of a signal for strong first-order phase transition in neutron stars, such as twin configurations or sharp kinks in the mass-radius diagram, and a limit on stellar central density (as discussed above) would give valuable new insight into nuclear physics. This is because a strong phase order phase transition in the regime reached in neutron stars is usually associated with deconfinement to quark matter, which so far has not been constrained at low temperature, either in nature or in density range.

Our analysis clearly reveals that non-smooth structure in the speed of sound should be included in the analysis of future X-ray and gravitational wave data. This can be done through the parameterized models introduced in this paper. With such models, it would be straightforward to carry out a Bayesian analysis of gravitational wave data to determine whether the data prefers the presence of non-trivial structure. Such a study will be the focus of future investigations. 

\acknowledgements
The authors would like to thank Hank Lamm, Mauricio Hippert, Deep Chatterjee, Alejandro C\'ardenas-Avenda\~no, and Abhishek Hegde for useful discussions related to this work  J.N.H. and T.D. acknowledge the support from the US-DOE Nuclear Science Grant No. DE-SC0020633. H.~T.~and N.~Y.~acknowledge support from NASA Grants No. NNX16AB98G, 80NSSC17M0041 and 80NSSC18K1352 and NSF Award No. 1759615. V.D acknowledges support from the National Science Foundation under grant PHY-1748621 and PHAROS (COST Action CA16214). The authors also acknowledge support from the Illinois Campus Cluster, a computing resource that is operated by the Illinois Campus Cluster Program (ICCP) in conjunction with the National Center for Supercomputing Applications (NCSA), and which is supported by funds from the University of Illinois at Urbana-Champaign.


\bibliography{BIG,NOTinspire}

\begin{thebibliography}{166}%
\makeatletter
\providecommand \@ifxundefined [1]{%
 \@ifx{#1\undefined}
}%
\providecommand \@ifnum [1]{%
 \ifnum #1\expandafter \@firstoftwo
 \else \expandafter \@secondoftwo
 \fi
}%
\providecommand \@ifx [1]{%
 \ifx #1\expandafter \@firstoftwo
 \else \expandafter \@secondoftwo
 \fi
}%
\providecommand \natexlab [1]{#1}%
\providecommand \enquote  [1]{``#1''}%
\providecommand \bibnamefont  [1]{#1}%
\providecommand \bibfnamefont [1]{#1}%
\providecommand \citenamefont [1]{#1}%
\providecommand \href@noop [0]{\@secondoftwo}%
\providecommand \href [0]{\begingroup \@sanitize@url \@href}%
\providecommand \@href[1]{\@@startlink{#1}\@@href}%
\providecommand \@@href[1]{\endgroup#1\@@endlink}%
\providecommand \@sanitize@url [0]{\catcode `\\12\catcode `\$12\catcode
  `\&12\catcode `\#12\catcode `\^12\catcode `\_12\catcode `\%12\relax}%
\providecommand \@@startlink[1]{}%
\providecommand \@@endlink[0]{}%
\providecommand \url  [0]{\begingroup\@sanitize@url \@url }%
\providecommand \@url [1]{\endgroup\@href {#1}{\urlprefix }}%
\providecommand \urlprefix  [0]{URL }%
\providecommand \Eprint [0]{\href }%
\providecommand \doibase [0]{http://dx.doi.org/}%
\providecommand \selectlanguage [0]{\@gobble}%
\providecommand \bibinfo  [0]{\@secondoftwo}%
\providecommand \bibfield  [0]{\@secondoftwo}%
\providecommand \translation [1]{[#1]}%
\providecommand \BibitemOpen [0]{}%
\providecommand \bibitemStop [0]{}%
\providecommand \bibitemNoStop [0]{.\EOS\space}%
\providecommand \EOS [0]{\spacefactor3000\relax}%
\providecommand \BibitemShut  [1]{\csname bibitem#1\endcsname}%
\let\auto@bib@innerbib\@empty
\bibitem [{\citenamefont {Glendenning}(1982)}]{Glendenning:1982nc}%
  \BibitemOpen
  \bibfield  {author} {\bibinfo {author} {\bibfnamefont {N.~K.}\ \bibnamefont
  {Glendenning}},\ }\href {\doibase 10.1016/0370-2693(82)90078-8} {\bibfield
  {journal} {\bibinfo  {journal} {Phys. Lett. B}\ }\textbf {\bibinfo {volume}
  {114}},\ \bibinfo {pages} {392} (\bibinfo {year} {1982})}\BibitemShut
  {NoStop}%
\bibitem [{\citenamefont {Bednarek}\ \emph {et~al.}(2012)\citenamefont
  {Bednarek}, \citenamefont {Haensel}, \citenamefont {Zdunik}, \citenamefont
  {Bejger},\ and\ \citenamefont {Manka}}]{Bednarek:2011gd}%
  \BibitemOpen
  \bibfield  {author} {\bibinfo {author} {\bibfnamefont {I.}~\bibnamefont
  {Bednarek}}, \bibinfo {author} {\bibfnamefont {P.}~\bibnamefont {Haensel}},
  \bibinfo {author} {\bibfnamefont {J.~L.}\ \bibnamefont {Zdunik}}, \bibinfo
  {author} {\bibfnamefont {M.}~\bibnamefont {Bejger}}, \ and\ \bibinfo {author}
  {\bibfnamefont {R.}~\bibnamefont {Manka}},\ }\href {\doibase
  10.1051/0004-6361/201118560} {\bibfield  {journal} {\bibinfo  {journal}
  {Astron. Astrophys.}\ }\textbf {\bibinfo {volume} {543}},\ \bibinfo {pages}
  {A157} (\bibinfo {year} {2012})},\ \Eprint {http://arxiv.org/abs/1111.6942}
  {arXiv:1111.6942 [astro-ph.SR]} \BibitemShut {NoStop}%
\bibitem [{\citenamefont {Fortin}\ \emph {et~al.}(2015)\citenamefont {Fortin},
  \citenamefont {Zdunik}, \citenamefont {Haensel},\ and\ \citenamefont
  {Bejger}}]{Fortin:2014mya}%
  \BibitemOpen
  \bibfield  {author} {\bibinfo {author} {\bibfnamefont {M.}~\bibnamefont
  {Fortin}}, \bibinfo {author} {\bibfnamefont {J.~L.}\ \bibnamefont {Zdunik}},
  \bibinfo {author} {\bibfnamefont {P.}~\bibnamefont {Haensel}}, \ and\
  \bibinfo {author} {\bibfnamefont {M.}~\bibnamefont {Bejger}},\ }\href
  {\doibase 10.1051/0004-6361/201424800} {\bibfield  {journal} {\bibinfo
  {journal} {Astron. Astrophys.}\ }\textbf {\bibinfo {volume} {576}},\ \bibinfo
  {pages} {A68} (\bibinfo {year} {2015})},\ \Eprint
  {http://arxiv.org/abs/1408.3052} {arXiv:1408.3052 [astro-ph.SR]} \BibitemShut
  {NoStop}%
\bibitem [{\citenamefont {Buballa}\ \emph {et~al.}(2014)\citenamefont {Buballa}
  \emph {et~al.}}]{Buballa:2014jta}%
  \BibitemOpen
  \bibfield  {author} {\bibinfo {author} {\bibfnamefont {M.}~\bibnamefont
  {Buballa}} \emph {et~al.},\ }\href {\doibase 10.1088/0954-3899/41/12/123001}
  {\bibfield  {journal} {\bibinfo  {journal} {J. Phys. G}\ }\textbf {\bibinfo
  {volume} {41}},\ \bibinfo {pages} {123001} (\bibinfo {year} {2014})},\
  \Eprint {http://arxiv.org/abs/1402.6911} {arXiv:1402.6911 [astro-ph.HE]}
  \BibitemShut {NoStop}%
\bibitem [{\citenamefont {Annala}\ \emph {et~al.}(2020)\citenamefont {Annala},
  \citenamefont {Gorda}, \citenamefont {Kurkela}, \citenamefont {N\"attil\"a},\
  and\ \citenamefont {Vuorinen}}]{Annala:2019puf}%
  \BibitemOpen
  \bibfield  {author} {\bibinfo {author} {\bibfnamefont {E.}~\bibnamefont
  {Annala}}, \bibinfo {author} {\bibfnamefont {T.}~\bibnamefont {Gorda}},
  \bibinfo {author} {\bibfnamefont {A.}~\bibnamefont {Kurkela}}, \bibinfo
  {author} {\bibfnamefont {J.}~\bibnamefont {N\"attil\"a}}, \ and\ \bibinfo
  {author} {\bibfnamefont {A.}~\bibnamefont {Vuorinen}},\ }\href {\doibase
  10.1038/s41567-020-0914-9} {\bibfield  {journal} {\bibinfo  {journal} {Nature
  Phys.}\ }\textbf {\bibinfo {volume} {16}},\ \bibinfo {pages} {907} (\bibinfo
  {year} {2020})},\ \Eprint {http://arxiv.org/abs/1903.09121} {arXiv:1903.09121
  [astro-ph.HE]} \BibitemShut {NoStop}%
\bibitem [{\citenamefont {Ivanenko}\ and\ \citenamefont
  {Kurdgelaidze}(1965)}]{Ivanenko:1965dg}%
  \BibitemOpen
  \bibfield  {author} {\bibinfo {author} {\bibfnamefont {D.~D.}\ \bibnamefont
  {Ivanenko}}\ and\ \bibinfo {author} {\bibfnamefont {D.~F.}\ \bibnamefont
  {Kurdgelaidze}},\ }\href {\doibase 10.1007/BF01042830} {\bibfield  {journal}
  {\bibinfo  {journal} {Astrophysics}\ }\textbf {\bibinfo {volume} {1}},\
  \bibinfo {pages} {251} (\bibinfo {year} {1965})}\BibitemShut {NoStop}%
\bibitem [{\citenamefont {Keister}\ and\ \citenamefont
  {Kisslinger}(1976)}]{Keister:1976gc}%
  \BibitemOpen
  \bibfield  {author} {\bibinfo {author} {\bibfnamefont {B.~D.}\ \bibnamefont
  {Keister}}\ and\ \bibinfo {author} {\bibfnamefont {L.~S.}\ \bibnamefont
  {Kisslinger}},\ }\href {\doibase 10.1016/0370-2693(76)90370-1} {\bibfield
  {journal} {\bibinfo  {journal} {Phys. Lett. B}\ }\textbf {\bibinfo {volume}
  {64}},\ \bibinfo {pages} {117} (\bibinfo {year} {1976})}\BibitemShut
  {NoStop}%
\bibitem [{\citenamefont {Grassi}(1988)}]{Grassi:1987bu}%
  \BibitemOpen
  \bibfield  {author} {\bibinfo {author} {\bibfnamefont {F.}~\bibnamefont
  {Grassi}},\ }\href {\doibase 10.1007/BF01574554} {\bibfield  {journal}
  {\bibinfo  {journal} {Z. Phys. C}\ }\textbf {\bibinfo {volume} {38}},\
  \bibinfo {pages} {307} (\bibinfo {year} {1988})}\BibitemShut {NoStop}%
\bibitem [{\citenamefont {Abbott}\ \emph {et~al.}(2020)\citenamefont {Abbott}
  \emph {et~al.}}]{Abbott:2020khf}%
  \BibitemOpen
  \bibfield  {author} {\bibinfo {author} {\bibfnamefont {R.}~\bibnamefont
  {Abbott}} \emph {et~al.} (\bibinfo {collaboration} {LIGO Scientific,
  Virgo}),\ }\href {\doibase 10.3847/2041-8213/ab960f} {\bibfield  {journal}
  {\bibinfo  {journal} {Astrophys. J. Lett.}\ }\textbf {\bibinfo {volume}
  {896}},\ \bibinfo {pages} {L44} (\bibinfo {year} {2020})},\ \Eprint
  {http://arxiv.org/abs/2006.12611} {arXiv:2006.12611 [astro-ph.HE]}
  \BibitemShut {NoStop}%
\bibitem [{\citenamefont {Tan}\ \emph {et~al.}(2020)\citenamefont {Tan},
  \citenamefont {Noronha-Hostler},\ and\ \citenamefont {Yunes}}]{Tan:2020ics}%
  \BibitemOpen
  \bibfield  {author} {\bibinfo {author} {\bibfnamefont {H.}~\bibnamefont
  {Tan}}, \bibinfo {author} {\bibfnamefont {J.}~\bibnamefont
  {Noronha-Hostler}}, \ and\ \bibinfo {author} {\bibfnamefont {N.}~\bibnamefont
  {Yunes}},\ }\href {\doibase 10.1103/PhysRevLett.125.261104} {\bibfield
  {journal} {\bibinfo  {journal} {Phys. Rev. Lett.}\ }\textbf {\bibinfo
  {volume} {125}},\ \bibinfo {pages} {261104} (\bibinfo {year} {2020})},\
  \Eprint {http://arxiv.org/abs/2006.16296} {arXiv:2006.16296 [astro-ph.HE]}
  \BibitemShut {NoStop}%
\bibitem [{\citenamefont {Dexheimer}\ \emph
  {et~al.}(2021{\natexlab{a}})\citenamefont {Dexheimer}, \citenamefont {Gomes},
  \citenamefont {Kl\"ahn}, \citenamefont {Han},\ and\ \citenamefont
  {Salinas}}]{Dexheimer:2020rlp}%
  \BibitemOpen
  \bibfield  {author} {\bibinfo {author} {\bibfnamefont {V.}~\bibnamefont
  {Dexheimer}}, \bibinfo {author} {\bibfnamefont {R.~O.}\ \bibnamefont
  {Gomes}}, \bibinfo {author} {\bibfnamefont {T.}~\bibnamefont {Kl\"ahn}},
  \bibinfo {author} {\bibfnamefont {S.}~\bibnamefont {Han}}, \ and\ \bibinfo
  {author} {\bibfnamefont {M.}~\bibnamefont {Salinas}},\ }\href {\doibase
  10.1103/PhysRevC.103.025808} {\bibfield  {journal} {\bibinfo  {journal}
  {Phys. Rev. C}\ }\textbf {\bibinfo {volume} {103}},\ \bibinfo {pages}
  {025808} (\bibinfo {year} {2021}{\natexlab{a}})},\ \Eprint
  {http://arxiv.org/abs/2007.08493} {arXiv:2007.08493 [astro-ph.HE]}
  \BibitemShut {NoStop}%
\bibitem [{\citenamefont {Demircik}\ \emph {et~al.}(2021)\citenamefont
  {Demircik}, \citenamefont {Ecker},\ and\ \citenamefont
  {J\"arvinen}}]{Demircik:2020jkc}%
  \BibitemOpen
  \bibfield  {author} {\bibinfo {author} {\bibfnamefont {T.}~\bibnamefont
  {Demircik}}, \bibinfo {author} {\bibfnamefont {C.}~\bibnamefont {Ecker}}, \
  and\ \bibinfo {author} {\bibfnamefont {M.}~\bibnamefont {J\"arvinen}},\
  }\href {\doibase 10.3847/2041-8213/abd853} {\bibfield  {journal} {\bibinfo
  {journal} {Astrophys. J. Lett.}\ }\textbf {\bibinfo {volume} {907}},\
  \bibinfo {pages} {L37} (\bibinfo {year} {2021})},\ \Eprint
  {http://arxiv.org/abs/2009.10731} {arXiv:2009.10731 [astro-ph.HE]}
  \BibitemShut {NoStop}%
\bibitem [{\citenamefont {Christian}\ and\ \citenamefont
  {Schaffner-Bielich}(2021)}]{Christian:2020xwz}%
  \BibitemOpen
  \bibfield  {author} {\bibinfo {author} {\bibfnamefont {J.-E.}\ \bibnamefont
  {Christian}}\ and\ \bibinfo {author} {\bibfnamefont {J.}~\bibnamefont
  {Schaffner-Bielich}},\ }\href {\doibase 10.1103/PhysRevD.103.063042}
  {\bibfield  {journal} {\bibinfo  {journal} {Phys. Rev. D}\ }\textbf {\bibinfo
  {volume} {103}},\ \bibinfo {pages} {063042} (\bibinfo {year} {2021})},\
  \Eprint {http://arxiv.org/abs/2011.01001} {arXiv:2011.01001 [astro-ph.HE]}
  \BibitemShut {NoStop}%
\bibitem [{\citenamefont {Blaschke}\ and\ \citenamefont
  {Cierniak}(2021)}]{Blaschke:2020vuy}%
  \BibitemOpen
  \bibfield  {author} {\bibinfo {author} {\bibfnamefont {D.}~\bibnamefont
  {Blaschke}}\ and\ \bibinfo {author} {\bibfnamefont {M.}~\bibnamefont
  {Cierniak}},\ }\href {\doibase 10.1002/asna.202113909} {\bibfield  {journal}
  {\bibinfo  {journal} {Astron. Nachr.}\ }\textbf {\bibinfo {volume} {342}},\
  \bibinfo {pages} {227} (\bibinfo {year} {2021})},\ \Eprint
  {http://arxiv.org/abs/2012.15785} {arXiv:2012.15785 [astro-ph.HE]}
  \BibitemShut {NoStop}%
\bibitem [{\citenamefont {Ayriyan}\ \emph {et~al.}(2021)\citenamefont
  {Ayriyan}, \citenamefont {Blaschke}, \citenamefont {Grunfeld}, \citenamefont
  {Alvarez-Castillo}, \citenamefont {Grigorian},\ and\ \citenamefont
  {Abgaryan}}]{Ayriyan:2021prr}%
  \BibitemOpen
  \bibfield  {author} {\bibinfo {author} {\bibfnamefont {A.}~\bibnamefont
  {Ayriyan}}, \bibinfo {author} {\bibfnamefont {D.}~\bibnamefont {Blaschke}},
  \bibinfo {author} {\bibfnamefont {A.~G.}\ \bibnamefont {Grunfeld}}, \bibinfo
  {author} {\bibfnamefont {D.}~\bibnamefont {Alvarez-Castillo}}, \bibinfo
  {author} {\bibfnamefont {H.}~\bibnamefont {Grigorian}}, \ and\ \bibinfo
  {author} {\bibfnamefont {V.}~\bibnamefont {Abgaryan}},\ }\href@noop {} {\
  (\bibinfo {year} {2021})},\ \Eprint {http://arxiv.org/abs/2102.13485}
  {arXiv:2102.13485 [astro-ph.HE]} \BibitemShut {NoStop}%
\bibitem [{\citenamefont {Li}\ \emph {et~al.}(2020{\natexlab{a}})\citenamefont
  {Li}, \citenamefont {Zhu}, \citenamefont {Zhou}, \citenamefont {Dong},
  \citenamefont {Hu},\ and\ \citenamefont {Xia}}]{Li:2020dst}%
  \BibitemOpen
  \bibfield  {author} {\bibinfo {author} {\bibfnamefont {A.}~\bibnamefont
  {Li}}, \bibinfo {author} {\bibfnamefont {Z.~Y.}\ \bibnamefont {Zhu}},
  \bibinfo {author} {\bibfnamefont {E.~P.}\ \bibnamefont {Zhou}}, \bibinfo
  {author} {\bibfnamefont {J.~M.}\ \bibnamefont {Dong}}, \bibinfo {author}
  {\bibfnamefont {J.~N.}\ \bibnamefont {Hu}}, \ and\ \bibinfo {author}
  {\bibfnamefont {C.~J.}\ \bibnamefont {Xia}},\ }\href {\doibase
  10.1016/j.jheap.2020.07.001} {\bibfield  {journal} {\bibinfo  {journal}
  {JHEAp}\ }\textbf {\bibinfo {volume} {28}},\ \bibinfo {pages} {19} (\bibinfo
  {year} {2020}{\natexlab{a}})},\ \Eprint {http://arxiv.org/abs/2007.05116}
  {arXiv:2007.05116 [nucl-th]} \BibitemShut {NoStop}%
\bibitem [{\citenamefont {Otto}\ \emph {et~al.}(2020)\citenamefont {Otto},
  \citenamefont {Oertel},\ and\ \citenamefont {Schaefer}}]{Otto:2020hoz}%
  \BibitemOpen
  \bibfield  {author} {\bibinfo {author} {\bibfnamefont {K.}~\bibnamefont
  {Otto}}, \bibinfo {author} {\bibfnamefont {M.}~\bibnamefont {Oertel}}, \ and\
  \bibinfo {author} {\bibfnamefont {B.-J.}\ \bibnamefont {Schaefer}},\ }\href
  {\doibase 10.1140/epjst/e2020-000155-y} {\bibfield  {journal} {\bibinfo
  {journal} {Eur. Phys. J. ST}\ }\textbf {\bibinfo {volume} {229}},\ \bibinfo
  {pages} {3629} (\bibinfo {year} {2020})},\ \Eprint
  {http://arxiv.org/abs/2007.07394} {arXiv:2007.07394 [hep-ph]} \BibitemShut
  {NoStop}%
\bibitem [{\citenamefont {Nandi}\ and\ \citenamefont
  {Pal}(2020)}]{Nandi:2020luz}%
  \BibitemOpen
  \bibfield  {author} {\bibinfo {author} {\bibfnamefont {R.}~\bibnamefont
  {Nandi}}\ and\ \bibinfo {author} {\bibfnamefont {S.}~\bibnamefont {Pal}},\
  }\href {\doibase 10.1140/epjs/s11734-021-00004-4} {\  (\bibinfo {year}
  {2020}),\ 10.1140/epjs/s11734-021-00004-4},\ \Eprint
  {http://arxiv.org/abs/2008.10943} {arXiv:2008.10943 [astro-ph.HE]}
  \BibitemShut {NoStop}%
\bibitem [{\citenamefont {Ferreira}\ \emph {et~al.}(2020)\citenamefont
  {Ferreira}, \citenamefont {C\^amara~Pereira},\ and\ \citenamefont
  {Provid\^encia}}]{Ferreira:2020kvu}%
  \BibitemOpen
  \bibfield  {author} {\bibinfo {author} {\bibfnamefont {M.}~\bibnamefont
  {Ferreira}}, \bibinfo {author} {\bibfnamefont {R.}~\bibnamefont
  {C\^amara~Pereira}}, \ and\ \bibinfo {author} {\bibfnamefont
  {C.}~\bibnamefont {Provid\^encia}},\ }\href {\doibase
  10.1103/PhysRevD.102.083030} {\bibfield  {journal} {\bibinfo  {journal}
  {Phys. Rev. D}\ }\textbf {\bibinfo {volume} {102}},\ \bibinfo {pages}
  {083030} (\bibinfo {year} {2020})},\ \Eprint
  {http://arxiv.org/abs/2008.12563} {arXiv:2008.12563 [nucl-th]} \BibitemShut
  {NoStop}%
\bibitem [{\citenamefont {Most}\ \emph {et~al.}(2020)\citenamefont {Most},
  \citenamefont {Papenfort}, \citenamefont {Weih},\ and\ \citenamefont
  {Rezzolla}}]{Most:2020bba}%
  \BibitemOpen
  \bibfield  {author} {\bibinfo {author} {\bibfnamefont {E.~R.}\ \bibnamefont
  {Most}}, \bibinfo {author} {\bibfnamefont {L.~J.}\ \bibnamefont {Papenfort}},
  \bibinfo {author} {\bibfnamefont {L.~R.}\ \bibnamefont {Weih}}, \ and\
  \bibinfo {author} {\bibfnamefont {L.}~\bibnamefont {Rezzolla}},\ }\href
  {\doibase 10.1093/mnrasl/slaa168} {\bibfield  {journal} {\bibinfo  {journal}
  {Mon. Not. Roy. Astron. Soc.}\ }\textbf {\bibinfo {volume} {499}},\ \bibinfo
  {pages} {L82} (\bibinfo {year} {2020})},\ \Eprint
  {http://arxiv.org/abs/2006.14601} {arXiv:2006.14601 [astro-ph.HE]}
  \BibitemShut {NoStop}%
\bibitem [{\citenamefont {Broadhurst}\ \emph {et~al.}(2020)\citenamefont
  {Broadhurst}, \citenamefont {Diego},\ and\ \citenamefont
  {Smoot}}]{Broadhurst:2020cvm}%
  \BibitemOpen
  \bibfield  {author} {\bibinfo {author} {\bibfnamefont {T.}~\bibnamefont
  {Broadhurst}}, \bibinfo {author} {\bibfnamefont {J.~M.}\ \bibnamefont
  {Diego}}, \ and\ \bibinfo {author} {\bibfnamefont {G.~F.}\ \bibnamefont
  {Smoot}},\ }\href@noop {} {\  (\bibinfo {year} {2020})},\ \Eprint
  {http://arxiv.org/abs/2006.13219} {arXiv:2006.13219 [astro-ph.CO]}
  \BibitemShut {NoStop}%
\bibitem [{\citenamefont {Fishbach}\ \emph {et~al.}(2020)\citenamefont
  {Fishbach}, \citenamefont {Essick},\ and\ \citenamefont
  {Holz}}]{Fishbach:2020ryj}%
  \BibitemOpen
  \bibfield  {author} {\bibinfo {author} {\bibfnamefont {M.}~\bibnamefont
  {Fishbach}}, \bibinfo {author} {\bibfnamefont {R.}~\bibnamefont {Essick}}, \
  and\ \bibinfo {author} {\bibfnamefont {D.~E.}\ \bibnamefont {Holz}},\ }\href
  {\doibase 10.3847/2041-8213/aba7b6} {\bibfield  {journal} {\bibinfo
  {journal} {Astrophys. J. Lett.}\ }\textbf {\bibinfo {volume} {899}},\
  \bibinfo {pages} {L8} (\bibinfo {year} {2020})},\ \Eprint
  {http://arxiv.org/abs/2006.13178} {arXiv:2006.13178 [astro-ph.HE]}
  \BibitemShut {NoStop}%
\bibitem [{\citenamefont {Nathanail}\ \emph {et~al.}(2021)\citenamefont
  {Nathanail}, \citenamefont {Most},\ and\ \citenamefont
  {Rezzolla}}]{Nathanail:2021tay}%
  \BibitemOpen
  \bibfield  {author} {\bibinfo {author} {\bibfnamefont {A.}~\bibnamefont
  {Nathanail}}, \bibinfo {author} {\bibfnamefont {E.~R.}\ \bibnamefont {Most}},
  \ and\ \bibinfo {author} {\bibfnamefont {L.}~\bibnamefont {Rezzolla}},\
  }\href {\doibase 10.3847/2041-8213/abdfc6} {\bibfield  {journal} {\bibinfo
  {journal} {Astrophys. J. Lett.}\ }\textbf {\bibinfo {volume} {908}},\
  \bibinfo {pages} {L28} (\bibinfo {year} {2021})},\ \Eprint
  {http://arxiv.org/abs/2101.01735} {arXiv:2101.01735 [astro-ph.HE]}
  \BibitemShut {NoStop}%
\bibitem [{\citenamefont {Rezzolla}\ \emph {et~al.}(2018)\citenamefont
  {Rezzolla}, \citenamefont {Most},\ and\ \citenamefont
  {Weih}}]{Rezzolla:2017aly}%
  \BibitemOpen
  \bibfield  {author} {\bibinfo {author} {\bibfnamefont {L.}~\bibnamefont
  {Rezzolla}}, \bibinfo {author} {\bibfnamefont {E.~R.}\ \bibnamefont {Most}},
  \ and\ \bibinfo {author} {\bibfnamefont {L.~R.}\ \bibnamefont {Weih}},\
  }\href {\doibase 10.3847/2041-8213/aaa401} {\bibfield  {journal} {\bibinfo
  {journal} {Astrophys. J. Lett.}\ }\textbf {\bibinfo {volume} {852}},\
  \bibinfo {pages} {L25} (\bibinfo {year} {2018})},\ \Eprint
  {http://arxiv.org/abs/1711.00314} {arXiv:1711.00314 [astro-ph.HE]}
  \BibitemShut {NoStop}%
\bibitem [{\citenamefont {Khadkikar}\ \emph {et~al.}(2021)\citenamefont
  {Khadkikar}, \citenamefont {Raduta}, \citenamefont {Oertel},\ and\
  \citenamefont {Sedrakian}}]{Khadkikar:2021yrj}%
  \BibitemOpen
  \bibfield  {author} {\bibinfo {author} {\bibfnamefont {S.}~\bibnamefont
  {Khadkikar}}, \bibinfo {author} {\bibfnamefont {A.~R.}\ \bibnamefont
  {Raduta}}, \bibinfo {author} {\bibfnamefont {M.}~\bibnamefont {Oertel}}, \
  and\ \bibinfo {author} {\bibfnamefont {A.}~\bibnamefont {Sedrakian}},\ }\href
  {\doibase 10.1103/PhysRevC.103.055811} {\bibfield  {journal} {\bibinfo
  {journal} {Phys. Rev. C}\ }\textbf {\bibinfo {volume} {103}},\ \bibinfo
  {pages} {055811} (\bibinfo {year} {2021})},\ \Eprint
  {http://arxiv.org/abs/2102.00988} {arXiv:2102.00988 [astro-ph.HE]}
  \BibitemShut {NoStop}%
\bibitem [{\citenamefont {Alford}\ \emph {et~al.}(2018)\citenamefont {Alford},
  \citenamefont {Bovard}, \citenamefont {Hanauske}, \citenamefont {Rezzolla},\
  and\ \citenamefont {Schwenzer}}]{Alford:2017rxf}%
  \BibitemOpen
  \bibfield  {author} {\bibinfo {author} {\bibfnamefont {M.~G.}\ \bibnamefont
  {Alford}}, \bibinfo {author} {\bibfnamefont {L.}~\bibnamefont {Bovard}},
  \bibinfo {author} {\bibfnamefont {M.}~\bibnamefont {Hanauske}}, \bibinfo
  {author} {\bibfnamefont {L.}~\bibnamefont {Rezzolla}}, \ and\ \bibinfo
  {author} {\bibfnamefont {K.}~\bibnamefont {Schwenzer}},\ }\href {\doibase
  10.1103/PhysRevLett.120.041101} {\bibfield  {journal} {\bibinfo  {journal}
  {Phys. Rev. Lett.}\ }\textbf {\bibinfo {volume} {120}},\ \bibinfo {pages}
  {041101} (\bibinfo {year} {2018})},\ \Eprint
  {http://arxiv.org/abs/1707.09475} {arXiv:1707.09475 [gr-qc]} \BibitemShut
  {NoStop}%
\bibitem [{\citenamefont {Alford}\ \emph {et~al.}(2019)\citenamefont {Alford},
  \citenamefont {Harutyunyan},\ and\ \citenamefont
  {Sedrakian}}]{Alford:2019kdw}%
  \BibitemOpen
  \bibfield  {author} {\bibinfo {author} {\bibfnamefont {M.}~\bibnamefont
  {Alford}}, \bibinfo {author} {\bibfnamefont {A.}~\bibnamefont {Harutyunyan}},
  \ and\ \bibinfo {author} {\bibfnamefont {A.}~\bibnamefont {Sedrakian}},\
  }\href {\doibase 10.1103/PhysRevD.100.103021} {\bibfield  {journal} {\bibinfo
   {journal} {Phys. Rev. D}\ }\textbf {\bibinfo {volume} {100}},\ \bibinfo
  {pages} {103021} (\bibinfo {year} {2019})},\ \Eprint
  {http://arxiv.org/abs/1907.04192} {arXiv:1907.04192 [astro-ph.HE]}
  \BibitemShut {NoStop}%
\bibitem [{\citenamefont {Routray}\ \emph {et~al.}(2020)\citenamefont
  {Routray}, \citenamefont {Pattnaik}, \citenamefont {Gonzalez-Boquera},
  \citenamefont {Vi\~nas}, \citenamefont {Centelles},\ and\ \citenamefont
  {Behera}}]{Routray:2020zkf}%
  \BibitemOpen
  \bibfield  {author} {\bibinfo {author} {\bibfnamefont {T.~R.}\ \bibnamefont
  {Routray}}, \bibinfo {author} {\bibfnamefont {S.~P.}\ \bibnamefont
  {Pattnaik}}, \bibinfo {author} {\bibfnamefont {C.}~\bibnamefont
  {Gonzalez-Boquera}}, \bibinfo {author} {\bibfnamefont {X.}~\bibnamefont
  {Vi\~nas}}, \bibinfo {author} {\bibfnamefont {M.}~\bibnamefont {Centelles}},
  \ and\ \bibinfo {author} {\bibfnamefont {B.}~\bibnamefont {Behera}},\
  }\href@noop {} {\  (\bibinfo {year} {2020})},\ \Eprint
  {http://arxiv.org/abs/2006.15430} {arXiv:2006.15430 [nucl-th]} \BibitemShut
  {NoStop}%
\bibitem [{\citenamefont {Parotto}\ \emph {et~al.}(2020)\citenamefont
  {Parotto}, \citenamefont {Bluhm}, \citenamefont {Mroczek}, \citenamefont
  {Nahrgang}, \citenamefont {Noronha-Hostler}, \citenamefont {Rajagopal},
  \citenamefont {Ratti}, \citenamefont {Sch\"afer},\ and\ \citenamefont
  {Stephanov}}]{Parotto:2018pwx}%
  \BibitemOpen
  \bibfield  {author} {\bibinfo {author} {\bibfnamefont {P.}~\bibnamefont
  {Parotto}}, \bibinfo {author} {\bibfnamefont {M.}~\bibnamefont {Bluhm}},
  \bibinfo {author} {\bibfnamefont {D.}~\bibnamefont {Mroczek}}, \bibinfo
  {author} {\bibfnamefont {M.}~\bibnamefont {Nahrgang}}, \bibinfo {author}
  {\bibfnamefont {J.}~\bibnamefont {Noronha-Hostler}}, \bibinfo {author}
  {\bibfnamefont {K.}~\bibnamefont {Rajagopal}}, \bibinfo {author}
  {\bibfnamefont {C.}~\bibnamefont {Ratti}}, \bibinfo {author} {\bibfnamefont
  {T.}~\bibnamefont {Sch\"afer}}, \ and\ \bibinfo {author} {\bibfnamefont
  {M.}~\bibnamefont {Stephanov}},\ }\href {\doibase
  10.1103/PhysRevC.101.034901} {\bibfield  {journal} {\bibinfo  {journal}
  {Phys. Rev. C}\ }\textbf {\bibinfo {volume} {101}},\ \bibinfo {pages}
  {034901} (\bibinfo {year} {2020})},\ \Eprint
  {http://arxiv.org/abs/1805.05249} {arXiv:1805.05249 [hep-ph]} \BibitemShut
  {NoStop}%
\bibitem [{\citenamefont {Grefa}\ \emph {et~al.}(2021)\citenamefont {Grefa},
  \citenamefont {Noronha}, \citenamefont {Noronha-Hostler}, \citenamefont
  {Portillo}, \citenamefont {Ratti},\ and\ \citenamefont
  {Rougemont}}]{Grefa:2021qvt}%
  \BibitemOpen
  \bibfield  {author} {\bibinfo {author} {\bibfnamefont {J.}~\bibnamefont
  {Grefa}}, \bibinfo {author} {\bibfnamefont {J.}~\bibnamefont {Noronha}},
  \bibinfo {author} {\bibfnamefont {J.}~\bibnamefont {Noronha-Hostler}},
  \bibinfo {author} {\bibfnamefont {I.}~\bibnamefont {Portillo}}, \bibinfo
  {author} {\bibfnamefont {C.}~\bibnamefont {Ratti}}, \ and\ \bibinfo {author}
  {\bibfnamefont {R.}~\bibnamefont {Rougemont}},\ }\href@noop {} {\  (\bibinfo
  {year} {2021})},\ \Eprint {http://arxiv.org/abs/2102.12042} {arXiv:2102.12042
  [nucl-th]} \BibitemShut {NoStop}%
\bibitem [{\citenamefont {Stafford}\ \emph {et~al.}(2021)\citenamefont
  {Stafford}, \citenamefont {Mroczek}, \citenamefont {Nava~Acuna},
  \citenamefont {Noronha-Hostler}, \citenamefont {Parotto}, \citenamefont
  {Price},\ and\ \citenamefont {Ratti}}]{Stafford:2021wik}%
  \BibitemOpen
  \bibfield  {author} {\bibinfo {author} {\bibfnamefont {J.~M.}\ \bibnamefont
  {Stafford}}, \bibinfo {author} {\bibfnamefont {D.}~\bibnamefont {Mroczek}},
  \bibinfo {author} {\bibfnamefont {A.~R.}\ \bibnamefont {Nava~Acuna}},
  \bibinfo {author} {\bibfnamefont {J.}~\bibnamefont {Noronha-Hostler}},
  \bibinfo {author} {\bibfnamefont {P.}~\bibnamefont {Parotto}}, \bibinfo
  {author} {\bibfnamefont {D.~R.~P.}\ \bibnamefont {Price}}, \ and\ \bibinfo
  {author} {\bibfnamefont {C.}~\bibnamefont {Ratti}},\ }\href@noop {} {\
  (\bibinfo {year} {2021})},\ \Eprint {http://arxiv.org/abs/2103.08146}
  {arXiv:2103.08146 [hep-ph]} \BibitemShut {NoStop}%
\bibitem [{\citenamefont {Aoki}\ \emph {et~al.}(2006)\citenamefont {Aoki},
  \citenamefont {Endrodi}, \citenamefont {Fodor}, \citenamefont {Katz},\ and\
  \citenamefont {Szabo}}]{Aoki:2006we}%
  \BibitemOpen
  \bibfield  {author} {\bibinfo {author} {\bibfnamefont {Y.}~\bibnamefont
  {Aoki}}, \bibinfo {author} {\bibfnamefont {G.}~\bibnamefont {Endrodi}},
  \bibinfo {author} {\bibfnamefont {Z.}~\bibnamefont {Fodor}}, \bibinfo
  {author} {\bibfnamefont {S.~D.}\ \bibnamefont {Katz}}, \ and\ \bibinfo
  {author} {\bibfnamefont {K.~K.}\ \bibnamefont {Szabo}},\ }\href {\doibase
  10.1038/nature05120} {\bibfield  {journal} {\bibinfo  {journal} {Nature}\
  }\textbf {\bibinfo {volume} {443}},\ \bibinfo {pages} {675} (\bibinfo {year}
  {2006})},\ \Eprint {http://arxiv.org/abs/hep-lat/0611014}
  {arXiv:hep-lat/0611014} \BibitemShut {NoStop}%
\bibitem [{\citenamefont {Bors\'anyi}\ \emph {et~al.}(2021)\citenamefont
  {Bors\'anyi}, \citenamefont {Fodor}, \citenamefont {Guenther}, \citenamefont
  {Kara}, \citenamefont {Katz}, \citenamefont {Parotto}, \citenamefont
  {P\'asztor}, \citenamefont {Ratti},\ and\ \citenamefont
  {Szab\'o}}]{Borsanyi:2021sxv}%
  \BibitemOpen
  \bibfield  {author} {\bibinfo {author} {\bibfnamefont {S.}~\bibnamefont
  {Bors\'anyi}}, \bibinfo {author} {\bibfnamefont {Z.}~\bibnamefont {Fodor}},
  \bibinfo {author} {\bibfnamefont {J.~N.}\ \bibnamefont {Guenther}}, \bibinfo
  {author} {\bibfnamefont {R.}~\bibnamefont {Kara}}, \bibinfo {author}
  {\bibfnamefont {S.~D.}\ \bibnamefont {Katz}}, \bibinfo {author}
  {\bibfnamefont {P.}~\bibnamefont {Parotto}}, \bibinfo {author} {\bibfnamefont
  {A.}~\bibnamefont {P\'asztor}}, \bibinfo {author} {\bibfnamefont
  {C.}~\bibnamefont {Ratti}}, \ and\ \bibinfo {author} {\bibfnamefont {K.~K.}\
  \bibnamefont {Szab\'o}},\ }\href@noop {} {\  (\bibinfo {year} {2021})},\
  \Eprint {http://arxiv.org/abs/2102.06660} {arXiv:2102.06660 [hep-lat]}
  \BibitemShut {NoStop}%
\bibitem [{\citenamefont {Asakawa}\ and\ \citenamefont
  {Yazaki}(1989)}]{Asakawa:1989bq}%
  \BibitemOpen
  \bibfield  {author} {\bibinfo {author} {\bibfnamefont {M.}~\bibnamefont
  {Asakawa}}\ and\ \bibinfo {author} {\bibfnamefont {K.}~\bibnamefont
  {Yazaki}},\ }\href {\doibase 10.1016/0375-9474(89)90002-X} {\bibfield
  {journal} {\bibinfo  {journal} {Nucl. Phys. A}\ }\textbf {\bibinfo {volume}
  {504}},\ \bibinfo {pages} {668} (\bibinfo {year} {1989})}\BibitemShut
  {NoStop}%
\bibitem [{\citenamefont {Berges}\ and\ \citenamefont
  {Rajagopal}(1999)}]{Berges:1998rc}%
  \BibitemOpen
  \bibfield  {author} {\bibinfo {author} {\bibfnamefont {J.}~\bibnamefont
  {Berges}}\ and\ \bibinfo {author} {\bibfnamefont {K.}~\bibnamefont
  {Rajagopal}},\ }\href {\doibase 10.1016/S0550-3213(98)00620-8} {\bibfield
  {journal} {\bibinfo  {journal} {Nucl. Phys. B}\ }\textbf {\bibinfo {volume}
  {538}},\ \bibinfo {pages} {215} (\bibinfo {year} {1999})},\ \Eprint
  {http://arxiv.org/abs/hep-ph/9804233} {arXiv:hep-ph/9804233} \BibitemShut
  {NoStop}%
\bibitem [{\citenamefont {Halasz}\ \emph {et~al.}(1998)\citenamefont {Halasz},
  \citenamefont {Jackson}, \citenamefont {Shrock}, \citenamefont {Stephanov},\
  and\ \citenamefont {Verbaarschot}}]{Halasz:1998qr}%
  \BibitemOpen
  \bibfield  {author} {\bibinfo {author} {\bibfnamefont {A.~M.}\ \bibnamefont
  {Halasz}}, \bibinfo {author} {\bibfnamefont {A.~D.}\ \bibnamefont {Jackson}},
  \bibinfo {author} {\bibfnamefont {R.~E.}\ \bibnamefont {Shrock}}, \bibinfo
  {author} {\bibfnamefont {M.~A.}\ \bibnamefont {Stephanov}}, \ and\ \bibinfo
  {author} {\bibfnamefont {J.~J.~M.}\ \bibnamefont {Verbaarschot}},\ }\href
  {\doibase 10.1103/PhysRevD.58.096007} {\bibfield  {journal} {\bibinfo
  {journal} {Phys. Rev. D}\ }\textbf {\bibinfo {volume} {58}},\ \bibinfo
  {pages} {096007} (\bibinfo {year} {1998})},\ \Eprint
  {http://arxiv.org/abs/hep-ph/9804290} {arXiv:hep-ph/9804290} \BibitemShut
  {NoStop}%
\bibitem [{\citenamefont {Adam}\ \emph {et~al.}(2021)\citenamefont {Adam} \emph
  {et~al.}}]{Adam:2020unf}%
  \BibitemOpen
  \bibfield  {author} {\bibinfo {author} {\bibfnamefont {J.}~\bibnamefont
  {Adam}} \emph {et~al.} (\bibinfo {collaboration} {STAR}),\ }\href {\doibase
  10.1103/PhysRevLett.126.092301} {\bibfield  {journal} {\bibinfo  {journal}
  {Phys. Rev. Lett.}\ }\textbf {\bibinfo {volume} {126}},\ \bibinfo {pages}
  {092301} (\bibinfo {year} {2021})},\ \Eprint
  {http://arxiv.org/abs/2001.02852} {arXiv:2001.02852 [nucl-ex]} \BibitemShut
  {NoStop}%
\bibitem [{\citenamefont {Dexheimer}\ \emph
  {et~al.}(2021{\natexlab{b}})\citenamefont {Dexheimer}, \citenamefont
  {Noronha}, \citenamefont {Noronha-Hostler}, \citenamefont {Ratti},\ and\
  \citenamefont {Yunes}}]{Dexheimer:2020zzs}%
  \BibitemOpen
  \bibfield  {author} {\bibinfo {author} {\bibfnamefont {V.}~\bibnamefont
  {Dexheimer}}, \bibinfo {author} {\bibfnamefont {J.}~\bibnamefont {Noronha}},
  \bibinfo {author} {\bibfnamefont {J.}~\bibnamefont {Noronha-Hostler}},
  \bibinfo {author} {\bibfnamefont {C.}~\bibnamefont {Ratti}}, \ and\ \bibinfo
  {author} {\bibfnamefont {N.}~\bibnamefont {Yunes}},\ }\href {\doibase
  10.1088/1361-6471/abe104} {\bibfield  {journal} {\bibinfo  {journal} {J.
  Phys. G}\ }\textbf {\bibinfo {volume} {48}},\ \bibinfo {pages} {073001}
  (\bibinfo {year} {2021}{\natexlab{b}})},\ \Eprint
  {http://arxiv.org/abs/2010.08834} {arXiv:2010.08834 [nucl-th]} \BibitemShut
  {NoStop}%
\bibitem [{\citenamefont {Ratti}(2018)}]{Ratti:2018ksb}%
  \BibitemOpen
  \bibfield  {author} {\bibinfo {author} {\bibfnamefont {C.}~\bibnamefont
  {Ratti}},\ }\href {\doibase 10.1088/1361-6633/aabb97} {\bibfield  {journal}
  {\bibinfo  {journal} {Rept. Prog. Phys.}\ }\textbf {\bibinfo {volume} {81}},\
  \bibinfo {pages} {084301} (\bibinfo {year} {2018})},\ \Eprint
  {http://arxiv.org/abs/1804.07810} {arXiv:1804.07810 [hep-lat]} \BibitemShut
  {NoStop}%
\bibitem [{\citenamefont {Bzdak}\ \emph {et~al.}(2020)\citenamefont {Bzdak},
  \citenamefont {Esumi}, \citenamefont {Koch}, \citenamefont {Liao},
  \citenamefont {Stephanov},\ and\ \citenamefont {Xu}}]{Bzdak:2019pkr}%
  \BibitemOpen
  \bibfield  {author} {\bibinfo {author} {\bibfnamefont {A.}~\bibnamefont
  {Bzdak}}, \bibinfo {author} {\bibfnamefont {S.}~\bibnamefont {Esumi}},
  \bibinfo {author} {\bibfnamefont {V.}~\bibnamefont {Koch}}, \bibinfo {author}
  {\bibfnamefont {J.}~\bibnamefont {Liao}}, \bibinfo {author} {\bibfnamefont
  {M.}~\bibnamefont {Stephanov}}, \ and\ \bibinfo {author} {\bibfnamefont
  {N.}~\bibnamefont {Xu}},\ }\href {\doibase 10.1016/j.physrep.2020.01.005}
  {\bibfield  {journal} {\bibinfo  {journal} {Phys. Rept.}\ }\textbf {\bibinfo
  {volume} {853}},\ \bibinfo {pages} {1} (\bibinfo {year} {2020})},\ \Eprint
  {http://arxiv.org/abs/1906.00936} {arXiv:1906.00936 [nucl-th]} \BibitemShut
  {NoStop}%
\bibitem [{\citenamefont {Monnai}\ \emph {et~al.}(2021)\citenamefont {Monnai},
  \citenamefont {Schenke},\ and\ \citenamefont {Shen}}]{Monnai:2021kgu}%
  \BibitemOpen
  \bibfield  {author} {\bibinfo {author} {\bibfnamefont {A.}~\bibnamefont
  {Monnai}}, \bibinfo {author} {\bibfnamefont {B.}~\bibnamefont {Schenke}}, \
  and\ \bibinfo {author} {\bibfnamefont {C.}~\bibnamefont {Shen}},\ }\href
  {\doibase 10.1142/S0217751X21300076} {\bibfield  {journal} {\bibinfo
  {journal} {Int. J. Mod. Phys. A}\ }\textbf {\bibinfo {volume} {36}},\
  \bibinfo {pages} {2130007} (\bibinfo {year} {2021})},\ \Eprint
  {http://arxiv.org/abs/2101.11591} {arXiv:2101.11591 [nucl-th]} \BibitemShut
  {NoStop}%
\bibitem [{\citenamefont {Gerlach}(1968)}]{Gerlach:1968zz}%
  \BibitemOpen
  \bibfield  {author} {\bibinfo {author} {\bibfnamefont {U.~H.}\ \bibnamefont
  {Gerlach}},\ }\href {\doibase 10.1103/PhysRev.172.1325} {\bibfield  {journal}
  {\bibinfo  {journal} {Phys. Rev.}\ }\textbf {\bibinfo {volume} {172}},\
  \bibinfo {pages} {1325} (\bibinfo {year} {1968})}\BibitemShut {NoStop}%
\bibitem [{\citenamefont {Kampfer}(1981)}]{Kampfer:1981yr}%
  \BibitemOpen
  \bibfield  {author} {\bibinfo {author} {\bibfnamefont {B.}~\bibnamefont
  {Kampfer}},\ }\href {\doibase 10.1088/0305-4470/14/11/009} {\bibfield
  {journal} {\bibinfo  {journal} {J. Phys. A}\ }\textbf {\bibinfo {volume}
  {14}},\ \bibinfo {pages} {L471} (\bibinfo {year} {1981})}\BibitemShut
  {NoStop}%
\bibitem [{\citenamefont {Han}\ and\ \citenamefont
  {Prakash}(2020)}]{Han:2020adu}%
  \BibitemOpen
  \bibfield  {author} {\bibinfo {author} {\bibfnamefont {S.}~\bibnamefont
  {Han}}\ and\ \bibinfo {author} {\bibfnamefont {M.}~\bibnamefont {Prakash}},\
  }\href {\doibase 10.3847/1538-4357/aba3c7} {\bibfield  {journal} {\bibinfo
  {journal} {Astrophys. J.}\ }\textbf {\bibinfo {volume} {899}},\ \bibinfo
  {pages} {164} (\bibinfo {year} {2020})},\ \Eprint
  {http://arxiv.org/abs/2006.02207} {arXiv:2006.02207 [astro-ph.HE]}
  \BibitemShut {NoStop}%
\bibitem [{\citenamefont {Pang}\ \emph {et~al.}(2020)\citenamefont {Pang},
  \citenamefont {Dietrich}, \citenamefont {Tews},\ and\ \citenamefont {Van
  Den~Broeck}}]{Pang:2020ilf}%
  \BibitemOpen
  \bibfield  {author} {\bibinfo {author} {\bibfnamefont {P.~T.~H.}\
  \bibnamefont {Pang}}, \bibinfo {author} {\bibfnamefont {T.}~\bibnamefont
  {Dietrich}}, \bibinfo {author} {\bibfnamefont {I.}~\bibnamefont {Tews}}, \
  and\ \bibinfo {author} {\bibfnamefont {C.}~\bibnamefont {Van Den~Broeck}},\
  }\href {\doibase 10.1103/PhysRevResearch.2.033514} {\bibfield  {journal}
  {\bibinfo  {journal} {Phys. Rev. Res.}\ }\textbf {\bibinfo {volume} {2}},\
  \bibinfo {pages} {033514} (\bibinfo {year} {2020})},\ \Eprint
  {http://arxiv.org/abs/2006.14936} {arXiv:2006.14936 [astro-ph.HE]}
  \BibitemShut {NoStop}%
\bibitem [{\citenamefont {Jakobus}\ \emph {et~al.}(2021)\citenamefont
  {Jakobus}, \citenamefont {Motornenko}, \citenamefont {Gomes}, \citenamefont
  {Steinheimer},\ and\ \citenamefont {Stoecker}}]{Jakobus:2020nxw}%
  \BibitemOpen
  \bibfield  {author} {\bibinfo {author} {\bibfnamefont {P.}~\bibnamefont
  {Jakobus}}, \bibinfo {author} {\bibfnamefont {A.}~\bibnamefont {Motornenko}},
  \bibinfo {author} {\bibfnamefont {R.~O.}\ \bibnamefont {Gomes}}, \bibinfo
  {author} {\bibfnamefont {J.}~\bibnamefont {Steinheimer}}, \ and\ \bibinfo
  {author} {\bibfnamefont {H.}~\bibnamefont {Stoecker}},\ }\href {\doibase
  10.1140/epjc/s10052-020-08779-x} {\bibfield  {journal} {\bibinfo  {journal}
  {Eur. Phys. J. C}\ }\textbf {\bibinfo {volume} {81}},\ \bibinfo {pages} {41}
  (\bibinfo {year} {2021})},\ \Eprint {http://arxiv.org/abs/2004.07026}
  {arXiv:2004.07026 [nucl-th]} \BibitemShut {NoStop}%
\bibitem [{\citenamefont {Glendenning}(1992)}]{Glendenning:1992vb}%
  \BibitemOpen
  \bibfield  {author} {\bibinfo {author} {\bibfnamefont {N.~K.}\ \bibnamefont
  {Glendenning}},\ }\href {\doibase 10.1103/PhysRevD.46.1274} {\bibfield
  {journal} {\bibinfo  {journal} {Phys. Rev. D}\ }\textbf {\bibinfo {volume}
  {46}},\ \bibinfo {pages} {1274} (\bibinfo {year} {1992})}\BibitemShut
  {NoStop}%
\bibitem [{\citenamefont {Baym}(1979)}]{Baym:1979etb}%
  \BibitemOpen
  \bibfield  {author} {\bibinfo {author} {\bibfnamefont {G.}~\bibnamefont
  {Baym}},\ }\href {\doibase 10.1016/0378-4371(79)90200-0} {\bibfield
  {journal} {\bibinfo  {journal} {Physica A}\ }\textbf {\bibinfo {volume}
  {96}},\ \bibinfo {pages} {131} (\bibinfo {year} {1979})}\BibitemShut
  {NoStop}%
\bibitem [{\citenamefont {Dexheimer}\ \emph {et~al.}(2015)\citenamefont
  {Dexheimer}, \citenamefont {Negreiros},\ and\ \citenamefont
  {Schramm}}]{Dexheimer:2014pea}%
  \BibitemOpen
  \bibfield  {author} {\bibinfo {author} {\bibfnamefont {V.}~\bibnamefont
  {Dexheimer}}, \bibinfo {author} {\bibfnamefont {R.}~\bibnamefont
  {Negreiros}}, \ and\ \bibinfo {author} {\bibfnamefont {S.}~\bibnamefont
  {Schramm}},\ }\href {\doibase 10.1103/PhysRevC.91.055808} {\bibfield
  {journal} {\bibinfo  {journal} {Phys. Rev. C}\ }\textbf {\bibinfo {volume}
  {91}},\ \bibinfo {pages} {055808} (\bibinfo {year} {2015})},\ \Eprint
  {http://arxiv.org/abs/1411.4623} {arXiv:1411.4623 [astro-ph.HE]} \BibitemShut
  {NoStop}%
\bibitem [{\citenamefont {Dutra}\ \emph {et~al.}(2016)\citenamefont {Dutra},
  \citenamefont {Louren\c{c}o},\ and\ \citenamefont {Menezes}}]{Dutra:2015hxa}%
  \BibitemOpen
  \bibfield  {author} {\bibinfo {author} {\bibfnamefont {M.}~\bibnamefont
  {Dutra}}, \bibinfo {author} {\bibfnamefont {O.}~\bibnamefont {Louren\c{c}o}},
  \ and\ \bibinfo {author} {\bibfnamefont {D.~P.}\ \bibnamefont {Menezes}},\
  }\href {\doibase 10.1103/PhysRevC.93.025806} {\bibfield  {journal} {\bibinfo
  {journal} {Phys. Rev. C}\ }\textbf {\bibinfo {volume} {93}},\ \bibinfo
  {pages} {025806} (\bibinfo {year} {2016})},\ \bibinfo {note} {[Erratum:
  Phys.Rev.C 94, 049901 (2016)]},\ \Eprint {http://arxiv.org/abs/1510.02060}
  {arXiv:1510.02060 [astro-ph.HE]} \BibitemShut {NoStop}%
\bibitem [{\citenamefont {McLerran}\ and\ \citenamefont
  {Reddy}(2019)}]{McLerran:2018hbz}%
  \BibitemOpen
  \bibfield  {author} {\bibinfo {author} {\bibfnamefont {L.}~\bibnamefont
  {McLerran}}\ and\ \bibinfo {author} {\bibfnamefont {S.}~\bibnamefont
  {Reddy}},\ }\href {\doibase 10.1103/PhysRevLett.122.122701} {\bibfield
  {journal} {\bibinfo  {journal} {Phys. Rev. Lett.}\ }\textbf {\bibinfo
  {volume} {122}},\ \bibinfo {pages} {122701} (\bibinfo {year} {2019})},\
  \Eprint {http://arxiv.org/abs/1811.12503} {arXiv:1811.12503 [nucl-th]}
  \BibitemShut {NoStop}%
\bibitem [{\citenamefont {Alford}\ and\ \citenamefont
  {Sedrakian}(2017)}]{Alford:2017qgh}%
  \BibitemOpen
  \bibfield  {author} {\bibinfo {author} {\bibfnamefont {M.~G.}\ \bibnamefont
  {Alford}}\ and\ \bibinfo {author} {\bibfnamefont {A.}~\bibnamefont
  {Sedrakian}},\ }\href {\doibase 10.1103/PhysRevLett.119.161104} {\bibfield
  {journal} {\bibinfo  {journal} {Phys. Rev. Lett.}\ }\textbf {\bibinfo
  {volume} {119}},\ \bibinfo {pages} {161104} (\bibinfo {year} {2017})},\
  \Eprint {http://arxiv.org/abs/1706.01592} {arXiv:1706.01592 [astro-ph.HE]}
  \BibitemShut {NoStop}%
\bibitem [{\citenamefont {Zacchi}\ \emph {et~al.}(2016)\citenamefont {Zacchi},
  \citenamefont {Hanauske},\ and\ \citenamefont
  {Schaffner-Bielich}}]{Zacchi:2015oma}%
  \BibitemOpen
  \bibfield  {author} {\bibinfo {author} {\bibfnamefont {A.}~\bibnamefont
  {Zacchi}}, \bibinfo {author} {\bibfnamefont {M.}~\bibnamefont {Hanauske}}, \
  and\ \bibinfo {author} {\bibfnamefont {J.}~\bibnamefont
  {Schaffner-Bielich}},\ }\href {\doibase 10.1103/PhysRevD.93.065011}
  {\bibfield  {journal} {\bibinfo  {journal} {Phys. Rev. D}\ }\textbf {\bibinfo
  {volume} {93}},\ \bibinfo {pages} {065011} (\bibinfo {year} {2016})},\
  \Eprint {http://arxiv.org/abs/1510.00180} {arXiv:1510.00180 [nucl-th]}
  \BibitemShut {NoStop}%
\bibitem [{\citenamefont {Alvarez-Castillo}\ \emph {et~al.}(2019)\citenamefont
  {Alvarez-Castillo}, \citenamefont {Blaschke}, \citenamefont {Grunfeld},\ and\
  \citenamefont {Pagura}}]{Alvarez-Castillo:2018pve}%
  \BibitemOpen
  \bibfield  {author} {\bibinfo {author} {\bibfnamefont {D.~E.}\ \bibnamefont
  {Alvarez-Castillo}}, \bibinfo {author} {\bibfnamefont {D.~B.}\ \bibnamefont
  {Blaschke}}, \bibinfo {author} {\bibfnamefont {A.~G.}\ \bibnamefont
  {Grunfeld}}, \ and\ \bibinfo {author} {\bibfnamefont {V.~P.}\ \bibnamefont
  {Pagura}},\ }\href {\doibase 10.1103/PhysRevD.99.063010} {\bibfield
  {journal} {\bibinfo  {journal} {Phys. Rev. D}\ }\textbf {\bibinfo {volume}
  {99}},\ \bibinfo {pages} {063010} (\bibinfo {year} {2019})},\ \Eprint
  {http://arxiv.org/abs/1805.04105} {arXiv:1805.04105 [hep-ph]} \BibitemShut
  {NoStop}%
\bibitem [{\citenamefont {Li}\ \emph {et~al.}(2020{\natexlab{b}})\citenamefont
  {Li}, \citenamefont {Sedrakian},\ and\ \citenamefont {Alford}}]{Li:2019fqe}%
  \BibitemOpen
  \bibfield  {author} {\bibinfo {author} {\bibfnamefont {J.~J.}\ \bibnamefont
  {Li}}, \bibinfo {author} {\bibfnamefont {A.}~\bibnamefont {Sedrakian}}, \
  and\ \bibinfo {author} {\bibfnamefont {M.}~\bibnamefont {Alford}},\ }\href
  {\doibase 10.1103/PhysRevD.101.063022} {\bibfield  {journal} {\bibinfo
  {journal} {Phys. Rev. D}\ }\textbf {\bibinfo {volume} {101}},\ \bibinfo
  {pages} {063022} (\bibinfo {year} {2020}{\natexlab{b}})},\ \Eprint
  {http://arxiv.org/abs/1911.00276} {arXiv:1911.00276 [astro-ph.HE]}
  \BibitemShut {NoStop}%
\bibitem [{\citenamefont {Wang}\ \emph {et~al.}(2019)\citenamefont {Wang},
  \citenamefont {Shi}, \citenamefont {Yan},\ and\ \citenamefont
  {Zong}}]{Wang:2019npj}%
  \BibitemOpen
  \bibfield  {author} {\bibinfo {author} {\bibfnamefont {Q.-w.}\ \bibnamefont
  {Wang}}, \bibinfo {author} {\bibfnamefont {C.}~\bibnamefont {Shi}}, \bibinfo
  {author} {\bibfnamefont {Y.}~\bibnamefont {Yan}}, \ and\ \bibinfo {author}
  {\bibfnamefont {H.-S.}\ \bibnamefont {Zong}},\ }\href@noop {} {\  (\bibinfo
  {year} {2019})},\ \Eprint {http://arxiv.org/abs/1912.02312} {arXiv:1912.02312
  [hep-ph]} \BibitemShut {NoStop}%
\bibitem [{\citenamefont {Bitaghsir~Fadafan}\ \emph {et~al.}(2020)\citenamefont
  {Bitaghsir~Fadafan}, \citenamefont {Cruz~Rojas},\ and\ \citenamefont
  {Evans}}]{Fadafa:2019euu}%
  \BibitemOpen
  \bibfield  {author} {\bibinfo {author} {\bibfnamefont {K.}~\bibnamefont
  {Bitaghsir~Fadafan}}, \bibinfo {author} {\bibfnamefont {J.}~\bibnamefont
  {Cruz~Rojas}}, \ and\ \bibinfo {author} {\bibfnamefont {N.}~\bibnamefont
  {Evans}},\ }\href {\doibase 10.1103/PhysRevD.101.126005} {\bibfield
  {journal} {\bibinfo  {journal} {Phys. Rev. D}\ }\textbf {\bibinfo {volume}
  {101}},\ \bibinfo {pages} {126005} (\bibinfo {year} {2020})},\ \Eprint
  {http://arxiv.org/abs/1911.12705} {arXiv:1911.12705 [hep-ph]} \BibitemShut
  {NoStop}%
\bibitem [{\citenamefont {Xia}\ \emph {et~al.}(2021)\citenamefont {Xia},
  \citenamefont {Zhu}, \citenamefont {Zhou},\ and\ \citenamefont
  {Li}}]{Xia:2019xax}%
  \BibitemOpen
  \bibfield  {author} {\bibinfo {author} {\bibfnamefont {C.}~\bibnamefont
  {Xia}}, \bibinfo {author} {\bibfnamefont {Z.}~\bibnamefont {Zhu}}, \bibinfo
  {author} {\bibfnamefont {X.}~\bibnamefont {Zhou}}, \ and\ \bibinfo {author}
  {\bibfnamefont {A.}~\bibnamefont {Li}},\ }\href {\doibase
  10.1088/1674-1137/abea0d} {\bibfield  {journal} {\bibinfo  {journal} {Chin.
  Phys. C}\ }\textbf {\bibinfo {volume} {45}},\ \bibinfo {pages} {055104}
  (\bibinfo {year} {2021})},\ \Eprint {http://arxiv.org/abs/1906.00826}
  {arXiv:1906.00826 [nucl-th]} \BibitemShut {NoStop}%
\bibitem [{\citenamefont {Yazdizadeh}\ and\ \citenamefont
  {Bordbar}(2019)}]{Yazdizadeh:2019ivy}%
  \BibitemOpen
  \bibfield  {author} {\bibinfo {author} {\bibfnamefont {T.}~\bibnamefont
  {Yazdizadeh}}\ and\ \bibinfo {author} {\bibfnamefont {G.~H.}\ \bibnamefont
  {Bordbar}},\ }\href {\doibase 10.1007/s40995-019-00731-3} {\bibfield
  {journal} {\bibinfo  {journal} {Iran. J. Sci. Technol. A}\ }\textbf {\bibinfo
  {volume} {43}},\ \bibinfo {pages} {2691} (\bibinfo {year} {2019})},\ \Eprint
  {http://arxiv.org/abs/1906.00175} {arXiv:1906.00175 [nucl-th]} \BibitemShut
  {NoStop}%
\bibitem [{\citenamefont {Shahrbaf}\ \emph {et~al.}(2020)\citenamefont
  {Shahrbaf}, \citenamefont {Blaschke}, \citenamefont {Grunfeld},\ and\
  \citenamefont {Moshfegh}}]{Shahrbaf:2019vtf}%
  \BibitemOpen
  \bibfield  {author} {\bibinfo {author} {\bibfnamefont {M.}~\bibnamefont
  {Shahrbaf}}, \bibinfo {author} {\bibfnamefont {D.}~\bibnamefont {Blaschke}},
  \bibinfo {author} {\bibfnamefont {A.~G.}\ \bibnamefont {Grunfeld}}, \ and\
  \bibinfo {author} {\bibfnamefont {H.~R.}\ \bibnamefont {Moshfegh}},\ }\href
  {\doibase 10.1103/PhysRevC.101.025807} {\bibfield  {journal} {\bibinfo
  {journal} {Phys. Rev. C}\ }\textbf {\bibinfo {volume} {101}},\ \bibinfo
  {pages} {025807} (\bibinfo {year} {2020})},\ \Eprint
  {http://arxiv.org/abs/1908.04740} {arXiv:1908.04740 [nucl-th]} \BibitemShut
  {NoStop}%
\bibitem [{\citenamefont {Zacchi}\ and\ \citenamefont
  {Schaffner-Bielich}(2019)}]{Zacchi:2019ayh}%
  \BibitemOpen
  \bibfield  {author} {\bibinfo {author} {\bibfnamefont {A.}~\bibnamefont
  {Zacchi}}\ and\ \bibinfo {author} {\bibfnamefont {J.}~\bibnamefont
  {Schaffner-Bielich}},\ }\href {\doibase 10.1103/PhysRevD.100.123024}
  {\bibfield  {journal} {\bibinfo  {journal} {Phys. Rev. D}\ }\textbf {\bibinfo
  {volume} {100}},\ \bibinfo {pages} {123024} (\bibinfo {year} {2019})},\
  \Eprint {http://arxiv.org/abs/1909.12071} {arXiv:1909.12071 [nucl-th]}
  \BibitemShut {NoStop}%
\bibitem [{\citenamefont {Zhao}\ and\ \citenamefont
  {Lattimer}(2020)}]{Zhao:2020dvu}%
  \BibitemOpen
  \bibfield  {author} {\bibinfo {author} {\bibfnamefont {T.}~\bibnamefont
  {Zhao}}\ and\ \bibinfo {author} {\bibfnamefont {J.~M.}\ \bibnamefont
  {Lattimer}},\ }\href {\doibase 10.1103/PhysRevD.102.023021} {\bibfield
  {journal} {\bibinfo  {journal} {Phys. Rev. D}\ }\textbf {\bibinfo {volume}
  {102}},\ \bibinfo {pages} {023021} (\bibinfo {year} {2020})},\ \Eprint
  {http://arxiv.org/abs/2004.08293} {arXiv:2004.08293 [astro-ph.HE]}
  \BibitemShut {NoStop}%
\bibitem [{\citenamefont {Lopes}\ and\ \citenamefont
  {Menezes}(2021)}]{Lopes:2020rqn}%
  \BibitemOpen
  \bibfield  {author} {\bibinfo {author} {\bibfnamefont {L.~L.}\ \bibnamefont
  {Lopes}}\ and\ \bibinfo {author} {\bibfnamefont {D.~P.}\ \bibnamefont
  {Menezes}},\ }\href {\doibase 10.1016/j.nuclphysa.2021.122171} {\bibfield
  {journal} {\bibinfo  {journal} {Nucl. Phys. A}\ }\textbf {\bibinfo {volume}
  {1009}},\ \bibinfo {pages} {122171} (\bibinfo {year} {2021})},\ \Eprint
  {http://arxiv.org/abs/2004.07909} {arXiv:2004.07909 [astro-ph.HE]}
  \BibitemShut {NoStop}%
\bibitem [{\citenamefont {Blaschke}\ \emph {et~al.}(2020)\citenamefont
  {Blaschke}, \citenamefont {Grigorian},\ and\ \citenamefont
  {R\"opke}}]{Blaschke:2020qrs}%
  \BibitemOpen
  \bibfield  {author} {\bibinfo {author} {\bibfnamefont {D.}~\bibnamefont
  {Blaschke}}, \bibinfo {author} {\bibfnamefont {H.}~\bibnamefont {Grigorian}},
  \ and\ \bibinfo {author} {\bibfnamefont {G.}~\bibnamefont {R\"opke}},\ }\href
  {\doibase 10.3390/particles3020033} {\bibfield  {journal} {\bibinfo
  {journal} {Particles}\ }\textbf {\bibinfo {volume} {3}},\ \bibinfo {pages}
  {477} (\bibinfo {year} {2020})},\ \Eprint {http://arxiv.org/abs/2005.10218}
  {arXiv:2005.10218 [nucl-th]} \BibitemShut {NoStop}%
\bibitem [{\citenamefont {Duarte}\ \emph {et~al.}(2020)\citenamefont {Duarte},
  \citenamefont {Hernandez-Ortiz},\ and\ \citenamefont
  {Jeong}}]{Duarte:2020xsp}%
  \BibitemOpen
  \bibfield  {author} {\bibinfo {author} {\bibfnamefont {D.~C.}\ \bibnamefont
  {Duarte}}, \bibinfo {author} {\bibfnamefont {S.}~\bibnamefont
  {Hernandez-Ortiz}}, \ and\ \bibinfo {author} {\bibfnamefont {K.~S.}\
  \bibnamefont {Jeong}},\ }\href {\doibase 10.1103/PhysRevC.102.025203}
  {\bibfield  {journal} {\bibinfo  {journal} {Phys. Rev. C}\ }\textbf {\bibinfo
  {volume} {102}},\ \bibinfo {pages} {025203} (\bibinfo {year} {2020})},\
  \Eprint {http://arxiv.org/abs/2003.02362} {arXiv:2003.02362 [nucl-th]}
  \BibitemShut {NoStop}%
\bibitem [{\citenamefont {Rho}(2020)}]{Rho:2020eqo}%
  \BibitemOpen
  \bibfield  {author} {\bibinfo {author} {\bibfnamefont {M.}~\bibnamefont
  {Rho}},\ }\href@noop {} {\  (\bibinfo {year} {2020})},\ \Eprint
  {http://arxiv.org/abs/2004.09082} {arXiv:2004.09082 [nucl-th]} \BibitemShut
  {NoStop}%
\bibitem [{\citenamefont {Marczenko}(2020)}]{Marczenko:2020wlc}%
  \BibitemOpen
  \bibfield  {author} {\bibinfo {author} {\bibfnamefont {M.}~\bibnamefont
  {Marczenko}},\ }\href {\doibase 10.1140/epjst/e2020-000093-3} {\bibfield
  {journal} {\bibinfo  {journal} {Eur. Phys. J. ST}\ }\textbf {\bibinfo
  {volume} {229}},\ \bibinfo {pages} {3651} (\bibinfo {year} {2020})},\ \Eprint
  {http://arxiv.org/abs/2005.14535} {arXiv:2005.14535 [nucl-th]} \BibitemShut
  {NoStop}%
\bibitem [{\citenamefont {Minamikawa}\ \emph {et~al.}(2021)\citenamefont
  {Minamikawa}, \citenamefont {Kojo},\ and\ \citenamefont
  {Harada}}]{Minamikawa:2020jfj}%
  \BibitemOpen
  \bibfield  {author} {\bibinfo {author} {\bibfnamefont {T.}~\bibnamefont
  {Minamikawa}}, \bibinfo {author} {\bibfnamefont {T.}~\bibnamefont {Kojo}}, \
  and\ \bibinfo {author} {\bibfnamefont {M.}~\bibnamefont {Harada}},\ }\href
  {\doibase 10.1103/PhysRevC.103.045205} {\bibfield  {journal} {\bibinfo
  {journal} {Phys. Rev. C}\ }\textbf {\bibinfo {volume} {103}},\ \bibinfo
  {pages} {045205} (\bibinfo {year} {2021})},\ \Eprint
  {http://arxiv.org/abs/2011.13684} {arXiv:2011.13684 [nucl-th]} \BibitemShut
  {NoStop}%
\bibitem [{\citenamefont {Hippert}\ \emph {et~al.}(2021)\citenamefont
  {Hippert}, \citenamefont {Fraga},\ and\ \citenamefont
  {Noronha}}]{Hippert:2021gfs}%
  \BibitemOpen
  \bibfield  {author} {\bibinfo {author} {\bibfnamefont {M.}~\bibnamefont
  {Hippert}}, \bibinfo {author} {\bibfnamefont {E.~S.}\ \bibnamefont {Fraga}},
  \ and\ \bibinfo {author} {\bibfnamefont {J.}~\bibnamefont {Noronha}},\
  }\href@noop {} {\  (\bibinfo {year} {2021})},\ \Eprint
  {http://arxiv.org/abs/2105.04535} {arXiv:2105.04535 [nucl-th]} \BibitemShut
  {NoStop}%
\bibitem [{\citenamefont {Pisarski}(2021)}]{Pisarski:2021aoz}%
  \BibitemOpen
  \bibfield  {author} {\bibinfo {author} {\bibfnamefont {R.~D.}\ \bibnamefont
  {Pisarski}},\ }\href {\doibase 10.1103/PhysRevD.103.L071504} {\bibfield
  {journal} {\bibinfo  {journal} {Phys. Rev. D}\ }\textbf {\bibinfo {volume}
  {103}},\ \bibinfo {pages} {L071504} (\bibinfo {year} {2021})},\ \Eprint
  {http://arxiv.org/abs/2101.05813} {arXiv:2101.05813 [nucl-th]} \BibitemShut
  {NoStop}%
\bibitem [{\citenamefont {Sen}\ and\ \citenamefont
  {Sivertsen}(2020)}]{Sen:2020qcd}%
  \BibitemOpen
  \bibfield  {author} {\bibinfo {author} {\bibfnamefont {S.}~\bibnamefont
  {Sen}}\ and\ \bibinfo {author} {\bibfnamefont {L.}~\bibnamefont
  {Sivertsen}},\ }\href@noop {} {\  (\bibinfo {year} {2020})},\ \Eprint
  {http://arxiv.org/abs/2011.04681} {arXiv:2011.04681 [astro-ph.HE]}
  \BibitemShut {NoStop}%
\bibitem [{\citenamefont {Stone}\ \emph
  {et~al.}(2021{\natexlab{a}})\citenamefont {Stone}, \citenamefont {Dexheimer},
  \citenamefont {Guichon}, \citenamefont {Thomas},\ and\ \citenamefont
  {Typel}}]{Stone:2021ngh}%
  \BibitemOpen
  \bibfield  {author} {\bibinfo {author} {\bibfnamefont {J.~R.}\ \bibnamefont
  {Stone}}, \bibinfo {author} {\bibfnamefont {V.}~\bibnamefont {Dexheimer}},
  \bibinfo {author} {\bibfnamefont {P.~A.~M.}\ \bibnamefont {Guichon}},
  \bibinfo {author} {\bibfnamefont {A.~W.}\ \bibnamefont {Thomas}}, \ and\
  \bibinfo {author} {\bibfnamefont {S.}~\bibnamefont {Typel}},\ }\href
  {\doibase 10.1093/mnras/staa4006} {\bibfield  {journal} {\bibinfo  {journal}
  {Mon. Not. Roy. Astron. Soc.}\ }\textbf {\bibinfo {volume} {502}},\ \bibinfo
  {pages} {3476} (\bibinfo {year} {2021}{\natexlab{a}})},\ \Eprint
  {http://arxiv.org/abs/1906.11100} {arXiv:1906.11100 [nucl-th]} \BibitemShut
  {NoStop}%
\bibitem [{\citenamefont {Kapusta}\ and\ \citenamefont
  {Welle}(2021)}]{Kapusta:2021ney}%
  \BibitemOpen
  \bibfield  {author} {\bibinfo {author} {\bibfnamefont {J.~I.}\ \bibnamefont
  {Kapusta}}\ and\ \bibinfo {author} {\bibfnamefont {T.}~\bibnamefont
  {Welle}},\ }\href@noop {} {\  (\bibinfo {year} {2021})},\ \Eprint
  {http://arxiv.org/abs/2103.16633} {arXiv:2103.16633 [nucl-th]} \BibitemShut
  {NoStop}%
\bibitem [{\citenamefont {Somasundaram}\ and\ \citenamefont
  {Margueron}(2021)}]{Somasundaram:2021ljr}%
  \BibitemOpen
  \bibfield  {author} {\bibinfo {author} {\bibfnamefont {R.}~\bibnamefont
  {Somasundaram}}\ and\ \bibinfo {author} {\bibfnamefont {J.}~\bibnamefont
  {Margueron}},\ }\href@noop {} {\  (\bibinfo {year} {2021})},\ \Eprint
  {http://arxiv.org/abs/2104.13612} {arXiv:2104.13612 [astro-ph.HE]}
  \BibitemShut {NoStop}%
\bibitem [{\citenamefont {Motornenko}\ \emph {et~al.}(2020)\citenamefont
  {Motornenko}, \citenamefont {Steinheimer}, \citenamefont {Vovchenko},
  \citenamefont {Schramm},\ and\ \citenamefont
  {Stoecker}}]{Motornenko:2019arp}%
  \BibitemOpen
  \bibfield  {author} {\bibinfo {author} {\bibfnamefont {A.}~\bibnamefont
  {Motornenko}}, \bibinfo {author} {\bibfnamefont {J.}~\bibnamefont
  {Steinheimer}}, \bibinfo {author} {\bibfnamefont {V.}~\bibnamefont
  {Vovchenko}}, \bibinfo {author} {\bibfnamefont {S.}~\bibnamefont {Schramm}},
  \ and\ \bibinfo {author} {\bibfnamefont {H.}~\bibnamefont {Stoecker}},\
  }\href {\doibase 10.1103/PhysRevC.101.034904} {\bibfield  {journal} {\bibinfo
   {journal} {Phys. Rev. C}\ }\textbf {\bibinfo {volume} {101}},\ \bibinfo
  {pages} {034904} (\bibinfo {year} {2020})},\ \Eprint
  {http://arxiv.org/abs/1905.00866} {arXiv:1905.00866 [hep-ph]} \BibitemShut
  {NoStop}%
\bibitem [{\citenamefont {Baym}\ \emph {et~al.}(2019)\citenamefont {Baym},
  \citenamefont {Furusawa}, \citenamefont {Hatsuda}, \citenamefont {Kojo},\
  and\ \citenamefont {Togashi}}]{Baym:2019iky}%
  \BibitemOpen
  \bibfield  {author} {\bibinfo {author} {\bibfnamefont {G.}~\bibnamefont
  {Baym}}, \bibinfo {author} {\bibfnamefont {S.}~\bibnamefont {Furusawa}},
  \bibinfo {author} {\bibfnamefont {T.}~\bibnamefont {Hatsuda}}, \bibinfo
  {author} {\bibfnamefont {T.}~\bibnamefont {Kojo}}, \ and\ \bibinfo {author}
  {\bibfnamefont {H.}~\bibnamefont {Togashi}},\ }\href {\doibase
  10.3847/1538-4357/ab441e} {\bibfield  {journal} {\bibinfo  {journal}
  {Astrophys. J.}\ }\textbf {\bibinfo {volume} {885}},\ \bibinfo {pages} {42}
  (\bibinfo {year} {2019})},\ \Eprint {http://arxiv.org/abs/1903.08963}
  {arXiv:1903.08963 [astro-ph.HE]} \BibitemShut {NoStop}%
\bibitem [{\citenamefont {Sen}\ and\ \citenamefont
  {Warrington}(2021)}]{Sen:2020peq}%
  \BibitemOpen
  \bibfield  {author} {\bibinfo {author} {\bibfnamefont {S.}~\bibnamefont
  {Sen}}\ and\ \bibinfo {author} {\bibfnamefont {N.~C.}\ \bibnamefont
  {Warrington}},\ }\href {\doibase 10.1016/j.nuclphysa.2020.122059} {\bibfield
  {journal} {\bibinfo  {journal} {Nucl. Phys. A}\ }\textbf {\bibinfo {volume}
  {1006}},\ \bibinfo {pages} {122059} (\bibinfo {year} {2021})},\ \Eprint
  {http://arxiv.org/abs/2002.11133} {arXiv:2002.11133 [nucl-th]} \BibitemShut
  {NoStop}%
\bibitem [{\citenamefont {Abbott}\ \emph {et~al.}(2017)\citenamefont {Abbott}
  \emph {et~al.}}]{TheLIGOScientific:2017qsa}%
  \BibitemOpen
  \bibfield  {author} {\bibinfo {author} {\bibfnamefont {B.~P.}\ \bibnamefont
  {Abbott}} \emph {et~al.} (\bibinfo {collaboration} {LIGO Scientific,
  Virgo}),\ }\href {\doibase 10.1103/PhysRevLett.119.161101} {\bibfield
  {journal} {\bibinfo  {journal} {Phys. Rev. Lett.}\ }\textbf {\bibinfo
  {volume} {119}},\ \bibinfo {pages} {161101} (\bibinfo {year} {2017})},\
  \Eprint {http://arxiv.org/abs/1710.05832} {arXiv:1710.05832 [gr-qc]}
  \BibitemShut {NoStop}%
\bibitem [{\citenamefont {Riley}\ \emph {et~al.}(2019)\citenamefont {Riley}
  \emph {et~al.}}]{Riley:2019yda}%
  \BibitemOpen
  \bibfield  {author} {\bibinfo {author} {\bibfnamefont {T.~E.}\ \bibnamefont
  {Riley}} \emph {et~al.},\ }\href {\doibase 10.3847/2041-8213/ab481c}
  {\bibfield  {journal} {\bibinfo  {journal} {Astrophys. J. Lett.}\ }\textbf
  {\bibinfo {volume} {887}},\ \bibinfo {pages} {L21} (\bibinfo {year}
  {2019})},\ \Eprint {http://arxiv.org/abs/1912.05702} {arXiv:1912.05702
  [astro-ph.HE]} \BibitemShut {NoStop}%
\bibitem [{\citenamefont {Miller}\ \emph {et~al.}(2019)\citenamefont {Miller}
  \emph {et~al.}}]{Miller:2019cac}%
  \BibitemOpen
  \bibfield  {author} {\bibinfo {author} {\bibfnamefont {M.~C.}\ \bibnamefont
  {Miller}} \emph {et~al.},\ }\href {\doibase 10.3847/2041-8213/ab50c5}
  {\bibfield  {journal} {\bibinfo  {journal} {Astrophys. J. Lett.}\ }\textbf
  {\bibinfo {volume} {887}},\ \bibinfo {pages} {L24} (\bibinfo {year}
  {2019})},\ \Eprint {http://arxiv.org/abs/1912.05705} {arXiv:1912.05705
  [astro-ph.HE]} \BibitemShut {NoStop}%
\bibitem [{\citenamefont {Jayasinghe}\ \emph {et~al.}(2021)\citenamefont
  {Jayasinghe} \emph {et~al.}}]{Jayasinghe:2021uqb}%
  \BibitemOpen
  \bibfield  {author} {\bibinfo {author} {\bibfnamefont {T.}~\bibnamefont
  {Jayasinghe}} \emph {et~al.},\ }\href {\doibase 10.1093/mnras/stab907} {\
  (\bibinfo {year} {2021}),\ 10.1093/mnras/stab907},\ \Eprint
  {http://arxiv.org/abs/2101.02212} {arXiv:2101.02212 [astro-ph.SR]}
  \BibitemShut {NoStop}%
\bibitem [{\citenamefont {Baym}\ \emph {et~al.}(2018)\citenamefont {Baym},
  \citenamefont {Hatsuda}, \citenamefont {Kojo}, \citenamefont {Powell},
  \citenamefont {Song},\ and\ \citenamefont {Takatsuka}}]{Baym:2017whm}%
  \BibitemOpen
  \bibfield  {author} {\bibinfo {author} {\bibfnamefont {G.}~\bibnamefont
  {Baym}}, \bibinfo {author} {\bibfnamefont {T.}~\bibnamefont {Hatsuda}},
  \bibinfo {author} {\bibfnamefont {T.}~\bibnamefont {Kojo}}, \bibinfo {author}
  {\bibfnamefont {P.~D.}\ \bibnamefont {Powell}}, \bibinfo {author}
  {\bibfnamefont {Y.}~\bibnamefont {Song}}, \ and\ \bibinfo {author}
  {\bibfnamefont {T.}~\bibnamefont {Takatsuka}},\ }\href {\doibase
  10.1088/1361-6633/aaae14} {\bibfield  {journal} {\bibinfo  {journal} {Rept.
  Prog. Phys.}\ }\textbf {\bibinfo {volume} {81}},\ \bibinfo {pages} {056902}
  (\bibinfo {year} {2018})},\ \Eprint {http://arxiv.org/abs/1707.04966}
  {arXiv:1707.04966 [astro-ph.HE]} \BibitemShut {NoStop}%
\bibitem [{\citenamefont {Rather}\ \emph
  {et~al.}(2021{\natexlab{a}})\citenamefont {Rather}, \citenamefont {Usmani},\
  and\ \citenamefont {Patra}}]{Rather:2020gja}%
  \BibitemOpen
  \bibfield  {author} {\bibinfo {author} {\bibfnamefont {I.~A.}\ \bibnamefont
  {Rather}}, \bibinfo {author} {\bibfnamefont {A.~A.}\ \bibnamefont {Usmani}},
  \ and\ \bibinfo {author} {\bibfnamefont {S.~K.}\ \bibnamefont {Patra}},\
  }\href {\doibase 10.1016/j.nuclphysa.2021.122189} {\bibfield  {journal}
  {\bibinfo  {journal} {Nucl. Phys. A}\ }\textbf {\bibinfo {volume} {1010}},\
  \bibinfo {pages} {122189} (\bibinfo {year} {2021}{\natexlab{a}})},\ \Eprint
  {http://arxiv.org/abs/2009.12613} {arXiv:2009.12613 [nucl-th]} \BibitemShut
  {NoStop}%
\bibitem [{\citenamefont {Ferreira}\ and\ \citenamefont
  {Provid\^encia}(2020)}]{Ferreira:2020zzy}%
  \BibitemOpen
  \bibfield  {author} {\bibinfo {author} {\bibfnamefont {M.}~\bibnamefont
  {Ferreira}}\ and\ \bibinfo {author} {\bibfnamefont {C.}~\bibnamefont
  {Provid\^encia}},\ }\href {\doibase 10.1103/PhysRevD.102.103003} {\bibfield
  {journal} {\bibinfo  {journal} {Phys. Rev. D}\ }\textbf {\bibinfo {volume}
  {102}},\ \bibinfo {pages} {103003} (\bibinfo {year} {2020})},\ \Eprint
  {http://arxiv.org/abs/2010.05588} {arXiv:2010.05588 [nucl-th]} \BibitemShut
  {NoStop}%
\bibitem [{\citenamefont {Riley}\ \emph {et~al.}(2021)\citenamefont {Riley}
  \emph {et~al.}}]{Riley:2021pdl}%
  \BibitemOpen
  \bibfield  {author} {\bibinfo {author} {\bibfnamefont {T.~E.}\ \bibnamefont
  {Riley}} \emph {et~al.},\ }\href@noop {} {\  (\bibinfo {year} {2021})},\
  \Eprint {http://arxiv.org/abs/2105.06980} {arXiv:2105.06980 [astro-ph.HE]}
  \BibitemShut {NoStop}%
\bibitem [{\citenamefont {Miller}\ \emph {et~al.}(2021)\citenamefont {Miller}
  \emph {et~al.}}]{Miller:2021qha}%
  \BibitemOpen
  \bibfield  {author} {\bibinfo {author} {\bibfnamefont {M.~C.}\ \bibnamefont
  {Miller}} \emph {et~al.},\ }\href@noop {} {\  (\bibinfo {year} {2021})},\
  \Eprint {http://arxiv.org/abs/2105.06979} {arXiv:2105.06979 [astro-ph.HE]}
  \BibitemShut {NoStop}%
\bibitem [{\citenamefont {Yagi}\ and\ \citenamefont
  {Yunes}(2017{\natexlab{a}})}]{Yagi:2016bkt}%
  \BibitemOpen
  \bibfield  {author} {\bibinfo {author} {\bibfnamefont {K.}~\bibnamefont
  {Yagi}}\ and\ \bibinfo {author} {\bibfnamefont {N.}~\bibnamefont {Yunes}},\
  }\href {\doibase 10.1016/j.physrep.2017.03.002} {\bibfield  {journal}
  {\bibinfo  {journal} {Phys. Rept.}\ }\textbf {\bibinfo {volume} {681}},\
  \bibinfo {pages} {1} (\bibinfo {year} {2017}{\natexlab{a}})},\ \Eprint
  {http://arxiv.org/abs/1608.02582} {arXiv:1608.02582 [gr-qc]} \BibitemShut
  {NoStop}%
\bibitem [{\citenamefont {Martinon}\ \emph {et~al.}(2014)\citenamefont
  {Martinon}, \citenamefont {Maselli}, \citenamefont {Gualtieri},\ and\
  \citenamefont {Ferrari}}]{Martinon:2014uua}%
  \BibitemOpen
  \bibfield  {author} {\bibinfo {author} {\bibfnamefont {G.}~\bibnamefont
  {Martinon}}, \bibinfo {author} {\bibfnamefont {A.}~\bibnamefont {Maselli}},
  \bibinfo {author} {\bibfnamefont {L.}~\bibnamefont {Gualtieri}}, \ and\
  \bibinfo {author} {\bibfnamefont {V.}~\bibnamefont {Ferrari}},\ }\href
  {\doibase 10.1103/PhysRevD.90.064026} {\bibfield  {journal} {\bibinfo
  {journal} {Phys. Rev. D}\ }\textbf {\bibinfo {volume} {90}},\ \bibinfo
  {pages} {064026} (\bibinfo {year} {2014})},\ \Eprint
  {http://arxiv.org/abs/1406.7661} {arXiv:1406.7661 [gr-qc]} \BibitemShut
  {NoStop}%
\bibitem [{\citenamefont {Alford}\ \emph {et~al.}(2017)\citenamefont {Alford},
  \citenamefont {Harris},\ and\ \citenamefont {Sachdeva}}]{Alford:2017vca}%
  \BibitemOpen
  \bibfield  {author} {\bibinfo {author} {\bibfnamefont {M.~G.}\ \bibnamefont
  {Alford}}, \bibinfo {author} {\bibfnamefont {S.~P.}\ \bibnamefont {Harris}},
  \ and\ \bibinfo {author} {\bibfnamefont {P.~S.}\ \bibnamefont {Sachdeva}},\
  }\href {\doibase 10.3847/1538-4357/aa8509} {\bibfield  {journal} {\bibinfo
  {journal} {Astrophys. J.}\ }\textbf {\bibinfo {volume} {847}},\ \bibinfo
  {pages} {109} (\bibinfo {year} {2017})},\ \Eprint
  {http://arxiv.org/abs/1705.09880} {arXiv:1705.09880 [astro-ph.HE]}
  \BibitemShut {NoStop}%
\bibitem [{\citenamefont {Pereira}\ \emph {et~al.}(2018)\citenamefont
  {Pereira}, \citenamefont {Flores},\ and\ \citenamefont
  {Lugones}}]{Pereira:2017rmp}%
  \BibitemOpen
  \bibfield  {author} {\bibinfo {author} {\bibfnamefont {J.~P.}\ \bibnamefont
  {Pereira}}, \bibinfo {author} {\bibfnamefont {C.~V.}\ \bibnamefont {Flores}},
  \ and\ \bibinfo {author} {\bibfnamefont {G.}~\bibnamefont {Lugones}},\ }\href
  {\doibase 10.3847/1538-4357/aabfbf} {\bibfield  {journal} {\bibinfo
  {journal} {Astrophys. J.}\ }\textbf {\bibinfo {volume} {860}},\ \bibinfo
  {pages} {12} (\bibinfo {year} {2018})},\ \Eprint
  {http://arxiv.org/abs/1706.09371} {arXiv:1706.09371 [gr-qc]} \BibitemShut
  {NoStop}%
\bibitem [{\citenamefont {Alford}\ \emph
  {et~al.}(2015{\natexlab{a}})\citenamefont {Alford}, \citenamefont {Han},\
  and\ \citenamefont {Schwenzer}}]{Alford:2014jha}%
  \BibitemOpen
  \bibfield  {author} {\bibinfo {author} {\bibfnamefont {M.~G.}\ \bibnamefont
  {Alford}}, \bibinfo {author} {\bibfnamefont {S.}~\bibnamefont {Han}}, \ and\
  \bibinfo {author} {\bibfnamefont {K.}~\bibnamefont {Schwenzer}},\ }\href
  {\doibase 10.1103/PhysRevC.91.055804} {\bibfield  {journal} {\bibinfo
  {journal} {Phys. Rev. C}\ }\textbf {\bibinfo {volume} {91}},\ \bibinfo
  {pages} {055804} (\bibinfo {year} {2015}{\natexlab{a}})},\ \Eprint
  {http://arxiv.org/abs/1404.5279} {arXiv:1404.5279 [astro-ph.SR]} \BibitemShut
  {NoStop}%
\bibitem [{\citenamefont {Abbott}\ \emph {et~al.}(2018)\citenamefont {Abbott}
  \emph {et~al.}}]{Abbott:2018exr}%
  \BibitemOpen
  \bibfield  {author} {\bibinfo {author} {\bibfnamefont {B.~P.}\ \bibnamefont
  {Abbott}} \emph {et~al.} (\bibinfo {collaboration} {LIGO Scientific,
  Virgo}),\ }\href {\doibase 10.1103/PhysRevLett.121.161101} {\bibfield
  {journal} {\bibinfo  {journal} {Phys. Rev. Lett.}\ }\textbf {\bibinfo
  {volume} {121}},\ \bibinfo {pages} {161101} (\bibinfo {year} {2018})},\
  \Eprint {http://arxiv.org/abs/1805.11581} {arXiv:1805.11581 [gr-qc]}
  \BibitemShut {NoStop}%
\bibitem [{\citenamefont {Landry}\ and\ \citenamefont
  {Essick}(2019)}]{Landry:2018prl}%
  \BibitemOpen
  \bibfield  {author} {\bibinfo {author} {\bibfnamefont {P.}~\bibnamefont
  {Landry}}\ and\ \bibinfo {author} {\bibfnamefont {R.}~\bibnamefont
  {Essick}},\ }\href {\doibase 10.1103/PhysRevD.99.084049} {\bibfield
  {journal} {\bibinfo  {journal} {Phys. Rev. D}\ }\textbf {\bibinfo {volume}
  {99}},\ \bibinfo {pages} {084049} (\bibinfo {year} {2019})},\ \Eprint
  {http://arxiv.org/abs/1811.12529} {arXiv:1811.12529 [gr-qc]} \BibitemShut
  {NoStop}%
\bibitem [{\citenamefont {Essick}\ \emph {et~al.}(2020)\citenamefont {Essick},
  \citenamefont {Landry},\ and\ \citenamefont {Holz}}]{Essick:2019ldf}%
  \BibitemOpen
  \bibfield  {author} {\bibinfo {author} {\bibfnamefont {R.}~\bibnamefont
  {Essick}}, \bibinfo {author} {\bibfnamefont {P.}~\bibnamefont {Landry}}, \
  and\ \bibinfo {author} {\bibfnamefont {D.~E.}\ \bibnamefont {Holz}},\ }\href
  {\doibase 10.1103/PhysRevD.101.063007} {\bibfield  {journal} {\bibinfo
  {journal} {Phys. Rev. D}\ }\textbf {\bibinfo {volume} {101}},\ \bibinfo
  {pages} {063007} (\bibinfo {year} {2020})},\ \Eprint
  {http://arxiv.org/abs/1910.09740} {arXiv:1910.09740 [astro-ph.HE]}
  \BibitemShut {NoStop}%
\bibitem [{\citenamefont {Landry}\ \emph {et~al.}(2020)\citenamefont {Landry},
  \citenamefont {Essick},\ and\ \citenamefont
  {Chatziioannou}}]{Landry:2020vaw}%
  \BibitemOpen
  \bibfield  {author} {\bibinfo {author} {\bibfnamefont {P.}~\bibnamefont
  {Landry}}, \bibinfo {author} {\bibfnamefont {R.}~\bibnamefont {Essick}}, \
  and\ \bibinfo {author} {\bibfnamefont {K.}~\bibnamefont {Chatziioannou}},\
  }\href {\doibase 10.1103/PhysRevD.101.123007} {\bibfield  {journal} {\bibinfo
   {journal} {Phys. Rev. D}\ }\textbf {\bibinfo {volume} {101}},\ \bibinfo
  {pages} {123007} (\bibinfo {year} {2020})},\ \Eprint
  {http://arxiv.org/abs/2003.04880} {arXiv:2003.04880 [astro-ph.HE]}
  \BibitemShut {NoStop}%
\bibitem [{\citenamefont {Yagi}\ and\ \citenamefont
  {Yunes}(2016)}]{Yagi:2015pkc}%
  \BibitemOpen
  \bibfield  {author} {\bibinfo {author} {\bibfnamefont {K.}~\bibnamefont
  {Yagi}}\ and\ \bibinfo {author} {\bibfnamefont {N.}~\bibnamefont {Yunes}},\
  }\href {\doibase 10.1088/0264-9381/33/13/13LT01} {\bibfield  {journal}
  {\bibinfo  {journal} {Class. Quant. Grav.}\ }\textbf {\bibinfo {volume}
  {33}},\ \bibinfo {pages} {13LT01} (\bibinfo {year} {2016})},\ \Eprint
  {http://arxiv.org/abs/1512.02639} {arXiv:1512.02639 [gr-qc]} \BibitemShut
  {NoStop}%
\bibitem [{\citenamefont {Yagi}\ and\ \citenamefont
  {Yunes}(2017{\natexlab{b}})}]{Yagi:2016qmr}%
  \BibitemOpen
  \bibfield  {author} {\bibinfo {author} {\bibfnamefont {K.}~\bibnamefont
  {Yagi}}\ and\ \bibinfo {author} {\bibfnamefont {N.}~\bibnamefont {Yunes}},\
  }\href {\doibase 10.1088/1361-6382/34/1/015006} {\bibfield  {journal}
  {\bibinfo  {journal} {Class. Quant. Grav.}\ }\textbf {\bibinfo {volume}
  {34}},\ \bibinfo {pages} {015006} (\bibinfo {year} {2017}{\natexlab{b}})},\
  \Eprint {http://arxiv.org/abs/1608.06187} {arXiv:1608.06187 [gr-qc]}
  \BibitemShut {NoStop}%
\bibitem [{\citenamefont {Kurkela}\ \emph {et~al.}(2010)\citenamefont
  {Kurkela}, \citenamefont {Romatschke},\ and\ \citenamefont
  {Vuorinen}}]{Kurkela:2009gj}%
  \BibitemOpen
  \bibfield  {author} {\bibinfo {author} {\bibfnamefont {A.}~\bibnamefont
  {Kurkela}}, \bibinfo {author} {\bibfnamefont {P.}~\bibnamefont {Romatschke}},
  \ and\ \bibinfo {author} {\bibfnamefont {A.}~\bibnamefont {Vuorinen}},\
  }\href {\doibase 10.1103/PhysRevD.81.105021} {\bibfield  {journal} {\bibinfo
  {journal} {Phys. Rev. D}\ }\textbf {\bibinfo {volume} {81}},\ \bibinfo
  {pages} {105021} (\bibinfo {year} {2010})},\ \Eprint
  {http://arxiv.org/abs/0912.1856} {arXiv:0912.1856 [hep-ph]} \BibitemShut
  {NoStop}%
\bibitem [{\citenamefont {Kurkela}\ \emph {et~al.}(2014)\citenamefont
  {Kurkela}, \citenamefont {Fraga}, \citenamefont {Schaffner-Bielich},\ and\
  \citenamefont {Vuorinen}}]{Kurkela:2014vha}%
  \BibitemOpen
  \bibfield  {author} {\bibinfo {author} {\bibfnamefont {A.}~\bibnamefont
  {Kurkela}}, \bibinfo {author} {\bibfnamefont {E.~S.}\ \bibnamefont {Fraga}},
  \bibinfo {author} {\bibfnamefont {J.}~\bibnamefont {Schaffner-Bielich}}, \
  and\ \bibinfo {author} {\bibfnamefont {A.}~\bibnamefont {Vuorinen}},\ }\href
  {\doibase 10.1088/0004-637X/789/2/127} {\bibfield  {journal} {\bibinfo
  {journal} {Astrophys. J.}\ }\textbf {\bibinfo {volume} {789}},\ \bibinfo
  {pages} {127} (\bibinfo {year} {2014})},\ \Eprint
  {http://arxiv.org/abs/1402.6618} {arXiv:1402.6618 [astro-ph.HE]} \BibitemShut
  {NoStop}%
\bibitem [{\citenamefont {Hebeler}\ \emph {et~al.}(2013)\citenamefont
  {Hebeler}, \citenamefont {Lattimer}, \citenamefont {Pethick},\ and\
  \citenamefont {Schwenk}}]{Hebeler:2013nza}%
  \BibitemOpen
  \bibfield  {author} {\bibinfo {author} {\bibfnamefont {K.}~\bibnamefont
  {Hebeler}}, \bibinfo {author} {\bibfnamefont {J.~M.}\ \bibnamefont
  {Lattimer}}, \bibinfo {author} {\bibfnamefont {C.~J.}\ \bibnamefont
  {Pethick}}, \ and\ \bibinfo {author} {\bibfnamefont {A.}~\bibnamefont
  {Schwenk}},\ }\href {\doibase 10.1088/0004-637X/773/1/11} {\bibfield
  {journal} {\bibinfo  {journal} {Astrophys. J.}\ }\textbf {\bibinfo {volume}
  {773}},\ \bibinfo {pages} {11} (\bibinfo {year} {2013})},\ \Eprint
  {http://arxiv.org/abs/1303.4662} {arXiv:1303.4662 [astro-ph.SR]} \BibitemShut
  {NoStop}%
\bibitem [{\citenamefont {Dutra}\ \emph {et~al.}(2012)\citenamefont {Dutra},
  \citenamefont {Lourenco}, \citenamefont {Sa~Martins}, \citenamefont
  {Delfino}, \citenamefont {Stone},\ and\ \citenamefont
  {Stevenson}}]{Dutra:2012mb}%
  \BibitemOpen
  \bibfield  {author} {\bibinfo {author} {\bibfnamefont {M.}~\bibnamefont
  {Dutra}}, \bibinfo {author} {\bibfnamefont {O.}~\bibnamefont {Lourenco}},
  \bibinfo {author} {\bibfnamefont {J.~S.}\ \bibnamefont {Sa~Martins}},
  \bibinfo {author} {\bibfnamefont {A.}~\bibnamefont {Delfino}}, \bibinfo
  {author} {\bibfnamefont {J.~R.}\ \bibnamefont {Stone}}, \ and\ \bibinfo
  {author} {\bibfnamefont {P.~D.}\ \bibnamefont {Stevenson}},\ }\href {\doibase
  10.1103/PhysRevC.85.035201} {\bibfield  {journal} {\bibinfo  {journal} {Phys.
  Rev. C}\ }\textbf {\bibinfo {volume} {85}},\ \bibinfo {pages} {035201}
  (\bibinfo {year} {2012})},\ \Eprint {http://arxiv.org/abs/1202.3902}
  {arXiv:1202.3902 [nucl-th]} \BibitemShut {NoStop}%
\bibitem [{\citenamefont {Tsang}\ \emph {et~al.}(2012)\citenamefont {Tsang}
  \emph {et~al.}}]{Tsang:2012se}%
  \BibitemOpen
  \bibfield  {author} {\bibinfo {author} {\bibfnamefont {M.~B.}\ \bibnamefont
  {Tsang}} \emph {et~al.},\ }\href {\doibase 10.1103/PhysRevC.86.015803}
  {\bibfield  {journal} {\bibinfo  {journal} {Phys. Rev. C}\ }\textbf {\bibinfo
  {volume} {86}},\ \bibinfo {pages} {015803} (\bibinfo {year} {2012})},\
  \Eprint {http://arxiv.org/abs/1204.0466} {arXiv:1204.0466 [nucl-ex]}
  \BibitemShut {NoStop}%
\bibitem [{\citenamefont {Fattoyev}\ \emph {et~al.}(2013)\citenamefont
  {Fattoyev}, \citenamefont {Carvajal}, \citenamefont {Newton},\ and\
  \citenamefont {Li}}]{Fattoyev:2012uu}%
  \BibitemOpen
  \bibfield  {author} {\bibinfo {author} {\bibfnamefont {F.~J.}\ \bibnamefont
  {Fattoyev}}, \bibinfo {author} {\bibfnamefont {J.}~\bibnamefont {Carvajal}},
  \bibinfo {author} {\bibfnamefont {W.~G.}\ \bibnamefont {Newton}}, \ and\
  \bibinfo {author} {\bibfnamefont {B.-A.}\ \bibnamefont {Li}},\ }\href
  {\doibase 10.1103/PhysRevC.87.015806} {\bibfield  {journal} {\bibinfo
  {journal} {Phys. Rev. C}\ }\textbf {\bibinfo {volume} {87}},\ \bibinfo
  {pages} {015806} (\bibinfo {year} {2013})},\ \Eprint
  {http://arxiv.org/abs/1210.3402} {arXiv:1210.3402 [nucl-th]} \BibitemShut
  {NoStop}%
\bibitem [{\citenamefont {Weissenborn}\ \emph {et~al.}(2012)\citenamefont
  {Weissenborn}, \citenamefont {Chatterjee},\ and\ \citenamefont
  {Schaffner-Bielich}}]{Weissenborn:2011kb}%
  \BibitemOpen
  \bibfield  {author} {\bibinfo {author} {\bibfnamefont {S.}~\bibnamefont
  {Weissenborn}}, \bibinfo {author} {\bibfnamefont {D.}~\bibnamefont
  {Chatterjee}}, \ and\ \bibinfo {author} {\bibfnamefont {J.}~\bibnamefont
  {Schaffner-Bielich}},\ }\href {\doibase 10.1016/j.nuclphysa.2012.02.012}
  {\bibfield  {journal} {\bibinfo  {journal} {Nucl. Phys. A}\ }\textbf
  {\bibinfo {volume} {881}},\ \bibinfo {pages} {62} (\bibinfo {year} {2012})},\
  \Eprint {http://arxiv.org/abs/1111.6049} {arXiv:1111.6049 [astro-ph.HE]}
  \BibitemShut {NoStop}%
\bibitem [{\citenamefont {Li}\ \emph {et~al.}(2018)\citenamefont {Li},
  \citenamefont {Sedrakian},\ and\ \citenamefont {Weber}}]{Li:2018qaw}%
  \BibitemOpen
  \bibfield  {author} {\bibinfo {author} {\bibfnamefont {J.~J.}\ \bibnamefont
  {Li}}, \bibinfo {author} {\bibfnamefont {A.}~\bibnamefont {Sedrakian}}, \
  and\ \bibinfo {author} {\bibfnamefont {F.}~\bibnamefont {Weber}},\ }\href
  {\doibase 10.1016/j.physletb.2018.06.051} {\bibfield  {journal} {\bibinfo
  {journal} {Phys. Lett. B}\ }\textbf {\bibinfo {volume} {783}},\ \bibinfo
  {pages} {234} (\bibinfo {year} {2018})},\ \Eprint
  {http://arxiv.org/abs/1803.03661} {arXiv:1803.03661 [nucl-th]} \BibitemShut
  {NoStop}%
\bibitem [{\citenamefont {Dexheimer}\ \emph
  {et~al.}(2008{\natexlab{a}})\citenamefont {Dexheimer}, \citenamefont
  {Schramm},\ and\ \citenamefont {Zschiesche}}]{Dexheimer:2007tn}%
  \BibitemOpen
  \bibfield  {author} {\bibinfo {author} {\bibfnamefont {V.}~\bibnamefont
  {Dexheimer}}, \bibinfo {author} {\bibfnamefont {S.}~\bibnamefont {Schramm}},
  \ and\ \bibinfo {author} {\bibfnamefont {D.}~\bibnamefont {Zschiesche}},\
  }\href {\doibase 10.1103/PhysRevC.77.025803} {\bibfield  {journal} {\bibinfo
  {journal} {Phys. Rev. C}\ }\textbf {\bibinfo {volume} {77}},\ \bibinfo
  {pages} {025803} (\bibinfo {year} {2008}{\natexlab{a}})},\ \Eprint
  {http://arxiv.org/abs/0710.4192} {arXiv:0710.4192 [nucl-th]} \BibitemShut
  {NoStop}%
\bibitem [{\citenamefont {Moustakidis}\ \emph {et~al.}(2017)\citenamefont
  {Moustakidis}, \citenamefont {Gaitanos}, \citenamefont {Margaritis},\ and\
  \citenamefont {Lalazissis}}]{Moustakidis:2016sab}%
  \BibitemOpen
  \bibfield  {author} {\bibinfo {author} {\bibfnamefont {C.~C.}\ \bibnamefont
  {Moustakidis}}, \bibinfo {author} {\bibfnamefont {T.}~\bibnamefont
  {Gaitanos}}, \bibinfo {author} {\bibfnamefont {C.}~\bibnamefont
  {Margaritis}}, \ and\ \bibinfo {author} {\bibfnamefont {G.~A.}\ \bibnamefont
  {Lalazissis}},\ }\href {\doibase 10.1103/PhysRevC.95.045801} {\bibfield
  {journal} {\bibinfo  {journal} {Phys. Rev. C}\ }\textbf {\bibinfo {volume}
  {95}},\ \bibinfo {pages} {045801} (\bibinfo {year} {2017})},\ \bibinfo {note}
  {[Erratum: Phys.Rev.C 95, 059904 (2017)]},\ \Eprint
  {http://arxiv.org/abs/1608.00344} {arXiv:1608.00344 [nucl-th]} \BibitemShut
  {NoStop}%
\bibitem [{\citenamefont {Bedaque}\ and\ \citenamefont
  {Steiner}(2015)}]{Bedaque:2014sqa}%
  \BibitemOpen
  \bibfield  {author} {\bibinfo {author} {\bibfnamefont {P.}~\bibnamefont
  {Bedaque}}\ and\ \bibinfo {author} {\bibfnamefont {A.~W.}\ \bibnamefont
  {Steiner}},\ }\href {\doibase 10.1103/PhysRevLett.114.031103} {\bibfield
  {journal} {\bibinfo  {journal} {Phys. Rev. Lett.}\ }\textbf {\bibinfo
  {volume} {114}},\ \bibinfo {pages} {031103} (\bibinfo {year} {2015})},\
  \Eprint {http://arxiv.org/abs/1408.5116} {arXiv:1408.5116 [nucl-th]}
  \BibitemShut {NoStop}%
\bibitem [{\citenamefont {Malfatti}\ \emph {et~al.}(2020)\citenamefont
  {Malfatti}, \citenamefont {Orsaria}, \citenamefont {Ranea-Sandoval},
  \citenamefont {Contrera},\ and\ \citenamefont {Weber}}]{Malfatti:2020onm}%
  \BibitemOpen
  \bibfield  {author} {\bibinfo {author} {\bibfnamefont {G.}~\bibnamefont
  {Malfatti}}, \bibinfo {author} {\bibfnamefont {M.~G.}\ \bibnamefont
  {Orsaria}}, \bibinfo {author} {\bibfnamefont {I.~F.}\ \bibnamefont
  {Ranea-Sandoval}}, \bibinfo {author} {\bibfnamefont {G.~A.}\ \bibnamefont
  {Contrera}}, \ and\ \bibinfo {author} {\bibfnamefont {F.}~\bibnamefont
  {Weber}},\ }\href {\doibase 10.1103/PhysRevD.102.063008} {\bibfield
  {journal} {\bibinfo  {journal} {Phys. Rev. D}\ }\textbf {\bibinfo {volume}
  {102}},\ \bibinfo {pages} {063008} (\bibinfo {year} {2020})},\ \Eprint
  {http://arxiv.org/abs/2008.06459} {arXiv:2008.06459 [astro-ph.HE]}
  \BibitemShut {NoStop}%
\bibitem [{\citenamefont {Guichon}\ \emph {et~al.}(1996)\citenamefont
  {Guichon}, \citenamefont {Saito}, \citenamefont {Rodionov},\ and\
  \citenamefont {Thomas}}]{Guichon:1995ue}%
  \BibitemOpen
  \bibfield  {author} {\bibinfo {author} {\bibfnamefont {P.~A.~M.}\
  \bibnamefont {Guichon}}, \bibinfo {author} {\bibfnamefont {K.}~\bibnamefont
  {Saito}}, \bibinfo {author} {\bibfnamefont {E.~N.}\ \bibnamefont {Rodionov}},
  \ and\ \bibinfo {author} {\bibfnamefont {A.~W.}\ \bibnamefont {Thomas}},\
  }\href {\doibase 10.1016/0375-9474(96)00033-4} {\bibfield  {journal}
  {\bibinfo  {journal} {Nucl. Phys. A}\ }\textbf {\bibinfo {volume} {601}},\
  \bibinfo {pages} {349} (\bibinfo {year} {1996})},\ \Eprint
  {http://arxiv.org/abs/nucl-th/9509034} {arXiv:nucl-th/9509034} \BibitemShut
  {NoStop}%
\bibitem [{\citenamefont {Dexheimer}\ and\ \citenamefont
  {Schramm}(2008)}]{Dexheimer:2008ax}%
  \BibitemOpen
  \bibfield  {author} {\bibinfo {author} {\bibfnamefont {V.}~\bibnamefont
  {Dexheimer}}\ and\ \bibinfo {author} {\bibfnamefont {S.}~\bibnamefont
  {Schramm}},\ }\href {\doibase 10.1086/589735} {\bibfield  {journal} {\bibinfo
   {journal} {Astrophys. J.}\ }\textbf {\bibinfo {volume} {683}},\ \bibinfo
  {pages} {943} (\bibinfo {year} {2008})},\ \Eprint
  {http://arxiv.org/abs/0802.1999} {arXiv:0802.1999 [astro-ph]} \BibitemShut
  {NoStop}%
\bibitem [{\citenamefont {Dexheimer}\ and\ \citenamefont
  {Schramm}(2010)}]{Dexheimer:2009hi}%
  \BibitemOpen
  \bibfield  {author} {\bibinfo {author} {\bibfnamefont {V.~A.}\ \bibnamefont
  {Dexheimer}}\ and\ \bibinfo {author} {\bibfnamefont {S.}~\bibnamefont
  {Schramm}},\ }\href {\doibase 10.1103/PhysRevC.81.045201} {\bibfield
  {journal} {\bibinfo  {journal} {Phys. Rev. C}\ }\textbf {\bibinfo {volume}
  {81}},\ \bibinfo {pages} {045201} (\bibinfo {year} {2010})},\ \Eprint
  {http://arxiv.org/abs/0901.1748} {arXiv:0901.1748 [astro-ph.SR]} \BibitemShut
  {NoStop}%
\bibitem [{\citenamefont {Fortin}\ \emph {et~al.}(2016)\citenamefont {Fortin},
  \citenamefont {Providencia}, \citenamefont {Raduta}, \citenamefont
  {Gulminelli}, \citenamefont {Zdunik}, \citenamefont {Haensel},\ and\
  \citenamefont {Bejger}}]{Fortin:2016hny}%
  \BibitemOpen
  \bibfield  {author} {\bibinfo {author} {\bibfnamefont {M.}~\bibnamefont
  {Fortin}}, \bibinfo {author} {\bibfnamefont {C.}~\bibnamefont {Providencia}},
  \bibinfo {author} {\bibfnamefont {A.~R.}\ \bibnamefont {Raduta}}, \bibinfo
  {author} {\bibfnamefont {F.}~\bibnamefont {Gulminelli}}, \bibinfo {author}
  {\bibfnamefont {J.~L.}\ \bibnamefont {Zdunik}}, \bibinfo {author}
  {\bibfnamefont {P.}~\bibnamefont {Haensel}}, \ and\ \bibinfo {author}
  {\bibfnamefont {M.}~\bibnamefont {Bejger}},\ }\href {\doibase
  10.1103/PhysRevC.94.035804} {\bibfield  {journal} {\bibinfo  {journal} {Phys.
  Rev. C}\ }\textbf {\bibinfo {volume} {94}},\ \bibinfo {pages} {035804}
  (\bibinfo {year} {2016})},\ \Eprint {http://arxiv.org/abs/1604.01944}
  {arXiv:1604.01944 [astro-ph.SR]} \BibitemShut {NoStop}%
\bibitem [{\citenamefont {Abbott}\ \emph {et~al.}(2019)\citenamefont {Abbott}
  \emph {et~al.}}]{Abbott:2018wiz}%
  \BibitemOpen
  \bibfield  {author} {\bibinfo {author} {\bibfnamefont {B.~P.}\ \bibnamefont
  {Abbott}} \emph {et~al.} (\bibinfo {collaboration} {LIGO Scientific,
  Virgo}),\ }\href {\doibase 10.1103/PhysRevX.9.011001} {\bibfield  {journal}
  {\bibinfo  {journal} {Phys. Rev. X}\ }\textbf {\bibinfo {volume} {9}},\
  \bibinfo {pages} {011001} (\bibinfo {year} {2019})},\ \Eprint
  {http://arxiv.org/abs/1805.11579} {arXiv:1805.11579 [gr-qc]} \BibitemShut
  {NoStop}%
\bibitem [{jak()}]{jakigit}%
  \BibitemOpen
  \href@noop {} {}\bibinfo {note}
  {\href{https://github.com/jnoronhahostler/Neutron_Star_EOS}{https://github.com/jnoronhahostler/Neutron\_Star\_EOS}}\BibitemShut
  {NoStop}%
\bibitem [{\citenamefont {Chabanat}\ \emph {et~al.}(1998)\citenamefont
  {Chabanat}, \citenamefont {Bonche}, \citenamefont {Haensel}, \citenamefont
  {Meyer},\ and\ \citenamefont {Schaeffer}}]{Chabanat:1997un}%
  \BibitemOpen
  \bibfield  {author} {\bibinfo {author} {\bibfnamefont {E.}~\bibnamefont
  {Chabanat}}, \bibinfo {author} {\bibfnamefont {P.}~\bibnamefont {Bonche}},
  \bibinfo {author} {\bibfnamefont {P.}~\bibnamefont {Haensel}}, \bibinfo
  {author} {\bibfnamefont {J.}~\bibnamefont {Meyer}}, \ and\ \bibinfo {author}
  {\bibfnamefont {R.}~\bibnamefont {Schaeffer}},\ }\href {\doibase
  10.1016/S0375-9474(98)00180-8} {\bibfield  {journal} {\bibinfo  {journal}
  {Nucl. Phys. A}\ }\textbf {\bibinfo {volume} {635}},\ \bibinfo {pages} {231}
  (\bibinfo {year} {1998})},\ \bibinfo {note} {[Erratum: Nucl.Phys.A 643,
  441--441 (1998)]}\BibitemShut {NoStop}%
\bibitem [{\citenamefont {Douchin}\ \emph {et~al.}(2000)\citenamefont
  {Douchin}, \citenamefont {Haensel},\ and\ \citenamefont
  {Meyer}}]{Douchin:2000kad}%
  \BibitemOpen
  \bibfield  {author} {\bibinfo {author} {\bibfnamefont {F.}~\bibnamefont
  {Douchin}}, \bibinfo {author} {\bibfnamefont {P.}~\bibnamefont {Haensel}}, \
  and\ \bibinfo {author} {\bibfnamefont {J.}~\bibnamefont {Meyer}},\ }\href
  {\doibase 10.1016/S0375-9474(99)00397-8} {\bibfield  {journal} {\bibinfo
  {journal} {Nucl. Phys. A}\ }\textbf {\bibinfo {volume} {665}},\ \bibinfo
  {pages} {419} (\bibinfo {year} {2000})}\BibitemShut {NoStop}%
\bibitem [{\citenamefont {Douchin}\ and\ \citenamefont
  {Haensel}(2000)}]{Douchin:2000kx}%
  \BibitemOpen
  \bibfield  {author} {\bibinfo {author} {\bibfnamefont {F.}~\bibnamefont
  {Douchin}}\ and\ \bibinfo {author} {\bibfnamefont {P.}~\bibnamefont
  {Haensel}},\ }\href {\doibase 10.1016/S0370-2693(00)00672-9} {\bibfield
  {journal} {\bibinfo  {journal} {Phys. Lett. B}\ }\textbf {\bibinfo {volume}
  {485}},\ \bibinfo {pages} {107} (\bibinfo {year} {2000})},\ \Eprint
  {http://arxiv.org/abs/astro-ph/0006135} {arXiv:astro-ph/0006135} \BibitemShut
  {NoStop}%
\bibitem [{\citenamefont {Douchin}\ and\ \citenamefont
  {Haensel}(2001)}]{Douchin:2001sv}%
  \BibitemOpen
  \bibfield  {author} {\bibinfo {author} {\bibfnamefont {F.}~\bibnamefont
  {Douchin}}\ and\ \bibinfo {author} {\bibfnamefont {P.}~\bibnamefont
  {Haensel}},\ }\href {\doibase 10.1051/0004-6361:20011402} {\bibfield
  {journal} {\bibinfo  {journal} {Astron. Astrophys.}\ }\textbf {\bibinfo
  {volume} {380}},\ \bibinfo {pages} {151} (\bibinfo {year} {2001})},\ \Eprint
  {http://arxiv.org/abs/astro-ph/0111092} {arXiv:astro-ph/0111092} \BibitemShut
  {NoStop}%
\bibitem [{\citenamefont {Togashi}\ \emph {et~al.}(2017)\citenamefont
  {Togashi}, \citenamefont {Nakazato}, \citenamefont {Takehara}, \citenamefont
  {Yamamuro}, \citenamefont {Suzuki},\ and\ \citenamefont
  {Takano}}]{Togashi:2017mjp}%
  \BibitemOpen
  \bibfield  {author} {\bibinfo {author} {\bibfnamefont {H.}~\bibnamefont
  {Togashi}}, \bibinfo {author} {\bibfnamefont {K.}~\bibnamefont {Nakazato}},
  \bibinfo {author} {\bibfnamefont {Y.}~\bibnamefont {Takehara}}, \bibinfo
  {author} {\bibfnamefont {S.}~\bibnamefont {Yamamuro}}, \bibinfo {author}
  {\bibfnamefont {H.}~\bibnamefont {Suzuki}}, \ and\ \bibinfo {author}
  {\bibfnamefont {M.}~\bibnamefont {Takano}},\ }\href {\doibase
  10.1016/j.nuclphysa.2017.02.010} {\bibfield  {journal} {\bibinfo  {journal}
  {Nucl. Phys. A}\ }\textbf {\bibinfo {volume} {961}},\ \bibinfo {pages} {78}
  (\bibinfo {year} {2017})},\ \Eprint {http://arxiv.org/abs/1702.05324}
  {arXiv:1702.05324 [nucl-th]} \BibitemShut {NoStop}%
\bibitem [{\citenamefont {Gulminelli}\ and\ \citenamefont
  {Raduta}(2015)}]{Gulminelli:2015csa}%
  \BibitemOpen
  \bibfield  {author} {\bibinfo {author} {\bibfnamefont {F.}~\bibnamefont
  {Gulminelli}}\ and\ \bibinfo {author} {\bibfnamefont {A.~R.}\ \bibnamefont
  {Raduta}},\ }\href {\doibase 10.1103/PhysRevC.92.055803} {\bibfield
  {journal} {\bibinfo  {journal} {Phys. Rev. C}\ }\textbf {\bibinfo {volume}
  {92}},\ \bibinfo {pages} {055803} (\bibinfo {year} {2015})},\ \Eprint
  {http://arxiv.org/abs/1504.04493} {arXiv:1504.04493 [nucl-th]} \BibitemShut
  {NoStop}%
\bibitem [{\citenamefont {Tews}\ \emph
  {et~al.}(2018{\natexlab{a}})\citenamefont {Tews}, \citenamefont {Carlson},
  \citenamefont {Gandolfi},\ and\ \citenamefont {Reddy}}]{Tews:2018kmu}%
  \BibitemOpen
  \bibfield  {author} {\bibinfo {author} {\bibfnamefont {I.}~\bibnamefont
  {Tews}}, \bibinfo {author} {\bibfnamefont {J.}~\bibnamefont {Carlson}},
  \bibinfo {author} {\bibfnamefont {S.}~\bibnamefont {Gandolfi}}, \ and\
  \bibinfo {author} {\bibfnamefont {S.}~\bibnamefont {Reddy}},\ }\href
  {\doibase 10.3847/1538-4357/aac267} {\bibfield  {journal} {\bibinfo
  {journal} {Astrophys. J.}\ }\textbf {\bibinfo {volume} {860}},\ \bibinfo
  {pages} {149} (\bibinfo {year} {2018}{\natexlab{a}})},\ \Eprint
  {http://arxiv.org/abs/1801.01923} {arXiv:1801.01923 [nucl-th]} \BibitemShut
  {NoStop}%
\bibitem [{\citenamefont {Yagi}\ \emph {et~al.}(2014)\citenamefont {Yagi},
  \citenamefont {Kyutoku}, \citenamefont {Pappas}, \citenamefont {Yunes},\ and\
  \citenamefont {Apostolatos}}]{Yagi:2014bxa}%
  \BibitemOpen
  \bibfield  {author} {\bibinfo {author} {\bibfnamefont {K.}~\bibnamefont
  {Yagi}}, \bibinfo {author} {\bibfnamefont {K.}~\bibnamefont {Kyutoku}},
  \bibinfo {author} {\bibfnamefont {G.}~\bibnamefont {Pappas}}, \bibinfo
  {author} {\bibfnamefont {N.}~\bibnamefont {Yunes}}, \ and\ \bibinfo {author}
  {\bibfnamefont {T.~A.}\ \bibnamefont {Apostolatos}},\ }\href {\doibase
  10.1103/PhysRevD.89.124013} {\bibfield  {journal} {\bibinfo  {journal} {Phys.
  Rev. D}\ }\textbf {\bibinfo {volume} {89}},\ \bibinfo {pages} {124013}
  (\bibinfo {year} {2014})},\ \Eprint {http://arxiv.org/abs/1403.6243}
  {arXiv:1403.6243 [gr-qc]} \BibitemShut {NoStop}%
\bibitem [{\citenamefont {Jim\'enez}\ and\ \citenamefont
  {Fraga}(2021)}]{Jimenez:2021wil}%
  \BibitemOpen
  \bibfield  {author} {\bibinfo {author} {\bibfnamefont {J.~C.}\ \bibnamefont
  {Jim\'enez}}\ and\ \bibinfo {author} {\bibfnamefont {E.~S.}\ \bibnamefont
  {Fraga}},\ }\href@noop {} {\  (\bibinfo {year} {2021})},\ \Eprint
  {http://arxiv.org/abs/2104.13480} {arXiv:2104.13480 [hep-ph]} \BibitemShut
  {NoStop}%
\bibitem [{\citenamefont {Zhang}\ and\ \citenamefont
  {Li}(2020)}]{Zhang:2020zsc}%
  \BibitemOpen
  \bibfield  {author} {\bibinfo {author} {\bibfnamefont {N.-B.}\ \bibnamefont
  {Zhang}}\ and\ \bibinfo {author} {\bibfnamefont {B.-A.}\ \bibnamefont {Li}},\
  }\href {\doibase 10.3847/1538-4357/abb470} {\bibfield  {journal} {\bibinfo
  {journal} {Astrophys. J.}\ }\textbf {\bibinfo {volume} {902}},\ \bibinfo
  {pages} {38} (\bibinfo {year} {2020})},\ \Eprint
  {http://arxiv.org/abs/2007.02513} {arXiv:2007.02513 [astro-ph.HE]}
  \BibitemShut {NoStop}%
\bibitem [{\citenamefont {Rather}\ \emph
  {et~al.}(2021{\natexlab{b}})\citenamefont {Rather}, \citenamefont {Rahaman},
  \citenamefont {Imran}, \citenamefont {Das}, \citenamefont {Usmani},\ and\
  \citenamefont {Patra}}]{Rather:2021yxo}%
  \BibitemOpen
  \bibfield  {author} {\bibinfo {author} {\bibfnamefont {I.~A.}\ \bibnamefont
  {Rather}}, \bibinfo {author} {\bibfnamefont {U.}~\bibnamefont {Rahaman}},
  \bibinfo {author} {\bibfnamefont {M.}~\bibnamefont {Imran}}, \bibinfo
  {author} {\bibfnamefont {H.~C.}\ \bibnamefont {Das}}, \bibinfo {author}
  {\bibfnamefont {A.~A.}\ \bibnamefont {Usmani}}, \ and\ \bibinfo {author}
  {\bibfnamefont {S.~K.}\ \bibnamefont {Patra}},\ }\href {\doibase
  10.1103/PhysRevC.103.055814} {\bibfield  {journal} {\bibinfo  {journal}
  {Phys. Rev. C}\ }\textbf {\bibinfo {volume} {103}},\ \bibinfo {pages}
  {055814} (\bibinfo {year} {2021}{\natexlab{b}})},\ \Eprint
  {http://arxiv.org/abs/2102.04067} {arXiv:2102.04067 [nucl-th]} \BibitemShut
  {NoStop}%
\bibitem [{\citenamefont {Hernandez-Vivanco}\ \emph {et~al.}(2021)\citenamefont
  {Hernandez-Vivanco}, \citenamefont {Lasky}, \citenamefont {Thrane},
  \citenamefont {Smith}, \citenamefont {Chatterjee}, \citenamefont {Banik},
  \citenamefont {Motta},\ and\ \citenamefont
  {Thomas}}]{Hernandez-Vivanco:2021urj}%
  \BibitemOpen
  \bibfield  {author} {\bibinfo {author} {\bibfnamefont {F.}~\bibnamefont
  {Hernandez-Vivanco}}, \bibinfo {author} {\bibfnamefont {P.~D.}\ \bibnamefont
  {Lasky}}, \bibinfo {author} {\bibfnamefont {E.}~\bibnamefont {Thrane}},
  \bibinfo {author} {\bibfnamefont {R.}~\bibnamefont {Smith}}, \bibinfo
  {author} {\bibfnamefont {D.}~\bibnamefont {Chatterjee}}, \bibinfo {author}
  {\bibfnamefont {S.}~\bibnamefont {Banik}}, \bibinfo {author} {\bibfnamefont
  {T.}~\bibnamefont {Motta}}, \ and\ \bibinfo {author} {\bibfnamefont
  {A.}~\bibnamefont {Thomas}},\ }\href@noop {} {\  (\bibinfo {year} {2021})},\
  \Eprint {http://arxiv.org/abs/2101.04782} {arXiv:2101.04782 [astro-ph.HE]}
  \BibitemShut {NoStop}%
\bibitem [{\citenamefont {Sen}(2021)}]{Sen:2021bms}%
  \BibitemOpen
  \bibfield  {author} {\bibinfo {author} {\bibfnamefont {D.}~\bibnamefont
  {Sen}},\ }\href {\doibase 10.1103/PhysRevC.103.045804} {\bibfield  {journal}
  {\bibinfo  {journal} {Phys. Rev. C}\ }\textbf {\bibinfo {volume} {103}},\
  \bibinfo {pages} {045804} (\bibinfo {year} {2021})},\ \Eprint
  {http://arxiv.org/abs/2103.14136} {arXiv:2103.14136 [nucl-th]} \BibitemShut
  {NoStop}%
\bibitem [{\citenamefont {Rather}\ \emph
  {et~al.}(2021{\natexlab{c}})\citenamefont {Rather}, \citenamefont {Rahaman},
  \citenamefont {Dexheimer}, \citenamefont {Usmani},\ and\ \citenamefont
  {Patra}}]{Rather:2021azv}%
  \BibitemOpen
  \bibfield  {author} {\bibinfo {author} {\bibfnamefont {I.~A.}\ \bibnamefont
  {Rather}}, \bibinfo {author} {\bibfnamefont {U.}~\bibnamefont {Rahaman}},
  \bibinfo {author} {\bibfnamefont {V.}~\bibnamefont {Dexheimer}}, \bibinfo
  {author} {\bibfnamefont {A.~A.}\ \bibnamefont {Usmani}}, \ and\ \bibinfo
  {author} {\bibfnamefont {S.~K.}\ \bibnamefont {Patra}},\ }\href@noop {} {\
  (\bibinfo {year} {2021}{\natexlab{c}})},\ \Eprint
  {http://arxiv.org/abs/2104.05950} {arXiv:2104.05950 [nucl-th]} \BibitemShut
  {NoStop}%
\bibitem [{\citenamefont {Tews}\ \emph
  {et~al.}(2018{\natexlab{b}})\citenamefont {Tews}, \citenamefont {Margueron},\
  and\ \citenamefont {Reddy}}]{Tews:2018iwm}%
  \BibitemOpen
  \bibfield  {author} {\bibinfo {author} {\bibfnamefont {I.}~\bibnamefont
  {Tews}}, \bibinfo {author} {\bibfnamefont {J.}~\bibnamefont {Margueron}}, \
  and\ \bibinfo {author} {\bibfnamefont {S.}~\bibnamefont {Reddy}},\ }\href
  {\doibase 10.1103/PhysRevC.98.045804} {\bibfield  {journal} {\bibinfo
  {journal} {Phys. Rev. C}\ }\textbf {\bibinfo {volume} {98}},\ \bibinfo
  {pages} {045804} (\bibinfo {year} {2018}{\natexlab{b}})},\ \Eprint
  {http://arxiv.org/abs/1804.02783} {arXiv:1804.02783 [nucl-th]} \BibitemShut
  {NoStop}%
\bibitem [{\citenamefont {Han}\ \emph {et~al.}(2021)\citenamefont {Han},
  \citenamefont {Jiang}, \citenamefont {Tang},\ and\ \citenamefont
  {Fan}}]{Han:2021kjx}%
  \BibitemOpen
  \bibfield  {author} {\bibinfo {author} {\bibfnamefont {M.-Z.}\ \bibnamefont
  {Han}}, \bibinfo {author} {\bibfnamefont {J.-L.}\ \bibnamefont {Jiang}},
  \bibinfo {author} {\bibfnamefont {S.-P.}\ \bibnamefont {Tang}}, \ and\
  \bibinfo {author} {\bibfnamefont {Y.-Z.}\ \bibnamefont {Fan}},\ }\href@noop
  {} {\  (\bibinfo {year} {2021})},\ \Eprint {http://arxiv.org/abs/2103.05408}
  {arXiv:2103.05408 [hep-ph]} \BibitemShut {NoStop}%
\bibitem [{\citenamefont {Ducoin}\ \emph {et~al.}(2011)\citenamefont {Ducoin},
  \citenamefont {Margueron}, \citenamefont {Providencia},\ and\ \citenamefont
  {Vidana}}]{Ducoin:2011fy}%
  \BibitemOpen
  \bibfield  {author} {\bibinfo {author} {\bibfnamefont {C.}~\bibnamefont
  {Ducoin}}, \bibinfo {author} {\bibfnamefont {J.}~\bibnamefont {Margueron}},
  \bibinfo {author} {\bibfnamefont {C.}~\bibnamefont {Providencia}}, \ and\
  \bibinfo {author} {\bibfnamefont {I.}~\bibnamefont {Vidana}},\ }\href
  {\doibase 10.1103/PhysRevC.83.045810} {\bibfield  {journal} {\bibinfo
  {journal} {Phys. Rev. C}\ }\textbf {\bibinfo {volume} {83}},\ \bibinfo
  {pages} {045810} (\bibinfo {year} {2011})},\ \Eprint
  {http://arxiv.org/abs/1102.1283} {arXiv:1102.1283 [nucl-th]} \BibitemShut
  {NoStop}%
\bibitem [{\citenamefont {Gamba}\ \emph {et~al.}(2020)\citenamefont {Gamba},
  \citenamefont {Read},\ and\ \citenamefont {Wade}}]{Gamba:2019kwu}%
  \BibitemOpen
  \bibfield  {author} {\bibinfo {author} {\bibfnamefont {R.}~\bibnamefont
  {Gamba}}, \bibinfo {author} {\bibfnamefont {J.~S.}\ \bibnamefont {Read}}, \
  and\ \bibinfo {author} {\bibfnamefont {L.~E.}\ \bibnamefont {Wade}},\ }\href
  {\doibase 10.1088/1361-6382/ab5ba4} {\bibfield  {journal} {\bibinfo
  {journal} {Class. Quant. Grav.}\ }\textbf {\bibinfo {volume} {37}},\ \bibinfo
  {pages} {025008} (\bibinfo {year} {2020})},\ \Eprint
  {http://arxiv.org/abs/1902.04616} {arXiv:1902.04616 [gr-qc]} \BibitemShut
  {NoStop}%
\bibitem [{\citenamefont {Li}\ \emph {et~al.}(2021)\citenamefont {Li},
  \citenamefont {Miao}, \citenamefont {Han},\ and\ \citenamefont
  {Zhang}}]{Li:2021crp}%
  \BibitemOpen
  \bibfield  {author} {\bibinfo {author} {\bibfnamefont {A.}~\bibnamefont
  {Li}}, \bibinfo {author} {\bibfnamefont {Z.}~\bibnamefont {Miao}}, \bibinfo
  {author} {\bibfnamefont {S.}~\bibnamefont {Han}}, \ and\ \bibinfo {author}
  {\bibfnamefont {B.}~\bibnamefont {Zhang}},\ }\href {\doibase
  10.3847/1538-4357/abf355} {\bibfield  {journal} {\bibinfo  {journal}
  {Astrophys. J.}\ }\textbf {\bibinfo {volume} {913}},\ \bibinfo {pages} {27}
  (\bibinfo {year} {2021})},\ \Eprint {http://arxiv.org/abs/2103.15119}
  {arXiv:2103.15119 [astro-ph.HE]} \BibitemShut {NoStop}%
\bibitem [{\citenamefont {Zhu}\ \emph {et~al.}(2018)\citenamefont {Zhu},
  \citenamefont {Zhou},\ and\ \citenamefont {Li}}]{Zhu:2018ona}%
  \BibitemOpen
  \bibfield  {author} {\bibinfo {author} {\bibfnamefont {Z.-Y.}\ \bibnamefont
  {Zhu}}, \bibinfo {author} {\bibfnamefont {E.-P.}\ \bibnamefont {Zhou}}, \
  and\ \bibinfo {author} {\bibfnamefont {A.}~\bibnamefont {Li}},\ }\href
  {\doibase 10.3847/1538-4357/aacc28} {\bibfield  {journal} {\bibinfo
  {journal} {Astrophys. J.}\ }\textbf {\bibinfo {volume} {862}},\ \bibinfo
  {pages} {98} (\bibinfo {year} {2018})},\ \Eprint
  {http://arxiv.org/abs/1802.05510} {arXiv:1802.05510 [nucl-th]} \BibitemShut
  {NoStop}%
\bibitem [{\citenamefont {Hu}\ \emph {et~al.}(2020)\citenamefont {Hu},
  \citenamefont {Bao}, \citenamefont {Zhang}, \citenamefont {Nakazato},
  \citenamefont {Sumiyoshi},\ and\ \citenamefont {Shen}}]{Hu:2020ujf}%
  \BibitemOpen
  \bibfield  {author} {\bibinfo {author} {\bibfnamefont {J.}~\bibnamefont
  {Hu}}, \bibinfo {author} {\bibfnamefont {S.}~\bibnamefont {Bao}}, \bibinfo
  {author} {\bibfnamefont {Y.}~\bibnamefont {Zhang}}, \bibinfo {author}
  {\bibfnamefont {K.}~\bibnamefont {Nakazato}}, \bibinfo {author}
  {\bibfnamefont {K.}~\bibnamefont {Sumiyoshi}}, \ and\ \bibinfo {author}
  {\bibfnamefont {H.}~\bibnamefont {Shen}},\ }\href {\doibase
  10.1093/ptep/ptaa016} {\bibfield  {journal} {\bibinfo  {journal} {PTEP}\
  }\textbf {\bibinfo {volume} {2020}},\ \bibinfo {pages} {043D01} (\bibinfo
  {year} {2020})},\ \Eprint {http://arxiv.org/abs/2002.00562} {arXiv:2002.00562
  [nucl-th]} \BibitemShut {NoStop}%
\bibitem [{\citenamefont {Balberg}\ \emph {et~al.}(1999)\citenamefont
  {Balberg}, \citenamefont {Lichtenstadt},\ and\ \citenamefont
  {Cook}}]{Balberg:1998ug}%
  \BibitemOpen
  \bibfield  {author} {\bibinfo {author} {\bibfnamefont {S.}~\bibnamefont
  {Balberg}}, \bibinfo {author} {\bibfnamefont {I.}~\bibnamefont
  {Lichtenstadt}}, \ and\ \bibinfo {author} {\bibfnamefont {G.~B.}\
  \bibnamefont {Cook}},\ }\href {\doibase 10.1086/313196} {\bibfield  {journal}
  {\bibinfo  {journal} {Astrophys. J. Suppl.}\ }\textbf {\bibinfo {volume}
  {121}},\ \bibinfo {pages} {515} (\bibinfo {year} {1999})},\ \Eprint
  {http://arxiv.org/abs/astro-ph/9810361} {arXiv:astro-ph/9810361} \BibitemShut
  {NoStop}%
\bibitem [{\citenamefont {Haensel}\ \emph {et~al.}(2002)\citenamefont
  {Haensel}, \citenamefont {Levenfish},\ and\ \citenamefont
  {Yakovlev}}]{Haensel:2002qw}%
  \BibitemOpen
  \bibfield  {author} {\bibinfo {author} {\bibfnamefont {P.}~\bibnamefont
  {Haensel}}, \bibinfo {author} {\bibfnamefont {K.~P.}\ \bibnamefont
  {Levenfish}}, \ and\ \bibinfo {author} {\bibfnamefont {D.~G.}\ \bibnamefont
  {Yakovlev}},\ }\href {\doibase 10.1051/0004-6361:20021112} {\bibfield
  {journal} {\bibinfo  {journal} {Astron. Astrophys.}\ }\textbf {\bibinfo
  {volume} {394}},\ \bibinfo {pages} {213} (\bibinfo {year} {2002})},\ \Eprint
  {http://arxiv.org/abs/astro-ph/0208078} {arXiv:astro-ph/0208078} \BibitemShut
  {NoStop}%
\bibitem [{\citenamefont {Dexheimer}\ \emph
  {et~al.}(2008{\natexlab{b}})\citenamefont {Dexheimer}, \citenamefont
  {Vasconcellos},\ and\ \citenamefont {Bodmann}}]{Dexheimer:2007mt}%
  \BibitemOpen
  \bibfield  {author} {\bibinfo {author} {\bibfnamefont {V.~A.}\ \bibnamefont
  {Dexheimer}}, \bibinfo {author} {\bibfnamefont {C.~A.~Z.}\ \bibnamefont
  {Vasconcellos}}, \ and\ \bibinfo {author} {\bibfnamefont {B.~E.~J.}\
  \bibnamefont {Bodmann}},\ }\href {\doibase 10.1103/PhysRevC.77.065803}
  {\bibfield  {journal} {\bibinfo  {journal} {Phys. Rev. C}\ }\textbf {\bibinfo
  {volume} {77}},\ \bibinfo {pages} {065803} (\bibinfo {year}
  {2008}{\natexlab{b}})},\ \Eprint {http://arxiv.org/abs/0708.0131}
  {arXiv:0708.0131 [astro-ph]} \BibitemShut {NoStop}%
\bibitem [{\citenamefont {Casali}\ and\ \citenamefont
  {Menezes}(2010)}]{CASALI2010}%
  \BibitemOpen
  \bibfield  {author} {\bibinfo {author} {\bibfnamefont {R.~H.}\ \bibnamefont
  {Casali}}\ and\ \bibinfo {author} {\bibfnamefont {D.~P.}\ \bibnamefont
  {Menezes}},\ }\href@noop {} {\bibfield  {journal} {\bibinfo  {journal}
  {Brazilian Journal of Physics}\ }\textbf {\bibinfo {volume} {40}},\ \bibinfo
  {pages} {166 } (\bibinfo {year} {2010})}\BibitemShut {NoStop}%
\bibitem [{\citenamefont {Miyatsu}\ \emph {et~al.}(2017)\citenamefont
  {Miyatsu}, \citenamefont {Kambe},\ and\ \citenamefont
  {Saito}}]{Miyatsu:2017teh}%
  \BibitemOpen
  \bibfield  {author} {\bibinfo {author} {\bibfnamefont {T.}~\bibnamefont
  {Miyatsu}}, \bibinfo {author} {\bibfnamefont {T.}~\bibnamefont {Kambe}}, \
  and\ \bibinfo {author} {\bibfnamefont {K.}~\bibnamefont {Saito}},\ }\href
  {\doibase 10.22323/1.281.0135} {\bibfield  {journal} {\bibinfo  {journal}
  {PoS}\ }\textbf {\bibinfo {volume} {INPC2016}},\ \bibinfo {pages} {135}
  (\bibinfo {year} {2017})},\ \Eprint {http://arxiv.org/abs/1702.05871}
  {arXiv:1702.05871 [nucl-th]} \BibitemShut {NoStop}%
\bibitem [{\citenamefont {Stone}\ \emph
  {et~al.}(2021{\natexlab{b}})\citenamefont {Stone}, \citenamefont {Dexheimer},
  \citenamefont {Guichon}, \citenamefont {Thomas},\ and\ \citenamefont
  {Typel}}]{Stone:2019abq}%
  \BibitemOpen
  \bibfield  {author} {\bibinfo {author} {\bibfnamefont {J.~R.}\ \bibnamefont
  {Stone}}, \bibinfo {author} {\bibfnamefont {V.}~\bibnamefont {Dexheimer}},
  \bibinfo {author} {\bibfnamefont {P.~A.~M.}\ \bibnamefont {Guichon}},
  \bibinfo {author} {\bibfnamefont {A.~W.}\ \bibnamefont {Thomas}}, \ and\
  \bibinfo {author} {\bibfnamefont {S.}~\bibnamefont {Typel}},\ }\href
  {\doibase 10.1093/mnras/staa4006} {\bibfield  {journal} {\bibinfo  {journal}
  {Mon. Not. Roy. Astron. Soc.}\ }\textbf {\bibinfo {volume} {502}},\ \bibinfo
  {pages} {3476} (\bibinfo {year} {2021}{\natexlab{b}})},\ \Eprint
  {http://arxiv.org/abs/1906.11100} {arXiv:1906.11100 [nucl-th]} \BibitemShut
  {NoStop}%
\bibitem [{\citenamefont {Motta}\ \emph {et~al.}(2021)\citenamefont {Motta},
  \citenamefont {Guichon},\ and\ \citenamefont {Thomas}}]{Motta:2020fvm}%
  \BibitemOpen
  \bibfield  {author} {\bibinfo {author} {\bibfnamefont {T.~F.}\ \bibnamefont
  {Motta}}, \bibinfo {author} {\bibfnamefont {P.~A.~M.}\ \bibnamefont
  {Guichon}}, \ and\ \bibinfo {author} {\bibfnamefont {A.~W.}\ \bibnamefont
  {Thomas}},\ }\href {\doibase 10.1016/j.nuclphysa.2021.122157} {\bibfield
  {journal} {\bibinfo  {journal} {Nucl. Phys. A}\ }\textbf {\bibinfo {volume}
  {1009}},\ \bibinfo {pages} {122157} (\bibinfo {year} {2021})},\ \Eprint
  {http://arxiv.org/abs/2009.10908} {arXiv:2009.10908 [nucl-th]} \BibitemShut
  {NoStop}%
\bibitem [{\citenamefont {Schneider}\ \emph {et~al.}(2019)\citenamefont
  {Schneider}, \citenamefont {Constantinou}, \citenamefont {Muccioli},\ and\
  \citenamefont {Prakash}}]{Schneider:2019vdm}%
  \BibitemOpen
  \bibfield  {author} {\bibinfo {author} {\bibfnamefont {A.~S.}\ \bibnamefont
  {Schneider}}, \bibinfo {author} {\bibfnamefont {C.}~\bibnamefont
  {Constantinou}}, \bibinfo {author} {\bibfnamefont {B.}~\bibnamefont
  {Muccioli}}, \ and\ \bibinfo {author} {\bibfnamefont {M.}~\bibnamefont
  {Prakash}},\ }\href {\doibase 10.1103/PhysRevC.100.025803} {\bibfield
  {journal} {\bibinfo  {journal} {Phys. Rev. C}\ }\textbf {\bibinfo {volume}
  {100}},\ \bibinfo {pages} {025803} (\bibinfo {year} {2019})},\ \Eprint
  {http://arxiv.org/abs/1901.09652} {arXiv:1901.09652 [nucl-th]} \BibitemShut
  {NoStop}%
\bibitem [{\citenamefont {Alford}\ \emph {et~al.}(2013)\citenamefont {Alford},
  \citenamefont {Han},\ and\ \citenamefont {Prakash}}]{Alford:2013aca}%
  \BibitemOpen
  \bibfield  {author} {\bibinfo {author} {\bibfnamefont {M.~G.}\ \bibnamefont
  {Alford}}, \bibinfo {author} {\bibfnamefont {S.}~\bibnamefont {Han}}, \ and\
  \bibinfo {author} {\bibfnamefont {M.}~\bibnamefont {Prakash}},\ }\href
  {\doibase 10.1103/PhysRevD.88.083013} {\bibfield  {journal} {\bibinfo
  {journal} {Phys. Rev. D}\ }\textbf {\bibinfo {volume} {88}},\ \bibinfo
  {pages} {083013} (\bibinfo {year} {2013})},\ \Eprint
  {http://arxiv.org/abs/1302.4732} {arXiv:1302.4732 [astro-ph.SR]} \BibitemShut
  {NoStop}%
\bibitem [{\citenamefont {Alford}\ \emph
  {et~al.}(2015{\natexlab{b}})\citenamefont {Alford}, \citenamefont {Burgio},
  \citenamefont {Han}, \citenamefont {Taranto},\ and\ \citenamefont
  {Zappal\`a}}]{Alford:2015dpa}%
  \BibitemOpen
  \bibfield  {author} {\bibinfo {author} {\bibfnamefont {M.~G.}\ \bibnamefont
  {Alford}}, \bibinfo {author} {\bibfnamefont {G.~F.}\ \bibnamefont {Burgio}},
  \bibinfo {author} {\bibfnamefont {S.}~\bibnamefont {Han}}, \bibinfo {author}
  {\bibfnamefont {G.}~\bibnamefont {Taranto}}, \ and\ \bibinfo {author}
  {\bibfnamefont {D.}~\bibnamefont {Zappal\`a}},\ }\href {\doibase
  10.1103/PhysRevD.92.083002} {\bibfield  {journal} {\bibinfo  {journal} {Phys.
  Rev. D}\ }\textbf {\bibinfo {volume} {92}},\ \bibinfo {pages} {083002}
  (\bibinfo {year} {2015}{\natexlab{b}})},\ \Eprint
  {http://arxiv.org/abs/1501.07902} {arXiv:1501.07902 [nucl-th]} \BibitemShut
  {NoStop}%
\bibitem [{\citenamefont {Alford}\ and\ \citenamefont
  {Han}(2016)}]{Alford:2015gna}%
  \BibitemOpen
  \bibfield  {author} {\bibinfo {author} {\bibfnamefont {M.~G.}\ \bibnamefont
  {Alford}}\ and\ \bibinfo {author} {\bibfnamefont {S.}~\bibnamefont {Han}},\
  }\href {\doibase 10.1140/epja/i2016-16062-9} {\bibfield  {journal} {\bibinfo
  {journal} {Eur. Phys. J. A}\ }\textbf {\bibinfo {volume} {52}},\ \bibinfo
  {pages} {62} (\bibinfo {year} {2016})},\ \Eprint
  {http://arxiv.org/abs/1508.01261} {arXiv:1508.01261 [nucl-th]} \BibitemShut
  {NoStop}%
\bibitem [{\citenamefont {Ranea-Sandoval}\ \emph {et~al.}(2016)\citenamefont
  {Ranea-Sandoval}, \citenamefont {Han}, \citenamefont {Orsaria}, \citenamefont
  {Contrera}, \citenamefont {Weber},\ and\ \citenamefont
  {Alford}}]{Ranea-Sandoval:2015ldr}%
  \BibitemOpen
  \bibfield  {author} {\bibinfo {author} {\bibfnamefont {I.~F.}\ \bibnamefont
  {Ranea-Sandoval}}, \bibinfo {author} {\bibfnamefont {S.}~\bibnamefont {Han}},
  \bibinfo {author} {\bibfnamefont {M.~G.}\ \bibnamefont {Orsaria}}, \bibinfo
  {author} {\bibfnamefont {G.~A.}\ \bibnamefont {Contrera}}, \bibinfo {author}
  {\bibfnamefont {F.}~\bibnamefont {Weber}}, \ and\ \bibinfo {author}
  {\bibfnamefont {M.~G.}\ \bibnamefont {Alford}},\ }\href {\doibase
  10.1103/PhysRevC.93.045812} {\bibfield  {journal} {\bibinfo  {journal} {Phys.
  Rev. C}\ }\textbf {\bibinfo {volume} {93}},\ \bibinfo {pages} {045812}
  (\bibinfo {year} {2016})},\ \Eprint {http://arxiv.org/abs/1512.09183}
  {arXiv:1512.09183 [nucl-th]} \BibitemShut {NoStop}%
\bibitem [{\citenamefont {Han}\ and\ \citenamefont
  {Steiner}(2019)}]{Han:2018mtj}%
  \BibitemOpen
  \bibfield  {author} {\bibinfo {author} {\bibfnamefont {S.}~\bibnamefont
  {Han}}\ and\ \bibinfo {author} {\bibfnamefont {A.~W.}\ \bibnamefont
  {Steiner}},\ }\href {\doibase 10.1103/PhysRevD.99.083014} {\bibfield
  {journal} {\bibinfo  {journal} {Phys. Rev. D}\ }\textbf {\bibinfo {volume}
  {99}},\ \bibinfo {pages} {083014} (\bibinfo {year} {2019})},\ \Eprint
  {http://arxiv.org/abs/1810.10967} {arXiv:1810.10967 [nucl-th]} \BibitemShut
  {NoStop}%
\bibitem [{\citenamefont {Chatziioannou}\ and\ \citenamefont
  {Han}(2020)}]{Chatziioannou:2019yko}%
  \BibitemOpen
  \bibfield  {author} {\bibinfo {author} {\bibfnamefont {K.}~\bibnamefont
  {Chatziioannou}}\ and\ \bibinfo {author} {\bibfnamefont {S.}~\bibnamefont
  {Han}},\ }\href {\doibase 10.1103/PhysRevD.101.044019} {\bibfield  {journal}
  {\bibinfo  {journal} {Phys. Rev. D}\ }\textbf {\bibinfo {volume} {101}},\
  \bibinfo {pages} {044019} (\bibinfo {year} {2020})},\ \Eprint
  {http://arxiv.org/abs/1911.07091} {arXiv:1911.07091 [gr-qc]} \BibitemShut
  {NoStop}%
\bibitem [{\citenamefont {Mishustin}\ \emph {et~al.}(2003)\citenamefont
  {Mishustin}, \citenamefont {Hanauske}, \citenamefont {Bhattacharyya},
  \citenamefont {Satarov}, \citenamefont {Stoecker},\ and\ \citenamefont
  {Greiner}}]{Mishustin:2002xe}%
  \BibitemOpen
  \bibfield  {author} {\bibinfo {author} {\bibfnamefont {I.~N.}\ \bibnamefont
  {Mishustin}}, \bibinfo {author} {\bibfnamefont {M.}~\bibnamefont {Hanauske}},
  \bibinfo {author} {\bibfnamefont {A.}~\bibnamefont {Bhattacharyya}}, \bibinfo
  {author} {\bibfnamefont {L.~M.}\ \bibnamefont {Satarov}}, \bibinfo {author}
  {\bibfnamefont {H.}~\bibnamefont {Stoecker}}, \ and\ \bibinfo {author}
  {\bibfnamefont {W.}~\bibnamefont {Greiner}},\ }\href {\doibase
  10.1016/S0370-2693(02)03108-8} {\bibfield  {journal} {\bibinfo  {journal}
  {Phys. Lett. B}\ }\textbf {\bibinfo {volume} {552}},\ \bibinfo {pages} {1}
  (\bibinfo {year} {2003})},\ \Eprint {http://arxiv.org/abs/hep-ph/0210422}
  {arXiv:hep-ph/0210422} \BibitemShut {NoStop}%
\bibitem [{\citenamefont {Jokela}\ \emph {et~al.}(2021)\citenamefont {Jokela},
  \citenamefont {J\"arvinen}, \citenamefont {Nijs},\ and\ \citenamefont
  {Remes}}]{Jokela:2020piw}%
  \BibitemOpen
  \bibfield  {author} {\bibinfo {author} {\bibfnamefont {N.}~\bibnamefont
  {Jokela}}, \bibinfo {author} {\bibfnamefont {M.}~\bibnamefont {J\"arvinen}},
  \bibinfo {author} {\bibfnamefont {G.}~\bibnamefont {Nijs}}, \ and\ \bibinfo
  {author} {\bibfnamefont {J.}~\bibnamefont {Remes}},\ }\href {\doibase
  10.1103/PhysRevD.103.086004} {\bibfield  {journal} {\bibinfo  {journal}
  {Phys. Rev. D}\ }\textbf {\bibinfo {volume} {103}},\ \bibinfo {pages}
  {086004} (\bibinfo {year} {2021})},\ \Eprint
  {http://arxiv.org/abs/2006.01141} {arXiv:2006.01141 [hep-ph]} \BibitemShut
  {NoStop}%
\bibitem [{\citenamefont {Kojo}\ \emph {et~al.}(2020)\citenamefont {Kojo},
  \citenamefont {Hou}, \citenamefont {Okafor},\ and\ \citenamefont
  {Togashi}}]{Kojo:2020ztt}%
  \BibitemOpen
  \bibfield  {author} {\bibinfo {author} {\bibfnamefont {T.}~\bibnamefont
  {Kojo}}, \bibinfo {author} {\bibfnamefont {D.}~\bibnamefont {Hou}}, \bibinfo
  {author} {\bibfnamefont {J.}~\bibnamefont {Okafor}}, \ and\ \bibinfo {author}
  {\bibfnamefont {H.}~\bibnamefont {Togashi}},\ }\href@noop {} {\  (\bibinfo
  {year} {2020})},\ \Eprint {http://arxiv.org/abs/2012.01650} {arXiv:2012.01650
  [astro-ph.HE]} \BibitemShut {NoStop}%
\bibitem [{\citenamefont {Alford}\ \emph {et~al.}(2005)\citenamefont {Alford},
  \citenamefont {Braby}, \citenamefont {Paris},\ and\ \citenamefont
  {Reddy}}]{Alford:2004pf}%
  \BibitemOpen
  \bibfield  {author} {\bibinfo {author} {\bibfnamefont {M.}~\bibnamefont
  {Alford}}, \bibinfo {author} {\bibfnamefont {M.}~\bibnamefont {Braby}},
  \bibinfo {author} {\bibfnamefont {M.~W.}\ \bibnamefont {Paris}}, \ and\
  \bibinfo {author} {\bibfnamefont {S.}~\bibnamefont {Reddy}},\ }\href
  {\doibase 10.1086/430902} {\bibfield  {journal} {\bibinfo  {journal}
  {Astrophys. J.}\ }\textbf {\bibinfo {volume} {629}},\ \bibinfo {pages} {969}
  (\bibinfo {year} {2005})},\ \Eprint {http://arxiv.org/abs/nucl-th/0411016}
  {arXiv:nucl-th/0411016} \BibitemShut {NoStop}%
\bibitem [{\citenamefont {Christian}\ \emph {et~al.}(2019)\citenamefont
  {Christian}, \citenamefont {Zacchi},\ and\ \citenamefont
  {Schaffner-Bielich}}]{Christian:2018jyd}%
  \BibitemOpen
  \bibfield  {author} {\bibinfo {author} {\bibfnamefont {J.-E.}\ \bibnamefont
  {Christian}}, \bibinfo {author} {\bibfnamefont {A.}~\bibnamefont {Zacchi}}, \
  and\ \bibinfo {author} {\bibfnamefont {J.}~\bibnamefont
  {Schaffner-Bielich}},\ }\href {\doibase 10.1103/PhysRevD.99.023009}
  {\bibfield  {journal} {\bibinfo  {journal} {Phys. Rev. D}\ }\textbf {\bibinfo
  {volume} {99}},\ \bibinfo {pages} {023009} (\bibinfo {year} {2019})},\
  \Eprint {http://arxiv.org/abs/1809.03333} {arXiv:1809.03333 [astro-ph.HE]}
  \BibitemShut {NoStop}%
\bibitem [{\citenamefont {Miao}\ \emph {et~al.}(2020)\citenamefont {Miao},
  \citenamefont {Li}, \citenamefont {Zhu},\ and\ \citenamefont
  {Han}}]{Miao:2020yjk}%
  \BibitemOpen
  \bibfield  {author} {\bibinfo {author} {\bibfnamefont {Z.}~\bibnamefont
  {Miao}}, \bibinfo {author} {\bibfnamefont {A.}~\bibnamefont {Li}}, \bibinfo
  {author} {\bibfnamefont {Z.}~\bibnamefont {Zhu}}, \ and\ \bibinfo {author}
  {\bibfnamefont {S.}~\bibnamefont {Han}},\ }\href {\doibase
  10.3847/1538-4357/abbd41} {\bibfield  {journal} {\bibinfo  {journal}
  {Astrophys. J.}\ }\textbf {\bibinfo {volume} {904}},\ \bibinfo {pages} {103}
  (\bibinfo {year} {2020})},\ \Eprint {http://arxiv.org/abs/2006.00839}
  {arXiv:2006.00839 [nucl-th]} \BibitemShut {NoStop}%
\bibitem [{\citenamefont {Benitez}\ \emph {et~al.}(2021)\citenamefont
  {Benitez}, \citenamefont {Weller}, \citenamefont {Guedes}, \citenamefont
  {Chirenti},\ and\ \citenamefont {Miller}}]{Benitez:2020fup}%
  \BibitemOpen
  \bibfield  {author} {\bibinfo {author} {\bibfnamefont {E.}~\bibnamefont
  {Benitez}}, \bibinfo {author} {\bibfnamefont {J.}~\bibnamefont {Weller}},
  \bibinfo {author} {\bibfnamefont {V.}~\bibnamefont {Guedes}}, \bibinfo
  {author} {\bibfnamefont {C.}~\bibnamefont {Chirenti}}, \ and\ \bibinfo
  {author} {\bibfnamefont {M.~C.}\ \bibnamefont {Miller}},\ }\href {\doibase
  10.1103/PhysRevD.103.023007} {\bibfield  {journal} {\bibinfo  {journal}
  {Phys. Rev. D}\ }\textbf {\bibinfo {volume} {103}},\ \bibinfo {pages}
  {023007} (\bibinfo {year} {2021})},\ \Eprint
  {http://arxiv.org/abs/2010.02619} {arXiv:2010.02619 [astro-ph.HE]}
  \BibitemShut {NoStop}%
\bibitem [{\citenamefont {Bauswein}\ \emph {et~al.}(2021)\citenamefont
  {Bauswein}, \citenamefont {Blacker}, \citenamefont {Lioutas}, \citenamefont
  {Soultanis}, \citenamefont {Vijayan},\ and\ \citenamefont
  {Stergioulas}}]{Bauswein:2020xlt}%
  \BibitemOpen
  \bibfield  {author} {\bibinfo {author} {\bibfnamefont {A.}~\bibnamefont
  {Bauswein}}, \bibinfo {author} {\bibfnamefont {S.}~\bibnamefont {Blacker}},
  \bibinfo {author} {\bibfnamefont {G.}~\bibnamefont {Lioutas}}, \bibinfo
  {author} {\bibfnamefont {T.}~\bibnamefont {Soultanis}}, \bibinfo {author}
  {\bibfnamefont {V.}~\bibnamefont {Vijayan}}, \ and\ \bibinfo {author}
  {\bibfnamefont {N.}~\bibnamefont {Stergioulas}},\ }\href {\doibase
  10.1103/PhysRevD.103.123004} {\bibfield  {journal} {\bibinfo  {journal}
  {Phys. Rev. D}\ }\textbf {\bibinfo {volume} {103}},\ \bibinfo {pages}
  {123004} (\bibinfo {year} {2021})},\ \Eprint
  {http://arxiv.org/abs/2010.04461} {arXiv:2010.04461 [astro-ph.HE]}
  \BibitemShut {NoStop}%
\bibitem [{\citenamefont {Ayriyan}\ \emph {et~al.}(2018)\citenamefont
  {Ayriyan}, \citenamefont {Bastian}, \citenamefont {Blaschke}, \citenamefont
  {Grigorian}, \citenamefont {Maslov},\ and\ \citenamefont
  {Voskresensky}}]{Ayriyan:2017nby}%
  \BibitemOpen
  \bibfield  {author} {\bibinfo {author} {\bibfnamefont {A.}~\bibnamefont
  {Ayriyan}}, \bibinfo {author} {\bibfnamefont {N.~U.}\ \bibnamefont
  {Bastian}}, \bibinfo {author} {\bibfnamefont {D.}~\bibnamefont {Blaschke}},
  \bibinfo {author} {\bibfnamefont {H.}~\bibnamefont {Grigorian}}, \bibinfo
  {author} {\bibfnamefont {K.}~\bibnamefont {Maslov}}, \ and\ \bibinfo {author}
  {\bibfnamefont {D.~N.}\ \bibnamefont {Voskresensky}},\ }\href {\doibase
  10.1103/PhysRevC.97.045802} {\bibfield  {journal} {\bibinfo  {journal} {Phys.
  Rev. C}\ }\textbf {\bibinfo {volume} {97}},\ \bibinfo {pages} {045802}
  (\bibinfo {year} {2018})},\ \Eprint {http://arxiv.org/abs/1711.03926}
  {arXiv:1711.03926 [nucl-th]} \BibitemShut {NoStop}%
\bibitem [{\citenamefont {Baacke}(1977)}]{Baacke:1976jv}%
  \BibitemOpen
  \bibfield  {author} {\bibinfo {author} {\bibfnamefont {J.}~\bibnamefont
  {Baacke}},\ }\href@noop {} {\bibfield  {journal} {\bibinfo  {journal} {Acta
  Phys. Polon. B}\ }\textbf {\bibinfo {volume} {8}},\ \bibinfo {pages} {625}
  (\bibinfo {year} {1977})}\BibitemShut {NoStop}%
\bibitem [{\citenamefont {Steinheimer}\ \emph {et~al.}(2011)\citenamefont
  {Steinheimer}, \citenamefont {Schramm},\ and\ \citenamefont
  {Stocker}}]{Steinheimer:2010ib}%
  \BibitemOpen
  \bibfield  {author} {\bibinfo {author} {\bibfnamefont {J.}~\bibnamefont
  {Steinheimer}}, \bibinfo {author} {\bibfnamefont {S.}~\bibnamefont
  {Schramm}}, \ and\ \bibinfo {author} {\bibfnamefont {H.}~\bibnamefont
  {Stocker}},\ }\href {\doibase 10.1088/0954-3899/38/3/035001} {\bibfield
  {journal} {\bibinfo  {journal} {J. Phys. G}\ }\textbf {\bibinfo {volume}
  {38}},\ \bibinfo {pages} {035001} (\bibinfo {year} {2011})},\ \Eprint
  {http://arxiv.org/abs/1009.5239} {arXiv:1009.5239 [hep-ph]} \BibitemShut
  {NoStop}%
\bibitem [{\citenamefont {Satarov}\ \emph {et~al.}(2015)\citenamefont
  {Satarov}, \citenamefont {Bugaev},\ and\ \citenamefont
  {Mishustin}}]{Satarov:2014voa}%
  \BibitemOpen
  \bibfield  {author} {\bibinfo {author} {\bibfnamefont {L.~M.}\ \bibnamefont
  {Satarov}}, \bibinfo {author} {\bibfnamefont {K.~A.}\ \bibnamefont {Bugaev}},
  \ and\ \bibinfo {author} {\bibfnamefont {I.~N.}\ \bibnamefont {Mishustin}},\
  }\href {\doibase 10.1103/PhysRevC.91.055203} {\bibfield  {journal} {\bibinfo
  {journal} {Phys. Rev. C}\ }\textbf {\bibinfo {volume} {91}},\ \bibinfo
  {pages} {055203} (\bibinfo {year} {2015})},\ \Eprint
  {http://arxiv.org/abs/1411.0959} {arXiv:1411.0959 [nucl-th]} \BibitemShut
  {NoStop}%
\bibitem [{\citenamefont {Dexheimer}\ \emph {et~al.}(2013)\citenamefont
  {Dexheimer}, \citenamefont {Steinheimer}, \citenamefont {Negreiros},\ and\
  \citenamefont {Schramm}}]{Dexheimer:2012eu}%
  \BibitemOpen
  \bibfield  {author} {\bibinfo {author} {\bibfnamefont {V.}~\bibnamefont
  {Dexheimer}}, \bibinfo {author} {\bibfnamefont {J.}~\bibnamefont
  {Steinheimer}}, \bibinfo {author} {\bibfnamefont {R.}~\bibnamefont
  {Negreiros}}, \ and\ \bibinfo {author} {\bibfnamefont {S.}~\bibnamefont
  {Schramm}},\ }\href {\doibase 10.1103/PhysRevC.87.015804} {\bibfield
  {journal} {\bibinfo  {journal} {Phys. Rev. C}\ }\textbf {\bibinfo {volume}
  {87}},\ \bibinfo {pages} {015804} (\bibinfo {year} {2013})},\ \Eprint
  {http://arxiv.org/abs/1206.3086} {arXiv:1206.3086 [astro-ph.HE]} \BibitemShut
  {NoStop}%
\bibitem [{\citenamefont {Mukherjee}\ \emph {et~al.}(2017)\citenamefont
  {Mukherjee}, \citenamefont {Schramm}, \citenamefont {Steinheimer},\ and\
  \citenamefont {Dexheimer}}]{Mukherjee:2017jzi}%
  \BibitemOpen
  \bibfield  {author} {\bibinfo {author} {\bibfnamefont {A.}~\bibnamefont
  {Mukherjee}}, \bibinfo {author} {\bibfnamefont {S.}~\bibnamefont {Schramm}},
  \bibinfo {author} {\bibfnamefont {J.}~\bibnamefont {Steinheimer}}, \ and\
  \bibinfo {author} {\bibfnamefont {V.}~\bibnamefont {Dexheimer}},\ }\href
  {\doibase 10.1051/0004-6361/201731505} {\bibfield  {journal} {\bibinfo
  {journal} {Astron. Astrophys.}\ }\textbf {\bibinfo {volume} {608}},\ \bibinfo
  {pages} {A110} (\bibinfo {year} {2017})},\ \Eprint
  {http://arxiv.org/abs/1706.09191} {arXiv:1706.09191 [nucl-th]} \BibitemShut
  {NoStop}%
\bibitem [{\citenamefont {Yagi}\ and\ \citenamefont
  {Yunes}(2013{\natexlab{a}})}]{Yagi:2013awa}%
  \BibitemOpen
  \bibfield  {author} {\bibinfo {author} {\bibfnamefont {K.}~\bibnamefont
  {Yagi}}\ and\ \bibinfo {author} {\bibfnamefont {N.}~\bibnamefont {Yunes}},\
  }\href {\doibase 10.1103/PhysRevD.88.023009} {\bibfield  {journal} {\bibinfo
  {journal} {Phys. Rev. D}\ }\textbf {\bibinfo {volume} {88}},\ \bibinfo
  {pages} {023009} (\bibinfo {year} {2013}{\natexlab{a}})},\ \Eprint
  {http://arxiv.org/abs/1303.1528} {arXiv:1303.1528 [gr-qc]} \BibitemShut
  {NoStop}%
\bibitem [{\citenamefont {Yagi}\ and\ \citenamefont
  {Yunes}(2013{\natexlab{b}})}]{Yagi:2013bca}%
  \BibitemOpen
  \bibfield  {author} {\bibinfo {author} {\bibfnamefont {K.}~\bibnamefont
  {Yagi}}\ and\ \bibinfo {author} {\bibfnamefont {N.}~\bibnamefont {Yunes}},\
  }\href {\doibase 10.1126/science.1236462} {\bibfield  {journal} {\bibinfo
  {journal} {Science}\ }\textbf {\bibinfo {volume} {341}},\ \bibinfo {pages}
  {365} (\bibinfo {year} {2013}{\natexlab{b}})},\ \Eprint
  {http://arxiv.org/abs/1302.4499} {arXiv:1302.4499 [gr-qc]} \BibitemShut
  {NoStop}%
\end{thebibliography}%


%

\end{document}